\documentclass[11pt]{article}
\usepackage{eurosym}
\usepackage{amsfonts}
\usepackage{amssymb}
\usepackage{graphicx}
\usepackage{amsmath}
\usepackage{makeidx}
\usepackage{indentfirst}
\usepackage[T1]{fontenc}
\usepackage[utf8]{inputenc}

\newcounter{resultnum}[section]
\newcounter{conclusionnum}[section]
\newcounter{conditionnum}[section]
\newcounter{conjecturenum}[section]
\newcounter{examplenum}[section]
\newcounter{exercisenum}[section]
\newcounter{lemmanum}[section]
\newcounter{notationnum}[section]
\newcounter{theoremnum}[section]
\newcounter{definitionnum}[section]
\newcounter{corollarynum}[section]
\newcounter{remarknum}[section]
\newcounter{propositionnum}[section]
\newcounter{acknowledgementnum}[section]
\newcounter{algorithmnum}[section]
\newcounter{axiomnum}[section]
\newcounter{casenum}[section]
\newcounter{claimnum}[section]
\newcounter{summarynum}[section]
\newcounter{problemnum}[section]
\begin{document}

\title{Nonassociative geometry of nonholonomic phase spaces with\\
star R-flux string deformations and (non) symmetric metrics}
\date{June 19, 2025; uploaded an equivalent of the published version}
\author{ \vspace{.1 in} \textbf{Sergiu I. Vacaru} \thanks{
email: sergiu.vacaru@gmail.com \newline
\textit{Address for post correspondence in 2020-2021 as a visitor senior
researcher at YF CNU Ukraine is:\ } Yu. Gagarin street, 37-3, Chernivtsi,
Ukraine, 58008;\ authors' order reflects certain chronology of former and
futures research programs and involves equal contributions in obtaining new
results in this work} \\
{\small \textit{Physics Department, California State University at Fresno,
Fresno, CA 93740, USA; }}\\
{\small \textit{and Dep. Theoretical Physics and Computer Modelling,}}\\
{\small \textit{\ Yu. Fedkovych Chernivtsi National University}, 101
Storozhynetska street, Chernivtsi, 58029, Ukraine} \vspace{.2 in} \\
{\textbf{El\c{s}en Veli Veliev} \vspace{.1 in}} \thanks{%
email: elsen@kocaeli.edu.tr and elsenveli@hotmail.com}\\
{\small \textit{Department of Physics,\ Kocaeli University, 41380, Izmit,
Turkey}} \vspace{.2 in} \\
\vspace{.1 in} {\textbf{Lauren\c{t}iu Bubuianu}}\thanks{%
email: laurentiu.bubuianu@tvr.ro} \\
{\small \textit{SRTV - Studioul TVR Ia\c{s}i} and \textit{University
Appolonia}, 2 Muzicii street, Ia\c{s}i, 700399, Romania} }
\maketitle

\begin{abstract}
We elaborate on nonassociative differential geometry of phase spaces endowed
with nonholonomic (non-integrable) distributions and frames, nonlinear and
linear connections, symmetric and nonsymmetric metrics, and correspondingly
adapted quasi-Hopf algebra structures. The approach is based on the concept
of nonassociative star product introduced for describing closed strings
moving in a constant R-flux background. Generalized Moyal-Weyl deformations
are considered when, for nonassociative and noncommutative terms of star
deformations, there are used nonholonomic frames (bases) instead of local
partial derivatives. In such modified nonassociative and nonholonomic
spacetimes and associated complex/ real phase spaces, the coefficients of
geometric and physical objects depend both on base spacetime coordinates and
conventional (co) fiber velocity/ momentum variables like in (non)
commutative Finsler-Lagrange-Hamilton geometry. For nonassociative and (non)
commutative phase spaces modelled as total spaces of (co) tangent bundles on
Lorentz manifolds enabled with star products and nonholonomic frames, we
consider associated nonlinear connection, N-connection, structures
determining conventional horizontal and (co) vertical (for instance, 4+4)
splitting of dimensions and N-adapted decompositions of fundamental
geometric objects. There are defined and computed in abstract geometric and
N-adapted coefficient forms the torsion, curvature and Ricci tensors. We
extend certain methods of nonholonomic geometry in order to construct R-flux
deformations of vacuum Einstein equations for the case of N-adapted linear
connections and symmetric and nonsymmetric metric structures.

\vskip3pt

\end{abstract}

\tableofcontents

\section{Introduction, motivations and goals}

We initialize a research on \textsf{nonassociative nonholonomic geometry}
induced by R-flux backgrounds in string gravity and classical and quantum
models on phase spaces modelled as (co) tangent Lorentz bundles endowed with
nonholonomic (non-integrable, equivalently, anholonomic) structures and
nonsymmetric metrics. Such a nonholonomic geometric formalism was elaborated
for various types of (non) commutative gravity and matter field theories,
modified (relativistic) geometric flows, and classical and quantum
information theories, see details in a series of our previous works \cite%
{vacaru01,vacaru03,vacaru09a,vacaru16,bubuianu17b,bubuianu17,bubuianu19,
vacaru18, bubuianu18a,vacaru19,vacaru20} and references therein. In this
article, nonassociative geometric and physical models are formulated in a
form which is accessible to researchers in particle physics and modern
cosmology. There are provided nonholonomic generalizations of two recent
models of nonassociative geometry and vacuum gravity due to \cite%
{blumenhagen16} and, with more formalized mathematical constructions
involving quasi--Hopf algebraic structures, \cite{aschieri17}. Such works on
nonassociative geometry and gravity involve symmetric and nonsymmetric
generic off-diagonal metrics, generalized connections, and systems of
coupled nonlinear partial differential equations, PDEs, with small
parameters and complex momentum variables.

We focus on nonholonomic geometric methods which are used in our partner
works for investigating nonassociative gravitational theories with
(effective) matter sources and matter fields equations. Such physically
important systems of nonlinear PDEs can be written in nonholonomic dyadic
variables, with distortions of connections, and which allow general
decoupling and constructing exact and parametric solutions. For instance, we
can consider nonassociative generalizations of the Einstein-Eisenhart-Moffat
gravity with nonsymmetric metrics \cite{einstein25,einstein45,
eisenhart51,eisenhart52,moffat79,moffat95,moffat95a,moffat00}. Here we note
that it was possible to elaborate also on nonassociative generalizations of
the geometry of nonholonomic (co) tangent bundles for commutative
Einstein-Finsler-Lagrange gravity with nonsymmetric metrics and their flows 
\cite{vacaru08aa,vacaru08bb,vacaru08cc} and to extend the results for
noncommutative gravity and geometric flow theories \cite%
{vacaru01,vacaru03,vacaru08,bubuianu17b}. Such geometric and physical models
with nonholonomic and/or Finsler-Hamilton-Lagrange variables in string and
modified gravity and phase space theories were studied in \cite%
{vacaru18,bubuianu18a,bubuianu17,bubuianu19} and references therein. This
work provides a generalized nonassociative and nonholonomic quasi-Hopf
formalism which is used in our partner works for developing new geometric
and analytic methods for finding new classes of exact and parametric black
hole and cosmological solutions in nonassociative gravity theories, see \cite%
{vacaru03,vacaru09a,bubuianu17b,bubuianu17,bubuianu19} for results in
commutative and noncommutative modified gravity theories, MGTs, and
nonholonomic geometry. The nonassociative nonholonomic geometric methods are
used also for constructing certain models of nonassociative geometric flows
when there are considered possible implications in quantum information
theory, nonassociative and noncommutative locally anisotropic string gravity
and MGTs and quantum deformation models as extensions of \cite%
{vacaru96a,vacaru96b,bubuianu17b,
vacaru07a,vacaru08b,vacaru08a,vacaru07,vacaru07b}.

Noncommutativity and nonassociativity appear in quite natural ways in string
theory and various modifications of (quantum) gravity and quantum field
theories. For open strings, such structures are present due to a
non-vanishing background 2-form in the world volume of a D-brane. In closed
string theory, the existence of non-commutative/ -associative structures can
be a consequence of flux compactification with non-trivial three-form and
related non-geometric backgrounds\footnote{%
The term "non-geometric" is largely used in literature on string theory.
Nevertheless, we show in this work that at least a formal geometrization of
such models is possible using nonholonomic distributions and N-adapted star
deformations.}. Evidences for such structures and various alternative
approaches to noncommutative geometric flows, noncommutative and
nonassociative gravity and gauge theories, membranes and double field
theory, etc. were studied in \cite{vacaru01,
vacaru03,bouwknegt04,alvarez06,luest10,blumenhagen10,mylonas12,condeescu13,blumenhagen13, kupriyanov19,kupriyanov19a}
and references therein\footnote{%
we cite here a series of papers which are most related to the purposes of
this article and related partner works; we do not attempt to review the
subjects and provide a comprehensive list of references}. Such geometric
constructions can be re-defined for models with quasi-Hopf algebras
developed in mathematical form \cite{barnes14,barnes15,aschieri17}, see also
some important previous works \cite{mylonas12,mylonas13}. In both cases
(with or without quasi-Hopf structures, see respectively, \cite{aschieri17}
or \cite{blumenhagen16}), certain self-consistent approaches to
nonassociative gravity were formulated up to a level where a metric and a
linear connection have been introduced and respective torsion, Riemann and
Ricci tensors were defined and computed. The first problem is that in such
theories a star metric generically does not satisfy typical relations for
pulling up/ down indices. For linear connections, there is a second both
conceptual and technical problem: how to adapt self-consistently the
nonassociative metric and connection structure? To solve such issues a new
type composition \cite{mylonas13} was employed in order to define an
nonassociative analogous Levi-Civita, LC, connection. Certain theories with
different type Moyal-Weyl star product were elaborated for noncommutative/
nonassociative gravity theories and deformation quantization \cite%
{vacaru01,vacaru03,vacaru07a,vacaru08b,kupriyanov15,kupriyanov16,kupriyanov18, kupriyanov18a}%
. They correspond to different types of nonholonomic distributions and
deformations of geometric and physical objects defined on respective phase
spaces.

\subsection{Why nonassociative theories should be formulated in nonholonomic
variables?}

This article generalizes the approaches toward to the theory of
nonassociative gravity from \cite{blumenhagen16,aschieri17} to a formalism
when the geometric constructions are performed in nonholonomic forms. This
allows us to work on the entire phase space and formulate nonassociative
gravitational field equations using nonholonomic geometric methods developed
and outlined in \cite{vacaru08bb,vacaru03,vacaru18,bubuianu18a}. In our
partner works, we elaborate on geometric and analytic methods for
constructing exact and parametric solutions of nonassociative modified
Einstein equations and (generalized) geometric flow evolution equations.
Such nonholonomic / noncommutative / supersymmetric etc. geometric
techniques (called the anholonomic frame and connection deformation method,
AFCDM; in our previous works we wrote in brief AFDM) of constructing exact
solutions in (non) commutative/ supersymmetric/ string / massive and other
type gravity theories was developed in \cite%
{vacaru18,bubuianu18a,bubuianu17,bubuianu19}, see also references therein.
The main idea of the AFCDM is to define certain classes of frame, metric,
and (auxiliary) nonlinear and linear connection structures which allow us to
decouple and integrate in very general forms various types of physically
important nonlinear systems of PDEs.

In a series of works with applications of our (non) associative nonholonomic
geometric formalism, we consider certain classes of nonassociatve geometric
flow and/or gravitational and matter field equations which can be modelled
also as nonholonomic Ricci soliton structures. The geometric constructions
involve a canonical distinguished connection, d-connection, which is metric
compatible and defined by the symmetric and nonsymmetric parts of metric (in
N-adapted form, it is called a d-metric). There are also associated
canonical nonassociative and/or nonsymmetric distortions of canonical
d-connections\footnote{%
we say that some geometric objects are distinguished, d-objects, for
instance, d-vectors, d-tensors etc. if they are adapted to a prescribed
nonholonomic structure, which can be nonassociative, noncommutative, with
nontrivial nonlinear connection, N-connection structure; in next section, we
provide basic definitions and proofs and necessary references}. Such spaces
contain certain nonholonomically induced torsion structures, with N-adapted
or coordinate frame components determined by generic off-diagonal terms of
metrics and anholonomy coefficients. For the canonical d-connections (with
respective nonholonomic diadic decompositions in spaces of higher dimensions
than four, for a conventional splitting of dimensions in as 2+2+2+...),
various physically important and associated systems of nonlinear PDEs can be
decoupled and integrated in certain general forms. Resulting generic
off-diagonal metrics\footnote{%
which can not be diagonalized in a finite spactime/ phase space region by
coordinate transforms; nonsymmetric metrics can be decomposed in respective
symmetric and anti-symmetric parts} and generalized connections are with
coefficients determined by generating functions and (effective) generating
sources, integration functions and constants. Such geometric objects are
subjected to corresponding nonholonomic conditions defining commutative /
noncommutative / nonassociative / supersymmetric / spinor and other type
structures. For some classes of solutions, the coefficients of nonsymmetric
metrics (which may be complex or real) are subjected to nonholonomic
constraints, or related dynamically to nonsymmetric coefficients of a Ricci
d-tensor which for nonholonomic configurations can be also nonsymmetric.

It should be emphasized that it is not possible to prove a general
decoupling property for equations involving directly the Levi-Civita
connection, LC-connection, because the zero-torsion conditions for this
connection contract and mix the indices in related tensor values (for
instance, in the Ricci d-tensor). Decoupling of such systems of nonlinear
PDEs is possible only for some canonical deformations of linear connections,
determined by the same metric structure, encoding nonholonomic and star
product deformations of Einstein equations and various modifications. For
well--defined auxiliary and/or additional constraints on already constructed
general classes of solutions, it is possible at the end to extract large
subclasses of solutions with zero torsion configurations. To eliminate
completely the torsion structure is not always possible in certain
noncommutative / nonassociative models of geometric flows, string gravity
etc. if the (non) linear connection formalism is elaborated directly on the
total phase space and not for the lifts of geometric objects and equations
with decoupling on a base manifold.

We note that the AFCDM was formulated also as a geometric method for
constructing exact and parametric generic off-diagonal solutions (in
general, with nontrivial nonlinear connection, N-connection, structure) in
general relativity, GR, and various modified commutative and noncommutative
Finsler-Lagrange-Hamilton like theories \cite{vacaru16,vacaru09a}.\footnote{%
Readers may consult references therein and the Introduction, Conclusion and
Appendix B to \cite{vacaru18}, for a historical review and summary of more
than 20 directions of research activity in different countries. Here we also
motivate that we have to consider Finsler geometric variables because
certain models of nonassociative gravity are formulated on phase spaces
which involves dependence of geometric objects on velocity and/or momentum
type coordinates, see additional motivations in section \ref{ssectmotivfinsl}%
.} We proved that Finsler like d-connections and adapted to N-connection
distinguished geometric objects and metrics (respectively, d-objects and
d-metrics) can be introduced also in the GR theory. Such nonholonomic
formulations can be alternative or equivalent to standard approaches with
the LC-connection. They allowed to construct various classes of generic
off-diagonal solutions of nonholonomically deformed Einstein equations with
possible constraints to zero torsion configurations.

\subsection{On theories with nonsymmetric metrics}

Nonassociative star product deformations of GR and certain quantum
deformation theories are formulated as complex geometric models with
nonsymmetric metrics \cite{blumenhagen16,aschieri17,
vacaru08a,vacaru07,vacaru07a,vacaru08,vacaru16,vacaru09a}. Such
constructions were studied in string gravity or other type (non) commutative
MGTs. For real manifolds and tangent bundles, nonsymmetric gravity theories
were formulated in some variants \cite{einstein25,einstein45,
eisenhart51,eisenhart52,moffat79,moffat95,moffat95a,moffat00} with
variational and geometric proofs of fundamental field equations and when
applications in modern astrophysics and cosmology \cite%
{clayton96,clayton96a,janssen06, prokopec06,janssen07,janssen07a,vacaru08aa}
were analyzed. These series of works define some directions of research in
mathematical and nonstandard particle physics and MGTs with nonsymmetric
metrics. Here we cite also \cite{vacaru08bb,vacaru08cc} and references
therein on nonsymmetric Ricci flows and Lagrange--Finsler geometry.

Let us summarize and discuss most important results on elaborating geometric
and physical models with nonsymmetric real metrics. Such works motivate our
research program on nonassociative geometric evolution flows and gravity
theories with applications, and quantum information theories when the
generalized metrics can be with complex components and nonholonomic
deformation structures are involved:

\begin{description}
\item[a.] There were studied examples of nonholonomic (non) commutative
geometric flow and gravity theories with nonsymmetric metrics and
generalized connections formulated on generalized phase spaces endowed with
nonholonomic structures \cite{vacaru08,vacaru08cc,vacaru03,vacaru09a,
vacaru16}. Those geometric constructions can be naturally extended for
nonassociative geometric models on complex phase spaces  \cite%
{blumenhagen16,aschieri17}. It should be emphasized here that none physical
principle prohibits us to elaborate on such theories with nonsymmetric
metrics. We can always decompose an nonsymmetric metric into skew-symmetric
and symmetric parts and consider respective star deformations in order to
elaborate certain classes of commutative / noncommutative / supersymmetric
metric--affine and generalized Finsler--Lagrange-Hamilton geometries,
theories of (locally anisotropic, classical, or quantum) matter fields and
gravitational interactions, and Ricci flow theories.

\item[b.] A. Einstein attempted to unify gravity with electromagnetism \cite%
{einstein25} in a model when the nonsymmetric part of metric had been
identified with the electromagnetic field strength tensor. He elaborated
also a unified theory of physical fields by introducing a complex metric
field with Hermitian symmetry \cite{einstein45}. Here we note that
nonassociative gravity theories \cite{blumenhagen16,aschieri17} also involve
nonsymmetric metrics with complex terms but in certain forms determined by a
star product related to contributions from string theory.

\item[c.] Then, L. P. Eisenhart studied generalized Riemannian spaces with
nonsymmetric metrics when the symmetric part is non-degenerate \cite%
{eisenhart51,eisenhart52}. The so-called Eisenhart problem was to define
various classes of linear connections which are compatible with a general
(nonsymmetric) metric structure. There were analysed also some important
particular solutions for symmetric and generic off-diagonal backgrounds with
Riemannian and Finsler-Lagrange spaces, see \cite{vacaru08bb} and references
therein for a review of results and applications in modified gravity and
cosmology theories. The geometric arena of nonassociative gravity \cite%
{blumenhagen16,aschieri17} involves generalized phase spaces as certain (co)
tangent complex manifolds with complex nonsymmetric metrics and respective
star products depending on velocity/momentum like variables. Such
constructions can be defined also in so-called nonholonomic (non)
commutative Finsler-Lagrange-Hamilton variables  \cite%
{vacaru16,vacaru09a,vacaru18} which allow us to find exact solutions  \cite%
{vacaru03,bubuianu17,bubuianu17b,bubuianu19} and perform deformation
quantization \cite{vacaru07a,vacaru08a,vacaru08b,vacaru07,vacaru07b}.

\item[d.] After A. Einstein and L. P. Eisenhart fundamental works, a more
than twenty years of research activity on elaborating principles and modern
type physical theories with nonsymmetric metrics was due to J. Moffat and
co--authors \cite{moffat79,moffat95,
moffat95a,moffat00,clayton96,clayton96a,moffat96}. In this article we write
in brief, NSGT, even NGT was also used in many previous papers (to avoid
confusions with "noncommutative gravity theories"). Late versions of such
models are free of ghosts, tachions and higher-order poles in the propagator
in the linear approximation on Minkowski space. It was proven that
expansions of the general nonsymmetric metric about an arbitrary Einstein
background metric result in effective field equations involving terms which
are first order in the skew-symmetric part of metric being free of coupling
to un-physical (negative energy)  modes. The solutions of such systems of
PDEs have consistent asymptotic boundary conditions.

\item[e.] T. Janssen and T. Prokopec published a series of works \cite%
{janssen06, prokopec06,janssen07,janssen07a} devoted to cosmological
implications and the problem of instabilities in gravity theories with
nonsymmetric metrics. They agreed that there are viable models of NSGT with
a nonzero mass term for the nonsymmetric part of metric, see classical and
quantum variants in \cite{clayton96,clayton96a,moffat96, moffat00}. Such a
term can be treated as an absolutely symmetric torsion induced by an
effective B--field like in string gravity but in four dimensions. Recently,
such nonholonomic string gravity models were studied in \cite{bubuianu17b}.
An important result was that even, for a GR background, a small B--field for
the nonsymmetric part may quickly grow other fields. The theory can be
stabilized by introducing extra Lagrange multipliers and finding solutions
when the unstable modes dynamically vanish. That solved also the problems
formally created by absence of gauge invariance found by Damour, Deser and
McCarthy \cite{damour93}. Nevertheless, the general conclusion of Janssen
and Prokopec works was that instabilities in NSGTs should not be seen as a
relic of a linearized theory with nonsymmetric metrics and associated tensor
fields. They found that certain nonlinear configurations with a nontrivial
Einstein background (for instance, on a Schwarzschild spacetime) are
positively unstable.

\item[f.] The idea to use Lagrange multipliers and dynamical constraints in
order to solve certain instability problems in NSGTs \cite%
{clayton96,clayton96a,moffat96, moffat00} contains already a relation to the
nonholonomic geometry and classical and quantum field dynamics with imposed
non-integrable conditions. Such a nonholonomic and almost symplectic
geometric approach to theories with nonsymmetric metrics was studied in a
series of works \cite{vacaru08aa,vacaru08bb, vacaru08cc}. We proved that the
Einstein gravity theory can be reformulated in almost K\"{a}hler
(nonsymmetric) variables with an effective symplectic form and compatible
linear connections uniquely defined by a (pseudo) Riemannian metric
following certain well--defined geometric principles. A large class of NSGTs
can be formulated on manifolds enabled with nonholonomic distributions (i.e.
nonholonomic manifolds) and their (co) tangent bundles. There were analyzed
the conditions when the fundamental geometric and physical objects defined
by (non) symmetric metrics are determined by nonholonomic deformations in GR
or by contributions from Ricci flow theory, or a quantum gravity model. We 
had shown how to impose certain classes of nonholonomic constraints
generating effective Lagrangians for a NSGTs which do not develop
instabilities. It was elaborated also a linearization formalism for such
anholonomic nonsymmetric geometric flow and gravity theories and analyzed
the stability of stationary ellipsoidal solutions defining certain classes
of nonholonomic and/or nonsymmetric deformations of the Schwarzschild
metric. Finally, we concluded that instabilities do not consist a general
feature of NSGTs but a particular property of certain models and/or classes
of unconstrained solutions. For geometric flows of nonsymmetric metrics, the
existence of singularities is a general property at least for the symmetric
part, but we can always elaborate on procedures of eliminating singularities
and stabilization imposing respective types of nonholonomic constraints and
deformation of geometric objects.
\end{description}

We note that there exists the so--called generality problem in NSGTs when
certain guiding principles have to be formulated in order to select from
nine and more constants and extra terms in generalized Lagrangians. J.
Moffat's group considered at least 11 undetermined parameters which come
from the full theory and the decomposition of a nonsymmetric metric tensor 
\cite{moffat79,moffat95,moffat95a,moffat00,clayton96, clayton96a,moffat96}.
In our works, we exploited the idea that (non) symmetric corrections to
metrics and connections can be derived following certain geometric
principles in nonholonomic Ricci flow and/or deformation quantization
theories \cite{vacaru08aa,vacaru08bb,vacaru08cc}.

\subsection{Nonassociative gravity and Finsler-Lagrange-Hamilton geometry}

\label{ssectmotivfinsl}

We present here a series of important geometric and physical arguments
motivating our approach to nonassociative geometry and modifications of
gravity and particle physics theories formulated as a synthesis of
nonholonomic and Finsler like geometric methods with the star product
techniques:

\begin{enumerate}
\item The nonassociativity for a constant R-flux background is determined by
certain commutation relations for a quasi-Poisson coordinate algebra 
\begin{equation}
\lbrack x^{i},x^{j}]=\frac{i\mathit{l}_{s}^{4}}{3\hbar }R^{ija}p_{a},\
[x^{i},p_{a}]=i\hbar \delta _{a}^{i}\mbox{ and }\lbrack p_{a},p_{b}]=0,
\label{comrel}
\end{equation}%
for $(x,p)=(x^{i},p_{a})$, with physical constants $l_{s},\hbar $ and $%
R^{ija}$ being antisymmetric and when all indices are (are) fiber type ones;
where, for the complex unity, $i^{2}=-1$ (one should not confuse with a
horizontal index); and when $x^{i}$ and $p_{a}$ denote respective spacetime
coordinates and momentum variables on a phase space $T^{\ast }V$ (modelled
as a cotangent bundle, for instance, to a Lorentz manifold $V$).\footnote{%
\label{fnfluxind}In above formulas coordinate indices run values $%
i,j,...=1,2,...n$ and $a=n+1,n+2,...,n+n.$ In order to contract an index $%
i,j,...$ with a respective $a,b,...$ we should express $a=n+i,b=n+j,...$ and
then summarize correspondingly on some up-low repeating $i,j,..$}
Commutation rules (\ref{comrel}) together with other type nonholonomic
constraints (there are used equivalent terms like anholonomic, i.e.
non-integrable) are defined by respective distributions of geometric objects
on tangent bundle $TV,$ with local coordinates $(x,y)=(x^{i},y^{a}),$ and
its dual $T^{\ast }V,$ see more evidence in \cite%
{vacaru01,vacaru03,bouwknegt04,alvarez06,
luest10,blumenhagen10,mylonas12,condeescu13,blumenhagen13}.

\item For elaborating particle physics models with
nonassociative/noncommutative variable, respective nonholonomic
distributions are determined by modified dispersion relations, MDRs, when
locally 
\begin{equation}
c^{2}\overrightarrow{\mathbf{p}}^{2}-E^{2}+c^{4}m^{2}=\varpi (E,%
\overrightarrow{\mathbf{p}},m;l_{s}),  \label{mdr}
\end{equation}%
for an indicator of modifications $\varpi (x^{k},E,\overrightarrow{\mathbf{p}%
},m;l_{s})$. In general, such an indicator depends on spacetime coordinates $%
x^{i}$ and momenta $p_{a}=(\overrightarrow{\mathbf{p}},E)$ with an energy
parameter $E$. Commutative theories with MDRs can be formulated in an
axiomatic relativistic Finsler-Lagrange-Hamilton form \cite%
{vacaru18,bubuianu18a}. Such models encode possible violations of local
Lorentz invariance, LIV, or can be restricted to constructions on Lorentz
manifolds. Formula (\ref{mdr}) can be generalized by introducing
noncommutative/nonassociative variables and respective parameters defining
such theories (for instance, considering nonassociative star product
deformations, see next sections). Geometrically, we work with corresponding
classes of nonholonomic distributions on $T^{\ast }V$ when, for instance,
metrics $g_{ij}(x^{k},p_{a})$ depend both on spacetime and momentum type
coordinates, and certain star product deformations (see rigorous geometric
definitions in the next section). Nonlinear metric structures (for instance,
written in the form $g_{ij}(x^{k},y^{l})$) are considered, in Lagrange
geometry defined on $TV$ and, imposing homogeneity conditions on fiber
coordinates $y^{l}$, in Finsler geometry. For various types of geometric and
physical models, star product deformations on total bundle spaces, their
dual, and/or projections on base manifolds, can be constructed.

\item A special class of canonical d-connections (Finsler like ones being
metric compatible but not with homogeneity conditions) allows to decouple
and integrate in general form the gravitational and (effective) matter field
equations various types of MGTs and GR,  \cite%
{vacaru18,bubuianu18a,bubuianu17,bubuianu19}. For four dimensional, 4-d,
spacetime manifolds,conventional 2+2 splitting determined by a nonlinear
connection, N-connection, structure allow to find exact solutions (with
generic off-diagonal metrics when coefficients depend on all spacetime
coordinates) of physically important systems of nonlinear PDEs. In this
series of partner works (see, for instance, the second partner work \cite%
{partner02} and further ones), we prove that such geometric methods can be
extended to nonassociative gravity theories. When nonholonomic geometric
constructions are performed on generalized phase spaces, for instance, of
total dimension 8, the approach is developed with a nonholonomic shell
diadic splitting, s-splitting, of dimensions (2+2+2+2) determined by 
respective s-connection and s-metric structures. This defines geometrically
the AFCDM for generating exact and parametric solutions in noncommutative
Finsler and gauge like gravity theories \cite{vacaru03,vacaru09a,bubuianu17}%
. As explicit examples, we can construct modified black hole and
cosmological solutions and all geometric formalism can be extended to 
nonassociative gravity following methods of nonholonomic geometry and the
theory of nonlinear PDEs.

\item Rewriting commutative and noncommutative gravity theories in Finsler
like variables, it is possible to define canonical almost symplectic
geometries and perform models of deformation, A-brane and gauge like
quantization \cite{vacaru08a,vacaru07,vacaru07b}. We expect that such
methods can be generalized for respective star products in nonassociative
gravity and even in the A. Connes approach to noncommutative geometry \cite%
{vacaru08} and further nonassociative extensions.

\item Using the results on AFCDMs in constructing new classes of exact
solution in heterotic string gravity theories \cite{bubuianu17b}, there are
possibilities to investigate the problem how a nonassociative gravity based
on the concept of star diffeomorphism can be related self-consistently with
a string or double field theory and to find exact and parametric solutions
in such scenarios. Here we note that J. Moffat's group elaborated their
nonsymmetric gravity theories following a paradigm that MGTs should be
elaborated for 4-d generalized spacetime geometries and not for (super)
string gravities with extra dimensions. Considering models of string gravity
on higher order (co) tangent bundles and respective (non) associative /
commutative geometric objects \cite{vacaru96a,vacaru96b,aschieri15, vacaru18}%
, we can formulate an unified geometric formalism for all type of such
theories even the arenas for constructions and physical principles can be
very different. For instance, we can use nonholonomic deformations of 4-d
Lorentz manifolds, their (co) tangent bundles, or 10-d and 11-d spacetimes,
correspondingly, in string gravity and supergravity etc. and respective
nonassociative models.

\item Nonholonomic constraints on geometric flow evolution of (pseudo)
Riemannian metrics naturally result in nonsymmetric metric components and
generalized nonlinear and linear connection structures, see reviews of
results in \cite{vacaru08bb,vacaru18}. Such configurations can be described
in Finsler--Lagrange and/or almost K\"{a}hler variables. The constructions
can be extended on complex and real (co) tangent bundles for nonassociative
phase space geometries and respective models of nonassociative and/or
noncommutative Finsler-Lagrange-Hamilton geometry. For self-similar
configurations, we obtain nonholonomic Ricci solitons with nonassociataive
star products providing certain modifications of the Einstein equations for
geometric objects defined in \cite{blumenhagen16,aschieri17}.

\item Finally, we note that nonholonomic generalized Finsler methods are
important for developing in a series of partner works certain models of
nonassociative and noncommutative quantum information theories and geometric
flows.
\end{enumerate}



This paper is organized as follows:

Section \ref{sec2} contains an introduction into the geometry of
nonassociative star products and nonholonomic (co) tangent Lorentz bundles.
In geometric preliminaries, we define nonlinear connections, N-connections,
and N-adapted frames and related nonholonomic horizontal, h, and vertical,v,
or covertical, c, decompositions (h- and v- or c-decompositions) for
associative and commutative base Lorentz manifolds and nonholonomic phase
spaces. The formulas for nonassociative star products are considered in
N-adapted and coordinate base forms on nonholonomic phase spaces with real
and complex momentum like variables. We study properties of the star product
and develop on a nonholonomically generalized star tensor calculus and Lie
derivatives.

In section \ref{sec3}, there are formulated and compared two approaches
(without and with quasi-Hopf N-adapted structure) to nonassociative
nonholonomic differential geometry. There are defined nonassociative
distinguished connections, d-connections, and covariant derivatives with
N-adapted h- and c-splitting and computed their Riemannian, torsion and
Ricci d-tensors. There are analyzed main features of star deformed
nonsymmetric and symmetric metrics studied in N-adapted frame bases and
related nonassociative variants of canonical distinguished and Levi-Civita,
LC, connections.

Section \ref{sec4} is devoted to nonassociative nonholonomic modifications
of vacuum Einstein equations, in general, with nontrivial cosmological
constant. For nontrivial R-fluxes, such models involve both symmetric and
nonsymmetric metrics. In canonical nonholonomic variables, the nonsymmetric
components of Ricci tensors can be related to parametric decompositions of
the nonsymmetric part of metrics. There are computed nontrivial R-flux
contributions to horizontal components of the Ricci tensor depending both on
spacetime and momentum like variables. We speculate how to formulate
nonassociative variants of vacuum Einstein equations with a nontrivial
cosmological constant using star nonholonomic deformations. The main issue
is that such constructions involve nonsymmetric complex metric structures
and related types of N- and d-connections. We formulate such variants of
nonassociative gravitational field equations when the nonsymmetric
components of metrics are described by nonsymmetric d-tensor fields and
argue that such equations can be decoupled and integrated in certain general
forms using the AFCDM (proofs are given in our partner work \cite{partner02}%
).

Conclusions are drawn in section \ref{sec5}. In this paper, proofs are
sketched in brief forms which are accessible both to mathematicians and
physicists. We cite relevant papers containing similar associative geometric
results and various details and examples. It is used the abstract (frame and
coordinate free) geometric method but certain important coefficient
N-adapted formulas are presented because they are useful for our future
partner works.

\section{Nonassociative nonholonomic star products on (co) tangent Lorentz
bundles}

\label{sec2} A self-consistent approach to nonassociative gravity was
elaborated in \cite{blumenhagen16} for respective star products and star
diffeomorphisms on a configuration space $M$ with local coordinates $(x^{j})$
and coordinate frames. In that work, there are considered also geometric
constructions on full phase space $\mathcal{M}=T^{\ast }M,$ with local
coordinates of type $(x^{k},ip_{a}),$ which is a dual tangent bundle with
(co) fiber momentum like coordinates $ip_{a}$ multiplied to the complex
unity, $i^{2}=-1$, see footnotes below on conventions on coordinates and
indices. It is speculated on formulating a variational calculus for star
diffeomorphism invariant of the Einstein-Hilbert action with embedding of
configuration space into the phase space as the $p_{0}$ leaf. Then, a more
rigorous mathematical model of nonassociative geometry and gravity with
quasi-Hopf algebraic structures was developed in \cite{aschieri17}. In this
paper, we extend and modify those geometric approaches in nonholonomic
geometric forms which allows us to formulate certain variants of
nonassociative vacuum gravitational field equations with nonsymmetric
metrics, see section \ref{sec4}. Using such a formalism, we construct exact
solutions in nonassociative gravity and consider possible applications in
modern cosmology, astrophysics and (quantum) information theory in a series
of partner works, see reviews of previous results for (non) commutative
geometric and physical models in \cite%
{vacaru09a,vacaru20,bubuianu17,bubuianu19}, and the second partner work \cite%
{partner02} on decoupling and integrability of nonassociative vacuum
gravitational equations.

In \cite{vacaru18,bubuianu18a}, a study of physical principles and axiomatic
approaches to (non) commutative MGTs with MDRs and LIVs of type (\ref{mdr})
and possible generalizations to models with nonsymmetric metrics is
provided. It was emphasized that a subclass of nonstandard physical models
of Finsler-Lagrange-Hamilton type can be formulated as causal extensions of
GR and standard particle theories of physics if certain canonical
nonholonomic variables are used for geometric constructions performed on
tangent and cotangent Lorentz bundles, respectively, $TV$ and $T^{\ast }V.$
Such total (phase) spaces can be modelled on a Lorentz manifold $V$ (base
spacetime). A similar geometric techniques can be applied for constructing
exact and parametric solutions in GR when a base spacetime $\mathbf{V}$ is
modeled as a four dimensional, 4-d, nonholonomic Lorentz manifold enabled
with a pseudo-Riemannian metric structure $g_{ij}(x^{k})$ (for $%
i,j,k,...=1,2,3,4$; for instance, of local signature $(+++-)$), and certain
classes of nonholonomic distributions with respect to which geometric
objects can be adapted.\footnote{%
On total (phase) spaces of dimension 8, the coefficients of geometric
objects (metrics, connections, curvatures etc.) depend both on base and
(fiber) coordinates. For instance, we consider theories when a metric
structure $g_{\alpha \beta }(x^{k},v^{a}),$ or $g_{\alpha%
\beta}(x^{k},p_{a}), $ is determined by MDRs in some forms depending
additionally on velocity/ momentum like (co) fiber coordinates, when indices 
$\alpha =(i,a),\beta =(j,b)$ etc. run values $1,2,...,7,8.$ Such MDRs and
commutation relations (\ref{comrel}) also determine a class of frames which
are linear with respect to coefficients of a nonlinear connection
(N-connection) structure $\mathbf{N}=\{N_{i}^{a}\}$ and/or $\ ^{\shortmid}%
\mathbf{N}=\{\ ^{\shortmid }N_{ia}\}, $ see definitions below.} We can
always introduce some type of nonholonomic variables on $TV$ and $T^{\ast }V$
and formulate various geometric and physical theories on such phase spaces
as certain relativistic Finsler-Lagrange-Hamilton geometries which can be
generalized to complex variables, for nonassociative and/or noncommutative
spaces, supersymmetric models etc. To formulate such theories in Finsler
like variables (they can be introduced even in GR) is important because they
allow a straightforward reformulation in almost symplectic variables which
can be used for performing deformation, A-brane and gauge like quantization 
\cite{vacaru07a,vacaru08b,vacaru07,vacaru01,vacaru03,vacaru08}. Similar
higher dimension constructions were considered for the heterotic string
gravity in \cite{bubuianu17b,vacaru96a,vacaru96b}.

In this section, we provide basic definitions and results on N-connection
geometry with nonhlonomic h- and v-/c-splitting of type 4+4 on $TV$ and $%
T^{\ast }V$, which are necessary for developing the anholonomic frame and
connection deformation method, AFCDM, for constructing exact and parametric
solutions in nonassociative gravity \cite%
{vacaru18,bubuianu17,bubuianu19,vacaru20}. Then, we study certain important
properties of nonassociative star products adapted to corresponding
nonlinear connection splitting and state that there are two important
operators which, respectively, control when the product is noncommutative
and nonassociative. To perform a nonholonomic generalization of
nonassociative gravity models from \cite{blumenhagen16,aschieri17} we
consider phase spaces with complexified momentum type variables of $%
T^{\ast}V.$

\subsection{Geometric preliminaries:\ Nonlinear connections in commutative
spaces}

We summarize main concepts and definitions from nonholonomic geometry of
relativistic phase spaces with real spacetime and complex velocity/momentum
type variables. For convenience, there will be provided some important
formulas with real and complex (co) fiber coordinates:%
\begin{eqnarray}
\ ^{\shortparallel }u &=&(x,\ ^{\shortparallel }p)=\{\ ^{\shortparallel
}u^{\alpha }=(u^{k}=x^{k},\ ^{\shortparallel }p_{a}=(i\hbar )^{-1}p_{a})\}%
\mbox{ on }T_{\shortparallel }^{\ast }\mathbf{V,}\mbox{ and }
\label{loccordph} \\
\ ^{\shortmid }u &=&(x,\ ^{\shortmid }p)=\{\ ^{\shortmid }u^{\alpha
}=(u^{k}=x^{k},\ ^{\shortmid }p_{a}=p_{a})\}\mbox{ on }T^{\ast }\mathbf{V}. 
\notag
\end{eqnarray}%
This defines a phase space model $T_{\shortparallel }^{\ast }\mathbf{V}$
with complex cofibers associated to a real cotangent bundle $T^{\ast }%
\mathbf{V.}$ Such notations are different from those introduced for formula
(3.1) in \cite{blumenhagen16}, where $X_{K}=(\frac{p_{k}}{i\hbar },x^{k})$
is written instead of $\ ^{\shortparallel }u^{\alpha }.$\footnote{$\hbar
=h/2\pi $ is the Planck constant} In this work, an up (or low, on
convenience), label "$\ ^{\shortparallel }$" is used in order to distinguish
such total "momentum-complexified" coordinates from real ones $\
^{\shortmid}u^{\alpha }=(x^{k},p_{a})$ on $T^{\ast }\mathbf{V}$ considered
in Finsler-Lagrange-Hamilton geometry \cite{vacaru18,bubuianu18a}.

\paragraph{Nonlinear connection structures on commutative phase spaces: 
\newline
}

Nonlinear connection, N--connection, structures on $TV,\ T^{\ast }V$ and $%
T_{\shortparallel }^{\ast }V$ can be defined in global forms by respective
Whitney sums $\oplus $ of conventional horizontal and vertical
distributions, $h$- and $v$--distributions, and horizontal and covertical
distributions, $h$ and $c$--distributions, when 
\begin{equation}
\mathbf{N}:\ T\mathbf{TV}=hTV\oplus vTV,{\quad }\ ^{\shortmid }\mathbf{N}:\ T%
\mathbf{T}^{\ast }\mathbf{V}=hT^{\ast }V\oplus cvT^{\ast }V\mbox{ and  }\ \
^{\shortparallel }\mathbf{N}:\ T\mathbf{T}_{\shortparallel }^{\ast }\mathbf{V%
}=hT_{\shortparallel }^{\ast }V\oplus cvT_{\shortparallel }^{\ast }V.
\label{ncon}
\end{equation}%
For a Lorentz base manifold, $\mathbf{V,}$ such a N--connection splitting is
with a $4+4$ decomposition of the total space dimension. In local forms, the
formulas (\ref{ncon}) are written respectively: 
\begin{equation*}
\mathbf{N}=N_{i}^{a}\frac{\partial }{\partial v^{a}}\otimes dx^{i},\
^{\shortmid }\mathbf{N}=\ ^{\shortmid }N_{ia}\frac{\partial }{\partial p_{a}}%
\otimes dx^{i}\mbox{ and }\ \ ^{\shortparallel }\mathbf{N}=\
^{\shortparallel }N_{ia}\frac{\partial }{\partial \ ^{\shortparallel }p_{a}}%
\otimes dx^{i},
\end{equation*}%
with corresponding N-connection coefficients $\mathbf{N}=\{N_{i}^{a}\},\
^{\shortmid}\mathbf{N}= \{\ ^{\shortmid }N_{ia}\},$ and $\ ^{\shortparallel }%
\mathbf{N}=\{\ ^{\shortparallel }N_{ia}\},$ for $i,j,k,...=1,2,3,4$ (used
for base coordinates) and $a,b,c,...=5,6,7,8$ (used for (co) fiber
coordinates). The spacetime and phase spaces and respective geometric
objects on $\mathbf{V},T\mathbf{V},T^{\ast }\mathbf{V,T}_{\shortparallel}^{%
\ast }\mathbf{V}$ are labeled by "bold face" symbols if they are enabled
with a N-connection structures and/or written in N-adapted form. An up, or
law (on convenience), label bar "$\ ^{\shortmid }$" is used in order to
emphasize that certain geometric objects are defined on cotangent bundles
and, respectively, labels with a double bar "$\ ^{\shortparallel }$" for
geometric objects on full phase spaces with co-vertical coordinates
containing the complex unity before momenta.

\paragraph{N-adapted commutative (co) frame structures: \newline
}

If a N-connection structure (\ref{ncon}) is introduced for a real tangent
Lorentz bundle, we can define respective classes of N-elongated
(equivalently, N-adapted) bases and dual bases (cobases), which are linear
on N-connection coefficients and written in the form: 
\begin{eqnarray}
\mathbf{e}_{\alpha } &=&(\mathbf{e}_{i}=\frac{\partial }{\partial x^{i}}%
-N_{i}^{a}(x,v)\frac{\partial }{\partial v^{a}},e_{b}=\partial _{b}=\frac{%
\partial }{\partial v^{b}}),\mbox{ on }\ T\mathbf{TV;}  \label{nadap} \\
\mathbf{e}^{\alpha } &=&(e^{i}=dx^{i},\mathbf{e}%
^{a}=dv^{a}+N_{i}^{a}(x,v)dx^{i}),\mbox{ on }T^{\ast }\mathbf{TV.}  \notag
\end{eqnarray}%
A local basis $\mathbf{e}_{\alpha }$ is nonholonomic if the commutators 
\begin{equation}
\mathbf{e}_{[\alpha }\mathbf{e}_{\beta ]}:=\mathbf{e}_{\alpha }\mathbf{e}%
_{\beta }-\mathbf{e}_{\beta }\mathbf{e}_{\alpha }=w_{\alpha \beta }^{\gamma
}(u)\mathbf{e}_{\gamma }  \label{anhrel}
\end{equation}%
contain nontrivial anholonomy coefficients $w_{\alpha \beta
}^{\gamma}=\{w_{ia}^{b}= \partial _{a}N_{i}^{b},w_{ji}^{a}=\mathbf{e}%
_{j}N_{i}^{a}-\mathbf{e}_{i}N_{j}^{a}\}.$ If $w_{\alpha \beta }^{\gamma }=0,$
such a basis is holonomic (integrable) and can be transformed via coordinate
transforms into a coordinated basis, $\mathbf{e}_{\alpha }\rightarrow 
\mathbf{\partial }_{\alpha }=\partial /\partial u^{\alpha }$. Nonholonomic
frames and coframes of type (\ref{nadap}) are considered in Einstein and
Finsler-Lagrange geometry and gravity theories \cite{vacaru18,bubuianu18a}
with respective 2+2 or 4+4 splitting. For Finsler variables, there are
imposed certain homogeneity conditions on velocity type fiber variables. In
principle, we can consider models with complex velocity type variables, but
such constructions are not largely used in modern mathematical and
theoretical particle physics.

Similarly, we can define N-elongated dual bases and cobases for cotangent
bundles, 
\begin{eqnarray}
\ ^{\shortmid }\mathbf{e}_{\alpha } &=&(\ ^{\shortmid }\mathbf{e}_{i}=\frac{%
\partial }{\partial x^{i}}-\ ^{\shortmid }N_{ia}(x,p)\frac{\partial }{%
\partial p_{a}},\ ^{\shortmid }e^{b}=\frac{\partial }{\partial p_{b}}),%
\mbox{ on }\ T\mathbf{T}^{\ast }\mathbf{V;}  \label{nadapd} \\
\ \ ^{\shortmid }\mathbf{e}^{\alpha } &=&(\ ^{\shortmid }e^{i}=dx^{i},\
^{\shortmid }\mathbf{e}_{a}=dp_{a}+\ ^{\shortmid }N_{ia}(x,p)dx^{i}),%
\mbox{
on }T^{\ast }\mathbf{T}^{\ast }\mathbf{V,}  \notag
\end{eqnarray}%
which, in general, are characterized by anholonomy coefficients 
\begin{equation}
\ ^{\shortmid }w_{\alpha \beta }^{\gamma }=\{\ ^{\shortmid }w_{ib}^{\ a}=\
^{\shortmid }e^{a}\ ^{\shortmid }N_{ib},\ ^{\shortmid }w_{jia}=\ ^{\shortmid
}\mathbf{e}_{j}\ ^{\shortmid }N_{ia}-\ ^{\shortmid }\mathbf{e}_{i}\
^{\shortmid }N_{ja}\}.  \label{anhrelcd}
\end{equation}%
For a N--connection on $\mathbf{TV,}$ or $\mathbf{T}^{\ast }\mathbf{V,}$ we
can introduce and compute corresponding coefficients of N--connection
curvature (called also Neijenhuis tensors) 
\begin{equation}
\Omega _{ij}^{a}=\frac{\partial N_{i}^{a}}{\partial x^{j}}-\frac{\partial
N_{j}^{a}}{\partial x^{i}}+N_{i}^{b}\frac{\partial N_{j}^{a}}{\partial v^{b}}%
-N_{j}^{b}\frac{\partial N_{i}^{a}}{\partial v^{b}},\mbox{\ or\ }\ \ \mathbf{%
\ ^{\shortmid }}\Omega _{ija}=\frac{\partial \mathbf{\ ^{\shortmid }}N_{ia}}{%
\partial x^{j}}-\frac{\partial \mathbf{\ ^{\shortmid }}N_{ja}}{\partial x^{i}%
}+\ \mathbf{^{\shortmid }}N_{ib}\frac{\partial \mathbf{\ ^{\shortmid }}N_{ja}%
}{\partial p_{b}}-\mathbf{\ ^{\shortmid }}N_{jb}\frac{\partial \mathbf{\
^{\shortmid }}N_{ia}}{\partial p_{b}}.  \label{neijtc}
\end{equation}%
Such curvatures are different from the Riemannian curvature for linear
connections. In brief, we shall write also $\partial _{i}:=\partial
/\partial x^{i}, \ ^{\shortmid } \partial ^{b}:= \partial /\partial
p_{b},\partial _{\alpha }:=(\partial _{i}, \partial _{a}),\ ^{\shortmid
}\partial _{\alpha }:=(\partial _{i}, \ ^{\shortmid}\partial ^{a})$ etc.

Formulas (\ref{nadapd}), (\ref{anhrelcd}) and (\ref{neijtc}) can be
generalized on a (complex) phase space $T_{\shortparallel }^{\ast }\mathbf{V}
$ for 
\begin{eqnarray}
\ ^{\shortparallel }\mathbf{e}_{\alpha } &=&(\ ^{\shortparallel }\mathbf{e}%
_{i}=\frac{\partial }{\partial x^{i}}-\ ^{\shortparallel }N_{ia}(x,\
^{\shortparallel }p)\frac{\partial }{\partial \ \ ^{\shortparallel }p_{a}},\
\ ^{\shortparallel }e^{b}=\frac{\partial }{\partial \ \ ^{\shortparallel
}p_{b}}),\mbox{ on }\ T\mathbf{T}_{\shortparallel }^{\ast }\mathbf{V;}
\label{nadapdc} \\
\ ^{\shortparallel }\mathbf{e}^{\alpha } &=&(\ ^{\shortparallel
}e^{i}=dx^{i},\ ^{\shortparallel }\mathbf{e}_{a}=d\ \ ^{\shortparallel
}p_{a}+\ \ ^{\shortparallel }N_{ia}(x,\ \ ^{\shortparallel }p)dx^{i}),\mbox{
on }T^{\ast }\mathbf{T}_{\shortparallel }^{\ast }\mathbf{V}.  \notag
\end{eqnarray}%
Various versions of such N-adapted (N-elongated) nonholonomic bases (frames)
and their frame transforms can be used for variables $\
^{\shortparallel}p_{a}=\ ^{\shortparallel }p_{a}=(i\hbar )^{-1}p_{a}$ and $\
^{\shortparallel }e^{b}=\partial /\partial \ ^{\shortparallel }p_{b}=i\hbar
\ ^{\shortmid }e^{b},$ in order to elaborate on generalizations of star
products and respective nonassociative calculus, see below formulas (\ref%
{starpn}). The coefficients of N-connection $\ ^{\shortparallel }\mathbf{N}%
=\{\ ^{\shortparallel }N_{ia}\}$ can be chosen in such forms that the
N-adapted frames (\ref{nadapdc}) are real bases even the cofiber coordinates
contain the complex unity $i,$ when $i^{2}=-1.$

\paragraph{General and N-adapted frame transforms in nonholonomic
commutative phase spaces: \newline
}

On phase spaces $T_{\shortparallel }^{\ast }V$ and/or $\mathbf{T}
_{\shortparallel }^{\ast }\mathbf{V,}$ we can consider arbitrary frame
transforms from coordinate (holonomic) bases and cobases, respectively, to
nonholonomic ones, which is written in the form. 
\begin{equation}
\ ^{\shortparallel }\partial _{\underline{\alpha }}=\partial /\partial \
^{\shortparallel }u^{\underline{\alpha }}\ \rightarrow \ ^{\shortparallel
}e_{\alpha }=\ ^{\shortparallel }e_{\alpha }^{\ \underline{\alpha }}(\
^{\shortparallel }u^{\underline{\beta }})\ ^{\shortparallel }\partial _{%
\underline{\alpha }}\mbox{ and }\ ^{\shortparallel }e^{\underline{\gamma }%
}=d\ ^{\shortparallel }u^{\underline{\gamma }}\rightarrow \ ^{\shortparallel
}e^{\gamma }=\ ^{\shortparallel }e_{\ \underline{\gamma }}^{\gamma }(\
^{\shortparallel }u^{\underline{\beta }})\ ^{\shortparallel }e^{\underline{%
\gamma }}  \label{nhframtr0}
\end{equation}%
where undrelined indices are used for objects with coefficients determined
with respect to a coordinate local (co) base and matrices $\
^{\shortparallel}e_{\alpha }^{\ \underline{\alpha }}$ and $e_{\ \underline{%
\gamma }}^{\gamma}$ are called vierbeinds. For (co) vector/tangent bundles,
there is a natural local splitting of total space indices into h- and
c-indices, for instance, $\underline{\beta }=(\underline{j},\underline{c}%
),\beta =(j,c),$ $\ ^{\shortparallel }u= (\ x,\ ^{\shortparallel }p)=\{\
^{\shortparallel }u^{\underline{\gamma }}= (x^{\underline{k}},\
^{\shortparallel }p_{\underline{c}})\}$ etc. There is a subclass of
so-called N-adapted frame transforms which are linear on N-connection
coefficients and transform local coordinate (co) bases into N-adapted ones (%
\ref{nadapdc}). They are parameterized in the form%
\begin{equation}
\ ^{\shortparallel }\mathbf{e}_{\alpha }^{\ \underline{\alpha }}=\left[ 
\begin{array}{cc}
\ ^{\shortparallel }e_{i}^{\ \underline{i}}(\ ^{\shortparallel }u) & \ -\
^{\shortparallel }N_{ib}\ ^{\shortparallel }e_{\ \underline{a}}^{b}(\
^{\shortparallel }u) \\ 
0 & ^{\shortparallel }e_{\ \underline{a}}^{a}(\ ^{\shortparallel }u)%
\end{array}%
\right] \mbox{ and }\ ^{\shortparallel }\mathbf{e}_{\ \underline{\gamma }%
}^{\gamma }=\left[ 
\begin{array}{cc}
\ ^{\shortparallel }e_{\ \underline{j}}^{j}(\ ^{\shortparallel }u) & \ \
^{\shortparallel }N_{kb}\ ^{\shortparallel }e_{\ \underline{k}}^{k}(\
^{\shortparallel }u) \\ 
0 & \ ^{\shortparallel }e_{a}^{\ \underline{a}}(\ ^{\shortparallel }u)%
\end{array}%
\right] ,  \label{nhframtr}
\end{equation}%
where $\ ^{\shortparallel }\mathbf{e}_{\alpha }=\ ^{\shortparallel }\mathbf{e%
}_{\alpha }^{\ \underline{\alpha }}(\ ^{\shortparallel }u^{\underline{\beta }%
})\ ^{\shortparallel }\partial _{\underline{\alpha }}$ and $\
^{\shortparallel }\mathbf{e}^{\gamma }=\ ^{\shortparallel }\mathbf{e}_{\ 
\underline{\gamma }}^{\gamma }(\ ^{\shortparallel }u^{\underline{\beta }})\
^{\shortparallel }e^{\underline{\gamma }}.$ We can consider N-adapted frame
transforms of type $\ ^{\shortparallel }\mathbf{e}_{\alpha ^{\prime }}=\
^{\shortparallel }\mathbf{e}_{\alpha ^{\prime }}^{\ \alpha }(\
^{\shortparallel }u^{\underline{\beta }})\ \ ^{\shortparallel }\mathbf{e}%
_{\alpha }$ and $\ ^{\shortparallel }\mathbf{e}^{\gamma ^{\prime }}=\
^{\shortparallel }\mathbf{e}_{\ \gamma }^{\gamma ^{\prime }}(\
^{\shortparallel }u^{\underline{\beta }})\ \ ^{\shortparallel }\mathbf{e}%
^{\gamma },$ or any general (not obligatory N-adapted, i.e. without boldface
symbols) $\ ^{\shortparallel }e_{\alpha ^{\prime }}=\ ^{\shortparallel
}e_{\alpha ^{\prime }}^{\ \alpha }(\ ^{\shortparallel }u^{\underline{\beta }%
})\ \ ^{\shortparallel }e_{\alpha }$ and $\ ^{\shortparallel }e^{\gamma
^{\prime }}=\ ^{\shortparallel }e_{\ \gamma }^{\gamma ^{\prime }}(\
^{\shortparallel }u^{\underline{\beta }})\ \ ^{\shortparallel }e^{\gamma }.$

Because we work on a (co) tangent bundle, we have the possibility to
contract not only horizontal (vertical) up with vertical (horizontal) low
indices, for instance, $A^{i}B_{i}$ ($A^{a}B_{a}$), but we can also
introduce a convention how to contract horizontal indices with vertical
ones, for instance, $A^{i}B_{n+i}\sim A^{i}B_{a},$ where $\dim V=n$ and the
same dimension is taken for the (co) fiber coordinates. This convention is
used for R-fluxes in string theory, when such a rule in stated in the
definition of star product (see next section). In the bulk of such works,
the geometric objects and star products are defined for coordinate (co)
bases, $\ ^{\shortparallel }\partial _{\underline{\alpha }}$ and $\
^{\shortparallel}e^{\underline{\gamma }}$, but the constructions became well
defined geometrically only if a general covariance principle (when the
geometric/physical models do not depend on frame/coordinate transforms) is
formulated. A self-consistent geometric approach has to be elaborated for
general frames $\ ^{\shortparallel }e_{\alpha }$ and $\ ^{\shortparallel
}e^{\gamma }$ for commutative phase spaces with possible star deformations
to nonassociative ones on tensor products. A subclass of such N-adapted
frames $\ ^{\shortparallel }\mathbf{e}_{\alpha }$ and $\ ^{\shortparallel }%
\mathbf{e}^{\gamma }$ will be defined in such a form which will allow
certain general decoupling and integration of physically important systems
of nonlinear PDEs.

\subsection{Nonassociative star product and N--connection structures}

We extend the definition of nonassociative star product introduced and
studied in \cite{mylonas12,mylonas13} to the case of nonholonomic phase
spaces and spacetime manifolds enabled with N-connection structure. The
approaches elaborated is section 2 of \cite{blumenhagen16} and section 2 of 
\cite{aschieri17} for configuration spaces can be developed for the case of
trivial N-connection structure if we follow the geometric formalism outlined
in previous subsection. In certain models of noncommutative gauge gravity
and generalized Finsler geometry and their deformation quantization, similar
N-adapted Moyal--Weyl star products were considered in \cite%
{vacaru01,vacaru03,vacaru07a,vacaru08b,vacaru16,vacaru09a,vacaru08a,vacaru07b}%
. Our motivation to consider such general frame and N-adapted frame
formulations is that the GR is a classical theory valid at the macroscopic
level at least for astrophysical and not very long cosmological distances.
The Einstein and matter field equations can be formulated in a general
covariant form when necessary type frames and coordinates are used for
constructing and analyzing properties of certain classes of solutions and
modeling classical field interactions. A general formalism of quantum and
string star deformations should not depend on the type of frames/
coordinates are used for the definition of the star product and respective
quantum/ string corrections determined by Planck and/or string length
scales. Such a geometrization is possible if we use general nonholonomic
frames which can be with N-connection splitting for certain purposes to
geometrize phase space constructions and/or solve systems of nonlinear PDEs.
For instance, one of the aims is to consider some primary classical generic
off-diagonal metrics and generalized connections and then elaborate on
self-consistent nonassociative and noncommutative modifications into certain
classes of target solutions, which are covariant on commutative base
manifolds but with star deformed symmetries and nonassociative geometric
objects on the total phase space.

We begin with the definition of nonassociative star products in
h-coordinates for h-coefficients in order to familiarize readers with
nonholonomic modifications of coordinate frame constructions in \cite%
{blumenhagen16,mylonas12,mylonas13}. In subsection \ref{ssectstarphase}, the
formalism of nonassociative star product is developed in N--adapted form on
the full phase space $T_{\shortparallel }^{\ast }\mathbf{V}$. Then (in
further subsections), the constructions are adapted also to quasi-Hopf
structures with N-connection splitting in order to elaborated on geometric
models with star deformations of coordinate diffeomorphism.

\subsubsection{The universal N-adapted R-matrix and the associator}

\label{ssconv}In our works, the geometric and physical constructions are
elaborated for general (co) frame structures both on base and phase (tangent
and cotangent bundles) spaces, i.e. for some $\ ^{\shortparallel }e_{\alpha
} $ and $\ ^{\shortparallel }e^{\gamma }$ and respective frame transforms (%
\ref{nhframtr0}). For a nontrivial N-connection structure, it is optimal to
work with bases $\ ^{\shortparallel }\mathbf{e}_{\alpha }$ and $\
^{\shortparallel }\mathbf{e}^{\gamma }$ (\ref{nadapdc}) and respective
vielbeinds (\ref{nhframtr}).

\paragraph{Nonassociative N-adapted star product: \newline
}

Let us consider a full phase space $\mathcal{M}$ containing a spacetime
direction determined by a N-adapted direction $\ ^{\shortmid }\mathbf{e}_{i}$
and a momentum like cofiber directions $\ ^{\shortmid }e^{b}$ as in (\ref%
{nadapd}). For any pair of functions $\ z(x,p)$ and $\ q(x,p),$ we define a
N-adapted star product $\star _{N}$: 
\begin{eqnarray}
z\star _{N}q:= &&\cdot \lbrack \mathcal{F}_{N}^{-1}(z,q)]  \label{starpn} \\
&=&\cdot \lbrack \exp \left( -\frac{1}{2}i\hbar (\ ^{\shortmid }\mathbf{e}%
_{i}\otimes \ ^{\shortmid }e^{i}-\ ^{\shortmid }e^{i}\otimes \ ^{\shortmid }%
\mathbf{e}_{i})\right) +\frac{i\mathit{\ell }_{s}^{4}}{12\hbar }%
R^{ija}(p_{a}\ ^{\shortmid }\mathbf{e}_{i}\otimes \ ^{\shortmid }\mathbf{e}%
_{j}-\ ^{\shortmid }\mathbf{e}_{j}\otimes p_{a}\ ^{\shortmid }\mathbf{e}%
_{i})]z\otimes q  \notag \\
&=&z\cdot q-\frac{i}{2}\hbar \lbrack (\ ^{\shortmid }\mathbf{e}_{i}z)(\
^{\shortmid }e^{i}q)-(\ ^{\shortmid }e^{i}z)(\ ^{\shortmid }\mathbf{e}%
_{i}q)]+\frac{i\mathit{\ell }_{s}^{4}}{6\hbar }R^{ija}p_{a}(\ ^{\shortmid }%
\mathbf{e}_{i}z)(\ ^{\shortmid }\mathbf{e}_{j}q)+\ldots ,  \notag
\end{eqnarray}%
where the constant $\mathit{\ell }$ and totally antisymmetric $R^{ija}$
capture the presence of a flux background in string theory, with h- and
v-indices contracted following the convention from footnote \ref{fnfluxind}.
In these formulas, the tensor product $\otimes $ is used in a form
indicating on which factor of $z\otimes q$ the N-adapted derivatives act
with the dot form, when eventually the tensor products turn into usual
multiplications. In coordinate frames and for holonomic frames determined by
trivial N-connection structures, the star product (\ref{starpn}) coincides
with that considered in \cite{blumenhagen16}. In such cases, we do not write
boldface symbols and do not use the label "N" for star products an related
geometric objects.

The properties of the twist distinguished operator, d-operator, $\mathcal{F}%
^{-1}$ (\ref{starpn}) will be discussed bellow. It is important to decide on
possible conventions on definition of nonassociative star products and
constructing models of curved phase phases:\ In flat spaces, we can work
always with partial derivatives $\ ^{\shortmid }\partial _{\alpha}=(\partial
_{i}, \ ^{\shortmid }\partial ^{a})$ identifying indices for coordinates on
base manifold with (co) fiber indices on (co) tangent bundle. In such cases,
we can define the nonassociative star product as in formulas (2.1) and (2.2)
of \cite{blumenhagen16}. For curved base spaces enabled with generalized
anholonomic (co) frames as commutative frame transforms of the partial
derivatives and differentials and involving respective frame decompositions
of the metric structure and certain covariant derivatives (and related
nonlinear and linear connections), the constructions are not straightforward
and depend on certain conventions on constructing respective nonassociative
geometric and curved spacetime models. There are two possibilities:

\begin{itemize}
\item \textbf{Convention 1:} We can introduce the concept of nonassociative
star product using partial derivatives on a conventional flat phase space
with contributions of a totally antisymmetric $R^{ija}.$ Then, we try to
elaborate on curved nonassociative phase space generalizations when in the
total space there are considered nonassociative frames/ metrics /
connections etc. with both star and nonholonomic deformations of geometric
data, 
\begin{equation}
(\ ^{\shortmid }\partial _{\alpha },\eta _{\beta \gamma })\rightarrow (\star
,\ ^{\shortmid }\partial _{\alpha },\eta _{\beta \gamma })\rightarrow (\star
;\ ^{\shortmid }\mathbf{e}_{\alpha }[\star ],\ ^{\shortmid }\mathbf{g}%
_{\beta \gamma }[\star ],\ ^{\shortmid }\mathbf{D}_{\alpha }[\star ]).
\label{conv1}
\end{equation}%
Following a geometric approach, the definition of various types of curve
nonassociative phase spaces with respective nonholonomic, $\ ^{\shortmid }%
\mathbf{e}_{\alpha }[\star ],$ metric, $\ ^{\shortmid }\mathbf{g}_{\beta
\gamma }[\star ],$ and linear connection, $\ ^{\shortmid }\mathbf{D}_{\alpha
}[\star ],$ structures depends explicitly on the type of star product we fix
for partial derivatives on a flat phase space. This way, we consider star
generalized frame transforms and generalized metric and connection
structures deforming nonholonomically a flat $TV$ to a curved $\mathcal{M}%
:=T_{\shortparallel}^{\ast }\mathbf{V}$, with nonassociative mixing of
coordinate and (co) frame indices. Here we note that in noncommutative and
nonassociative geometry we can define an infinite number of local
differential calculi because there are different definitions of star
products, various types of adapting to nonholonomic structures and different
concepts of connections, symmetric and nonsymmetric metrics etc. In such
cases, there are a number of ambiguities for elaborating equivalent
geometric models and selecting physically important theories from an
infinite number of nonholonomic generalizations in the total space with $%
TV\rightarrow \mathcal{M}$. It is also a problem how to find unique
constraints/ limits, for instance, to GR and commutative standard particle
theories.

\item \textbf{Convention 2: } In a self-consistent geometric form, we can
consider a commutative $TV$ with curved spacetime and phase space geometric
data $(\ ^{\shortmid }\mathbf{e}_{\alpha },\ ^{\shortmid }\mathbf{g}_{\beta
\gamma },\ ^{\shortmid }\mathbf{D}_{\alpha })$ determined as a lift of
geometric objects from a spacetime $V$ (for instance, in GR) to its (co)
tangent bundle. This way, we can construct nonassociative deformations%
\begin{eqnarray}
&&(\ ^{\shortmid }\mathbf{e}_{\alpha },\ ^{\shortmid }\mathbf{g}_{\beta
\gamma },\ ^{\shortmid }\mathbf{D}_{\alpha };  \label{conv2} \\
&&%
\mbox{and solutions of standard or modified commutative Einstein eqs
in nonholonomic variables})\rightarrow  \notag \\
&&(\star _{N}:\mbox{ star deformations  of nonholonomic geometric
objects }\ ^{\shortmid }\mathbf{e}_{\alpha },\ ^{\shortmid }\mathbf{g}%
_{\beta \gamma },\ ^{\shortmid }\mathbf{D}_{\alpha };  \notag \\
&&\mbox{ and solutions of nonassociative gravitational eqs}  \notag \\
&&%
\mbox{ which are nonassociative deformations of solutions of commutative gravitational
equations}).  \notag
\end{eqnarray}%
In brief, we elaborate on star deformed geometric models on phase spaces
following another procedure than (\ref{conv1}), 
\begin{equation*}
(\ ^{\shortparallel }\partial _{\alpha },\eta _{\beta \gamma })\rightarrow
(\ \ ^{\shortparallel }\mathbf{e}_{\alpha },\ \ ^{\shortparallel }\mathbf{g}%
_{\beta \gamma },\ \ ^{\shortparallel }\mathbf{D}_{\alpha })\rightarrow
(\star ;\ \ ^{\shortparallel }\mathbf{e}_{\alpha },\ \ ^{\shortparallel }%
\mathbf{g}_{\beta \gamma }[\star ],\ \ ^{\shortparallel }\mathbf{D}_{\alpha
}[\star ]),
\end{equation*}%
where we consider nonholonomic deforms from a flat phase space to a curve
commutative nonholonomic phase space and then we consider star deforms
using, for instance, a N-adapted nonholonomic structure of type (\ref%
{nhframtr}). In such a geometric approach, we firstly take a well defined
commutative geometric/ physical model with respective (co) frame / metric /
connection / physical objects and equations, and corresponding lifts and
nonholonomic deformations in commutative (co) tangent bundles. Then,
secondly, we $\star _{N}$-transform all types of nonholonomic commutative
geometric/ physical data in a self-consistent form generating a
nonassociative nonholonomic phase space $\mathcal{M}$. On the base manifold,
we can consider horizontal frame transforms $\ ^{\shortmid }\mathbf{e}_{i}=\
^{\shortmid }\mathbf{e}_{i}^{i^{\prime }}(x,\ ^{\shortmid }p)\partial
_{i^{\prime }}$ with $\ ^{\shortmid }\mathbf{e}_{i}^{i^{\prime }}\
^{\shortmid }\mathbf{e}_{i^{\prime }}^{j}=\delta _{i}^{j},$ when 
\begin{equation*}
\ ^{\shortmid }\mathbf{e}_{i}\otimes \ ^{\shortmid }e^{i}=\ ^{\shortmid }%
\mathbf{e}_{i}^{i^{\prime }}\ ^{\shortmid }\mathbf{e}_{i^{\prime }}\otimes \
^{\shortmid }e^{i},\ R^{ija}p_{a}\ ^{\shortmid }\mathbf{e}_{i}\otimes \
^{\shortmid }\mathbf{e}_{j}=R^{i^{\prime }j^{\prime }a}p_{a}\ ^{\shortmid }%
\mathbf{\partial }_{i^{\prime }}\otimes \ ^{\shortmid }\mathbf{\partial }%
_{j^{\prime }},
\end{equation*}%
and re-define (\ref{starpn}) in not N-adapted form involving local
coordinate frames in respective forms written similarly to formulas (2.1)
and (2.2) of \cite{blumenhagen16}. In general N-elongated frames $\
^{\shortmid }\mathbf{e}_{\alpha }$ and/or $\ ^{\shortparallel }\mathbf{e}%
_{\alpha },$ we have non-trivial anholonomy coefficients (\ref{anhrelcd})
which contribute with additional terms in various N-adapted coefficient
formulas for linear connections and respective torsion, curvature, Ricci and
other type d-tensors. Such nonholonomic commutative geometric effects are
studied in details in Refs. \cite{vacaru18,bubuianu18a}. Similar formulas
can be reproduced by analogy on $T\mathbf{V,}$ $T^{\ast }\mathbf{V}$ and $%
T_{\shortparallel }^{\ast }\mathbf{V.}$
\end{itemize}

In our works, we follow the Convention 2, when the fundamental geometric
objects and physically important equations and their solutions are defined
in N-adapted form and then we star deform all commutative geometric/
physical data into respective nonassociative geometric / equations /
solutions on the total space of $\mathcal{M}$. In this approach, $R^{ija}$
are taken for a string theory, or other type model. They determine star
deformations on a base commutative nonholonomic manifold $V$ and together
with lifts and nonholonomic structures on $T\mathbf{V}$ and $T^{\ast }%
\mathbf{V}$ generate a nonassociative differential geometry of phase space $%
\mathcal{M}$. We emphasize that the sign in exponential terms in (\ref%
{starpn}) is taken as in coordinate bases of \cite{aschieri17} which is
opposite to the sign used in \cite{blumenhagen16}. Such an approach allows
us to consider geometrically "by analogy" both types of coordinate and/or
N-adapted nonholonomic and star deformed formulas.

For $z=x^{i}$ and $q=p_{i},$ the first part of the star product with trivial
N-connection structure, when $z\star _{N}q=z\star q,$ the formulas (\ref%
{starpn}) give the non-trivial commutator (\ref{comrel}). Such products are
nonassociative and violate the Jacobi identity. To guaranty compatibility
with the exterior derivative we can follow a proposal from \cite{mylonas13}
suggesting to replace partial derivatives by Lie derivatives $\mathcal{L}_{\
^{\shortmid }\mathbf{e}_{i}}$ and $\mathcal{L}_{\ ^{\shortmid }e^{i}}$ on
respective directions $\ ^{\shortmid }\mathbf{e}_{i}$ and $\
^{\shortmid}e^{i}.$ On nonholonomic manifolds, all derivatives can be
considered in N-adapted form, when the star product (\ref{starpn}) is
defined respectively as%
\begin{eqnarray}
z\star _{N}q &:=& z\cdot q-\frac{i}{2}\hbar \lbrack (\mathcal{L}_{\
^{\shortmid }\mathbf{e}_{i}}z)(\mathcal{L}_{\ ^{\shortmid }e^{i}}q)+(%
\mathcal{L}_{\ ^{\shortmid }e^{i}}z)(\mathcal{L}_{\ ^{\shortmid }\mathbf{e}%
_{i}}q)]  \label{starplie} \\
&&+\frac{i\mathit{\ell }_{s}^{4}}{12\hbar }R^{ija}[(\mathcal{L}_{p_{a}\
^{\shortmid }\mathbf{e}_{i}}z)(\mathcal{L}_{\ ^{\shortmid }e_{j}}q)+(%
\mathcal{L}_{\ ^{\shortmid }e_{j}}z)(\mathcal{L}_{p_{a}\ ^{\shortmid }%
\mathbf{e}_{i}}q)]+\ldots  \notag
\end{eqnarray}%
A nonassociative Lie star product (\ref{starplie}) can be re-written in "non
N-adapted" forms when $\ ^{\shortmid }\mathbf{e}_{i}\rightarrow \partial
/\partial x^{i}$ and $\ ^{\shortmid}e^{a} \rightarrow dp^{a}$ etc. using
respective frame transforms\footnote{%
such formulas are 8-d cotangent Lorentz bundles variants of formula (2.4) in 
\cite{blumenhagen16}; we consider in certain cases corresponding coordinate
base constructions because they allow to understand how generalizations are
performed in this work;\ nevertheless, coordinate bases are not convenient
for decoupling and integrating nonlinear systems of physically important
PDEs and we shall always give priorities to N-adapted formulas which admit
application of the AFCDM.}.

\paragraph{Nonholonomic twist d-operators and N-adapted universal $\mathcal{R%
}$-matrices: \newline
}

The operator $\mathcal{F}_{N}^{-1}$ (\ref{starpn}) is called respectively
the N-twist which transforms in the usual twist operator $\mathcal{F}^{-1}$
if the formulas are redefined in terms of coordinate bases. In all cases
(with $N$, or without any label), it is satisfied the property $\mathcal{F}%
^{-1}(z,q)=\mathcal{F}(z,q).$ Using such values, we can define respective 
\textsf{permutation} d-operators of scalars in star products in the form%
\begin{eqnarray*}
z\star _{N}q &=&\cdot \lbrack \mathcal{F}_{N}^{-1}(z,q)]=\cdot \lbrack 
\mathcal{F}_{N}(q,z)]=\cdot \lbrack \mathcal{F}_{N}^{-1}\underbrace{\mathcal{%
F}_{N}\mathcal{F}_{N}}_{:=\overline{\mathcal{R}}_{N}}(q,z)]:=\overline{%
\mathcal{R}}_{N}(q)\star _{N}\overline{\mathcal{R}}_{N}(z); \\
z\star q &=&\cdot \lbrack \mathcal{F}^{-1}(z,q)]=\cdot \lbrack \mathcal{F}%
(q,z)]=\cdot \lbrack \mathcal{F}^{-1}\underbrace{\mathcal{FF}}_{:=\overline{%
\mathcal{R}}}(q,z)]:=\overline{\mathcal{R}}(q)\star \overline{\mathcal{R}}%
(z).
\end{eqnarray*}%
The notations in these formulas mean that respective $\overline{\mathcal{R}}$%
-matrix (permutation d-operator) acts fist on $q$ and then on $z,$ and after
that the corresponding star product is carried out. For instance, we have 
\begin{eqnarray*}
z\star _{N}q &=&zq+\frac{i\mathit{\ell }_{s}^{4}}{6\hbar }R^{ija}p_{a}(%
\mathbf{e}_{i}z)(\mathbf{e}_{j}q)+\ldots \mbox{ and } \\
\overline{\mathcal{R}}_{N}(q)\star _{N}\overline{\mathcal{R}}_{N}(z)
&=&q\star _{N}z-2\frac{i\mathit{\ell }_{s}^{4}}{6\hbar }R^{ija}p_{a}(\mathbf{%
e}_{i}q)\star _{N}(\mathbf{e}_{j}z)+\ldots \\
&&=qz+\frac{i\mathit{\ell }_{s}^{4}}{6\hbar }R^{ija}p_{a}(\mathbf{e}_{i}q)(%
\mathbf{e}_{j}z)-2\frac{i\mathit{\ell }_{s}^{4}}{6\hbar }R^{ija}p_{a}(%
\mathbf{e}_{i}q)\star _{N}(\mathbf{e}_{j}z)+\ldots \\
&&=zq+\frac{i\mathit{\ell }_{s}^{4}}{6\hbar }R^{ija}p_{a}(\mathbf{e}_{i}z)(%
\mathbf{e}_{j}q)+\ldots ;
\end{eqnarray*}%
\begin{eqnarray*}
z\star q &=&zq+\frac{i\mathit{\ell }_{s}^{4}}{6\hbar }R^{ija}p_{a}(\mathbf{%
\partial }_{i}z)(\mathbf{\partial }_{j}q)+\ldots \mbox{ and } \\
\overline{\mathcal{R}}(q)\star \overline{\mathcal{R}}_{N}(z) &=&q\star z-2%
\frac{i\mathit{\ell }_{s}^{4}}{6\hbar }R^{ija}p_{a}(\mathbf{\partial }%
_{i}q)\star (\mathbf{\partial }_{j}z)+\ldots \\
&&=qz+\frac{i\mathit{\ell }_{s}^{4}}{6\hbar }R^{ija}p_{a}(\mathbf{\partial }%
_{i}q)(\mathbf{\partial }_{j}z)-2\frac{i\mathit{\ell }_{s}^{4}}{6\hbar }%
R^{ija}p_{a}(\mathbf{\partial }_{i}q)\star (\mathbf{\partial }_{j}z)+\ldots
\\
&&=zq+\frac{i\mathit{\ell }_{s}^{4}}{6\hbar }R^{ija}p_{a}(\mathbf{\partial }%
_{i}z)(\mathbf{\partial }_{j}q)+\ldots ,
\end{eqnarray*}%
respectively for $\star _{N}$ and $\star ,$ and $z,q\in C^{\infty }(\mathcal{%
M}).$

\paragraph{Nonholonomic associators: \newline
}

An associator $\phi $ is a geometric object (d-object, for nontrivial
nonholonomic configurations) which reorders the brackets in star products of
three functions. It can be defined for $\star _{N}$ and $\star .$ For
simplicity, we shall present hereafter only formulas for $\star _{N}$ if
that will not result in ambiguities. \textsf{The d-operator $\phi $ acts on
the functions first and then the star product is executed. Such a star
product is determined by respective (non) holonomic (co) bases.} By
definition,%
\begin{equation}
\phi :(\ z\star _{N}\ q)\star _{N}\ v=\ _{\phi }z\star _{N}(\ _{\phi }q\star
_{N}\ _{\phi }v):=z\star _{N}\ _{\phi }(q\star _{N}\ v).  \label{threeprod}
\end{equation}%
Our system of notations is different from that in \cite{blumenhagen16} (for
instance, in that work it is used $q^{\phi }$ instead of $\ _{\phi }q$; we
shall put equivalently an up abstract label, like $\ ^{\phi }q$, if a low
one will be not convenient in certain formulas) because we consider
N-adapted constructions and follow a convention to use left up and low
labels for the symbols.

Similarly, we can introduce the \textsf{inverse associator} $\overline{\phi}$
shifting brackets to the left,%
\begin{equation*}
\overline{\phi }:\ z\star _{N}\ (q\star _{N}\ v)=(\ _{\overline{\phi }%
}z\star _{N}\ _{\overline{\phi }}q)\star _{N}\ _{\overline{\phi }}v:=\ _{%
\overline{\phi }}(z\star _{N}q)\star _{N}\ v.
\end{equation*}

The d-operators $\phi $ and $\overline{\phi }$ are central and commute
respectively with $\mathcal{F}$ and $\mathcal{R}.$ Explicit N-adapted and/or
coordinate calculus result in expressions%
\begin{equation}
\phi (z,q,k)=\exp \left( \frac{i\mathit{\ell }_{s}^{4}}{6\hbar }R^{ijk}(%
\mathcal{L}_{\mathbf{e}_{i}}z)(\mathcal{L}_{\mathbf{e}_{j}}q)(\mathcal{L}_{%
\mathbf{e}_{k}}v)\right) (\ z\star _{N}\ q\star _{N}\ v),  \label{assoc1}
\end{equation}%
where indices $i,j,k$ are taken from splitting $a=n+i,b=n+j,c=n+k.$ Using
N-adapted frames (\ref{nadapd}), when h-components $\ ^{\shortmid }\mathbf{e}%
_{i}$ are identified with $\mathbf{e}_{i},$ we express (\ref{threeprod}) as 
\begin{equation*}
(\ z\star _{N}\ q)\star _{N}\ k=\ z\star _{N}\ (q\star _{N}\
k)=\sum_{r=1}^{\infty }\frac{1}{r!}(\frac{\mathit{\ell }_{s}^{4}}{6}%
)^{s}R^{i^{1}j^{1}k^{1}}\ldots R^{i^{r}j^{r}k^{r}}(\mathbf{e}_{i^{1}}\ldots 
\mathbf{e}_{i^{r}}z)\star _{N}\left( (\mathbf{e}_{j^{1}}\ldots \mathbf{e}%
_{j^{r}}q)\star _{N}(\mathbf{e}_{k^{1}}\ldots \mathbf{e}_{k^{r}}v)\right) .
\end{equation*}

Similar formulas hold true for the inverse associator $\overline{\phi }$
following the rule that it switches the sign in the exponent and kipping in
mind that $R^{ija}$ is antisymmetric. This way we conclude that a
permutation of the arguments invests the twists in N-adapted or coordinate
frame forms, when $\phi (z,q,k)=\overline{\phi }(z,k,q).$ Using the $%
\mathcal{R}$-matrix, we can invert the associator, 
\begin{equation}
\ _{\phi }z\star _{N}(\ _{\phi }q\star _{N}\ _{\phi }k)=\ _{\overline{\phi }%
}z\star _{N}\left( \overline{\mathcal{R}}_{N}(\ _{\overline{\phi }}k)\star
_{N}\overline{\mathcal{R}}_{N}(\ _{\overline{\phi }}q)\right) .
\label{auxf1}
\end{equation}%
We conclude that the definition of nonassociative star product \cite%
{blumenhagen16} preserves its properties for N-adapting in the sense that it
admits in all cases two important d-operators: \textsf{the $\mathcal{R}$%
-matrix capture the effect of noncommutativity and the associator $\phi $
characterizes nonassociativity. Nonholonomic distributions allow such a
diadic decomposition of the coefficients of fundamental geometric objects
and physically important PDEs when it is possible to decouple and integrate
such systems in general forms,} for nonassociative gravity models (such
properties are proven in our partner works).

\paragraph{N-adapted nonassociative tri-products: \newline
}

The nonassociative star products studied in this work are different form the
standard one on the Moyal-Weyl plane because in our case there are involved
momentum/ velocity type coordinates. We elaborated on such theories in the
so-called Finsler extensions of string theory \cite{vacaru96a,vacaru96b} and
by applying generalized Finsler-Lagrange-Hamilton methods in heterotic
string gravity and modifications \cite%
{vacaru03,vacaru16,vacaru18,bubuianu18a}. In \cite{aschieri15}, it was
studied the relation between the star and tri-products in relation to
configuration spaces and conformal field theories. Using the formalism of
N--connections and adapted star products (\ref{starpn}), we can define
so-called \textsf{tri-products for functions on configuration spaces} via 
\begin{eqnarray*}
q_{[1]}\bigtriangleup q_{[2]}\bigtriangleup \ldots \bigtriangleup q_{[r]}
&=&q_{[1]}\star _{N}\left( q_{[2]}\left( \star _{N}\left( \ldots
q_{[r-1]}\star _{N}q_{[r]}\right) \ldots \right) \right) _{\mid p_{8}=0} \\
&=&\cdot \left[ \exp \left( -\frac{\mathit{\ell }_{s}^{4}}{12}\sum_{1\leq 
\check{a}<\check{b}<\check{c}\leq \lbrack r]}R^{ijk}(\mathbf{e}_{i}^{\check{a%
}}\otimes \mathbf{e}_{j}^{\check{b}}\otimes \mathbf{e}_{k}^{\check{c}%
})\right) \left( q_{[1]}\otimes q_{[2]}\otimes \ldots \otimes q_{[r]}\right) %
\right]
\end{eqnarray*}%
for integer labeling values $[r],\check{a},\check{b},\check{c},$ where $%
p_{8} $ is energy type coordinate in momentum space. So, we first evaluate
the nested star product and then restrict the values to the so-called $%
p_{8}=0$ leaf. For instance, this implies that $q_{[1]}\bigtriangleup
q_{[2]}=q_{[1]}q_{[2]},$ when tri-products have the peculiar property for
all type of N-adapted partial derivatives all R-flux dependent corrections
become total derivatives. We obtain such simplifications:%
\begin{equation*}
\int d^{4}xq_{[1]}\bigtriangleup q_{[2]}\bigtriangleup \ldots \bigtriangleup
q_{[r]}=\int d^{4}xq_{[1]}q_{[2]}\ldots q_{[r]}.
\end{equation*}%
All formulas with nonassociative tri-products involving nonholonomic
structures in coordinate frames transform into similar constructions with
membrane corrections of deviations from $p_{8}=0,$ see \cite%
{blumenhagen13,aschieri15}.

\subsubsection{Nonholonomic star scalars and the Leibniz rule}

The Leibniz rule for star product \cite{mylonas12,mylonas13}, see also \cite%
{blumenhagen16}, can be considered for nonholonomic manifolds and bundle
spaces.

We define a star scalar $z$ is a d-object that transforms under star
diffemorphism N-adapted forms $\xi =\xi ^{i}\star _{N}\mathbf{e}_{i}$ (here
we consider only h-components and total space constructions and v-components
will be considered below) with the star Lie N-adapted derivative, $\delta
_{\xi }z=\mathcal{L}_{\xi }^{\star }z:=\xi ^{i}\star _{N}\mathbf{e}_{i}z$.
Than we demand that the star product of two star scalars should be also a
star scalar when 
\begin{equation*}
\delta _{\xi }(q\star _{N}z)=\mathcal{L}_{\xi }^{\star }(q\star _{N}z)=\xi
^{i}\star _{N}\mathbf{e}_{i}(q\star _{N}z)=\xi ^{i}\star _{N}(\mathbf{e}%
_{i}q\star _{N}z)+\xi ^{i}\star _{N}(q\star _{N}\mathbf{e}_{i}z).
\end{equation*}%
Using the $\mathcal{R}$-matrix and the associator $\phi ,$ we formulate a
generalized Leibnitz rule, 
\begin{equation*}
\xi ^{i}\star _{N}(q\star _{N}\mathbf{e}_{i}z)=(\xi ^{i}\star _{N}q)\star
_{N}\mathbf{e}_{i}z=\ _{\overline{\phi }}\left( \overline{\mathcal{R}}%
_{N}(q)\star _{N}\overline{\mathcal{R}}_{N}(\xi ^{i})\right) \star _{N}\ _{%
\overline{\phi }}(\mathbf{e}_{i}z)=\overline{\mathcal{R}}_{N}(q)\star
_{N}{}_{\phi ^{2}}(\overline{\mathcal{R}}_{N}(\xi ^{i})\star _{N}\ (\mathbf{e%
}_{i}z)).
\end{equation*}%
This permutation formula inverts the arguments as an associator similarly to
(\ref{auxf1}), 
\begin{equation}
\mathcal{L}_{\xi }^{\star }(q\star _{N}z)=\mathcal{L}_{\ _{\overline{\phi }%
}\xi }^{\star }(\ _{\overline{\phi }}q)\star _{N}(\ _{\overline{\phi }}z)+%
\overline{\mathcal{R}}_{N}(\ _{\phi ^{2}}q)\star _{N}\mathcal{L}_{\overline{%
\mathcal{R}}_{N}(\ _{\phi ^{2}}\xi )}^{\star }\ _{\phi ^{2}}z.  \label{leibr}
\end{equation}

Now, we can develop the theory in N-adapted form for nonassociative star
d-tensors.

\paragraph{Star--commutators of d-vectors and d-covectors: \newline
}

We define a star covector in N-adapted form $\omega _{i}=\mathbf{e}_{i}q$
with $\mathbf{e}_{i}$ (\ref{nadap}) (in similar forms we can consider (\ref%
{nadapd})). This is the h-part of a star distinguished 1-form which is
subjected to such transformation laws%
\begin{equation*}
\delta _{\xi }(\mathbf{e}_{i}q)=\mathbf{e}_{i}(\delta _{\xi }q)=\xi
^{j}\star _{N}(\mathbf{e}_{j}\mathbf{e}_{i}q)+(\mathbf{e}_{i}\xi ^{j})\star
_{N}(\mathbf{e}_{j}q)\mbox{ and }\delta _{\xi }\omega _{i}=\mathcal{L}_{\xi
}^{\star }\omega _{i}=\xi ^{j}\star _{N}(\mathbf{e}_{j}\omega _{i})+(\mathbf{%
e}_{i}\xi ^{j})\star _{N}\omega _{j}.
\end{equation*}%
In a compatible form with above formulas for covectors, a star d-vector $%
v^{i}$ is defined using formulas%
\begin{equation*}
\delta _{\xi }v^{i}=\mathcal{L}_{\xi }^{\star }v^{i}=\xi ^{j}\star _{N}(%
\mathbf{e}_{j}v^{i})-\overline{\mathcal{R}}_{N}(v^{j})\star _{N}\overline{%
\mathcal{R}}_{N}(\mathbf{e}_{j}\xi ^{i}),
\end{equation*}%
which guarantees that $z=v^{i}\star _{N}w_{i}$ transforms as a star scalar.

By definition, the star commutator of star d-vectors, $[,]_{\star
_{N}}:=[,]\circ $ $\mathcal{F}_{N}^{-1},$ which is equal to the Lie
derivative in N-adapted from, is constructed to be manifestly antisymmetric
and satisfy the horizontal, h, conditions%
\begin{equation}
\lbrack \ ^{h}v,\ ^{h}\omega ]_{\star _{N}}=v^{j}\star _{N}(\mathbf{e}%
_{j}\omega ^{i})-\overline{\mathcal{R}}_{N}(\omega ^{j})\star _{N}\overline{%
\mathcal{R}}_{N}(\mathbf{e}_{j}v^{i})=-[\overline{\mathcal{R}}_{N}(\ ^{h}v),%
\overline{\mathcal{R}}_{N}(\ ^{h}\omega )]_{\star _{N}}.  \label{hstcom}
\end{equation}

Similar transformation laws hold for $\star $ when $(\star _{N},\mathbf{e}%
_{i},\overline{\mathcal{R}}_{N})\leftrightarrow (\star ,\mathbf{\partial }%
_{i},\overline{\mathcal{R}}).$

\paragraph{Star d-tensors products: \newline
}

Let us explain how to construct star d-tensor products as generalizations of
definitions for d-vectors and d-covectors. To motivate the formulas we
consider a star tensor product of h-vectors and, respectively, a star
d-tensor,%
\begin{equation*}
\ ^{h}v\otimes _{\star _{N}}\ ^{h}z=\{v^{i}\star _{N}z^{j}\}\mbox{ and }\
^{h}b=\{b^{ij}\}.
\end{equation*}%
Extending the Leibniz rule (\ref{leibr}) to d-tensor products, we write down%
\begin{eqnarray*}
\mathcal{L}_{\xi }^{\star }(\ ^{h}v\otimes _{\star _{N}}\ ^{h}z) &=&\mathcal{%
L}_{\ _{\overline{\phi }}\xi }^{\star }(\ _{\overline{\phi }}^{h}v)\otimes
_{\star _{N}}(\ _{\overline{\phi }}z)+\overline{\mathcal{R}}_{N}(\ _{\phi
^{2}}^{h}v)\otimes _{\star _{N}}\left( \mathcal{L}_{\overline{\mathcal{R}}%
_{N}(\ _{\overline{\phi }}\xi )}^{\star }\ _{\phi ^{2}}^{h}z\right) \\
\mbox{
and }\mathcal{L}_{\xi }^{\star }b^{ij} &=&\xi ^{k}\star _{N}(\mathbf{e}%
_{k}b^{ij})-(\mathbf{e}_{k}\xi ^{i})\star _{N}b^{kj}-(\mathbf{e}_{k}\xi
^{j})\star _{N}b^{ik}.
\end{eqnarray*}%
In a similar form, we can consider higher order h-components of d-tensors
and redefinitions for $\star $ products.

\paragraph{N-adapted composition, closure of commutators and star Taylor
expansion: \newline
}

To eliminate breaking ambiguities with actions of two Lie derivatives on
d-objects, is convenient to introduce the commutator of such derivatives
with the closure property 
\begin{equation*}
\lbrack \mathcal{L}_{\xi }^{\star },\mathcal{L}_{\eta }^{\star }]_{\star
_{N}}\ ^{h}v=\mathcal{L}_{[\xi ,\eta ]_{\star _{N}}}^{\star }\ ^{h}v.
\end{equation*}%
This condition is satisfied if by definition 
\begin{equation*}
\lbrack \mathcal{L}_{\xi }^{\star },\mathcal{L}_{\eta }^{\star }]_{\star
_{N}}\ ^{h}v:=\mathcal{L}_{\ _{\phi }\xi }^{\star }(\mathcal{L}_{\ _{\phi
}\eta }^{\star }(\ _{\phi }^{h}v))-\mathcal{L}_{\overline{\mathcal{R}}_{N}(\
_{\phi }\eta )}^{\star }(\mathcal{L}_{\overline{\mathcal{R}}_{N}(\ _{\phi
}\xi )}^{\star }(\ _{\phi }^{h}v)).
\end{equation*}%
Such d-operators are similar to those for the Hopf algebras and can be
extended for d-algebras, see discussion in section 2.3 of \cite%
{blumenhagen16}. The composition $\bullet $ of operators $O,A,B$ and
commutators in N-adapted or local coordinate basis forms satisfy the rules%
\begin{equation}
(O\bullet O^{\prime })(\ ^{h}v)=O(O^{\prime }(\ ^{h}v))\mbox{ and }\lbrack
A,B]=A\bullet B-\overline{\mathcal{R}}(B)\bullet \overline{\mathcal{R}}(A).
\label{compcom}
\end{equation}

Finally, we note that every function $q(x)\in C^{\infty }(V)$ and/or
respective extensions on (co) tangent bundles are star scalars as it is
discussed in section 2.4 of \cite{blumenhagen16}. Nonholonomic structures on
such manifolds/ bundles do not modify properties of star Taylor expansions
like 
\begin{equation*}
q_{\star }(x)=\sum_{s_{1},s_{2},...,s_{k}=0}^{\infty
}q_{s_{1},s_{2},...,s_{k}}x_{1}^{s_{1}}\star (x_{2}^{s_{2}}\star (\ldots
\star x_{k}^{s_{k}})\ldots ),
\end{equation*}%
which are nonassociative analogs of commutative spacetime Taylor series 
\begin{equation*}
q(x)=\sum_{s_{1},s_{2},...,s_{k}=0}^{\infty
}q_{s_{1},s_{2},...,s_{k}}x_{1}^{s_{1}}x_{2}^{s_{2}}\ldots
x_{k}^{s_{k}}\ldots ,
\end{equation*}%
which holds for $x_{i}\star x_{i}=x_{i}x_{i},$ with linear functions $%
l(x)=x_{i}$ and a subset of those star scalars for which the star
multiplications in expansions act trivially.

\subsection{Nonassociativity, N-connections, and quasi-Hopf algebras}

\label{ssqhda} In \cite{aschieri17}, the general covariance under quasi-Hopf
algebra of deformed diffeomorphisms is implemented in order to formulate a
version of vacuum nonassociative Einstein equations. Authors of that paper
elaborated both an abstract and coordinate base formalism which is not
suitable for a general decoupling of nonassociative and/or (non) commutative
gravitational equations. In our approach, we use nonholonomic manifolds and
(co) tangent bundles, and their nonassociative generalizations, which allows
us to extend the AFCDM for constructing exact and parametric solutions of
nonassociative geometric flow and gravitational field (Ricci soliton)
equations. In N-adapted form, this involves distinguished algebras,
d-algebras, for a conventional nonholonomic h- and (c)v-splitting, and
respective quasi-Hopf d-algebras.

Let us consider the universal enveloping Hopf algebra $U\emph{Vec}(\mathcal{M%
})$ of the Lie algebra of vector fields $\emph{Vec}(\mathcal{M})$ on $%
\mathcal{M}.$ For non-trivial N-connection structures with h- and
v-splitting of phase space $\mathcal{M}=T_{\shortparallel }^{\ast }\mathbf{V,%
}$ we elaborate on a nonholonomic algebraic and geometric formalism with
d-vectors and d-tensors. A nonholonomic splitting $\emph{Vec}(\mathcal{M})=h%
\emph{Vec}(\mathcal{M})\oplus v\emph{Vec}(\mathcal{M})$ follows from the
definition of N-connection structure (\ref{ncon}). This defines a Hopf
d-algebra $U\emph{Vec}(\mathcal{M},N).$ We note that our results do not
depend on the signature of the spacetime metric but to study nonassociative
modifications of GR the signature of the basic spacetime manifolds must be
stated to be Lorentzian. The N-adapted cochain twist element $\mathcal{F}%
_{N} $ (\ref{starpn}) can be written in the form 
\begin{equation}
\ ^{\shortparallel }\mathcal{F}_{N}=\ ^{\shortparallel }\mathcal{\check{F}}%
_{N}\ ^{\shortmid }\mathcal{\check{F}}_{RN}=\ ^{\shortmid }\mathcal{\check{F}%
}_{RN}\ ^{\shortparallel }\mathcal{\check{F}}_{N},  \label{twisthopf}
\end{equation}%
for a nonholonomic Hopf 2-cocycle (determining a nonholonomic Moyal-Weyl
deformation of the phase space)%
\begin{equation*}
\ ^{\shortparallel }\mathcal{\check{F}}_{N}=\exp \left[ \left( \
^{\shortparallel }\mathbf{e}_{i}\otimes \ ^{\shortparallel }e^{n+i}-\
^{\shortparallel }e^{n+i}\otimes \ ^{\shortparallel }\mathbf{e}_{i}\right) %
\right]
\end{equation*}%
and a nonholonomic 2-cocycle for the so-called R-flux%
\begin{equation*}
\ ^{\shortmid }\mathcal{\check{F}}_{RN}=\exp \left[ \frac{i\kappa }{2}%
R^{jka}\left( p_{a}\ ^{\shortmid }\mathbf{e}_{k}\otimes \ ^{\shortmid }%
\mathbf{e}_{j}-\ ^{\shortmid }\mathbf{e}_{j}\otimes p_{a}\ ^{\shortmid }%
\mathbf{e}_{k}\right) ,\right]
\end{equation*}%
where $\kappa :=\mathit{\ell }_{s}^{3}/6\hbar $ and $\hbar $ are treated as
independent small deformation parameters (in some works, one consider as
parameters $\mathit{\ell }_{s}$ and $\hbar $). Such twist operators can be
parameterized in terms of certain elements $\mathfrak{f}^{\alpha },\mathfrak{%
f}_{\beta }\in $ $U\emph{Vec}(\mathcal{M})$ when 
\begin{eqnarray*}
\ ^{\shortparallel }\mathcal{\check{F}}_{N}&:=&\mathfrak{f}^{\alpha }\otimes 
\mathfrak{f}_{\alpha }=1\otimes 1+\mathit{O}(\hbar ,\mathit{\ell }_{s}^{3}),%
\mbox{ with inverse twist } \ ^{\shortparallel }\mathcal{\check{F}}%
_{N}^{-1}:=\overline{\mathfrak{f}}^{\alpha }\otimes \overline{\mathfrak{f}}%
_{\alpha }; \\
\ ^{\shortparallel }\mathcal{\check{F}}_{RN}&:= &\mathfrak{f}_{RN}^{\alpha
}\otimes \mathfrak{f}_{\alpha }^{RN}=1\otimes 1+\mathit{O}(\kappa ),
\end{eqnarray*}%
where summation on low-up repeating indices is understood.

A Hopf d-algebra $U\emph{Vec}(\mathcal{M},N)$ is determined by such
N-adapted structures: a) a co-product $\blacktriangle $ defined as $%
\blacktriangle (1)=1\otimes 1,\blacktriangle (\ ^{\shortparallel }\mathbf{e}%
_{\alpha })=1\otimes \ ^{\shortparallel }\mathbf{e}_{\alpha }+\
^{\shortparallel }\mathbf{e}_{\alpha }\otimes 1;$\ b) a co-unit $\epsilon $
defined as $\epsilon (1)=1,\epsilon (\ ^{\shortparallel }\mathbf{e}%
_{\alpha})=0;$ and c) and an antipode $\mathbf{S}$ defined as $\mathbf{S}%
(1)=1,\mathbf{S}(\ ^{\shortparallel }\mathbf{e}_{\alpha })=-\
^{\shortparallel }\mathbf{e}_{\alpha }$; such $\blacktriangle $ and $%
\epsilon $ d-operators are extended to the total space $U\emph{Vec}(\mathcal{%
M},N)$ as d-algebra homomorphisms and $\mathbf{S}$ extended as an d-algebra
anti-homomorphism (linear, anti-multiplicative and N-adapted). The
N-connection splitting states a h- and v-decomposition $\blacktriangle
=(h\blacktriangle ,v\blacktriangle ).$

A quasi-Hopf d-algebra $U\emph{Vec}^{\mathcal{F}}(\mathcal{M},N)$ is
generating as an extension in N-adapted form with power series in $\hbar $
and $\kappa $ of a Hopf d-algebra $U\emph{Vec}(\mathcal{M},N)$ using a twist 
$\ ^{\shortparallel }\mathcal{F}_{N}$ (\ref{twisthopf}). The algebraic Hopf
d-structure is preserved for respective $\mathcal{F}$-extended d-objects:
co-product $\blacktriangle _{\mathcal{F}}=\mathcal{F}_{N}\blacktriangle 
\mathcal{F}_{N}^{-1};$ quasi-antipode $\mathbf{S}_{\mathcal{F}}=\mathbf{S}$;
and co-unit $\epsilon _{\mathcal{F}}=\epsilon .$ Explicitly, there are
satisfied such important properties:%
\begin{eqnarray*}
\blacktriangle _{\mathcal{F}}(\ ^{\shortparallel }\mathbf{e}_{i})
&=&1\otimes \ ^{\shortparallel }\mathbf{e}_{i}+\ ^{\shortparallel }\mathbf{e}%
_{i}\otimes 1,\blacktriangle _{\mathcal{F}}(\ ^{\shortparallel
}e^{a})=1\otimes \ ^{\shortparallel }e^{a}+\ ^{\shortparallel }e^{a}\otimes
1+i\kappa R^{jka}\ ^{\shortparallel }\mathbf{e}_{j}\otimes \
^{\shortparallel }\mathbf{e}_{k}\mbox{ and } \\
\mathfrak{f}^{\alpha }\ \mathbf{S(}\mathfrak{f}_{\alpha }) &=&\mathfrak{f}%
_{RN}^{\alpha }\ \mathbf{S(}\mathfrak{f}_{\alpha }^{RN})=\overline{\mathfrak{%
f}}^{\alpha }\ \mathbf{S(}\overline{\mathfrak{f}}_{\alpha })=\overline{%
\mathfrak{f}}_{RN}^{\alpha }\ \mathbf{S(}\overline{\mathfrak{f}}%
_{\alpha}^{RN})=1.
\end{eqnarray*}%
As in holonomic bases, the N-adapted variant of twist $\ ^{\shortparallel }%
\mathcal{F}_{N}$ does not fulfill the 2-cocycle condition and we have to
consider the associator (\ref{threeprod}). For Hopf d-algebraic structures,
we use such an associator $\Phi $ (defining a Hopf 3-cocycle; see also
formula (\ref{assoc1})) and its inverse associator $\Phi ^{-1},$%
\begin{eqnarray*}
\Phi _{N} &=&\exp \left( \frac{\mathit{\ell }_{s}^{3}}{6}R^{n+j\ n+k\ n+i}\
^{\shortparallel }\mathbf{e}_{j}\otimes \ ^{\shortparallel }\mathbf{e}%
_{k}\otimes \ ^{\shortparallel }\mathbf{e}_{i}\right) =\ _{1}\phi \otimes \
_{2}\phi \otimes \ _{3}\phi =1\otimes 1\otimes 1+O(\mathit{\ell }_{s}^{3}),
\\
&& \mbox{ and }\Phi _{N}^{-1} = \ _{1}\overline{\phi }\otimes \ _{2}%
\overline{\phi }\otimes \ _{3}\overline{\phi }.
\end{eqnarray*}%
The co-product $\blacktriangle _{\mathcal{F}}$ is not co-associative because
the 2-cocycle condition is not hold true, when 
\begin{equation*}
\Phi _{N}(\ ^{\shortparallel }\mathcal{F}_{N}\otimes 1)(\blacktriangle
id\otimes 1)\ ^{\shortparallel }\mathcal{F}_{N}=(1\otimes \ ^{\shortparallel
}\mathcal{F}_{N})(\blacktriangle id\otimes 1)\ ^{\shortparallel }\mathcal{F}%
_{N}
\end{equation*}%
results in a quasi-associativity condition 
\begin{equation*}
\Phi _{N}(\blacktriangle _{\mathcal{F}}id\otimes 1)\blacktriangle _{\mathcal{%
F}}(\xi )=(id\otimes \blacktriangle _{\mathcal{F}})\blacktriangle _{\mathcal{%
F}}(\xi )\Phi _{N},\forall \xi \in U\emph{Vec}(\mathcal{M},N).
\end{equation*}

We conclude that a sextuple $\left( U\emph{Vec}(\mathcal{M}%
,N),\bullet,\blacktriangle _{\mathcal{F}},\Phi _{N},\mathbf{S,}\epsilon
\right) $ defines on the d-vector space $U\emph{Vec}(\mathcal{M},N),$ i.e.
on $T\mathcal{M}=TT_{\shortparallel }^{\ast }\mathbf{V,}$ a structure of
quasi-Hopf d-algebra $U\emph{Vec}^{\mathcal{F}}(\mathcal{M},N).$ Such
geometric d-objects and nonholonomic structures can be defined, for
instance, on $TT_{\shortparallel }\mathbf{V}$ which can be used for
constructing models of nonassociative Finsler-Lagrange geometry.

In above formulas, we use boldface d-operators in order to emphasize that we
work with N-adapted algebraic and geometric structures. For $\
^{\shortparallel }\mathbf{e}_{\alpha }\rightarrow \ ^{\shortparallel }%
\mathbf{\partial }_{\alpha }$ and respective $\blacktriangle _{\mathcal{F}%
}\rightarrow \triangle _{\mathcal{F}},\Phi _{N}\rightarrow \Phi ,\mathbf{S}%
\rightarrow S$ etc., we obtain the definition of standard Hopf algebra $U%
\emph{Vec}(\mathcal{M})$ and coordinate base constructions as in \cite%
{drinf,aschieri17,mylonas13}. The nonholonomic geometric formalism
reproduces, in principle, all results from those papers and \cite%
{blumenhagen16} if we follow the Convention 2 and re-define the symbols and
indices of geometric objects. In this article, we do not repeat such
abstract algebraic and geometric calculations but provide certain important
N-adapted coefficient formulas in order to familiarize readers with a
nonholonomic (non) associtative/ commutative geometric background for the
AFCDM which is used in our partner works.

\subsection{The star product on nonholonomic phase spaces}

\label{ssectstarphase}The N-adapted star product (\ref{starpn}) can be
naturally considered on total spaces of nonholonomic (co) tangent Lorentz
bundles. On $\mathcal{M}$, we can use local coordinates $\
^{\shortparallel}u=(x,\ ^{\shortparallel }p)$ (\ref{loccordph}), for a phase
space with complex cofibers on $T_{\shortparallel }^{\ast }\mathbf{V)}$,
which are associated to $\ ^{\shortmid }u=(x,\ ^{\shortmid }p)$ for real
cofibers in $T^{\ast }\mathbf{V}$.

\subsubsection{Momentum like nonholonomic variables on phase spaces with
complex cofibers}

A vector on $T_{\shortparallel }^{\ast }\mathbf{V}$ can be written in a
d-vector in N-adapted form,%
\begin{eqnarray*}
\ ^{\shortparallel }\mathbf{A}(\ ^{\shortparallel }u) &=&(A(x,\
^{\shortparallel }p),\ ^{\shortparallel }A(x,\ ^{\shortparallel }p))=\
^{\shortparallel }\mathbf{A}^{\alpha }(\ ^{\shortparallel }u)\star _{N}\
^{\shortparallel }\mathbf{e}_{\alpha }=A^{i}(\ ^{\shortparallel }u)\star _{N}%
\mathbf{e}_{i}+\ ^{\shortparallel }A_{b}(\ ^{\shortparallel }u)\star
_{N}i\hbar \ \ ^{\shortmid }e^{b}, \\
&&\mbox{ for }\ ^{\shortparallel }A_{b}(\ ^{\shortparallel }u)\star
_{N}i\hbar \ \ ^{\shortmid }e^{b}=\ ^{\shortparallel }A_{b}(\
^{\shortparallel }u)\star _{N}\ ^{\shortparallel }e^{b}; \\
&=&\ ^{\shortparallel }A^{\alpha }(\ ^{\shortparallel }u)\star \
^{\shortparallel }\partial _{\alpha }=A^{i}(\ ^{\shortparallel }u)\star
\partial _{i}+\ ^{\shortparallel }A_{b}(\ ^{\shortparallel }u)\star i\hbar \
\ ^{\shortmid }e^{b}, \\
&&\mbox{ for }\ ^{\shortparallel }A_{b}(\ ^{\shortparallel }u)\star i\hbar \
\ ^{\shortmid }e^{b}=\ ^{\shortparallel }A_{b}(\ ^{\shortparallel }u)\star \
^{\shortparallel }e^{b},
\end{eqnarray*}%
where nonholonomic bases are similar, respectively to (\ref{nadapd}) and
coordinate bases.

The N-adapted star products (\ref{starpn}) can be generalized to $%
T_{\shortparallel }^{\ast }\mathbf{V}$ in these respective forms: 
\begin{equation}
q\star _{N}z=\cdot \left[ \exp \frac{1}{2}\left( \left( P_{i}\otimes _{N}\
^{\shortparallel }P^{i}-\ ^{\shortparallel }P^{i}\otimes _{N}P_{i}\right)
+\left( \ ^{\shortparallel }M^{i}\otimes _{N}P_{i}-P_{i}\otimes _{N}\
^{\shortparallel }M^{i}\right) \right) q\otimes _{N}z\right] ,
\label{fstarp}
\end{equation}%
\begin{eqnarray}
\mbox{ where }\ ^{\shortparallel }\mathbf{P}^{\alpha } &=&(P_{i}=\mathcal{L}%
_{\mathbf{e}_{i}}^{\star },\ ^{\shortparallel }P^{a}=i\hbar \mathcal{L}_{\
^{\shortmid }e^{a}}^{\star })\mbox{ and }  \label{dmomentoper} \\
\ ^{\shortparallel }\mathbf{M}^{\gamma } &=&-\ ^{\shortparallel }F_{\ \alpha
\beta }^{\gamma }\ ^{\shortparallel }u^{\alpha }\ ^{\shortparallel }\mathbf{P%
}^{\beta }=(M_{i}=0,\ ^{\shortparallel }M^{i}=\frac{i\mathit{\ell }_{s}^{4}}{%
6\hbar }\mathcal{L}_{R^{ijb}p_{b}\ \mathbf{e}_{j}}^{\star }).  \notag
\end{eqnarray}%
In N-adapted form, we have such constraints on coefficients $\
^{\shortparallel }F_{\ \alpha \beta }^{\gamma }(\ ^{\shortparallel }u)$
which can be stated algebraically for any prescribed $R^{ija},p_{a}\ $and $\
^{\shortparallel }\mathbf{e}_{k}=\partial _{k}-\ ^{\shortparallel }N_{ak}(\
^{\shortparallel }u)\ ^{\shortparallel }\partial ^{a},$ see formulas (\ref%
{nadapdc}), in order to satisfy the conditions%
\begin{equation}
\ ^{\shortparallel }F_{\ i\alpha \beta }=0\mbox{ and any nontrivial }%
^{\shortparallel }F_{\ \alpha \beta }^{a}(\ ^{\shortparallel }u)\
^{\shortparallel }u^{\alpha }=-\frac{i\mathit{\ell }_{s}^{4}}{6\hbar }%
\mathcal{L}_{R^{ajk}p_{j}\ \mathbf{e}_{k}}^{\star }.  \label{fmomc}
\end{equation}

The nonassociative d-operators (\ref{dmomentoper}) implies a nontrivial
action on base 1-form $d\ ^{\shortmid }u^{\alpha }=(dx^{i},dp_{i})$ because
of Lie N-adapted derivatives. Considering a similar 1-form on $%
T_{\shortparallel }^{\ast }\mathbf{V,}$ when $d\ ^{\shortparallel }u^{\alpha
}=(dx^{i},(i\hbar )^{-1}dp_{i}),$ which is similar to $dX^{I}$ used in
formulas (3.5) in \ \cite{blumenhagen16}, and using $[\mathcal{L},d]=0,$ we
compute in coordinate bases%
\begin{equation*}
\ ^{\shortparallel }P^{\alpha }(d\ ^{\shortparallel }u^{\beta })=0%
\mbox{ and
}\ ^{\shortparallel }M^{\gamma }(d\ ^{\shortparallel }u^{\alpha })=\
^{\shortparallel }F_{\quad \beta }^{\gamma \alpha }\ d\ ^{\shortparallel
}u^{\beta }\mbox{ or }\ ^{\shortparallel }M^{a}(dx^{j})=-\frac{i\mathit{\ell 
}_{s}^{4}}{6\hbar }R^{ajc}p_{c}.
\end{equation*}%
With respect to N-elongated frames $\ ^{\shortparallel }\mathbf{e}_{\alpha}, 
$ we obtain 
\begin{equation}
\ ^{\shortparallel }\mathbf{P}^{\alpha }(\ ^{\shortparallel }\mathbf{e}%
^{\beta })=0\mbox{ and }\ ^{\shortparallel }\mathbf{M}^{\gamma }(\
^{\shortparallel }\mathbf{e}^{\alpha })=\ _{N}^{\shortparallel }F_{\quad
\beta }^{\gamma \alpha }\ \ ^{\shortparallel }\mathbf{e}^{\beta },
\label{dmoment1a}
\end{equation}%
see formulas \ (\ref{fmomc}). We can chose, for instance, such data when 
\begin{equation}
\ _{N}^{\shortparallel }F_{i\quad \beta }^{\ \alpha }=0,\ \
_{N}^{\shortparallel }F_{\quad kl}^{a}=0,\ ,\ _{N}^{\shortparallel }F_{\quad
ib}^{a}\ \delta _{8}^{b}=-\frac{i\mathit{\ell }_{s}^{4}}{6\hbar }%
R^{ajc}p_{c}.  \label{fmomc1}
\end{equation}%
Applying coordinate transforms adapted to a prescribed N-connection
structure, we can transform such values into some functionals $\
_{N}^{\shortparallel }F_{\quad \beta }^{\gamma \alpha }\
[R^{ajc},p_{c},^{\shortparallel }N_{ak}(\ ^{\shortparallel }u)].$

Acting on%
\begin{equation}
\delta _{\beta }^{\alpha }=\ ^{\shortparallel }\partial _{\beta }\star d\
^{\shortparallel }u^{\alpha }=\partial _{j}\star d\ x^{i}+\ ^{\shortparallel
}\partial ^{a}\star d\ ^{\shortparallel }p_{b}=\mathbf{e}_{j}\star _{N}%
\mathbf{e}^{i}+\ ^{\shortparallel }\mathbf{e}^{a}\star _{N}\
^{\shortparallel }\mathbf{e}_{b},  \label{dualcond}
\end{equation}%
where $\ ^{\shortparallel }\partial _{\beta }$ is considered as a coordinate
basis in $T_{\shortparallel }\mathbf{V}$ with complexified fiber
coordinates, we find%
\begin{eqnarray}
\ ^{\shortparallel }P^{\alpha }(\ ^{\shortparallel }\partial _{\beta }) &=&0%
\mbox{ and }\ ^{\shortparallel }M^{\gamma }(\ ^{\shortparallel }\partial
_{\alpha })=\ ^{\shortparallel }F_{\quad \alpha }^{\gamma \beta }\ \ \
^{\shortparallel }\partial _{\beta }\mbox{ or }\ ^{\shortparallel }M^{a}(\
^{\shortparallel }\partial ^{b})=\ ^{\shortparallel }M^{a}(i\hbar \
^{\shortmid }\partial ^{b})=-\frac{i\mathit{\ell }_{s}^{4}}{6}R^{abc}\ \
^{\shortparallel }\partial _{c};  \notag \\
\ ^{\shortparallel }\mathbf{P}^{\alpha }(\ \ ^{\shortparallel }\mathbf{e}%
_{\beta }) &=&0\mbox{ and }\ ^{\shortparallel }\mathbf{M}^{\gamma }(\
^{\shortparallel }\mathbf{e}_{\alpha })=\ _{N}^{\shortparallel }F_{\quad
\alpha }^{\gamma \beta }\ \ \ ^{\shortparallel }\mathbf{e}_{\beta }.
\label{dmoment1}
\end{eqnarray}%
Such formulas are similar to those for the relativistic Hamilton geometry on 
$T^{\ast }\mathbf{V,}$ see details in \cite{vacaru18,bubuianu18a}, where $\
^{\shortparallel }M^{a}$ defines a complex anholonomic structure determined
by even in coordinate frames on $R^{ijk}.$

There is a natural generalization of this formalism with nonholonomic star
products (\ref{starpn}) and (\ref{fstarp}) for Moyal-Wely spaces (see \cite%
{vacaru01,vacaru03}, Part III of \cite{vacaru05a} and \cite{aschieri10})
which can be also related to double field theory and appropriated curved
backgrounds \cite{blumenhagen10a,blumenhagen13}. Using a N-adapted basis $\
^{\shortparallel }\mathbf{e}_{\alpha }=(\ ^{\shortparallel }\mathbf{e}_{i},\
^{\shortparallel }e^{b})$ (\ref{nadapdc}), we define an antisymmetric
d-tensor%
\begin{equation}
\ ^{\shortparallel }\mathbf{e}_{ij}:=\ ^{\shortparallel }p_{n+i}\
^{\shortparallel }\mathbf{e}_{j}-\ ^{\shortparallel }p_{n+j}\
^{\shortparallel }\mathbf{e}_{i},  \label{antsmv}
\end{equation}
for some v-indices $a=n+i,$ $b=n+j$ and $c=n+k,$ defining a nonvanishing Lie
bracket, 
\begin{equation*}
\lbrack \ \ ^{\shortparallel }e^{n+k},\ ^{\shortparallel }\mathbf{e}%
_{ij}]=\delta _{\ i}^{k}\ ^{\shortparallel }\mathbf{e}_{j}-\delta _{\
j}^{k}\ ^{\shortparallel }\mathbf{e}_{i}.\ 
\end{equation*}%
A variant of cochain twist element $\ ^{\shortparallel }\mathcal{F}_{N}$ (%
\ref{twisthopf}) is defined by 
\begin{equation*}
\ ^{\shortparallel }\mathcal{F}_{N}^{c}=\exp \left[ -\frac{i\hbar }{2}\left(
\ ^{\shortparallel }\mathbf{e}_{j}\otimes _{N}\ ^{\shortparallel }e^{n+j}-\
^{\shortparallel }e^{n+j}\otimes _{N}\ ^{\shortparallel }\mathbf{e}%
_{j}\right) -\frac{i\kappa }{2}R^{jkn+i}\left( \ ^{\shortparallel }\mathbf{e}%
_{ki}\otimes _{N}\ ^{\shortparallel }\mathbf{e}_{j}-\ ^{\shortparallel }%
\mathbf{e}_{j}\otimes _{N}\ ^{\shortparallel }\mathbf{e}_{ki}\right) \right],
\end{equation*}%
which is an element of the Hopf d-algebra $U\mathfrak{iso}(2n)\subset U\emph{%
Vec}^{\mathcal{F}}(\mathcal{M},N).$

Such a Lie bracket relates in N-adapted form certain h- and v-components of $%
\ ^{\shortparallel }\mathbf{e}_{\alpha }$ and any collections of components
of a d-vector $\ ^{\shortparallel }\mathbf{e}_{\alpha }\rightarrow \
^{\shortparallel }\mathbf{X}_{\alpha }= (\ ^{\shortparallel }\mathbf{X}%
_{i},\ ^{\shortparallel }X^{b})$ and respective d-tensor $\ ^{\shortparallel
}\mathbf{e}_{ij}\rightarrow \ ^{\shortparallel }\mathbf{X}_{ij}:= \
^{\shortparallel }p_{n+i}\ ^{\shortparallel }\mathbf{X}_{j}-\
^{\shortparallel }p_{n+j}\ ^{\shortparallel }\mathbf{X}_{i}.$ This allows to
compute any $\ ^{\shortparallel }\mathcal{F}_{N}^{c}[\ ^{\shortparallel }%
\mathbf{X}_{\alpha }]$ and provides a nonassociative N-adapted deformation
of $\mathcal{M}.$ In the present paper, we shall work with N-adapted bases $%
\ ^{\shortparallel }\mathbf{e}_{\alpha }$ and cobases $\ ^{\shortparallel}%
\mathbf{e}^{\beta }$ and $\ ^{\shortparallel }\mathcal{F}_{N}$ (\ref%
{twisthopf}) if necessary in the form $\ ^{\shortparallel }\mathcal{F}%
_{N}^{c}$ for quasi-Hopf d-algebra configurations.

\subsubsection{Nonholonomic star commuting scalars and d-vectors}

Let us consider how to compute in N-adapted form the expression $q(\
^{\shortparallel }u)$ $\star\ ^{\shortparallel }\partial _{\alpha }$ given
in coordinate form. We note that for a function $q(\ ^{\shortmid}u)=q(x,\
^{\shortmid }p)=q(x,p),$ in contrast to $\partial _{i}q,$ the star
derivatives like $\partial _{i}\star q$ and $\ ^{\shortparallel }\mathbf{e}%
_{i}\star _{N}q$ act on $q$ encoding both the nonassociative and
nonholonomic structures.

\paragraph{The h-components in $T_{\shortparallel }^{\ast }\mathbf{V}$\ : 
\newline
}

We have 
\begin{equation*}
q\star \partial _{i}=q\cdot \partial _{i}\Longrightarrow q\star \partial
_{i}-\partial _{i}\star q=0,\mbox{ for }\ ^{\shortparallel }\partial
_{i}=\partial _{i}=\partial /\partial x^{i}.
\end{equation*}%
Considering $\ ^{\shortparallel }\mathbf{e}_{i}=\mathbf{e}_{i}=e_{\
i}^{\alpha ^{\prime }}(x,p)\partial _{\alpha ^{\prime }}=\
^{\shortparallel}e_{\ i}^{\alpha ^{\prime }}(x,p)\ ^{\shortparallel
}\partial _{\alpha ^{\prime }},$ see formula (\ref{nadapd}) for $%
p\rightarrow i\hbar p=\ ^{\shortparallel }p$, we treat any coefficient $\
^{\shortparallel }e_{\ i}^{\alpha ^{\prime }}(x,p)$ as a function similar to 
$q(x,p).$ In result, for any N-adapted construction%
\begin{equation*}
q\star _{N}\mathbf{e}_{i}=q\cdot \mathbf{e}_{i}\Longrightarrow q\star _{N}%
\mathbf{e}_{i}-\mathbf{e}_{i}\star _{N}q=0.
\end{equation*}

Additional terms arise when these h-formulas are extended on total phase
space $T_{\shortparallel }^{\ast }\mathbf{V}$ with star products of type (%
\ref{fstarp}). For an arbitrary $q(x,p),$ we compute 
\begin{eqnarray*}
q\star i\hbar \ ^{\shortmid }\partial ^{a} &=&q\cdot i\hbar \ ^{\shortmid
}\partial ^{a}-\frac{1}{2}P_{k}q\cdot \ ^{\shortparallel }M^{k}(i\hbar \
^{\shortmid }\partial ^{a})=q\cdot i\hbar \ ^{\shortmid }\partial ^{a}+\frac{%
i\mathit{\ell }_{s}^{4}}{12}R^{ajk}\ \partial _{j}q\star \partial _{k}, \\
&&\mbox{ i.e. }q\star i\hbar \ ^{\shortmid }\partial ^{a}-i\hbar \
^{\shortmid }\partial ^{a}\star q=\frac{\mathit{\ell }_{s}^{4}}{6}R^{ajk}\
\partial _{j}q\star \partial _{k}
\end{eqnarray*}%
For nonholonomic distributions, via frame transforms and N-elongating the
partial h-derivatives in the terms with $R^{...}$ and taking $\
^{\shortparallel }e^{a}= i\hbar $ $\ ^{\shortmid }e^{a}=i\hbar \
^{\shortmid}\partial ^{a},$ we can write 
\begin{equation*}
q\star _{N}i\hbar \ ^{\shortmid }e^{a}=q\cdot i\hbar \ ^{\shortmid }e^{a}+%
\frac{\mathit{\ell }_{s}^{4}}{12}R^{aij}\ \mathbf{e}_{j}q\star _{N}\mathbf{e}%
_{k}\mbox{
i.e. }q\star _{N}\ ^{\shortparallel }e^{a}-\ ^{\shortparallel }e^{a}\star
_{N}q=\frac{\mathit{\ell }_{s}^{4}}{6}R^{ajk}\ \mathbf{e}_{j}q\star _{N}%
\mathbf{e}_{k}.
\end{equation*}

In a more general context, we can consider such \textbf{principles of
N-adapted twist deformation}: Let $\mathcal{A}_{N}=(h\mathcal{A},v\mathcal{A}%
)$ be a d-algebra that carries a N-adapted representation of a Hopf
d-algebra $U\emph{Vec}(\mathcal{M},N)$ and a respective induced by a R-flux
twist deformation into a quasi-Hopf d-algebra $U\emph{Vec}^{\mathcal{F}}(%
\mathcal{M},N)$ as we defined in subsection \ref{ssqhda}. A d-vector $%
\vartheta =(h\vartheta ,v\vartheta )\in \emph{Vec}(\mathcal{M},N)$ acts on $%
\mathcal{A}$ $_{N}$ in such a form: 1) For every algebraic product $ab,$ we
have $\vartheta (ab)=u(a)b+au(b),$ i.e. we have a $U\emph{Vec}(\mathcal{M},N)
$-module d-algebra $\mathcal{A}_{N}.$ 2) Then, this multiplication is
deformed into a N-adapted star multiplication 
\begin{equation}
a\star _{N}b=\overline{\mathfrak{f}}^{\alpha }(a)\overline{\mathfrak{f}}%
_{\alpha }(b)=\overline{\mathfrak{f}}^{i}(a)\overline{\mathfrak{f}}_{i}(b)+\
^{\shortparallel }\overline{\mathfrak{f}}_{c}(a)\ ^{\shortparallel }%
\overline{\mathfrak{f}}^{c}(b),  \label{starmult}
\end{equation}%
which results in a noncommmutative and nonassociative d-algebra $\mathcal{A}%
_{N}^{\star }=(h\mathcal{A}^{\star },v\mathcal{A}^{\star }).$

$\mathcal{A}_{N}^{\star }$ carries a representation of the quasi-Hopf
d-algebra $U\emph{Vec}^{\mathcal{F}}(\mathcal{M},N)$ because for any element 
$\xi $ of this d-algebra we prove in N-adapted form that 
\begin{equation*}
\xi (a\star _{N}b)=\xi \left( \overline{\mathfrak{f}}^{\alpha }(a)\overline{%
\mathfrak{f}}_{\alpha }(b)\right) =\xi _{(1_{0})}\left( \overline{\mathfrak{f%
}}^{\alpha }(a))\xi _{(2_{0})}(\overline{\mathfrak{f}}_{\alpha }(b)\right) =%
\overline{\mathfrak{f}}^{\alpha }(\xi _{(1_{0})}(a))\xi _{(2_{0})}(\overline{%
\mathfrak{f}}_{\alpha }(b)).
\end{equation*}%
In these formulas, it is used a un-deformed N-adapted co-product $%
\blacktriangle (\xi )=\xi _{(1_{0})}\otimes _{N}\xi _{(2_{0})}$ when $%
\blacktriangle (\xi )\mathcal{F}_{N}^{-1}= \mathcal{F}_{N}^{-1}%
\blacktriangle _{\mathcal{F}}(\xi ).$

\paragraph{N-adapted star commuting scalars on total phase space $%
T_{\shortparallel }^{\ast }\mathbf{V}$: \newline
}

Above formulas can be written on total complexified Lorentz cobundles using (%
\ref{dmoment1a}) and (\ref{dmoment1}). They state that the star product acts
in the same way in coordinate frames and coframes using $\star _{N}$.
Respectively, we can consider 
\begin{equation}
q\star \ ^{\shortparallel }\partial _{\alpha }-\ ^{\shortparallel }\partial
_{\alpha }\star q=\ _{\partial }^{\shortparallel }F_{\alpha \quad }^{\quad
\gamma \beta }\ \ ^{\shortparallel }\partial _{\gamma }q\star \
^{\shortparallel }\partial _{\beta }\mbox{ and }q\star d\ ^{\shortparallel
}u^{\gamma }-d\ ^{\shortparallel }u^{\gamma }\star q=\ ^{\shortparallel
}F_{\quad \alpha }^{\gamma \beta }\ ^{\shortparallel }\partial _{\beta
}q\star d\ ^{\shortparallel }u^{\alpha }.  \label{aux02}
\end{equation}%
Here it should be noted that expressions of type $A^{\alpha }\star \
^{\shortparallel }\partial _{\alpha }$ and $\ ^{\shortparallel }\partial
_{\alpha }\star A^{\alpha }$ are not equivalent. The nontrivial coefficients
of two (in general, different but related) geometric objects $\ _{\partial
}^{\shortparallel }F_{\alpha \quad }^{\quad \gamma \beta }$ and $\
^{\shortparallel }F_{\quad \alpha }^{\gamma \beta }$ are determined by $%
R^{ajk}$. Our conventions are different from similar formulas (3.10) and
(3.11) in \cite{blumenhagen16} where there are used the values $\partial
_{I} $ and $\partial ^{I},$ but we accept the convention to use expressions
of type $A^{\alpha }\star \ ^{\shortparallel }\partial _{\alpha }$. This has
also consequences for the right and left multiplications of scalars and
vectors respective modifications of the Leibnitz rule. For instance, we
derive%
\begin{eqnarray}
q\star (\ ^{\shortparallel }A^{\alpha }\star \ ^{\shortparallel }\partial
_{\alpha }) &=&(q\star \ ^{\shortparallel }A^{\alpha })\star \
^{\shortparallel }\partial _{\alpha },  \label{aux03} \\
(\ ^{\shortparallel }A^{\alpha }\star \ ^{\shortparallel }\partial _{\alpha
})\star q &=&(\ ^{\shortparallel }A^{\alpha }\star q)\star \
^{\shortparallel }\partial _{\alpha }-\ _{\partial }^{\shortparallel
}F_{\alpha \quad }^{\quad \gamma \beta }\ \ ^{\shortparallel }\partial
_{\gamma }q\star \ ^{\shortparallel }\partial _{\beta },  \notag \\
\ ^{\shortparallel }\partial _{\alpha }(q\star z) &=&\ ^{\shortparallel
}\partial _{\alpha }q\star z+q\star \ ^{\shortparallel }\partial _{\alpha
}z-\ _{\partial }^{\shortparallel }F_{\alpha \quad }^{\quad \gamma \beta }\
\ ^{\shortparallel }\partial _{\gamma }q\star \ ^{\shortparallel }\partial
_{\beta }z.  \notag
\end{eqnarray}%
In explicit form for momentum d-operators (\ref{dmoment1a}) and (\ref%
{dmoment1}), we write%
\begin{equation}
\lbrack \ ^{\shortparallel }P^{\gamma },\ ^{\shortparallel }M^{\beta }]=\
^{\shortparallel }F_{\quad \alpha }^{\gamma \beta }\ ^{\shortparallel
}P^{\alpha }\mbox{ and }\ ^{\shortparallel }M^{\alpha }(q\star z)=\
^{\shortparallel }M^{\alpha }q\star z+q\star \ ^{\shortparallel }M^{\alpha
}z+\ _{M}^{\shortparallel }F_{\alpha \quad }^{\quad \gamma \beta }\ \
^{\shortparallel }\partial _{\gamma }q\star \ ^{\shortparallel }\partial
_{\beta }z,  \label{aux03a}
\end{equation}%
where the coefficients $\ ^{\shortparallel }F_{\quad \alpha }^{\gamma \beta
},\ _{\partial }^{\shortparallel }F_{\alpha \quad }^{\quad \gamma \beta },\
_{M}^{\shortparallel }F_{\alpha \quad }^{\quad \gamma \beta }$ are different
but determined by $R^{ajk}$ (we can prescribe values which are compatible to
analogous formulas (3.16) in \cite{blumenhagen16}).

In N-adapted forms, the formulas (\ref{aux02}) are written%
\begin{equation}
q\star _{N}\ ^{\shortparallel }\mathbf{e}_{\alpha }-\ ^{\shortparallel }%
\mathbf{e}_{\alpha }\star _{N}q=\ _{N}^{\shortparallel }F_{\alpha \quad
}^{\quad \gamma \beta }\ \ ^{\shortparallel }\mathbf{e}_{\gamma }q\star
_{N}\ ^{\shortparallel }\mathbf{e}_{\beta }\mbox{ and }q\star _{N}\
^{\shortparallel }\mathbf{e}^{\gamma }-\ ^{\shortparallel }\mathbf{e}\
^{\gamma }\star _{N}q=\ _{N}^{\shortparallel }F_{\quad \alpha }^{\gamma
\beta }\ \ ^{\shortparallel }\mathbf{e}_{\beta }q\star _{N}\
^{\shortparallel }\mathbf{e}^{\alpha }.  \label{aux04}
\end{equation}%
The multiplication and modification of Lebintz rules (\ref{aux03}) and (\ref%
{aux03a}) can be redefined with boldface symbols and $\star _{N}$ terms for
star commuting rules (\ref{aux04}).

\paragraph{$\mathcal{R}$ matrix action in $T_{\shortparallel }^{\ast }%
\mathbf{V}$: \newline
}

For a scalar and a d-vector, we have the relation%
\begin{equation}
\overline{\mathcal{R}}_{N}(q)\otimes _{N}\overline{\mathcal{R}}_{N}(\
^{\shortparallel }\mathbf{A}^{\alpha }\star _{N}\ ^{\shortparallel }\mathbf{e%
}_{\alpha })=\overline{\mathcal{R}}_{N}(q)\otimes _{N}\overline{\mathcal{R}}%
_{N}(\ ^{\shortparallel }\mathbf{A}^{\alpha })\star _{N}\ ^{\shortparallel }%
\mathbf{e}_{\alpha }+\ \ _{N}^{\shortparallel }F_{\quad \alpha }^{\gamma
\beta }\overline{\mathcal{R}}_{N}(\ ^{\shortparallel }\mathbf{e}_{\gamma
}q)\otimes _{N}\overline{\mathcal{R}}_{N}(\ ^{\shortparallel }\mathbf{A}%
^{\alpha })\star _{N}\ ^{\shortparallel }\mathbf{e}_{\beta }.  \label{aux03b}
\end{equation}%
By iteration, we can interchange two d-vectors following the rules%
\begin{eqnarray}
\overline{\mathcal{R}}_{N}(\ ^{\shortparallel }\mathbf{A}^{\alpha }\star
_{N}\ ^{\shortparallel }\mathbf{e}_{\alpha }) &\otimes _{N}&\overline{%
\mathcal{R}}_{N}(\ ^{\shortparallel }\mathbf{B}^{\beta }\star _{N}\
^{\shortparallel }\mathbf{e}_{\beta })=\overline{\mathcal{R}}_{N}(\
^{\shortparallel }\mathbf{A}^{\alpha })\star _{N}\ ^{\shortparallel }\mathbf{%
e}_{\alpha }\otimes _{N}\overline{\mathcal{R}}_{N}(\ ^{\shortparallel }%
\mathbf{B}^{\beta })\star _{N}\ ^{\shortparallel }\mathbf{e}_{\beta }
\label{aux03f} \\
&&-\ _{N}^{\shortparallel }F_{\quad \tau }^{\gamma \alpha }\overline{%
\mathcal{R}}_{N}(\ ^{\shortparallel }\mathbf{A}^{\tau })\star _{N}\
^{\shortparallel }\mathbf{e}_{\alpha }\otimes _{N}\overline{\mathcal{R}}%
_{N}(\ ^{\shortparallel }\mathbf{e}_{\gamma }\ ^{\shortparallel }\mathbf{B}%
^{\beta })\star _{N}\ ^{\shortparallel }\mathbf{e}_{\beta }  \notag \\
&&+\ _{N}^{\shortparallel }F_{\quad \tau }^{\gamma \beta }\overline{\mathcal{%
R}}_{N}(\ ^{\shortparallel }\mathbf{e}_{\gamma }\ ^{\shortparallel }\mathbf{A%
}^{\alpha })\star _{N}\ ^{\shortparallel }\mathbf{e}_{\alpha }\otimes _{N}%
\overline{\mathcal{R}}_{N}(\ ^{\shortparallel }\mathbf{B}^{\tau })\star
_{N}\ ^{\shortparallel }\mathbf{e}_{\beta }  \notag \\
&&-\ _{N}^{\shortparallel }F_{\quad \tau }^{\alpha \nu }\ \
_{N}^{\shortparallel }F_{\quad \varepsilon }^{\gamma \beta }\overline{%
\mathcal{R}}_{N}(\ ^{\shortparallel }\mathbf{e}_{\gamma }\ ^{\shortparallel }%
\mathbf{A}^{\tau })\star _{N}\ ^{\shortparallel }\mathbf{e}_{\nu }\otimes
_{N}\overline{\mathcal{R}}_{N}(\ ^{\shortparallel }\mathbf{e}_{\alpha }\
^{\shortparallel }\mathbf{B}^{\varepsilon })\star _{N}\ ^{\shortparallel }%
\mathbf{e}_{\beta }.  \notag
\end{eqnarray}%
Such properties were proven for arbitrary quasi-Hopf algebras \cite%
{barnes14,barnes15}. In this subsection, we performed a N-adapted (co)
tangent Lorentz bundle generalization of constructions from \cite%
{blumenhagen16}. Respectively, above formulas can be written for coordinate
frames with respective star $\star .$

\paragraph{Hopf d-algebras, N-adapted Lie derivatives, and $\mathcal{R}$
action on functions on $T_{\shortparallel }^{\ast }\mathbf{V}$: \newline
}

For a commutative d-algebra $\mathcal{A}_{N}$ and it N-adapted star
deformation $\mathcal{A}_{N}^{\star }$, we can control noncommutativity via
the action of $\mathcal{R}$-matrix, $\mathcal{R}_{N}=\mathcal{F}_{N}^{-2},$
when for the star product (\ref{starmult}) of two elements $a$ and $b$ is
computed 
\begin{eqnarray*}
a\star _{N}b &=&\overline{R}^{\gamma }(b)\star _{N}\overline{R}_{\gamma
}(a):=b^{\overline{\gamma }}\star _{N}a_{\overline{\gamma }}=b^{\overline{k}%
}\star _{N}a_{\overline{k}}+\ ^{\shortparallel }b_{_{\overline{c}}}\star
_{N}\ ^{\shortparallel }a^{_{\overline{c}}}, \\
&=&\overline{\mathfrak{f}}^{\alpha }(a)\overline{\mathfrak{f}}_{\alpha }(b)=%
\overline{\mathfrak{f}}_{\alpha }(b)\overline{\mathfrak{f}}^{\alpha }(a),%
\mbox{ where }b^{\overline{\gamma }}:=\overline{R}^{\gamma }(b)\mbox{ and }%
a_{\overline{\gamma }}:=\overline{R}_{\gamma }(a).
\end{eqnarray*}%
If the d-algebra $\mathcal{A}_{N}$ is associative, we can use the associator
d-operator $\Phi $ acting as%
\begin{equation}
\Phi :(a\star _{N}b)\star _{N}c=\ ^{\phi _{1}}a\star _{N}(\ ^{\phi
_{2}}b\star _{N}\ ^{\phi _{3}}c),  \label{aux13}
\end{equation}%
where, for instance, $\ ^{\phi _{1}}a=\phi _{1}(a).$ In coordinate bases, a
similar proof is provided in \cite{blumenhagen16}, see also references
therein.

Let us consider a simplest case of algebraic operations for a function $q\in
C^{\infty }(\mathcal{M},N)$, where the space of functions is $C^{\infty }(%
\mathcal{M)}$ where usual partial derivatives can be transformed into
N-adapted ones. The Lie d-derivative $\mathcal{L}_{\xi }(q):=\xi (q)$ is
defined by any d-vector field $\xi =\xi ^{\beta }\ ^{\shortparallel }\mathbf{%
e}_{\beta }.$ The action of the Lie d-algebra $\emph{Vec}(\mathcal{M},N)$ on
such functions can be extended to an action of the universal enveloping
d-algebra $U\emph{Vec}(\mathcal{M},N)$ if we define the Lie d-derivative on
products of d-vectors $\mathcal{L}_{\ ^{1}\xi \ ^{2}\xi ...\ ^{m}\xi }:=%
\mathcal{L}_{\ ^{1}\xi \ }\circ \mathcal{L}_{\ ^{2}\xi }\circ ...\circ 
\mathcal{L}_{\ ^{m}\xi }$ and by linearity. Then we deform\ with power
series on $\hbar $ and $\kappa $ the $U\emph{Vec}(\mathcal{M},N)$-module
d-algebra $C^{\infty }(\mathcal{M},N)$ into $U\emph{Vec}^{\mathcal{F}}(%
\mathcal{M},N)$-module d-algebra $\mathcal{A}_{N}^{\star }:=C_{\star
}^{\infty }(\mathcal{M},N)$ for which the d-vector space is considered the
same as $C^{\infty }(\mathcal{M},N)$ but with multiplication defined by a
N-adapted star product (\ref{starpn}), or (\ref{starplie}). For two
functions $q,z\in C^{\infty }(\mathcal{M},N),$ the N-adapted star product is
possess such properties:%
\begin{eqnarray}
q\star _{N}z &=&\overline{\mathfrak{f}}^{\alpha }(q)\cdot \overline{%
\mathfrak{f}}_{\alpha }(z)  \label{starpnh} \\
&=&\overline{R}^{\alpha }(z)\star _{N}\overline{R}_{\alpha }(q):=z^{%
\overline{\gamma }}\star _{N}q_{\overline{\gamma }}%
\mbox{ controls
noncommutativity };  \notag \\
\mbox{ and }\Phi &:&(z\star _{N}q)\star _{N}f=\ ^{\phi _{1}}z\star _{N}(\
^{\phi _{2}}q\star _{N}\ ^{\phi _{3}}f)\mbox{ controls nonassociativity}. 
\notag
\end{eqnarray}%
For such conventions on nonholonomic and nonassociative distributions, the
constant function $1$ on $\mathcal{M}$ is also the unit of the star
d-algebra $\mathcal{A}_{N}^{\star }$ which is stated by the property that $%
q\star _{N}1=q=1\star _{N}q.$

Similarly to (\ref{hstcom}), we can consider a N-adapted star commutator of
functions, $\left[ q,z\right] _{\star _{N}}=q\star _{N}z-z\star _{N}q,$ and
define on nonholonomic phase space $(\mathcal{M},N)$ a quasi-Poisson
coordinate d-algebra similar to (\ref{comrel}),%
\begin{equation*}
\left[ x^{j},x^{k}\right] _{\star _{N}}=2i\kappa R^{n+j\ n+k\ a}p_{a},\left[
x^{i},p_{n+j}\right] _{\star _{N}}=i\hbar \delta _{\ j}^{i}\mbox{ and }\left[
p_{n+i},p_{n+j}\right] _{\star _{N}}=0,
\end{equation*}%
for a parabolic R-fluc nonholonomic background, and a nontrivial Jacobiator%
\begin{equation*}
\left[ x^{i},x^{j},x^{k}\right] _{\star _{N}}=\mathit{\ell }_{s}^{3}R^{n+i\
n+j\ n+k}.
\end{equation*}%
We note that such formal coordinate functions on $T_{\shortparallel }^{\ast }%
\mathbf{V}$ are computed with indices determined with respect to $\
^{\shortparallel }\mathbf{e}_{\beta }.$

\subsubsection{Star pairing between d-vectors and d-forms}

Let us study the star pairing on nonholonomic phase spaces for N-adapted
star vectors and star forms. We begin with this convention on star product
of N-adapted bases and cobases in $T_{\shortparallel }^{\ast }\mathbf{V:}$%
\begin{equation}
\ ^{\shortparallel }\mathbf{e}_{\alpha }\star _{N}\ ^{\shortparallel }%
\mathbf{e}^{\beta }=\ ^{\shortparallel }\mathbf{e}^{\beta }\ \star _{N}\
^{\shortparallel }\mathbf{e}_{\alpha }=\delta _{\alpha }^{\beta }.
\label{aux03i}
\end{equation}%
The next two conventions are that the action of the N-adapted star product
on basis d-vectors must be on the right and for the N-adapted basis
d-covectors the cation is on the left, when 
\begin{equation*}
\ ^{\shortparallel }\mathbf{A=}\ ^{\shortparallel }\mathbf{A}^{\alpha }\star
_{N}\ ^{\shortparallel }\mathbf{e}_{\alpha }\mbox{ and }\ ^{\shortparallel }%
\mathbf{B=}\ ^{\shortparallel }\mathbf{B}_{\beta }\star _{N}\
^{\shortparallel }\mathbf{e}^{\beta }.
\end{equation*}%
We express%
\begin{eqnarray*}
\ ^{\shortparallel }\mathbf{A}\star _{N}\ ^{\shortparallel }\mathbf{B} &=&(\
^{\shortparallel }\mathbf{A}^{\alpha }\star _{N}\ ^{\shortparallel }\mathbf{e%
}_{\alpha })\star _{N}(\ ^{\shortparallel }\mathbf{B}_{\beta }\star _{N}\
^{\shortparallel }\mathbf{e}^{\beta })=\ ^{\shortparallel }\mathbf{A}%
^{\alpha }\star _{N}\delta _{\alpha }^{\beta }\mathbf{\star _{N}}\
^{\shortparallel }\mathbf{B}_{\beta }= \\
&&\ ^{\shortparallel }\mathbf{A}^{\alpha }\star _{N}\ ^{\shortparallel }%
\mathbf{B}_{\alpha }=\mathbf{A}^{i}\star _{N}\mathbf{B}_{i}+\
^{\shortparallel }\mathbf{A}_{a}\mathbf{\star }_{N}\ ^{\shortparallel }%
\mathbf{B}^{a}.
\end{eqnarray*}%
This way, we define the contraction of d-vectors and d-covectors as a
d-operator: 
\begin{eqnarray*}
<,>_{\mathbf{\star _{N}}}:\ T_{\shortparallel }\mathbf{V\otimes }_{\star
_{N}}T_{\shortparallel }^{\ast }\mathbf{V} &\rightarrow &\mathbb{R}, \\
\ ^{\shortparallel }\mathbf{A\otimes }_{\star _{N}}\ ^{\shortparallel }%
\mathbf{B} &\rightarrow &<,>_{N}\circ \mathcal{F}_{N}^{-1}(\
^{\shortparallel }\mathbf{A},\ ^{\shortparallel }\mathbf{B})=\
^{\shortparallel }\mathbf{A}^{\alpha }\star _{N}\ ^{\shortparallel }\mathbf{B%
}_{\alpha },
\end{eqnarray*}%
where the first entry in $<,>_{\mathbf{\star _{N}}}$ is reserved for a
d-vector and the second one is for a d-covector. Similar parings can be
defined using $\star .$

\subsection{The nonassociative star d-tensor calculus}

We elaborate a nonassociative star tensor calculus which can be performed in
N-adapted and coordinate form. The purpose is to show how this concretely
works for the nonassociative and nonholonomic star products (\ref{starpn}).
Such a calculus will be used in next sections for formulating explicit
models of nonassociative nonholonomic differential geometry and modified
gravity theories.

\subsubsection{N-adapted Lie-derivatives of star scalars}

By definition a star scalar $q(\ ^{\shortparallel }u)$ on $%
T_{\shortparallel}^{\ast }\mathbf{V}$ is a quantity which in N-adapted form
under diffeomorphisms generated by a d-vector $\ ^{\shortparallel }\mathbf{%
\xi }$ (choosing the convention $\ ^{\shortparallel }\xi =\ ^{\shortparallel}%
{\xi }^{\alpha }\star _{N}\ ^{\shortparallel }\mathbf{e}_{\alpha }$ instead
of $\ ^{\shortparallel }\mathbf{e}_{\alpha }\star _{N}\ ^{\shortparallel }%
\mathbf{\xi }^{\alpha })$ transforms under the rule%
\begin{equation*}
\delta _{\xi }q=\mathcal{L}_{\xi }^{\star }q=\ ^{\shortparallel }\mathbf{\xi 
}^{\alpha }\star _{N}\ ^{\shortparallel }\mathbf{e}_{\alpha }q=\
^{\shortparallel }\mathbf{\xi }^{\alpha }\star \ ^{\shortparallel }\mathbf{%
\partial }_{\alpha }q.
\end{equation*}%
It is also demanded that the star product of two scalars is also a star
scalar and that the switching partial derivatives obey the rule from (\ref%
{aux03}) and (\ref{aux03b}) (see also (\ref{auxf1})) 
\begin{eqnarray*}
\mathcal{L}_{\xi }^{\star }(q\star _{N}z) &=&\ ^{\shortparallel }{\xi}%
^{\alpha } \star _{N}\ ^{\shortparallel }\mathbf{e}_{\alpha }(q\star _{N}z)
\\
&=&\ ^{\shortparallel }\mathbf{\xi }^{\alpha }\star _{N}\ (\
^{\shortparallel }\mathbf{e}_{\alpha }q\star _{N}z)+\ ^{\shortparallel }%
\mathbf{\xi }^{\alpha }\star _{N}\ (q\star _{N}\ ^{\shortparallel }\mathbf{e}%
_{\alpha }z)-\ _{\partial }^{\shortparallel }F_{\alpha \quad }^{\quad \gamma
\beta }\ \ ^{\shortparallel }\mathbf{\xi }^{\alpha }\star _{N}\ (\
^{\shortparallel }\mathbf{e}_{\gamma }q\star _{N}\ ^{\shortparallel }\mathbf{%
e}_{\beta }z) \\
&=&\ _{\overline{\phi }}\left( (\mathcal{L}_{\xi }^{\star }q)\star
_{N}z\right) +\overline{\mathcal{R}}_{N}(q)\star _{N}\ _{\phi ^{2}}\left( 
\mathcal{L}_{\overline{\mathcal{R}}_{N}(\xi )}^{\star }z\right) .
\end{eqnarray*}%
These formulas can be written equivalently for respective frames and star
product $\star $ when the Leibniz rule does not change.

\subsubsection{Nonholonomic star Lie derivatives of d-vectors and d-covectors%
}

\paragraph{Twisting commutators for \textbf{d-vectors}: \newline
}

The formula for the horizontal star commutator (\ref{hstcom}) is extended on
total phase space with twisted N-adapted and/or coordinate basis commutators
which guarantees covariant behaviours under star multiplications,%
\begin{equation*}
\delta _{\xi }\ ^{\shortparallel }\mathbf{A}=\mathcal{L}_{\xi }^{\star }(\
^{\shortparallel }\mathbf{A})=[\xi ,\ ^{\shortparallel }\mathbf{A}]_{\star
_{N}}=[,]_{N}\circ \mathcal{F}_{N}^{-1}.
\end{equation*}%
This can be computed in component form which does not emphasize, or
(inversely) state explicitly, respective terms $\ ^{\shortparallel }F_{\quad
\alpha }^{\gamma \beta }$ and $\ _{\partial }^{\shortparallel }F_{\alpha
\quad }^{\quad \gamma \beta }$ correspondingly determined by $R^{ajk},$ 
\begin{eqnarray}
\mathcal{L}_{\xi }^{\star }(\ ^{\shortparallel }\mathbf{A}) &=&\xi (\
^{\shortparallel }\mathbf{A})-\overline{\mathcal{R}}_{N}(\ ^{\shortparallel }%
\mathbf{A})\overline{\mathcal{R}}_{N}(\xi )  \notag \\
&=&\ ^{\shortparallel }\mathbf{\xi }^{\alpha }\star _{N}\ ^{\shortparallel }%
\mathbf{e}_{\alpha }\ ^{\shortparallel }\mathbf{A}^{\beta }\star _{N}\
^{\shortparallel }\mathbf{e}_{\beta }-\overline{\mathcal{R}}_{N}(\
^{\shortparallel }\mathbf{A})^{\alpha }\star _{N}\ ^{\shortparallel }\mathbf{%
e}_{\alpha }\overline{\mathcal{R}}_{N}\ (\ ^{\shortparallel }\mathbf{\xi }%
)^{\beta }\star _{N}\ ^{\shortparallel }\mathbf{e}_{\beta },  \label{aux06}
\\
\mathcal{L}_{\xi }^{\star }(\ ^{\shortparallel }\mathbf{A}^{\alpha }) &=&\
^{\shortparallel }\mathbf{\xi }^{\beta }\star _{N}\ ^{\shortparallel }%
\mathbf{e}_{\beta }\ ^{\shortparallel }\mathbf{A}^{\alpha }-\overline{%
\mathcal{R}}_{N}(\ ^{\shortparallel }\mathbf{A}^{\beta })\star _{N}\
^{\shortparallel }\mathbf{e}_{\alpha }\overline{\mathcal{R}}_{N}\ (\
^{\shortparallel }\mathbf{\xi }^{\alpha })-\overline{\mathcal{R}}_{N}(\
^{\shortparallel }\mathbf{e}_{\mu }\ ^{\shortparallel }\mathbf{A}^{\beta
})\star _{N}{}^{\shortparallel }\mathbf{e}_{\beta }(\overline{\mathcal{R}}%
_{N}\ (\ ^{\shortparallel }\mathbf{\xi }^{\nu }))\ _{\partial
}^{\shortparallel }F_{\nu \quad }^{\ \mu \alpha }.  \notag
\end{eqnarray}

Extending on $T_{\shortparallel }^{\ast }\mathbf{V}$ the h-compositions (\ref%
{compcom}) for a N-connection structure, we compute%
\begin{eqnarray*}
(\xi \bullet _{N}\eta )q &=&\ _{\phi }\xi (\ _{\phi }\eta \ _{\phi }q)=\
^{\shortparallel }\mathbf{\xi }^{\alpha }\star _{N}\ ^{\shortparallel }%
\mathbf{e}_{\alpha }\ _{\phi }(\ ^{\shortparallel }\mathbf{\eta }^{\beta
}\star _{N}\ ^{\shortparallel }\mathbf{e}_{\beta }) \\
&=&(\ ^{\shortparallel }\mathbf{\xi }^{\alpha }\star _{N}\ ^{\shortparallel }%
\mathbf{\eta }^{\beta }\ )\star _{N}\ ^{\shortparallel }\mathbf{e}_{\alpha
}\ ^{\shortparallel }\mathbf{e}_{\beta }q+(\ ^{\shortparallel }\mathbf{\xi }%
^{\alpha }\star _{N}\ ^{\shortparallel }\mathbf{e}_{\alpha }\
^{\shortparallel }\mathbf{\eta }^{\beta })\star _{N}\ ^{\shortparallel }%
\mathbf{e}_{\beta }q \\
&&-\ _{\partial }^{\shortparallel }F_{\nu \quad }^{\quad \alpha \mu }(\
^{\shortparallel }\mathbf{\xi }^{\nu }\star _{N}\ ^{\shortparallel }\mathbf{e%
}_{\alpha }\ ^{\shortparallel }\mathbf{\eta }^{\beta })\star _{N}\
^{\shortparallel }\mathbf{e}_{\mu }\ ^{\shortparallel }\mathbf{e}_{\beta }q
\end{eqnarray*}%
\begin{eqnarray*}
\mbox{ and }(\overline{\mathcal{R}}_{N}\eta \bullet _{N}\overline{\mathcal{R}%
}_{N}\xi )q &=&(\ ^{\shortparallel }\mathbf{\xi }^{\alpha }\star _{N}\
^{\shortparallel }\mathbf{\eta }^{\beta }\ )\star _{N}\ ^{\shortparallel }%
\mathbf{e}_{\alpha }\ ^{\shortparallel }\mathbf{e}_{\beta }q+(\overline{%
\mathcal{R}}_{N}(\ ^{\shortparallel }\mathbf{\eta }^{\alpha })\star _{N}\
^{\shortparallel }\mathbf{e}_{\alpha }\overline{\mathcal{R}}_{N}(\
^{\shortparallel }\mathbf{\xi }^{\beta }))\star _{N}\ ^{\shortparallel }%
\mathbf{e}_{\beta }q\  \\
&&+(_{\partial }^{\shortparallel }F_{\nu \quad }^{\quad \alpha \mu }%
\overline{\mathcal{R}}_{N}(\ ^{\shortparallel }\mathbf{e}_{\alpha }\
^{\shortparallel }\mathbf{\eta }^{\beta })\star _{N}\ ^{\shortparallel }%
\mathbf{e}_{\beta }\overline{\mathcal{R}}_{N}(\ ^{\shortparallel }\mathbf{%
\xi }^{\nu }))\star _{N}\ ^{\shortparallel }\mathbf{e}_{\mu }q \\
&&+(\ _{\partial }^{\shortparallel }F_{\nu \quad }^{\quad \alpha \mu }\
^{\shortparallel }\mathbf{\xi }^{\nu }\star _{N}(\ ^{\shortparallel }\mathbf{%
e}_{\alpha }\ ^{\shortparallel }\mathbf{\eta }^{\beta }))\star _{N}\
^{\shortparallel }\mathbf{e}_{\beta }\ ^{\shortparallel }\mathbf{e}_{\mu }q.
\end{eqnarray*}%
Adding both last terms and after corresponding cancellations we obtain 
\begin{equation}
\lbrack \mathcal{L}_{\xi }^{\star },\mathcal{L}_{\eta }^{\star }]_{\star
_{N}}q=\ _{\phi }\xi (\ \eta (q))-(\overline{\mathcal{R}}_{N}\eta )\ _{\phi
}(\overline{\mathcal{R}}_{N}(\xi )(q))=\mathcal{L}_{[\xi ,\eta ]_{\star
_{N}}}^{\star }.  \label{twistvect1}
\end{equation}%
A tedious calculus in N-adapted or coordinate basis forms allow to verify
the Leibniz rule%
\begin{equation*}
\lbrack \ ^{\shortparallel }\mathbf{A},[\ ^{\shortparallel }\mathbf{B},\
^{\shortparallel }\mathbf{C}]_{\star _{N}}]_{\star _{N}}=\ _{\overline{\phi }%
}[[\ ^{\shortparallel }\mathbf{A},\ ^{\shortparallel }\mathbf{B]}_{\star
_{N}},\ ^{\shortparallel }\mathbf{C}]_{\star _{N}}+\ _{\phi ^{2}}[\overline{%
\mathcal{R}}_{N}(\ ^{\shortparallel }\mathbf{A)},[\overline{\mathcal{R}}%
_{N}(\ ^{\shortparallel }\mathbf{B)},\ ^{\shortparallel }\mathbf{C}]_{\star
_{N}}]_{\star _{N}}.
\end{equation*}%
The formula for twisting commutator (\ref{twistvect1}) can be written in
equivalent form using $\star .$

\paragraph{Twisting commutators for \textbf{d-covectors}: \newline
}

We consider respective N-adapted dualizations of formulas (\ref{aux02}), (%
\ref{aux03}) and (\ref{aux04}) using star d-covectors $\
^{\shortparallel}\omega =d\ ^{\shortparallel }u^{\alpha }\star \
^{\shortparallel }\omega _{\alpha }.$ Multiplying the formulas for $\
^{\shortparallel }\partial _{\alpha }(q\star z),$ one computes (see similar
details with cancellation of $F$-terms in (4.12) and the derivative operator 
$\partial :=dx^{I}\star \partial _{I}$ (4.13) from \cite{blumenhagen16}),%
\begin{eqnarray*}
d\ ^{\shortparallel }u^{\alpha }\star \ ^{\shortparallel }\partial _{\alpha
}(q\star z) &=&d\ ^{\shortparallel }u^{\alpha }\star \ ^{\shortparallel
}\partial _{\alpha }q\star z+q\star d\ ^{\shortparallel }u^{\alpha }\star \
^{\shortparallel }\partial _{\alpha }z,\mbox{ in local coordinate bases }; \\
\ ^{\shortparallel }\mathbf{e}^{\alpha }\star _{N}\ ^{\shortparallel }%
\mathbf{e}_{\alpha }(q\star _{N}z) &=&\ ^{\shortparallel }\mathbf{e}^{\alpha
}\star _{N}\ ^{\shortparallel }\mathbf{e}_{\alpha }q\star _{N}z+q\star _{N}\
^{\shortparallel }\mathbf{e}^{\alpha }\star _{N}\ ^{\shortparallel }\mathbf{e%
}_{\alpha }z,\mbox{ in N-adapted frames }.
\end{eqnarray*}%
We introduce the N-adapted derivative operator (it is not an exterior
derivative of type $d$ because the antisymmetrization is not involved) 
\begin{equation}
\ ^{\shortparallel }\mathbf{\partial }:=d\ ^{\shortparallel }u^{\alpha
}\star \ ^{\shortparallel }\partial _{\alpha },\ ^{\shortparallel }\mathbf{e}%
=\ ^{\shortparallel }\mathbf{e}^{\alpha }\star _{N}\ ^{\shortparallel }%
\mathbf{e}_{\alpha }.  \label{derivop}
\end{equation}%
Using above formulas, we can check that the Lorentz rule holds true, for
instance, 
\begin{equation}
\ ^{\shortparallel }\mathbf{e}(q\star _{N}z)=\ ^{\shortparallel }\mathbf{e(}%
q)\star _{N}z+\ ^{\shortparallel }q\star _{N}\ ^{\shortparallel }\mathbf{e}%
(z),  \label{aux01c}
\end{equation}%
when $\ ^{\shortparallel }\mathbf{\partial }$ and $\ ^{\shortparallel }%
\mathbf{e}$ from (\ref{derivop}) commutes with the star N-adapted product
and also the respective $\mathcal{R}$ matrixes.

From a star scalar, we can produce a 1-form, $\ ^{\shortparallel }\mathbf{e}%
q=\ ^{\shortparallel }\mathbf{e}^{\alpha }\star _{N}\ ^{\shortparallel}%
\mathbf{e}_{\alpha }q.$ So, we can interchange $\ ^{\shortparallel }\delta
_{\xi }^{\star }$ with $\ ^{\shortparallel }\mathbf{e}$ without producing
extra terms and use formulas%
\begin{equation*}
\ ^{\shortparallel }\delta _{\xi }^{\star }\ ^{\shortparallel }\mathbf{e}q=\
^{\shortparallel }\mathbf{e}\ ^{\shortparallel }\delta _{\xi }^{\star }q=\
^{\shortparallel }\mathbf{e(\ ^{\shortparallel }\xi }^{\alpha })\star _{N}\
^{\shortparallel }\mathbf{e}_{\alpha }q\mathbf{+\ ^{\shortparallel }\xi }%
^{\alpha }\star _{N}\ ^{\shortparallel }\mathbf{e}_{\alpha }\
^{\shortparallel }\mathbf{e}q.
\end{equation*}%
For a star 1-form 
\begin{equation*}
\ ^{\shortparallel }\omega =d\ ^{\shortparallel }u^{\alpha }\star \
^{\shortparallel }\omega _{\alpha }=\ ^{\shortparallel }\mathbf{e}^{\alpha
}\star _{N}\ ^{\shortparallel }\omega _{\alpha },
\end{equation*}%
we write in abstract and component forms, 
\begin{equation}
\mathcal{L}_{\xi }^{\star }\ ^{\shortparallel }\omega =\mathbf{\
^{\shortparallel }\xi }^{\alpha }\star _{N}\ ^{\shortparallel }\mathbf{e}%
_{\alpha }\ ^{\shortparallel }\omega +\ ^{\shortparallel }\mathbf{e\
^{\shortparallel }\xi }^{\beta }\star _{N}\ ^{\shortparallel }\omega _{\beta
}\mbox{ and }\mathcal{L}_{\xi }^{\star }\ ^{\shortparallel }\omega _{\gamma
}=\mathbf{\ ^{\shortparallel }\xi }^{\alpha }\star _{N}\ ^{\shortparallel }%
\mathbf{e}_{\alpha }\ ^{\shortparallel }\omega _{\gamma }+\ ^{\shortparallel
}\mathbf{e_{\gamma }\ ^{\shortparallel }\xi }^{\beta }\star _{N}\
^{\shortparallel }\omega _{\beta }.  \label{aux07}
\end{equation}%
These formulas prove that the star Lie N-adapted derivatives of d-vectors (%
\ref{aux06}) and d-covectors (\ref{aux07}) are compatible with the
contraction. We obtain a scalar if we contact a d-vector $\ ^{\shortparallel}%
\mathbf{A}^{\beta }$ and covector $\ ^{\shortparallel }\omega _{\gamma },$
when%
\begin{equation}
\mathcal{L}_{\xi }^{\star }(\ \ ^{\shortparallel }\mathbf{A}^{\beta }\star
_{N}\ ^{\shortparallel }\omega _{\beta })=\ _{\overline{\phi }}\left( 
\mathcal{L}_{\xi }^{\star }(\ \ ^{\shortparallel }\mathbf{A}^{\beta })\star
_{N}\ ^{\shortparallel }\omega _{\beta }\right) +\overline{\mathcal{R}}%
_{N}(\ \ ^{\shortparallel }\mathbf{A}^{\beta })\star _{N}\ _{\phi
^{2}}\left( \mathcal{L}_{\overline{\mathcal{R}}_{N}(\xi )}^{\star }(\
^{\shortparallel }\omega _{\beta })\right) .  \label{aux07c}
\end{equation}%
Such formulas hold true for $\star .$

\subsubsection{Star Lie N-adapted derivatives and d-tensor products}

We investigate in more details the notion of nonassociative star product for
d-tensors. We define respective star tensor products adapted to respective
coordinate or N-frame structures (usually, we shall write down N-adapted
formulas), respectively, $\otimes _{\star }:=\otimes \circ \mathcal{F}^{-1}$
and $\otimes _{\star N}:=\otimes _{N}\circ \mathcal{F}_{N}^{-1}$, which is
similar to the definition of star products and respective commutators. We
can prescribe any of such tensor products in $T^{\ast }\mathbf{V}$ and then
consider nonholonomic and star deformations on $T_{\shortparallel }^{\ast }%
\mathbf{V.}$

\paragraph{The Leibniz rule for star d-tensors: \newline
}

This rule for two d-tensors $\ ^{\shortparallel }\mathbf{Q}$ and $\
^{\shortparallel }\mathbf{K}$ can be written in the form 
\begin{equation*}
\mathcal{L}_{\xi }^{\star }(\ ^{\shortparallel }\mathbf{Q}\otimes _{\star
N}\ ^{\shortparallel }\mathbf{K})=\ _{\overline{\phi }}(\mathcal{L}_{\xi
}^{\star }(\ ^{\shortparallel }\mathbf{Q)}\otimes _{\star N}\
^{\shortparallel }\mathbf{K})+\overline{\mathcal{R}}_{N}(\ ^{\shortparallel }%
\mathbf{Q})\otimes _{\star N}\ _{\phi ^{2}}\left( \mathcal{L}_{\overline{%
\mathcal{R}}_{N}(\xi )}^{\star }\ ^{\shortparallel }\mathbf{K}\right) .
\end{equation*}

Let us consider, for instance, how to apply this rule to an element 
\begin{eqnarray}
\ ^{\shortparallel }\mathbf{Q} &=&\ ^{\shortparallel }\mathbf{A}_{\alpha
^{\prime }}\otimes _{\star N}\ ^{\shortparallel }\mathbf{B}^{\alpha ^{\prime
}}\in T_{\shortparallel }\mathbf{V}\otimes _{\star N}(T_{\shortparallel
}^{\ast }\mathbf{V}),  \label{aux05a} \\
&=&\ ^{\shortparallel }\mathbf{A}_{\alpha ^{\prime }}^{\alpha }\star _{N}\
^{\shortparallel }\mathbf{e}_{\alpha }\otimes _{\star N}\ ^{\shortparallel }%
\mathbf{B}^{\alpha ^{\prime }\beta }\star _{N}\ ^{\shortparallel }\mathbf{e}%
_{\beta },  \notag \\
&=&(\ ^{\shortparallel }\mathbf{A}_{\alpha ^{\prime }}^{\alpha }\star _{N}\
^{\shortparallel }\mathbf{B}^{\alpha ^{\prime }\beta }-\
_{N}^{\shortparallel }F_{\mu \quad }^{\quad \nu \beta }\ ^{\shortparallel }%
\mathbf{A}_{\alpha ^{\prime }}^{\mu }\star _{N}^{\shortparallel }\mathbf{e}%
_{\nu }\ ^{\shortparallel }\mathbf{B}^{\alpha ^{\prime }\alpha })\star
_{N}(\ ^{\shortparallel }\mathbf{e}_{\alpha }\otimes _{\star N}\
^{\shortparallel }\mathbf{e}_{\beta }),  \notag
\end{eqnarray}%
where for (co) tangent Lorentz bundles the abstract (primed) indices can be
moved up/down using the standard Lorentz metric $\eta _{\alpha
^{\prime}\beta ^{\prime }}=(\eta _{i^{\prime }j^{\prime }},\eta
^{a^{\prime}b^{\prime }})$ with $\eta _{i^{\prime }j^{\prime }}=(1,1,1,-1)$
and $\eta ^{a^{\prime }b^{\prime }}=(1,1,1,-1).$ Using (\ref{aux03f}), we
can compute the transpose $\ ^{\shortparallel }\mathbf{Q}^{T}$ which results
in interchange of indices related to corresponding cancellations,%
\begin{eqnarray}
\ ^{\shortparallel }\mathbf{Q}^{T} &=&\overline{\mathcal{R}}_{N}(\
^{\shortparallel }\mathbf{B}^{\alpha ^{\prime }})\otimes _{\star N}\overline{%
\mathcal{R}}_{N}(\ ^{\shortparallel }\mathbf{A}_{\alpha ^{\prime }})=%
\overline{\mathcal{R}}_{N}(\ ^{\shortparallel }\mathbf{A}_{\alpha ^{\prime
}}^{\alpha }\star _{N}\ ^{\shortparallel }\mathbf{e}_{\alpha })\otimes
_{\star N}\overline{\mathcal{R}}_{N}(\ \ ^{\shortparallel }\mathbf{B}%
^{\alpha ^{\prime }\beta }\star _{N}\ ^{\shortparallel }\mathbf{e}_{\beta }),
\notag \\
&=&(\ ^{\shortparallel }\mathbf{A}_{\alpha ^{\prime }}^{\alpha }\star _{N}\
^{\shortparallel }\mathbf{B}^{\alpha ^{\prime }\beta }-\
_{N}^{\shortparallel }F_{\mu \quad }^{\quad \nu \beta }\ ^{\shortparallel }%
\mathbf{A}_{\alpha ^{\prime }}^{\mu }\star _{N}\ ^{\shortparallel }\mathbf{e}%
_{\nu }\ ^{\shortparallel }\mathbf{B}^{\alpha ^{\prime }\alpha })\star
_{N}(\ ^{\shortparallel }\mathbf{e}_{\alpha }\otimes _{\star N}\
^{\shortparallel }\mathbf{e}_{\beta }).  \label{aux05b}
\end{eqnarray}%
We have that $\overline{\mathcal{R}}$--symmetric and $\overline{\mathcal{R}}$%
--antisymmetric d-tensors correspond respectively to symmetric and
antisymmetric matrices.

\paragraph{Contraction of star d-tensor products: \newline
}

The contraction of d-tensor products is done by multiplying the d-covectors
and d-vectors standing next to each other following the convention (\ref%
{aux03i}). We can also bring the basis d-vectors and d-covector to the
middle as in (\ref{aux05a}) and apply the rule $<\ ^{\shortparallel } 
\mathbf{e}_{\gamma }\otimes _{\star N}\ ^{\shortparallel }\mathbf{e}_{\tau
}, \ ^{\shortparallel }\mathbf{e}^{\alpha }\otimes _{\star N}\
^{\shortparallel }\mathbf{e}^{\beta}> =\delta _{\gamma }^{\beta }\delta
_{\tau }^{\alpha }$. Considering second rank d-tensors as respective
products of d-vectors and d-covectors, 
\begin{equation*}
\ ^{\shortparallel }\mathbf{A}\otimes _{\star N}\ ^{\shortparallel }\mathbf{B%
}\in T_{\shortparallel }^{\ast }\mathbf{V}\otimes _{\star
N}(T_{\shortparallel }^{\ast }\mathbf{V})\mbox{ and }\ \ ^{\shortparallel }%
\mathbf{\omega }\otimes _{\star N}\ ^{\shortparallel }\mathbf{\psi }\in
(T_{\shortparallel }^{\ast }\mathbf{V)}^{\ast }\otimes _{\star
N}(T_{\shortparallel }^{\ast }\mathbf{V})^{\ast },
\end{equation*}%
we write the product%
\begin{eqnarray*}
&&<\ ^{\shortparallel }\mathbf{A}\otimes _{\star _{N}}\ ^{\shortparallel }%
\mathbf{B,}\ \ ^{\shortparallel }\mathbf{\omega }\otimes _{\star _{N}}\
^{\shortparallel }\mathbf{\psi }>_{\star _{N}}=(\ ^{\shortparallel }\mathbf{A%
}^{\alpha }\star _{N}\ ^{\shortparallel }\mathbf{B}^{\beta }\mathbf{)\star }%
_{N}(\ \ ^{\shortparallel }\mathbf{\omega }_{\beta }\otimes _{\star _{N}}\
^{\shortparallel }\mathbf{\psi }_{\alpha }) \\
&&-\ _{N}^{\shortparallel }F_{\mu \quad }^{\quad \nu \beta }(\
^{\shortparallel }\mathbf{A}^{\mu }\star _{N}\mathbf{e}_{\nu }\
^{\shortparallel }\mathbf{B}^{\alpha })\star _{N}(\ ^{\shortparallel }%
\mathbf{\omega }_{\alpha }\otimes _{\star _{N}}\ ^{\shortparallel }\mathbf{%
\psi }_{\beta })-\ _{N}^{\shortparallel }F_{\mu \quad }^{\quad \nu \beta }(\
^{\shortparallel }\mathbf{A}^{\mu }\star _{N}\ ^{\shortparallel }\mathbf{B}%
^{\alpha })\star _{N}(\mathbf{e}_{\nu }\ ^{\shortparallel }\mathbf{\omega }%
_{\alpha }\otimes _{\star N}\ ^{\shortparallel }\mathbf{\psi }_{\beta }).
\end{eqnarray*}%
Formulas (\ref{aux05a}) and (\ref{aux05b}) can be applied in similar form to
d-forms, which allow to show the transposition symmetry of the previous
formula,%
\begin{equation}
<\ ^{\shortparallel }\mathbf{A}\otimes _{\star _{N}}\ ^{\shortparallel }%
\mathbf{B,}\ \ ^{\shortparallel }\mathbf{\omega }\otimes _{\star _{N}}\
^{\shortparallel }\mathbf{\psi }>_{\star _{N}}=<\ \overline{\mathcal{R}}%
_{N}(\ ^{\shortparallel }\mathbf{B)}\otimes _{\star _{N}}\overline{\mathcal{R%
}}_{N}(\ ^{\shortparallel }\mathbf{A),}\overline{\mathcal{R}}_{N}(\ \
^{\shortparallel }\psi \mathbf{)}\otimes _{\star _{N}}\overline{\mathcal{R}}%
_{N}(\ ^{\shortparallel }\omega \mathbf{)}>_{\star _{N}}.  \label{aux06a}
\end{equation}%
Such formulas can generalized in a straightforward forms to higher rank
d-tensors and $\star .$

\paragraph{N-adapted antisymmetric forms and star wedge products: \newline
}

Antisymmetric $k$-forms $\ ^{\shortparallel }\omega \in \wedge _{\star
_{N}}^{k}T_{\shortparallel }^{\ast }\mathbf{V}$ are defined as usual but
with additional requirements to be N-adapted in respective cobases, with $%
\mathcal{R}$-antisymmetry and adjusting to the star-wedge product to the $%
\wedge _{\star _{N}}=\wedge \circ \mathcal{F}_{N}^{-1}.$ For instance, the
star wedge product of two one forms $\ ^{\shortparallel }\gamma $ and $\
^{\shortparallel }\mu $ is defined in a clear $\mathcal{R}$-antisymmetric
form when%
\begin{equation*}
\ ^{\shortparallel }\gamma \wedge _{\star _{N}}\ ^{\shortparallel }\mu =\
^{\shortparallel }\gamma \otimes _{\star _{N}}\ ^{\shortparallel }\mu -%
\overline{\mathcal{R}}_{N}(\ ^{\shortparallel }\mu )\otimes _{\star _{N}}%
\overline{\mathcal{R}}_{N}(\ ^{\shortparallel }\gamma ).
\end{equation*}%
Using this formula and (\ref{aux06a}), we conclude that the star wedge
product projects out the antisymmetric part%
\begin{equation}
<\ ^{\shortparallel }\mathbf{A}\otimes _{\star _{N}}\ ^{\shortparallel }%
\mathbf{B,}\ ^{\shortparallel }\gamma \wedge _{\star _{N}}\ ^{\shortparallel
}\mu >_{\star _{N}}=<\ ^{\shortparallel }\mathbf{A}\otimes _{\star _{N}}\ \
^{\shortparallel }\mathbf{B,}\ \ ^{\shortparallel }\gamma \otimes _{\star
_{N}}\ ^{\shortparallel }\mu >_{\star _{N}}-<\overline{\mathcal{R}}_{N}(\
^{\shortparallel }\mathbf{B)}\otimes _{\star _{N}}\overline{\mathcal{R}}%
_{N}(\ ^{\shortparallel }\mathbf{A),}\ ^{\shortparallel }\gamma \otimes
_{\star _{N}}\ ^{\shortparallel }\mu >_{\star _{N}}.  \label{aux10}
\end{equation}

The exterior derivative is defined as the antisymmetrized partial derivative 
\begin{equation*}
\ ^{\shortparallel }\mathbf{d}:=\ _{\wedge }^{\shortparallel }\mathbf{e=}\
^{\shortparallel }\mathbf{e}^{\alpha }\wedge _{\star _{N}}\ ^{\shortparallel
}\mathbf{e}_{\alpha }=\ ^{\shortparallel }\mathbf{\partial }^{\alpha }\wedge
_{\star }\ ^{\shortparallel }\mathbf{\partial }_{\alpha }
\end{equation*}%
which is different from respective operators $^{\shortparallel }\mathbf{e}$ (%
\ref{derivop}). In this work, $\ ^{\shortparallel }\mathbf{d}$ is a
nonholonomic generalization of the exterior derivative $d=\mathbf{\partial }%
^{\wedge _{\star }}=dx^{I}\wedge _{\star }\mathbf{\partial }_{I}$ used, for
instance, in formula (4.28) from \cite{blumenhagen16}. In N-adapted form,
the exterior derivative $\ ^{\shortparallel }\mathbf{d}$ is invariant under
the $\mathcal{R}$-matrix, acting in a compatible forms with the Lie
derivative, $[\mathcal{L},\ ^{\shortparallel }\mathbf{d}]=0,$ where 
\begin{equation*}
\ ^{\shortparallel }\mathbf{d}(\ ^{\shortparallel }\gamma \otimes _{\star
_{N}}\ ^{\shortparallel }\mu )=\ ^{\shortparallel }\mathbf{d}(\
^{\shortparallel }\gamma )\otimes _{\star _{N}}\ ^{\shortparallel }\mu 
\mathbf{+}\ ^{\shortparallel }\gamma \otimes _{\star _{N}}\mathbf{\
^{\shortparallel }\mathbf{d}}(\ ^{\shortparallel }\mu ).
\end{equation*}%
For holonomic configurations, such nonassociative exterior differentials
were studied in \cite{mylonas12,mylonas13}.

\subsection{Forms and d-tensors for quasi-Hopf N-adapted structures}

In this subsection, we show that we can consider such types of N-adapted
nonholonomic distributions when the star deformed geometric constructions
with quasi-Hopf structure are similar to un-deformed analogs.

\subsubsection{Nonassociative deformations of star exterior products}

Let us consider the exterior algebra of differential forms $\Omega
^{\natural }(\mathcal{M},N)$ and study N-adapted nonassociative deformations
of star exterior products, i.e. respective $\Omega _{\star }^{\natural }(%
\mathcal{M},N).$ Stating for zero-forms, i.e. functions with star N-adapted
quasi-Hopf product (\ref{starpnh}), $\Omega _{\star }^{0}(\mathcal{M},N)=%
\mathcal{A}_{N}^{\star }$ we define the algebra of differential forms with
nonassociative product $\wedge _{\star _{N}}$ for any couple of 1-forms $\
^{\shortparallel }\omega ,\ ^{\shortparallel }\gamma \in $ $\Omega _{\star
}^{1}(\mathcal{M},N)$ when 
\begin{equation*}
\ ^{\shortparallel }\omega \wedge _{\star _{N}}\ ^{\shortparallel }\gamma =%
\overline{\mathfrak{f}}^{\alpha }(\ ^{\shortparallel }\omega )\cdot 
\overline{\mathfrak{f}}_{\alpha }(\ ^{\shortparallel }\gamma ).
\end{equation*}%
We use the N-adapted Lie derivative commuting with the exterior derivative
for actions with/on N-elongated (co) bases ($\ ^{\shortparallel }\mathbf{e}%
_{\alpha }$ and $\ ^{\shortparallel }\mathbf{e}^{\beta }$ from (\ref{nadapdc}%
)) and anti-symmetric h-vector$\ ^{\shortparallel }\mathbf{e}_{ij}$ (\ref%
{antsmv}), when%
\begin{equation}
\mathcal{L}_{\ ^{\shortparallel }\mathbf{e}_{\alpha }}^{\star }(\
^{\shortparallel }\mathbf{e}^{\beta })=0,\mathcal{L}_{R^{aij}\
^{\shortparallel }\mathbf{e}_{ij}}^{\star }(\ ^{\shortparallel }\mathbf{e}%
^{k})=2R^{abk}\ ^{\shortparallel }\mathbf{e}_{b}\mbox{ and },\mathcal{L}%
_{R^{aij}\ ^{\shortparallel }\mathbf{e}_{ij}}^{\star }(\ ^{\shortparallel }%
\mathbf{e}_{c})=0.  \label{aux14}
\end{equation}%
Iterating the commutativity of the exterior derivative in N-adapted form, $\
^{\shortparallel }\mathbf{d:\ }\Omega _{\star }^{\natural }(\mathcal{M}%
,N)\rightarrow \Omega _{\star }^{\natural +1}(\mathcal{M},N),$ we can work
with the un-deformed Leibniz rule,%
\begin{equation*}
\ ^{\shortparallel }\mathbf{d}(\ ^{\shortparallel }\omega \wedge _{\star
_{N}}\ ^{\shortparallel }\gamma )=\ ^{\shortparallel }\mathbf{d}\
^{\shortparallel }\omega \wedge _{\star _{N}}\ ^{\shortparallel }\gamma
+(-1)^{\mid \omega \mid }\ ^{\shortparallel }\omega \wedge _{\star _{N}}\
^{\shortparallel }\mathbf{d}\ ^{\shortparallel }\gamma ,
\end{equation*}%
where $\ ^{\shortparallel }\omega $ is a homogeneous form of degree $|\omega
|,$ and it is assumed that $\ ^{\shortparallel }\mathbf{d}\overline{%
\mathfrak{f}}^{\alpha }(\ ^{\shortparallel }\omega )=\overline{\mathfrak{f}}%
^{\alpha }(\ ^{\shortparallel }\mathbf{d}\ ^{\shortparallel }\omega )$ and $%
\ ^{\shortparallel }\mathbf{d}\overline{\mathfrak{f}}_{\alpha }(\
^{\shortparallel }\omega )=\overline{\mathfrak{f}}_{\alpha }(\
^{\shortparallel }\mathbf{d}\ ^{\shortparallel }\omega ).$

Both for N-adapted and coordinate cobases, we prove applying formulas (\ref%
{aux13}) and (\ref{aux14}) that their exterior products of such
orthonormalized 1-forms result in%
\begin{eqnarray*}
\ ^{\shortparallel }\mathbf{e}^{\alpha }\wedge _{\star _{N}}\
^{\shortparallel }\mathbf{e}^{\beta } &=&\ ^{\shortparallel }\mathbf{e}%
^{\alpha }\wedge _{_{N}}\ ^{\shortparallel }\mathbf{e}^{\beta }=-\
^{\shortparallel }\mathbf{e}^{\beta }\wedge _{_{N}}\ ^{\shortparallel }%
\mathbf{e}^{\alpha }=-\ ^{\shortparallel }\mathbf{e}^{\beta }\wedge _{\star
_{N}}\ ^{\shortparallel }\mathbf{e}^{\alpha }\mbox{ and } \\
(\ ^{\shortparallel }\mathbf{e}^{\alpha }\wedge _{\star _{N}}\
^{\shortparallel }\mathbf{e}^{\beta })\wedge _{\star _{N}}\ ^{\shortparallel
}\mathbf{e}^{\gamma } &=&\ _{\phi _{1}}^{\shortparallel }\mathbf{e}^{\alpha
}\wedge _{\star _{N}}(\ _{\phi _{2}}^{\shortparallel }\mathbf{e}^{\beta
}\wedge _{\star _{N}}\ _{\phi _{3}}^{\shortparallel }\mathbf{e}^{\gamma })=\
^{\shortparallel }\mathbf{e}^{\alpha }\wedge _{\star _{N}}(\
^{\shortparallel }\mathbf{e}^{\beta }\wedge _{\star _{N}}\ ^{\shortparallel }%
\mathbf{e}^{\gamma })=\ ^{\shortparallel }\mathbf{e}^{\alpha }\wedge
_{_{N}}\ ^{\shortparallel }\mathbf{e}^{\beta }\wedge _{_{N}}\
^{\shortparallel }\mathbf{e}^{\gamma },
\end{eqnarray*}%
where the associators act via Lie d-derivatives. Considering exterior
products between functions, i.e. 0-forms and 1-forms, and respective
N-adapted star deformations, we generate corresponding $C^{\infty }(\mathcal{%
M},N)$-bimodule and resulting $\mathcal{A}_{N}^{\star }$--bimodule structure
of spaces of 1-forms $\Omega _{N}^{1}(\mathcal{M)=}\Omega ^{1}(\mathcal{M}%
,N) $ and $\Omega _{\star N}^{1}(\mathcal{M)=}\Omega _{\star }^{1}(\mathcal{M%
},N).$ Such N-adapted coefficient formulas can be written in "packaged" form
using an antisymmetric d-tensor $\mathcal{R}_{\quad \gamma }^{\alpha \beta }$
with non-vanishing components $R^{jka}.$

We present here some important formulas when the h- and c-components 
\begin{eqnarray*}
f\star _{N}\ ^{\shortparallel }e^{j} &=&f\cdot \ ^{\shortparallel }e^{j}-%
\frac{\kappa }{2}R^{jkb}\mathbf{e}_{k}f\cdot \ ^{\shortparallel }\mathbf{e}%
_{b}=\ ^{\shortparallel }\mathbf{e}^{j}\star _{N}f-\ ^{\shortparallel }%
\mathbf{e}_{b}\star _{N}\kappa R^{jkb}\mathbf{e}_{k}f, \\
f\star _{N}\ ^{\shortparallel }\mathbf{e}_{a} &=&f\cdot \ ^{\shortparallel }%
\mathbf{e}_{a}=\ ^{\shortparallel }\mathbf{e}_{a}\star _{N}f
\end{eqnarray*}%
are written with generalized indices,%
\begin{equation*}
f\star _{N}\ ^{\shortparallel }\mathbf{e}^{\alpha }=\ ^{\shortparallel }%
\mathbf{e}^{\gamma }\star _{N}(f\delta _{\ \gamma }^{\alpha }-i\kappa 
\mathcal{R}_{\quad \gamma }^{\alpha \beta }\ ^{\shortparallel }\mathbf{e}%
_{\beta }f),
\end{equation*}%
when the absolute differential is star deformed in N-adapted form%
\begin{equation*}
\ ^{\shortparallel }\mathbf{d}f=\ ^{\shortparallel }\mathbf{e}_{\beta }f\
^{\shortparallel }\mathbf{e}^{\beta }=\ ^{\shortparallel }\mathbf{e}_{\beta
}f\star _{N}\ ^{\shortparallel }\mathbf{e}^{\beta }=\ ^{\shortparallel }%
\mathbf{e}^{\beta }\star _{N}\ ^{\shortparallel }\mathbf{e}_{\beta }f.
\end{equation*}%
Above formulas can be re-defined in local coordinate bases, 
\begin{eqnarray*}
f\star _{N}\ dx^{j} &=&f\cdot \ dx^{j}-\frac{i\kappa }{2}R^{jkb}\frac{%
\partial f}{\partial x^{k}}\cdot dp_{b}=\ dx^{j}\star _{N}f-\ dp_{b}\star
_{N}i\kappa R^{jkb}\frac{\partial f}{\partial x^{k}}, \\
f\star _{N}dp_{a} &=&f\cdot \ dp_{a}=dp_{a}\star _{N}f
\end{eqnarray*}

This formalism can be extended to d-tensors involving N-connections and
quasi-Hopf structures.

\subsubsection{d-tensor and quasi-Hopf star deformations}

\label{ssdtqh}The N-adapted tensor product $\otimes _{C^{\infty }(\mathcal{M}%
,N)}$over $C^{\infty }(\mathcal{M},N)$ is deformed to a d-tensor product $%
\otimes _{\star _{N}}$over d-algebra $\mathcal{A}_{N}^{\star }$ following
the convention 
\begin{equation*}
\ ^{\shortparallel }\mathbf{B}\otimes _{\star _{N}}\ ^{\shortparallel }%
\mathbf{Q}=\overline{\mathfrak{f}}^{\alpha }(\ ^{\shortparallel }\mathbf{B}%
)\otimes _{C^{\infty }(\mathcal{M},N)}\overline{\mathfrak{f}}_{\alpha }(\
^{\shortparallel }\mathbf{Q}).
\end{equation*}%
In this formula, the N-adapted twist on the d-tensors $\ ^{\shortparallel}%
\mathbf{B}$ and $\ ^{\shortparallel }\mathbf{Q}$ is taken via the Lie
d-derivative and the nonassociativity for any $a\in \mathcal{A}_{N}^{\star }$
is involved as 
\begin{equation*}
(\ ^{\shortparallel }\mathbf{B}\star _{N}a)\otimes _{\star _{N}}\
^{\shortparallel }\mathbf{Q=}\ _{\phi _{1}}^{\shortparallel }\mathbf{B}%
\otimes _{\star _{N}}(\ _{\phi _{2}}a\star _{N}\ _{\phi
_{3}}^{\shortparallel }\mathbf{Q).}
\end{equation*}%
The N-adapted star product $\star _{N}$ acts between functions and d-tensors
and states on the space of d-vector fields $\emph{Vec}(\mathcal{M},N)$ a
N-adapted $\mathcal{A}_{N}^{\star }$ --module structure, denoted by $\emph{%
Vec}_{\star _{N}}.$ Explicit coefficient formulas are defined by computing
Lie d-derivatives on N-elongated bases $\ ^{\shortparallel }\mathbf{e}
_{\beta }=(\ ^{\shortparallel }\mathbf{e}_{j},\ ^{\shortparallel }e^{b})$
and respective twists,%
\begin{equation*}
\mathcal{L}_{\ ^{\shortparallel }\mathbf{e}_{\alpha }}^{\star }(\
^{\shortparallel }\mathbf{e}_{\beta })=0,\mathcal{L}_{R^{aij}\
^{\shortparallel }\mathbf{e}_{ij}}^{\star }(\ ^{\shortparallel }\mathbf{e}%
_{k})=0\mbox{ and }\mathcal{L}_{R^{aij}\ ^{\shortparallel }\mathbf{e}%
_{ij}}^{\star }(\ ^{\shortparallel }e^{b})=2R^{abk}\ ^{\shortparallel }%
\mathbf{e}_{k}.
\end{equation*}%
The N-adapted star product of a function $f$ and h- and v-components of
partial derivatives 
\begin{eqnarray*}
f\star _{N}\ ^{\shortparallel }\mathbf{e}_{i} &=&f\cdot \ ^{\shortparallel }%
\mathbf{e}_{i}=\ ^{\shortparallel }\mathbf{e}_{i}\star _{N}f \\
f\star _{N}\ ^{\shortparallel }e^{a} &=&f\cdot \ ^{\shortparallel }e^{a}-%
\frac{i\kappa }{2}R^{jka}\mathbf{e}_{j}f\cdot \ ^{\shortparallel }\mathbf{e}%
_{k}=\ ^{\shortparallel }e^{a}\star _{N}f-\ ^{\shortparallel }\mathbf{e}%
_{j}\star _{N}i\kappa R^{ajk}\mathbf{e}_{k}f.
\end{eqnarray*}%
Here we nota that $\ ^{\shortparallel }\mathbf{e}_{i}\star _{N}f$ denotes
the right N-adapted \ $\mathcal{A}_{N}^{\star }$ --action on $\emph{Vec}%
_{\star _{N}}$ (this is not on action of $\ ^{\shortparallel }\mathbf{e}_{i}$
on the function $f$). In generalized indices,%
\begin{equation*}
f\star _{N}\ ^{\shortparallel }\mathbf{e}_{\alpha }=\ ^{\shortparallel }%
\mathbf{e}_{\gamma }\star _{N}(f\delta _{\ \alpha }^{\gamma }+i\kappa 
\mathcal{R}_{\quad \alpha }^{\gamma \beta }\ ^{\shortparallel }\mathbf{e}%
_{\beta }f),
\end{equation*}

Using the N-adapted star tensor product, we can extend the $\mathcal{A}%
_{N}^{\star }$ --bimodule $\emph{Vec}_{\star _{N}}$ of d-vector fields to
the $\Omega _{\star }^{\natural }$--bimodule $\emph{Vec}_{\star
_{N}}^{\natural }=\emph{Vec}_{\star _{N}}$ $\otimes _{\star _{N}}\mathbf{\ }%
\Omega _{\star }^{\natural }(\mathcal{M},N).$ An abstract differential form
calculus with left and right actions, duality, module homomorphisms, and
quantum Lie algebras of diffeomorphisms is elaborated in sections 3.5-3.7 of 
\cite{aschieri17}. All constructions can be formulated in a similar form for
N-adapted configurations. We omit such formulas in this work but present
certain important N-adapted coefficient formulas which are necessary for
formulating nonassociative gravitational field equations.

\section{Nonassociative differential geometry with N-connections \&
non\-symmet\-r\-ic met\-rics}

\label{sec3}

We define the geometric distinguished objects, d-objects, which are used for
elaborating on nonholonomic extensions of nonassociative differential
geometry models in \cite{blumenhagen16,aschieri17}. There are provided
abstract and N-adapted coefficient formulas for star deformed distinguished
connections, d-connections, their torsion and curvature distinguished
tensors, d-tensors. It is studied the important case when properties of such
d-objects are completely determined by symmetric and nonsymmetric star
deformations of distinguished metrics, d-metrics. At the end of respective
subsections, we provide coefficient formulas involving quasi-Hopf d-algebras.

\subsection{Nonassociative linear connections and d--connections}

The concepts of linear connection and associated covariant derivative in
nonassociative differential geometry can be introduced in a general and/or
N-adapted form, for distinguished connections, d-connections, not
considering metric structures.

\subsubsection{Covariant derivatives with d-connections}

There are two geometrically consistent notions of covariant derivatives on
nonassociative spaces which are determined by respective d-connections.

\paragraph{Left and right d-connections: \newline
}

On a (co) tangent Lorentz bundle enabled with respective real and/or complex
momentum like coordinates, a d-connection 
\begin{eqnarray}
\mathbf{D} &=&(\ _{h}D,\ _{v}D)=\{\mathbf{D}_{\alpha }=(D_{i},D_{a})\},\ 
\mathbf{\ \ ^{\shortmid }D}=(\ _{h}^{\shortmid }D,\ _{v}^{\shortmid }D)=\{%
\mathbf{\ ^{\shortmid }D}_{\alpha }=(\ ^{\shortmid }\mathbf{D}_{i},\
^{\shortmid }\mathbf{D}^{a})\},\mbox{ or }  \notag \\
\mathbf{\ \ ^{\shortparallel }D} &=&(\ _{h}^{\shortparallel }D,\
_{v}^{\shortparallel }D)=\{\ ^{\shortparallel }\mathbf{D}_{\alpha }=(\
^{\shortparallel }D_{i},\ ^{\shortparallel }D^{a})\}  \label{dcon}
\end{eqnarray}%
is a usual linear connection preserving under parallelism the h- and
(c)v-splitting (\ref{ncon}). For such a d-connection, we can define
N-adapted coefficients of nonassociative analogs of generalized Christoffel
symbols, which are determined by their coefficients $\ ^{\shortparallel
}\Gamma =\{\ ^{\shortparallel } \mathbf{\Gamma }_{\ \alpha \beta }^{\gamma
}\}$. On nonassociative phase spaces, we can introduce a similar N-adapted
covariant derivative as a d-operator $\ _{\flat }^{\shortparallel }\mathbf{D}%
:=\ ^{\shortparallel }\mathbf{e}- \ _{\flat }^{\shortparallel }\Gamma ,$ for 
$\ ^{\shortparallel }\mathbf{e}=\{\ ^{\shortparallel }\mathbf{e}_{\alpha }\}$
and compute such symbols using formula 
\begin{equation*}
\ _{\flat }^{\shortparallel }\mathbf{\Gamma }:=\ \ ^{\shortparallel }\mathbf{%
e}^{\alpha }\star _{N}\ ^{\shortparallel }\mathbf{e}^{\beta }\star _{N}\
_{\flat }^{\shortparallel }\mathbf{\Gamma }_{\ \alpha \beta }^{\gamma }\star
_{N}\ ^{\shortparallel }\mathbf{e}_{\gamma },
\end{equation*}%
where star products act with star contractions. In similar forms, we can
define and compute N-adapted coefficients $\Gamma =\{\Gamma _{\ \alpha
\beta}^{\gamma }\}$ and $\ ^{\shortmid }\Gamma =\{\ ^{\shortmid }\Gamma _{\
\alpha \beta }^{\gamma }\}$ for real (co) tangent Lorentz bundles. A label "$%
\flat $" will be used in order to distinguish certain geometric d-objects in
nonassociative geometry from similar ones when additionally will be adapted
to a quasi-Hopf d-structure (see next sections, when, for instance, it is
written$\ ^{\shortparallel}\Gamma ^{\star }$).

Dealing with a noncommutative star product, we have either let a covariant
d-operator to act from the left or from the right. This is related to the
ambiguity from which side we multiply a corresponding $\mathbf{\Gamma }$%
--operator, see details in \cite{blumenhagen16}. For any star distinguished
1--form 
\begin{equation*}
\omega =\mathbf{\ e}^{\alpha }\star _{N}\omega _{\alpha },\mathbf{\
^{\shortmid }}\omega =\mathbf{\ \ ^{\shortmid }e}^{\alpha }\star _{N}\mathbf{%
\ ^{\shortmid }}\omega _{\alpha },\mathbf{\mathbf{\mathbf{\mathbf{\
^{\shortparallel }}}}}\omega =\mathbf{\mathbf{\mathbf{\mathbf{\
^{\shortparallel }}}}e}^{\alpha }\star _{N}\mathbf{\mathbf{\mathbf{\mathbf{\
^{\shortparallel }}}}}\omega _{\alpha },
\end{equation*}%
we can introduce respectively: 
\begin{eqnarray}
\overrightarrow{\mathbf{D}}\omega &=&\ _{\flat }\mathbf{D}(\omega )=\mathbf{e%
}\omega -\ _{\flat }\mathbf{\Gamma }\star _{N}\omega \mbox{ or }%
\overleftarrow{\mathbf{D}}\omega =(\omega )\ _{\flat }\mathbf{D}=\mathbf{e}%
\omega -\omega \star _{N}\ _{\flat }\mathbf{\Gamma ;}  \label{leftrightcov}
\\
\mathbf{\ ^{\shortmid }}\overrightarrow{\mathbf{D}}\mathbf{\ \ ^{\shortmid }}%
\omega &=&\mathbf{\ \ _{\flat }^{\shortmid }D}(\mathbf{\ \ ^{\shortmid }}%
\omega )=\mathbf{\ \ ^{\shortmid }e\ \ ^{\shortmid }}\omega -\mathbf{\ \
_{\flat }^{\shortmid }\Gamma }\star _{N}\mathbf{\ \ ^{\shortmid }}\omega %
\mbox{ or }\mathbf{\ \ ^{\shortmid }}\overleftarrow{\mathbf{D}}\mathbf{\ \
^{\shortmid }}\omega =(\mathbf{\ \ ^{\shortmid }}\omega )\mathbf{\ \ _{\flat
}^{\shortmid }D}=\mathbf{\ \ ^{\shortmid }e\ \ ^{\shortmid }}\omega -\mathbf{%
\ \ ^{\shortmid }}\omega \star _{N}\mathbf{\ \ _{\flat }^{\shortmid }\Gamma ;%
}  \notag \\
\ ^{\shortparallel }\overrightarrow{\mathbf{D}}\mathbf{\mathbf{\mathbf{%
\mathbf{\ ^{\shortparallel }}}}}\omega &=&\ _{\flat }^{\shortparallel }%
\mathbf{D}(\mathbf{\mathbf{\mathbf{\mathbf{\ ^{\shortparallel }}}}}\omega
)=\ ^{\shortparallel }\mathbf{e\mathbf{\mathbf{\mathbf{\ ^{\shortparallel }}}%
}}\omega -\ _{\flat }^{\shortparallel }\mathbf{\Gamma }\star _{N}\mathbf{%
\mathbf{\mathbf{\mathbf{\ ^{\shortparallel }}}}}\omega \mbox{ or }\
^{\shortparallel }\overleftarrow{\mathbf{D}}\mathbf{\mathbf{\mathbf{\mathbf{%
\ ^{\shortparallel }}}}}\omega =(\mathbf{\mathbf{\mathbf{\mathbf{\
^{\shortparallel }}}}}\omega )\ _{\flat }^{\shortparallel }\mathbf{D}=\
^{\shortparallel }\mathbf{e\mathbf{\mathbf{\mathbf{\ ^{\shortparallel }}}}}%
\omega -\mathbf{\mathbf{\mathbf{\mathbf{\ ^{\shortparallel }}}}}\omega \star
_{N}\ _{\flat }^{\shortparallel }\Gamma .  \notag
\end{eqnarray}%
Here we note that equations of type (\ref{aux01c}) and (\ref{aux07c}) impose
the condition that to define d-operators of type $\overleftarrow{\mathbf{D}}$
we need to consider the $\mathcal{R}$-matrix.

It should be emphasized that both the left $\overrightarrow{\mathbf{D}}$ and
right \ $\overleftarrow{\mathbf{D}}$ covariant d-derivatives are consistent.
Using formulas (\ref{derivop}) -- (\ref{aux07c}), we can compute, for
instance, such \textsf{anomalous variations, } $\ ^{\shortparallel }\mathcal{%
\triangle }_{\xi }^{\star }:=\ ^{\shortparallel }\delta _{\xi }^{\star }-%
\mathcal{L}_{\xi }^{\star }$, of the second terms in (\ref{leftrightcov})
and find that both left and right choices compensate (for simplicity, we
consider star covectors) contributions of $\ ^{\shortparallel }\mathcal{%
\triangle }_{\xi }^{\star }$ $\ ^{\shortparallel }\mathbf{e}\
^{\shortparallel }\omega =$ $\ ^{\shortparallel }\mathbf{e}\
^{\shortparallel }\mathbf{e\ ^{\shortparallel }\xi }^{\beta }\star _{N}\
^{\shortparallel }\omega _{\beta }$ (we omit indices on d-vectors and
1-forms and write only one time the label "$\mathbf{\ ^{\shortparallel }}$"
in a term if this do not result in ambiguities). There are obtained the same
values if $\ ^{\shortparallel }\mathcal{\triangle }_{\xi }^{\star }\mathbf{%
\mathbf{\mathbf{\mathbf{\ _{\flat }^{\shortparallel }}}}\Gamma =}$ $\
^{\shortparallel }\mathbf{e}\ ^{\shortparallel }\mathbf{e\ ^{\shortparallel
}\xi :}$%
\begin{eqnarray}
\ ^{\shortparallel }\mathcal{\triangle }_{\xi }^{\star }(\mathbf{\mathbf{%
\mathbf{\mathbf{\ _{\flat }^{\shortparallel }}}}\Gamma }\star _{N}\mathbf{%
\mathbf{\mathbf{\mathbf{\ ^{\shortparallel }}}}}\omega ) &=&\
^{\shortparallel }\mathcal{\triangle }_{\xi }^{\star }\mathbf{\mathbf{%
\mathbf{\mathbf{\ _{\flat }^{\shortparallel }}}}\Gamma }\star _{N}\mathbf{%
\mathbf{\mathbf{\mathbf{\ ^{\shortparallel }}}}}\omega =\ ^{\shortparallel }%
\mathbf{e}\ ^{\shortparallel }\mathbf{e\ ^{\shortparallel }\xi }\star _{N}\
^{\shortparallel }\omega  \label{aux08} \\
\ ^{\shortparallel }\mathcal{\triangle }_{\xi }(\mathbf{\mathbf{\mathbf{%
\mathbf{\ ^{\shortparallel }}}}}\omega \star _{N}\mathbf{\mathbf{\mathbf{%
\mathbf{\ _{\flat }^{\shortparallel }}}}\Gamma }) &=&\overline{\mathcal{R}}%
_{N}(\mathbf{\mathbf{\mathbf{\mathbf{\ \ ^{\shortparallel }}}}}\omega )\star
_{N}\ \ ^{\shortparallel }\mathcal{\triangle }_{\overline{\mathcal{R}}%
_{N}(\xi )}\mathbf{\mathbf{\mathbf{\mathbf{\ _{\flat }^{\shortparallel }}}}%
\Gamma =}\overline{\mathcal{R}}_{N}(\mathbf{\mathbf{\mathbf{\mathbf{\ \
^{\shortparallel }}}}}\omega )\star _{N}\ ^{\shortparallel }\mathbf{e}\
^{\shortparallel }\mathbf{e}\overline{\mathcal{R}}_{N}\mathbf{(\
^{\shortparallel }\xi )}  \notag \\
&=&\overline{\mathcal{R}}_{N}(\mathbf{\mathbf{\mathbf{\mathbf{\ \
^{\shortparallel }}}}}\omega )\star _{N}\overline{\mathcal{R}}_{N}\mathbf{(}%
\ ^{\shortparallel }\mathbf{e}\ ^{\shortparallel }\mathbf{e\
^{\shortparallel }\xi )=}\ ^{\shortparallel }\mathbf{e}\ ^{\shortparallel }%
\mathbf{e\ ^{\shortparallel }\xi }\star _{N}\mathbf{\mathbf{\mathbf{\mathbf{%
\ \ ^{\shortparallel }}}}}\omega .  \notag
\end{eqnarray}%
Such formulas can be proven for $\star _{N}$ and $\star $ when there are
used respective 1-forms $\omega $ and $\ ^{\shortmid }\omega $ and related
d-vectors N-elongated (co) frames or coordinate ones.

\paragraph{An axiomatic viewpoint to covariant N-adapted derivatives: 
\newline
}

Geometrically, we can define a covariant d-derivative associated to a
d-connection $\ _{\flat }^{\shortparallel }\mathbf{D}$ on complex phase
space as a respective N-adapted map 
\begin{equation*}
T_{\shortparallel }^{\ast }\mathbf{V\rightarrow }T_{\shortparallel }^{\ast }%
\mathbf{V\otimes }_{\star _{N}}T_{\shortparallel }^{\ast }\mathbf{V}
\end{equation*}%
obeying the Leibniz rule when, for instance, 
\begin{equation*}
\ _{\flat }^{\shortparallel }\mathbf{D}(\ ^{\shortparallel }\omega \mathbf{%
\otimes }_{\star _{N}}\mathbf{\mathbf{\mathbf{\mathbf{\ ^{\shortparallel }}}}%
}\gamma )=\ _{\overline{\phi }}(\ _{\flat }^{\shortparallel }\mathbf{D}\
^{\shortparallel }\omega \mathbf{\otimes }_{\star _{N}}\mathbf{\mathbf{%
\mathbf{\mathbf{\ ^{\shortparallel }}}}}\gamma )+\ _{\phi ^{2}}\left( 
\overline{\mathcal{R}}_{N}(\ ^{\shortparallel }\omega )\star _{N}\overline{%
\mathcal{R}}_{N}(\ _{\flat }^{\shortparallel }\mathbf{D})\mathbf{\mathbf{%
\mathbf{\mathbf{\ ^{\shortparallel }}}}}\gamma \right) ,
\end{equation*}%
for some 1-forms (d-covectors) $\mathbf{\mathbf{\mathbf{\mathbf{\
^{\shortparallel }}}}}\omega $ and $\mathbf{\mathbf{\mathbf{\mathbf{\
^{\shortparallel }}}}}\gamma .$ Here we note that for a scalar $f$ we have $%
\ _{\flat }^{\shortparallel }\mathbf{D}f=\ ^{\shortparallel }\mathbf{e}f.$

The formulas for the covariant d-derivative can be considered for left and
right d-operators acting on a d-covector $\mathbf{\mathbf{\mathbf{\mathbf{\
^{\shortparallel }}}}}\omega $ (for simplicity, we consider only the cases
with $\star _{N}$-products),%
\begin{equation*}
\mathbf{\mathbf{\mathbf{\mathbf{\ ^{\shortparallel }}}}}\overrightarrow{%
\mathbf{D}}\mathbf{\mathbf{\mathbf{\mathbf{\ ^{\shortparallel }}}}}\omega =\
_{\flat }^{\shortparallel }\mathbf{D(\mathbf{\mathbf{\mathbf{\mathbf{\
^{\shortparallel }}}}e}}^{\alpha }\star _{N}\ ^{\shortparallel }\omega
_{\alpha }\mathbf{)=}\ _{\overline{\phi }}\left( \mathbf{\ _{\flat
}^{\shortparallel }\mathbf{D(\mathbf{\mathbf{\mathbf{\mathbf{\
^{\shortparallel }}}}e}}}^{\alpha }\mathbf{)}\star _{N}\ ^{\shortparallel
}\omega _{\alpha }\right) \mathbf{+}\overline{\mathcal{R}}_{N}\mathbf{%
\mathbf{(\mathbf{\mathbf{\mathbf{\mathbf{\ ^{\shortparallel }}}}e}}}^{\alpha
}\mathbf{)}\star _{N}\overline{\mathcal{R}}_{N}\mathbf{\mathbf{(\mathbf{%
\mathbf{\mathbf{\mathbf{\mathbf{\ ^{\shortparallel }}}}e)}}\ }}%
^{\shortparallel }\omega _{\alpha }\mathbf{=\mathbf{\mathbf{\mathbf{\
^{\shortparallel }}}}e\mathbf{\mathbf{\mathbf{\ ^{\shortparallel }}}}}\omega
-\mathbf{\mathbf{\mathbf{\mathbf{\ _{\flat }^{\shortparallel }}}}\Gamma }%
\star _{N}\mathbf{\mathbf{\mathbf{\mathbf{\ ^{\shortparallel }}}}}\omega .
\end{equation*}%
Let us consider how similar formulas can be written for a d-vector $\
^{\shortparallel }\mathbf{A}$ instead of a 1-form $\ ^{\shortparallel}\omega$%
. Using (\ref{aux08}), we compute 
\begin{equation*}
\ ^{\shortparallel }\mathcal{\triangle }_{\xi }^{\star }\mathbf{\mathbf{%
\mathbf{\mathbf{\ ^{\shortparallel }}}}e\mathbf{\mathbf{\mathbf{\
^{\shortparallel }}}A}}=-\overline{\mathcal{R}}_{N}\mathbf{\mathbf{(\mathbf{%
\mathbf{\mathbf{\mathbf{\ ^{\shortparallel }}}}}A})}\star _{N}\mathbf{%
\mathbf{\mathbf{\mathbf{\ ^{\shortparallel }}}}e\mathbf{\mathbf{\mathbf{\
^{\shortparallel }}}}e}\overline{\mathcal{R}}_{N}(\mathbf{\ ^{\shortparallel
}\xi })=-\mathbf{\mathbf{\mathbf{\mathbf{\ ^{\shortparallel }}}}e\mathbf{%
\mathbf{\mathbf{\ ^{\shortparallel }}}}e\ ^{\shortparallel }\xi }\star _{N}%
\mathbf{\mathbf{\mathbf{\mathbf{\mathbf{\mathbf{\ ^{\shortparallel }}}}}A,}}
\end{equation*}%
which means that such anomalous transforms can be compensated if $\mathbf{%
\mathbf{\mathbf{\mathbf{\ ^{\shortparallel }}}}}\overleftarrow{\mathbf{D}}$
and $\mathbf{\mathbf{\mathbf{\mathbf{\ ^{\shortparallel }}}}}\overrightarrow{%
\mathbf{D}}$ are defined with changed signs in front of $\mathbf{\mathbf{%
\mathbf{\mathbf{\ _{\flat }^{\shortparallel }}}}\Gamma }.$ From a more
axiomatic viewpoint (similar coordinate base conventions were introduced in 
\cite{aschieri05,blumenhagen16}), we can consider two consistent covariant
d-derivatives%
\begin{equation*}
\mathbf{\mathbf{\mathbf{\mathbf{\ ^{\shortparallel }}}}}\overrightarrow{%
\mathbf{D}}\mathbf{\mathbf{\mathbf{\mathbf{\ ^{\shortparallel }}}A}}=\
_{\flat }^{\shortparallel }\mathbf{D(\mathbf{\mathbf{\mathbf{\mathbf{\
^{\shortparallel }}}}}A})=\mathbf{\mathbf{\mathbf{\mathbf{\ ^{\shortparallel
}}}}e\mathbf{\mathbf{\mathbf{\mathbf{\ ^{\shortparallel }}}}}A+\mathbf{%
\mathbf{\mathbf{\ _{\flat }^{\shortparallel }}}}\Gamma }\star _{N}\
^{\shortparallel }\mathbf{A}\mbox{ and }\mathbf{\mathbf{\mathbf{\mathbf{\
^{\shortparallel }}}}}\overleftarrow{\mathbf{D}}\mathbf{\mathbf{\mathbf{%
\mathbf{\ ^{\shortparallel }}}A}}=\mathbf{(\mathbf{\mathbf{\mathbf{\mathbf{\
^{\shortparallel }}}}}A})\ _{\flat }^{\shortparallel }\mathbf{D}=\mathbf{%
\mathbf{\mathbf{\mathbf{\ ^{\shortparallel }}}}e\mathbf{\mathbf{\mathbf{%
\mathbf{\ ^{\shortparallel }}}}}A+\mathbf{\mathbf{\mathbf{\ ^{\shortparallel
}}}}A}\star _{N}\ _{\flat }^{\shortparallel }\mathbf{\Gamma .}
\end{equation*}%
It should be noted that all above type expressions simplify if we use $%
\mathbf{\mathbf{\mathbf{\mathbf{\ ^{\shortparallel }}}}}\overrightarrow{%
\mathbf{D}}$ for d-covectors and $\mathbf{\mathbf{\mathbf{\mathbf{\
^{\shortparallel }}}}}\overleftarrow{\mathbf{D}}$ for d-vectors. This
convention is compatible with the contraction if we follow the rule that
similarly to $\mathbf{\mathbf{\mathbf{\mathbf{\ ^{\shortparallel }}}}}%
\partial $ and $\mathbf{\mathbf{\mathbf{\mathbf{\ ^{\shortparallel }}}}e}$
the covariant d-derivative acts without an $\overline{\mathcal{R}}$-matrix
on respective products but such permutation matrices are used for left and
right covariant d-derivatives:%
\begin{eqnarray*}
\mathbf{\mathbf{\mathbf{\mathbf{\ ^{\shortparallel }}}}e(\mathbf{\mathbf{%
\mathbf{\mathbf{\ ^{\shortparallel }}}A}}}\star _{N}\ ^{\shortparallel
}\omega \mathbf{)} &=&\ _{\overline{\phi }}\left( \mathbf{\mathbf{\mathbf{%
\mathbf{\ ^{\shortparallel }}}}}\overleftarrow{\mathbf{D}}\mathbf{\mathbf{%
\mathbf{\mathbf{\ ^{\shortparallel }}}A}}\star _{N}\ ^{\shortparallel
}\omega \right) +\ _{\phi ^{2}}\left( \mathbf{\mathbf{\mathbf{\mathbf{\
^{\shortparallel }}}A}}\star _{N}\mathbf{\mathbf{\mathbf{\mathbf{\
^{\shortparallel }}}}}\overrightarrow{\mathbf{D}}\ ^{\shortparallel }\omega
\right) \\
&=&\mathbf{\mathbf{\mathbf{\mathbf{\ ^{\shortparallel }}}}e(\mathbf{\mathbf{%
\mathbf{\mathbf{\ ^{\shortparallel }}}A}}}\star _{N}\ ^{\shortparallel
}\omega \mathbf{)+}\ _{\phi }\left( (\mathbf{\mathbf{\mathbf{\mathbf{\
^{\shortparallel }}}}A}\star _{N}\ _{\flat }^{\shortparallel }\mathbf{\Gamma 
})\star _{N}\ ^{\shortparallel }\omega \right) -\ _{\phi ^{2}}\left( \
^{\shortparallel }\mathbf{A}\star _{N}\ (\mathbf{\mathbf{\mathbf{\mathbf{\
_{\flat }^{\shortparallel }}}}\Gamma }\star _{N}\ ^{\shortparallel }\mathbf{A%
})\right) ; \\
&=&\ ^{\shortparallel }\overrightarrow{\mathbf{D}}\mathbf{(\mathbf{\mathbf{%
\mathbf{\mathbf{\ ^{\shortparallel }}}A}}}\star _{N}\ ^{\shortparallel
}\omega )=\ _{\overline{\phi }}\mathbf{\mathbf{(\mathbf{\mathbf{\
^{\shortparallel }}}}}\overrightarrow{\mathbf{D}}\mathbf{\mathbf{\mathbf{%
\mathbf{\mathbf{\ ^{\shortparallel }}}A}}}\star _{N}\ ^{\shortparallel
}\omega \mathbf{)+}\overline{\mathcal{R}}_{N}\mathbf{\mathbf{(\mathbf{%
\mathbf{\mathbf{\mathbf{\ ^{\shortparallel }}}}}A})}\star _{N}\ _{\phi
^{2}}\left( \overline{\mathcal{R}}_{N}(\mathbf{\mathbf{\mathbf{\mathbf{\
^{\shortparallel }}}}}\overrightarrow{\mathbf{D}})\ ^{\shortparallel }\omega
\right) ; \\
&=&\ ^{\shortparallel }\overleftarrow{\mathbf{D}}(\mathbf{\mathbf{\
^{\shortparallel }A}}\star _{N}\ ^{\shortparallel }\omega )=\ \mathbf{%
\mathbf{\mathbf{\mathbf{\mathbf{\ ^{\shortparallel }}}A}}}\star _{N}\
^{\shortparallel }\omega \ _{\phi }(\mathbf{\ ^{\shortparallel }}%
\overleftarrow{\mathbf{D}})\mathbf{+\mathbf{\mathbf{\mathbf{\mathbf{\mathbf{%
\ ^{\shortparallel }}}}}A}}\overline{\mathcal{R}}_{N}(\mathbf{\mathbf{%
\mathbf{\mathbf{\ ^{\shortparallel }}}}}\overleftarrow{\mathbf{D}})\star
_{N}\ _{\overline{\phi }^{2}}\left( \overline{\mathcal{R}}_{N}(\
^{\shortparallel }\omega )\right) .
\end{eqnarray*}

Using corresponding labels, such formulas can be derived for N-adapted
splitting respectively on $T_{\shortparallel }\mathbf{V}$ and $%
T_{\shortparallel }^{\ast }\mathbf{V.}$

\paragraph{Directional covariant d-derivatives: \newline
}

The conventions for covariant d-derivatives are chosen in such forms that
for not adapted coordinate bases we obtain respective formulas from \cite%
{aschieri05,blumenhagen16} but also N-adapted formulas for torsion and
curvature on various (non) commutative nonholonomic spaces \cite%
{vacaru01,vacaru03,vacaru07a,vacaru08b,vacaru16,vacaru09a,vacaru08a,vacaru07b}%
. For instance, using $\star _{N}$ both in abstract and with N--adapted
coefficients, we write%
\begin{eqnarray}
\mathbf{\mathbf{\mathbf{\mathbf{\ _{\flat }^{\shortparallel }}}}D}\ _{%
\mathbf{\mathbf{\mathbf{\mathbf{\ ^{\shortparallel }}}}X}}\mathbf{\mathbf{%
\mathbf{\mathbf{\ ^{\shortparallel }}}}Y} &\mathbf{=}&\mathbf{\mathbf{%
\mathbf{\mathbf{\ ^{\shortparallel }}}}}\overleftarrow{\mathbf{D}}\ _{%
\mathbf{\mathbf{\mathbf{\mathbf{\ ^{\shortparallel }}}}}\overrightarrow{%
\mathbf{X}}}\mathbf{\mathbf{\mathbf{\mathbf{\ ^{\shortparallel }}}}Y:}=%
\mathbf{\mathbf{\mathbf{\mathbf{\ ^{\shortparallel }}}}X}\star _{N}\ _{\phi
}\left( \mathbf{\mathbf{\mathbf{\mathbf{\ ^{\shortparallel }}}}}%
\overleftarrow{\mathbf{D}}\ \mathbf{\mathbf{\mathbf{\mathbf{\
^{\shortparallel }}}}Y}\right) =\mathbf{\mathbf{\mathbf{\mathbf{\
^{\shortparallel }}}}X}\star _{N}(\mathbf{\mathbf{\mathbf{\mathbf{\
^{\shortparallel }}}}e\mathbf{\mathbf{\mathbf{\mathbf{\ ^{\shortparallel }}}}%
}Y)+(\mathbf{\mathbf{\mathbf{\ ^{\shortparallel }}}}X\otimes }_{\star _{N}}%
\mathbf{\mathbf{\mathbf{\mathbf{\mathbf{\ ^{\shortparallel }}}}}Y)}\star _{N}%
\mathbf{\mathbf{\mathbf{\mathbf{\ _{\flat }^{\shortparallel }}}}\Gamma } 
\notag \\
&=&\mathbf{\mathbf{\mathbf{\mathbf{\ ^{\shortparallel }}}}X}\star _{N}(%
\mathbf{\mathbf{\mathbf{\mathbf{\ ^{\shortparallel }}}}e\mathbf{\mathbf{%
\mathbf{\mathbf{\ ^{\shortparallel }}}}}Y)+\mathbf{\mathbf{\mathbf{\langle 
\mathbf{\mathbf{\mathbf{\mathbf{\ ^{\shortparallel }}}}X\otimes }_{\star
_{N}}\mathbf{\mathbf{\mathbf{\mathbf{\mathbf{\ ^{\shortparallel }}}}}Y,%
\mathbf{\mathbf{\ _{\flat }^{\shortparallel }}}}\Gamma \rangle _{\star _{N}}}%
}}}  \label{aux09a} \\
&=&\mathbf{\mathbf{\mathbf{\mathbf{\ ^{\shortparallel }}}}X}^{\alpha }\star
_{N}(\mathbf{\mathbf{\mathbf{\mathbf{\ ^{\shortparallel }}}}e}_{\alpha }%
\mathbf{\mathbf{\mathbf{\mathbf{\mathbf{\ ^{\shortparallel }}}}}Y}^{\beta })%
\mathbf{\star _{N}\mathbf{\mathbf{\mathbf{\ ^{\shortparallel }}}}e}_{\beta }%
\mathbf{+\mathbf{\mathbf{\mathbf{\ ^{\shortparallel }}}}X}^{\alpha }\star
_{N}\mathbf{\mathbf{\mathbf{\mathbf{\mathbf{\ ^{\shortparallel }}}}}Y}%
^{\beta }\mathbf{\ }\star _{N}\mathbf{\mathbf{\mathbf{\mathbf{\ _{\flat
}^{\shortparallel }}}}\Gamma }_{\ \alpha \beta }^{\gamma }\star _{N}\mathbf{%
\mathbf{\mathbf{\mathbf{\ ^{\shortparallel }}}}e}_{\gamma }  \notag \\
&&\mathbf{-}\ _{N}^{\shortparallel }F_{\mu \quad }^{\quad \nu \beta }(\
^{\shortparallel }\mathbf{X}^{\mu }\star _{N}\mathbf{\mathbf{\mathbf{\mathbf{%
\ ^{\shortparallel }}}}e}_{\nu }\mathbf{\mathbf{\mathbf{\mathbf{\mathbf{\
^{\shortparallel }}}}}Y}^{\alpha })\star _{N}\mathbf{\mathbf{\mathbf{\mathbf{%
\ _{\flat }^{\shortparallel }}}}\Gamma }_{\ \alpha \beta }^{\gamma }\star
_{N}\ ^{\shortparallel }\mathbf{e}_{\gamma }.  \notag
\end{eqnarray}%
Permutations of directional covariant d-derivatives via $\mathcal{R}$%
-matrices can be computed using formulas (\ref{aux05b}),%
\begin{eqnarray}
\mathbf{\mathbf{\mathbf{\mathbf{\ _{\flat }^{\shortparallel }}}}D}\ _{%
\overline{\mathcal{R}}_{N}(\mathbf{\mathbf{\mathbf{\mathbf{\
^{\shortparallel }}}}Y)}}\overline{\mathcal{R}}_{N}(\mathbf{\mathbf{\mathbf{%
\mathbf{\ ^{\shortparallel }}}}X)} &=&\overline{\mathcal{R}}_{N}\mathbf{(%
\mathbf{\mathbf{\mathbf{\ ^{\shortparallel }}}}Y})^{\alpha }\star _{N}(%
\mathbf{\mathbf{\mathbf{\mathbf{\ ^{\shortparallel }}}}e}_{\alpha }\overline{%
\mathcal{R}}_{N}(\mathbf{\mathbf{\mathbf{\mathbf{\mathbf{\ ^{\shortparallel }%
}}}}Y)}^{\beta })\mathbf{\star _{N}\mathbf{\mathbf{\mathbf{\
^{\shortparallel }}}}e}_{\beta }\mathbf{+\mathbf{\mathbf{\mathbf{\
^{\shortparallel }}}}X}^{\alpha }\star _{N}\mathbf{\mathbf{\mathbf{\mathbf{%
\mathbf{\ ^{\shortparallel }}}}}Y}^{\beta }\mathbf{\ }\star _{N}\mathbf{%
\mathbf{\mathbf{\mathbf{\ _{\flat }^{\shortparallel }}}}\Gamma }_{\ \alpha
\beta }^{\gamma }\star _{N}\mathbf{\mathbf{\mathbf{\mathbf{\
^{\shortparallel }}}}e}_{\gamma }  \notag \\
&&\mathbf{-}\ _{N}^{\shortparallel }F_{\mu \quad }^{\quad \nu \beta }(\
^{\shortparallel }\mathbf{X}^{\mu }\star _{N}\mathbf{\mathbf{\mathbf{\mathbf{%
\ ^{\shortparallel }}}}e}_{\nu }\mathbf{\mathbf{\mathbf{\mathbf{\mathbf{\
^{\shortparallel }}}}}Y}^{\alpha })\star _{N}\mathbf{\mathbf{\mathbf{\mathbf{%
\ _{\flat }^{\shortparallel }}}}\Gamma }_{\ \alpha \beta }^{\gamma }\star
_{N}\ ^{\shortparallel }\mathbf{e}_{\gamma }.  \label{aux09}
\end{eqnarray}%
Such formulas can be written for any geometric data $(\ ^{\shortparallel
}\partial ,\star ,\overline{\mathcal{R}},\ _{\flat }^{\shortparallel }D,\
^{\shortparallel }\overleftarrow{D},\ ^{\shortparallel }\overrightarrow{D})$
with coordinate frames $\ ^{\shortparallel }\partial $.

\subsubsection{N-adapted linear connections and quasi-Hopf structures}

\label{ssdconstar}The d-connection formalism introduced in previous
subsection can be extended for d-tensors and differential forms on the $%
\mathcal{A}_{N}^{\star }$--bimodule $\emph{Vec}_{\star _{N}}$ of d-vector
fields and the $\Omega _{\star }^{\natural }$--bimodule $\emph{Vec}_{\star
_{N}}^{\natural }=\emph{Vec}_{\star _{N}}$ $\otimes _{\star _{N}}\Omega
_{\star }^{\natural }(\mathcal{M},N)$ as we considered in section \ref%
{ssdtqh}.

\paragraph{Star d-connections as N-adapted linear maps: \newline
}

We define a star d-connection as a N-adapted linear map $h$%
\begin{eqnarray}
\mathbf{\mathbf{\mathbf{\mathbf{\ ^{\shortparallel }}}}D}^{\star }: &&%
\mathbf{\ }\emph{Vec}_{\star _{N}}\rightarrow \emph{Vec}_{\star _{N}}\otimes
_{\star _{N}}\mathbf{\ }\Omega _{\star }^{1},\mbox{  i. e. }\mathbf{v}%
\rightarrow \mathbf{\mathbf{\mathbf{\mathbf{\ ^{\shortparallel }}}}D}^{\star
}\mathbf{\mathbf{\mathbf{\mathbf{\ ^{\shortparallel }}}}v}=\mathbf{\mathbf{%
\mathbf{\mathbf{\ ^{\shortparallel }}}}v}^{\alpha }\otimes _{\star _{N}}%
\mathbf{\mathbf{\mathbf{\mathbf{\ ^{\shortparallel }}}}}\omega _{\alpha
}=(v^{i}\otimes _{\star _{N}}\omega _{i},\mathbf{\mathbf{\mathbf{\mathbf{\
^{\shortparallel }}}}}v_{a}\otimes _{\star _{N}}\mathbf{\mathbf{\mathbf{%
\mathbf{\ ^{\shortparallel }}}}}\omega ^{a});  \notag \\
&&\mbox{with  h- c-splitting }  \notag \\
\mathbf{\mathbf{\mathbf{\mathbf{\ ^{\shortparallel }}}}D}^{\star } &=&(\
_{h}^{\shortparallel }\mathbf{D}^{\star },\ _{c}^{\shortparallel }\mathbf{D}%
^{\star }):\ \emph{Vec}_{\star _{N}}=(h\emph{Vec}_{\star _{N}},c\emph{Vec}%
_{\star _{N}})\rightarrow (h\emph{Vec}_{\star _{N}},c\emph{Vec}_{\star
_{N}})\otimes _{\star _{N}}(\ h\Omega _{\star }^{\natural },\ c\Omega
_{\star }^{\natural }),  \label{dconhopf}
\end{eqnarray}%
for $\mathbf{\mathbf{\mathbf{\mathbf{\ ^{\shortparallel }}}}v}^{\alpha
}\otimes _{\star _{N}}\mathbf{\mathbf{\mathbf{\mathbf{\ ^{\shortparallel }}}}%
}\omega _{\alpha }\in \emph{Vec}_{\star _{N}}\otimes _{\star _{N}}\mathbf{\ }%
\Omega _{\star }^{1}.$ We put a star label for $\mathbf{\mathbf{\mathbf{%
\mathbf{\ ^{\shortparallel }}}}D}^{\star }$ in order to distinguish this
d-connection from $\mathbf{\mathbf{\mathbf{\mathbf{\ ^{\shortparallel }}}}D}$
(\ref{leftrightcov}) introduces for a geometric nonholonomic formalism not
involving Hopf d-algebras. For any element $a\in \mathcal{A}_{N}^{\star }$
and $\mathbf{\mathbf{\mathbf{\mathbf{\ ^{\shortparallel }}}}v\in }\emph{Vec}
_{\star _{N}}$ and respective associators, it is satisfied the right Leibniz
rule,%
\begin{equation}
\mathbf{\mathbf{\mathbf{\mathbf{\ ^{\shortparallel }}}}D}^{\star }(\mathbf{%
\mathbf{\mathbf{\mathbf{\ ^{\shortparallel }}}}v}\star _{N}a)=(\ _{\overline{%
\phi }_{1}}^{\shortparallel }\mathbf{D}^{\star }(\ _{\overline{\phi }%
_{2}}^{\shortparallel }\mathbf{v))}\star _{N}\ _{\overline{\phi }_{3}}a+%
\mathbf{\mathbf{\mathbf{\mathbf{\ ^{\shortparallel }}}}v}\star _{N}\
^{\shortparallel }\mathbf{d}a.  \label{leibr01}
\end{equation}

In a more general context, the adjoint action of a d-vector $\xi \in U\emph{%
Vec}^{\mathcal{F}}(\mathcal{M},N)$ also results in a N-adapted linear map $\
_{\xi }^{\shortparallel }\mathbf{D}^{\star }\mathbf{:\ }\emph{Vec}_{\star
_{N}}\rightarrow \emph{Vec}_{\star _{N}}\otimes _{\star _{N}}\Omega _{\star
}^{1},$ which is also a d-connection. This property exists for all types of
d-connections in nonholonomic (non) commutative and/or nonassociative
geometry.

The d-connection $\mathbf{\mathbf{\mathbf{\mathbf{\ ^{\shortparallel }}}}D}%
^{\star }$ (\ref{dconhopf}) can be uniquely extended to a N-adapted
covariant derivative for d-vector filed values in the exterior d-algebra $%
\emph{Vec}_{\star _{N}}^{\natural }=\emph{Vec}_{\star _{N}}\otimes _{\star
_{N}}\mathbf{\ }\Omega _{\star }^{\natural }$, when 
\begin{equation}
\ _{\mathbf{D}^{\star }}^{\shortparallel }\mathbf{d:\ }\emph{Vec}_{\star
_{N}}^{\natural }\rightarrow \emph{Vec}_{\star _{N}}^{\natural +1},%
\mbox{
i. e. }\mathbf{\mathbf{\mathbf{\mathbf{\ ^{\shortparallel }}}}v}^{\alpha
}\otimes _{\star _{N}}\mathbf{\mathbf{\mathbf{\mathbf{\ ^{\shortparallel }}}}%
}\omega _{\alpha }\rightarrow (\ _{\overline{\phi }_{1}}^{\shortparallel }%
\mathbf{D}^{\star }(\ _{\overline{\phi }_{2}}^{\shortparallel }\mathbf{%
v))\wedge }_{\star _{N}}\ _{\overline{\phi }_{3}}^{\shortparallel }\omega +%
\mathbf{\mathbf{\mathbf{\mathbf{\ ^{\shortparallel }}}}v}\otimes _{\star
_{N}}\ ^{\shortparallel }\mathbf{d\mathbf{\mathbf{\mathbf{\ ^{\shortparallel
}}}}}\omega .  \label{dcov01}
\end{equation}

We can consider the action of covariant d-derivative along a d-vector field $%
\mathbf{\mathbf{\mathbf{\mathbf{\ ^{\shortparallel }}}}z}^{\alpha }\in \emph{%
Vec}_{\star _{N}}$ using the pairing d-operator and respective associators,%
\begin{equation}
\mathbf{\mathbf{\mathbf{\mathbf{\ ^{\shortparallel }}}}D}_{\mathbf{z}%
}^{\star }\mathbf{\mathbf{\mathbf{\mathbf{\ ^{\shortparallel }}}}}\Psi
=\left\langle \mathbf{\mathbf{\mathbf{\mathbf{\ ^{\shortparallel }}}}D}%
^{\star }\mathbf{\mathbf{\mathbf{\mathbf{\ ^{\shortparallel }}}}}\Psi ,%
\mathbf{\mathbf{\mathbf{\mathbf{\ ^{\shortparallel }}}}z}\right\rangle
_{\star _{N}}=\left\langle \mathbf{\mathbf{\mathbf{\mathbf{\ }}}}(\mathbf{%
\mathbf{\mathbf{\mathbf{\ ^{\shortparallel }}}}v}^{\alpha }\otimes _{\star
_{N}}\mathbf{\mathbf{\mathbf{\mathbf{\ ^{\shortparallel }}}}}\omega _{\alpha
}\mathbf{)},\mathbf{\mathbf{\mathbf{\mathbf{\ ^{\shortparallel }}}}z}%
\right\rangle _{\star _{N}}=\mathbf{\mathbf{\mathbf{\mathbf{\ \mathbf{\ }}}}}%
\ _{\overline{\phi }_{1}}^{\shortparallel }\mathbf{v}^{\alpha }\star
_{N}\left\langle \mathbf{\mathbf{\mathbf{\mathbf{\mathbf{\ }}}}}\ _{%
\overline{\phi }_{2}}^{\shortparallel }\omega _{\alpha },\ _{\overline{\phi }%
_{3}}^{\shortparallel }\mathbf{z}\right\rangle _{\star _{N}},
\label{dcov01a}
\end{equation}%
for any $\mathbf{\mathbf{\mathbf{\mathbf{\ ^{\shortparallel }}}}}\Psi =$ $%
\mathbf{\mathbf{\mathbf{\mathbf{\ ^{\shortparallel }}}}v}^{\alpha }\otimes
_{\star _{N}}\mathbf{\mathbf{\mathbf{\mathbf{\ ^{\shortparallel }}}}}\omega
_{\alpha }\in \emph{Vec}_{\star _{N}}^{\natural }.$ These formulas can be
used for defining and computing the covariant d-derivative,%
\begin{equation*}
\ _{\mathbf{D}_{\mathbf{z}}^{\star }}^{\shortparallel }\mathbf{d\mathbf{%
\mathbf{\mathbf{\ ^{\shortparallel }}}}}\Psi :=\left\langle \mathbf{\mathbf{%
\mathbf{\mathbf{\ }}}}\ _{\mathbf{D}^{\star }}^{\shortparallel }\mathbf{d%
\mathbf{\mathbf{\mathbf{\ ^{\shortparallel }}}}}\Psi ,\mathbf{\mathbf{%
\mathbf{\mathbf{\ ^{\shortparallel }}}}z}\right\rangle _{\star _{N}}+\mathbf{%
\mathbf{\mathbf{\mathbf{\ }}}}\ _{\mathbf{D}^{\star }}^{\shortparallel }%
\mathbf{d}\left\langle \mathbf{\mathbf{\mathbf{\mathbf{\ ^{\shortparallel }}}%
}}\Psi ,\mathbf{\mathbf{\mathbf{\mathbf{\ ^{\shortparallel }}}}z}%
\right\rangle _{\star _{N}}.
\end{equation*}%
The d-operator $\ _{\mathbf{D}^{\star }}^{\shortparallel }\mathbf{d}$
satisfies in N-adapted form this graded right Leibnitz rule:%
\begin{equation*}
\ _{\mathbf{D}^{\star }}^{\shortparallel }\mathbf{d}(\mathbf{\mathbf{\mathbf{%
\mathbf{\ ^{\shortparallel }}}}}\Psi \mathbf{\wedge }_{\star _{N}}\mathbf{%
\mathbf{\mathbf{\mathbf{\ ^{\shortparallel }}}}}\omega )=\left( \ _{\mathbf{D%
}^{\star }}^{\overline{\phi }_{1}\shortparallel }\mathbf{d}(\mathbf{\mathbf{%
\mathbf{\mathbf{\ }}}}\ _{\overline{\phi }_{2}}^{\shortparallel }\Psi
)\right) \mathbf{\wedge }_{\star _{N}}\mathbf{\mathbf{\mathbf{\mathbf{\ }}}}%
\ _{\overline{\phi }_{3}}^{\shortparallel }\omega +(-1)^{|\Psi |}\mathbf{%
\mathbf{\mathbf{\mathbf{\ ^{\shortparallel }}}}}\Psi \mathbf{\wedge }_{\star
_{N}}\mathbf{\mathbf{\mathbf{\mathbf{\ }}}}\ _{\mathbf{D}^{\star
}}^{\shortparallel }\mathbf{d}\ ^{\shortparallel }\omega .
\end{equation*}%
In abstract geometric form, we can compute various star products of
d-tensors and multiplications to elements of d-algebras, scalars etc. using
respective (\ref{dcov01}) and (\ref{dcov01a}), pairing d-operators and
associators. For instance, we can consider $\mathbf{\mathbf{\mathbf{\mathbf{%
^{\shortparallel }}}}D}_{\mathbf{z}}^{\star }(\mathbf{\mathbf{\mathbf{%
\mathbf{\ ^{\shortparallel }}}}v}\star _{N}a)$ as a generalization of the
Leibniz rule (\ref{leibr01}). We omit such a tedious geometric calculus
which in not N-adapted form is similar to formula (4.8) in \cite{aschieri17}%
. In many cases, the constructions with quasi-Hopf algebras and respective
nonassociative covariant calculus can be extended symbolically in N-adapted
forms when (using notations from that work), for instance, $\partial $ $%
\rightarrow \mathbf{e,}\bigtriangledown _{z}^{\star }\rightarrow $ $\mathbf{%
\mathbf{\mathbf{\mathbf{^{\shortparallel }}}}D}_{\mathbf{z}}^{\star }$ and $%
d_{\bigtriangledown _{z}^{\star }}$ $\rightarrow \ _{\mathbf{D}%
^{\star}}^{\shortparallel }\mathbf{d.}$ In our approach, this can be
performed in a self-consistent geometric form using Convention 2 discussed
in section \ref{ssconv}.

\paragraph{N-adapted coefficient formulas for star deformed d-connections
and their h- and v-splitting : \newline
}

For a star deformed d-connection (\ref{dconhopf}), such coefficients $%
\mathbf{\mathbf{\mathbf{\mathbf{\ ^{\shortparallel }}}}D}^{\star }=\{\mathbf{%
\ ^{\shortparallel }\Gamma }_{\star \beta \gamma }^{\alpha }\in \mathcal{A}%
_{N}^{\star }\}$ are computed using the action of this d-operator on $\
^{\shortparallel }\mathbf{e}_{\alpha }$ and $\ ^{\shortparallel }\mathbf{e}%
^{\gamma }$ following above geometric rules, 
\begin{eqnarray}
\mathbf{\mathbf{\mathbf{\mathbf{\ ^{\shortparallel }}}}D}^{\star }\
^{\shortparallel }\mathbf{e}_{\alpha }&:= &\ ^{\shortparallel }\mathbf{e}%
_{\beta }\otimes _{\star _{N}}\mathbf{\ ^{\shortparallel }\Gamma }_{\star
\alpha }^{\beta }, \mbox{ where }  \label{aux300} \\
&&\mathbf{\ ^{\shortparallel }\Gamma }_{\star \beta }^{\alpha }=\mathbf{\
^{\shortparallel }\Gamma }_{\star \beta \gamma }^{\alpha }\star _{N}\
^{\shortparallel }\mathbf{e}^{\gamma }%
\mbox{ is a star
d-connection 1-form; in brief, }\mathbf{\ ^{\shortparallel }\Gamma }=\mathbf{%
\ _{\star }^{\shortparallel }\Gamma }_{\gamma }\ \star _{N}\
^{\shortparallel }\mathbf{e}^{\gamma }.  \notag
\end{eqnarray}%
Introducing $\mathbf{\mathbf{\mathbf{\mathbf{^{\shortparallel }}}}D}_{\
^{\shortparallel }\mathbf{e}_{\beta }}^{\star }:=\mathbf{\ ^{\shortparallel
}D}_{\ \beta }^{\star },$ we write%
\begin{eqnarray}
\mathbf{\mathbf{\mathbf{\mathbf{^{\shortparallel }}}}D}_{\ \alpha }^{\star
}\ ^{\shortparallel }\mathbf{e}_{\beta }:= &&\left\langle \mathbf{\mathbf{%
\mathbf{\mathbf{\ ^{\shortparallel }}}}D}^{\star }\ ^{\shortparallel }%
\mathbf{e}_{\beta },\mathbf{\mathbf{\mathbf{\mathbf{\ }}}}\ ^{\shortparallel
}\mathbf{e}_{\alpha }\right\rangle _{\star _{N}}=\left\langle \left( \
^{\shortparallel }\mathbf{e}_{\gamma }\otimes _{\star _{N}}(\mathbf{\
^{\shortparallel }\Gamma }_{\star \beta \tau }^{\gamma }\star _{N}\
^{\shortparallel }\mathbf{e}^{\tau })\right) ,\mathbf{\mathbf{\mathbf{%
\mathbf{\ }}}}\ ^{\shortparallel }\mathbf{e}_{\alpha }\right\rangle _{\star
_{N}}  \notag \\
&=&\left\langle (\ ^{\shortparallel }\mathbf{e}_{\gamma }\star _{N}\mathbf{\
^{\shortparallel }\Gamma }_{\star \beta \tau }^{\gamma })\otimes _{\star
_{N}}\ ^{\shortparallel }\mathbf{e}^{\tau },\mathbf{\mathbf{\mathbf{\mathbf{%
\ }}}}\ ^{\shortparallel }\mathbf{e}_{\alpha }\right\rangle _{\star _{N}}=\
^{\phi _{1}}(\ ^{\shortparallel }\mathbf{e}_{\gamma }\star _{N}\mathbf{\
^{\shortparallel }\Gamma }_{\star \beta \tau }^{\gamma })\star
_{N}\left\langle \mathbf{\mathbf{\mathbf{\mathbf{\ }}}}\ _{\phi
_{2}}^{\shortparallel }\mathbf{e}^{\tau },\mathbf{\mathbf{\mathbf{\mathbf{\ }%
}}}\ _{\phi _{3}}^{\shortparallel }\mathbf{e}_{\alpha }\right\rangle _{\star
_{N}}  \notag \\
&=&\ ^{\shortparallel }\mathbf{e}_{\gamma }\star _{N}\mathbf{\
^{\shortparallel }\Gamma }_{\star \beta \alpha }^{\gamma }.  \label{aux301}
\end{eqnarray}%
The d-connection coefficients $\mathbf{\ ^{\shortparallel }\Gamma }_{\star
\beta \gamma }^{\alpha }$ and respective 1-form $\mathbf{\ ^{\shortparallel
}\Gamma }_{\star }$ are used for computing N-adapted covariant derivatives:%
\begin{eqnarray}
\mathbf{\mathbf{\mathbf{\mathbf{\ ^{\shortparallel }}}}D}^{\star }\mathbf{\
^{\shortparallel }v} &=&\ ^{\shortparallel }\mathbf{e}_{\gamma }\otimes
_{\star _{N}}(\ ^{\shortparallel }\mathbf{d}\ ^{\shortparallel }\mathbf{v}%
^{\gamma }+\mathbf{\ ^{\shortparallel }\Gamma }_{\star \alpha }^{\gamma
}\star _{N}\mathbf{\ ^{\shortparallel }v}^{\alpha }),\mbox{ for
d-vector }\mathbf{\ ^{\shortparallel }v=\ ^{\shortparallel }\mathbf{e}}%
_{\alpha }\star _{N}\mathbf{\ ^{\shortparallel }v}^{\alpha },\mathbf{\
^{\shortparallel }v}^{\alpha }\in \mathcal{A}_{N}^{\star };  \notag \\
\ _{\mathbf{D}^{\star }}^{\shortparallel }\mathbf{d}(\ ^{\shortparallel }%
\mathbf{e}_{\gamma }\otimes _{\star _{N}}\ \mathbf{\mathbf{\mathbf{\mathbf{%
^{\shortparallel }}}}}\omega ^{\gamma }) &=&\ ^{\shortparallel }\mathbf{e}%
_{\gamma }\otimes _{\star _{N}}(\ ^{\shortparallel }\mathbf{d\mathbf{\mathbf{%
\mathbf{\ ^{\shortparallel }}}}}\omega ^{\gamma }+\ ^{\shortparallel }%
\mathbf{\Gamma }_{\star \alpha }^{\gamma }\star _{N}\mathbf{\mathbf{\mathbf{%
\mathbf{\ ^{\shortparallel }}}}}\omega ^{\alpha }),\mbox{
for }\mathbf{\mathbf{\mathbf{\mathbf{\ ^{\shortparallel }}}}}\omega ^{\alpha
}\in \mathbf{\ }\Omega _{\star }^{\natural }.  \label{aux302}
\end{eqnarray}

For commutative N-adapted (co) frames $^{\shortparallel }\mathbf{e}%
_{\alpha}= (\ ^{\shortparallel }\mathbf{e}_{i},\ ^{\shortparallel }e^{a})$ (%
\ref{nadapdc}), a d-connection is determined by h- and c-coefficients which
can be parameterized in the form $\ ^{\shortparallel }\mathbf{D}=\{\
^{\shortparallel }\widehat{\Gamma}_{\ \alpha \beta }^{\gamma }=(\
^{\shortparallel }L_{jk}^{i},\ ^{\shortparallel }L_{a\ k}^{\ b},\
^{\shortparallel }C_{\ j}^{i\ c},\ ^{\shortparallel }C_{ab}^{\quad c})\},$
see details in Refs. \cite{vacaru18,bubuianu18a}. In a similar form, we
define h-(c)v-components of a star deformed d-connection by introducing $\
^{\shortparallel }\mathbf{e}_{i}$ and $\ ^{\shortparallel }e^{b}$ in (\ref%
{aux301}). For instance, $\mathbf{\mathbf{\mathbf{\mathbf{^{\shortparallel }}%
}}D}_{\ k}^{\star }\ ^{\shortparallel }e^{b}=\ ^{\shortparallel }e^{a}\star
_{N}\mathbf{\ ^{\shortparallel }}L_{\star a\ k}^{\ b},$ which results in a
h- and c-decomposition 
\begin{equation}
\ \ ^{\shortparallel }\mathbf{D}^{\star }=\{\ \ ^{\shortparallel }\mathbf{%
\Gamma }_{\star \alpha \beta }^{\gamma }=(\ \ ^{\shortparallel }L_{\star
jk}^{i},\ \ ^{\shortparallel }L_{\star a\ k}^{\ b},\ \ ^{\shortparallel
}C_{\star \ j}^{i\ c},\ \ ^{\shortparallel }C_{\star ab}^{\quad c})\}.
\label{irevndecomdc}
\end{equation}%
Similar N-adapted coefficients can be computed for star deformations of $T%
\mathbf{TV}$ and $T\mathbf{T}^{\ast }\mathbf{V.}$

\paragraph{Dual star d-connections : \newline
}

For quasi-Hopf d-structures, we can also consider left actions of covariant
d-derivatives as in (\ref{aux09a}), which can be formulated in terms of the 
\textbf{dual star d-connection} $\mathbf{\mathbf{\mathbf{\mathbf{\
_{\star}^{\shortparallel }}}}D}$ which is different from $\mathbf{\mathbf{%
\mathbf{\mathbf{\ ^{\shortparallel }}}}D}^{\star }$ (\ref{dconhopf}) defined
without N-adapting to a quasi-Hopf d-algebra. Such d-connections are related
via pairing of actions on a distinguished 1-form $\mathbf{\mathbf{\mathbf{%
\mathbf{\ ^{\shortparallel }}}}}\omega $ and d-vector $\mathbf{\
^{\shortparallel }v,}$%
\begin{equation*}
\left\langle \mathbf{\mathbf{\mathbf{\mathbf{\ _{\star }^{\shortparallel }}}}%
D}\ ^{\shortparallel }\omega ,\mathbf{\mathbf{\mathbf{\mathbf{\ }}}}\
^{\shortparallel }\mathbf{v}\right\rangle _{\star _{N}}=\ _{^{\star }\mathbf{%
D}}^{\shortparallel }\mathbf{d\left\langle \mathbf{\mathbf{\mathbf{\mathbf{\
_{\star }^{\shortparallel }}}}D}\ ^{\shortparallel }\omega ,\mathbf{\mathbf{%
\mathbf{\mathbf{\ }}}}\ ^{\shortparallel }\mathbf{v}\right\rangle _{\star
_{N}}-}\left\langle \ _{\phi _{1}}^{\shortparallel }\omega ,\mathbf{\mathbf{%
\mathbf{\mathbf{\ }}}}\ _{\phi _{2}}^{\shortparallel }\mathbf{D}^{\star }%
\mathbf{(}\ _{\phi _{3}}^{\shortparallel }\mathbf{v)}\right\rangle _{\star
_{N}}.
\end{equation*}%
This defines a star d-connection on the dual N-adapted bimodule 
\begin{equation}
\mathbf{\mathbf{\mathbf{\mathbf{\ _{\star }^{\shortparallel }}}}D:\ }\Omega
_{\star }^{1}\rightarrow \mathbf{\ }\Omega _{\star }^{1}\otimes _{\star _{N}}%
\mathbf{\ }\Omega _{\star }^{1}.  \label{dualdcon}
\end{equation}%
Here we note that such a star d-connection acts from the right and from a
formal point of view we should write $(\mathbf{\mathbf{\mathbf{\mathbf{\
^{\shortparallel }}}}}\omega )\mathbf{\mathbf{\mathbf{\mathbf{\
_{\star}^{\shortparallel }}}}D}$ but following a convention from the bulk of
former works one nonassociative geometry, we use $\mathbf{\mathbf{\mathbf{%
\mathbf{\ _{\star }^{\shortparallel }}}}D}(\mathbf{\mathbf{\mathbf{\mathbf{\
^{\shortparallel }}}}}\omega ).$

The d-operator $\mathbf{\mathbf{\mathbf{\mathbf{\ _{\star }^{\shortparallel}}%
}}D}$ satisfies the left Leibniz rule,%
\begin{equation*}
\mathbf{\mathbf{\mathbf{\mathbf{\ _{\star }^{\shortparallel }}}}D}(a\star
_{N}\ ^{\shortparallel }\omega )=\ _{\phi _{1}}a\star _{N}\left( \ _{\phi
_{3}\star }^{\shortparallel }\mathbf{D}(\ _{\phi _{2}}^{\shortparallel
}\omega )\right) +\ _{^{\star }\mathbf{D}}^{\shortparallel }\mathbf{d}%
a\otimes _{\star _{N}}\mathbf{\mathbf{\mathbf{\mathbf{\ ^{\shortparallel }}}}%
}\omega ,\mbox{ for }a\in \mathcal{A}_{N}^{\star },\ ^{\shortparallel
}\omega \in \Omega _{\star }^{1},
\end{equation*}%
and has a unique lift to this type of d-connection acting on $\Omega _{\star
}^{\natural }$--bimodule $\emph{Vec}_{\star _{N}}^{\natural }=\emph{Vec}%
_{\star _{N}}$ $\otimes _{\star _{N}}\mathbf{\ }\Omega _{\star }^{\natural }$
when%
\begin{equation*}
\ _{^{\star }\mathbf{D}}^{\shortparallel }\mathbf{d:\ }\Omega _{\star
}^{\natural }\otimes _{\star _{N}}\Omega _{\star }^{1}\rightarrow \Omega
_{\star }^{\natural +1}\otimes _{\star _{N}}\Omega _{\star }^{1}.
\end{equation*}

We can compute the N-adapted coefficients of $\ _{\star }^{\shortparallel }%
\mathbf{D}= \{\ _{\star }^{\shortparallel}\Gamma _{\ \alpha \beta }^{\gamma
}\}$ if we take $\ ^{\shortparallel }\omega ^{\alpha }= \ ^{\shortparallel }%
\mathbf{e}^{\alpha }=(\ ^{\shortparallel }e^{i},\ ^{\shortparallel }\mathbf{e%
}_{b})$ and $\mathbf{\ ^{\shortparallel }v}_{\beta }=$ $^{\shortparallel }%
\mathbf{e}_{\beta }=(\ ^{\shortparallel }\mathbf{e}_{j},\ ^{\shortparallel
}e^{b})$ from (\ref{nadapdc}), when $\ _{^{\star }\mathbf{D}%
}^{\shortparallel }\mathbf{d}\left\langle \ ^{\shortparallel }\mathbf{e}%
^{\alpha },\mathbf{\mathbf{\mathbf{\mathbf{\ }}}}^{\shortparallel }\mathbf{e}%
_{\beta }\right\rangle _{\star _{N}}\mathbf{=}\ _{^{\star }\mathbf{D}%
}^{\shortparallel }\mathbf{d}\delta _{\beta }^{\alpha }.$ We can consider
the same N-adapted coefficients for the star d-connection as in (\ref{aux300}%
), when 
\begin{equation*}
\left\langle \mathbf{\mathbf{\mathbf{\mathbf{\ _{\star }^{\shortparallel }}}}%
D(}\ ^{\shortparallel }\mathbf{e}^{\alpha }),\mathbf{\mathbf{\mathbf{\mathbf{%
\ }}}}^{\shortparallel }\mathbf{e}_{\beta }\right\rangle _{\star _{N}}%
\mathbf{=-}\left\langle \ ^{\shortparallel }\mathbf{e}^{\alpha },\mathbf{%
\mathbf{\mathbf{\mathbf{\ }}}}\ ^{\shortparallel }\mathbf{D}^{\star }(%
\mathbf{\mathbf{\mathbf{\mathbf{\ }}}}^{\shortparallel }\mathbf{e}_{\beta
})\right\rangle _{\star _{N}}=-\left\langle \ ^{\shortparallel }\mathbf{e}%
^{\alpha },\mathbf{\mathbf{\mathbf{\mathbf{\ \ }}}}^{\shortparallel }\mathbf{%
e}_{\gamma }\otimes _{\star _{N}}\mathbf{\ ^{\shortparallel }\Gamma }_{\star
\beta }^{\gamma }\right\rangle _{\star _{N}}=-\mathbf{\ ^{\shortparallel
}\Gamma }_{\star \beta }^{\alpha },
\end{equation*}%
for $\mathbf{\mathbf{\mathbf{\mathbf{\ _{\star }^{\shortparallel }}}}D(}\
^{\shortparallel }\mathbf{e}^{\alpha })=-\mathbf{\ ^{\shortparallel }\Gamma }%
_{\star \beta }^{\alpha }\otimes _{\star _{N}}\ ^{\shortparallel }\mathbf{e}%
^{\beta }$ if we identify $\mathbf{\ ^{\shortparallel }\Gamma }_{\star \beta
\alpha }^{\gamma }=\mathbf{\ _{\star }^{\shortparallel }\Gamma }_{\ \beta
\alpha }^{\gamma }.$

In a similar N-adapted form, for $\ ^{\shortparallel }\omega =\
^{\shortparallel }\omega _{\alpha }\star _{N}\ ^{\shortparallel }\mathbf{e}%
^{\alpha }\in \Omega _{\star }^{1},$ we can compute%
\begin{eqnarray*}
\mathbf{\mathbf{\mathbf{\mathbf{\ _{\star }^{\shortparallel }}}}D(}\
^{\shortparallel }\omega ) &=&\mathbf{\mathbf{\mathbf{\mathbf{\ _{\star
}^{\shortparallel }}}}D(}\ ^{\shortparallel }\omega _{\alpha }\star _{N}\
^{\shortparallel }\mathbf{e}^{\alpha })=-\ ^{\shortparallel }\omega _{\alpha
}\star _{N}\mathbf{\mathbf{\mathbf{\mathbf{\ _{\star }^{\shortparallel }}}}D(%
}\ ^{\shortparallel }\mathbf{e}^{\alpha })+\ ^{\shortparallel }\mathbf{d}\
^{\shortparallel }\omega _{\alpha }\otimes _{\star _{N}}\ ^{\shortparallel }%
\mathbf{e}^{\alpha } \\
&=&\mathbf{-}\ ^{\shortparallel }\omega _{\alpha }\star _{N}\mathbf{\
^{\shortparallel }\Gamma }_{\star \beta }^{\alpha }\otimes _{\star _{N}}\
^{\shortparallel }\mathbf{e}^{\beta }+\ ^{\shortparallel }\mathbf{d}\
^{\shortparallel }\omega _{\alpha }\otimes _{\star _{N}}\ ^{\shortparallel }%
\mathbf{e}^{\alpha } \\
&=&(\ ^{\shortparallel }\mathbf{d}\ ^{\shortparallel }\omega _{\beta }-\
^{\shortparallel }\omega _{\alpha }\star _{N}\mathbf{\ ^{\shortparallel
}\Gamma }_{\star \beta }^{\alpha })\otimes _{\star _{N}}\ ^{\shortparallel }%
\mathbf{e}^{\beta }.
\end{eqnarray*}%
These formulas result in 
\begin{equation*}
\ _{^{\star }\mathbf{D}}^{\shortparallel }\mathbf{d(}\ ^{\shortparallel
}\omega _{\alpha }\otimes _{\star _{N}}{}^{\shortparallel }\mathbf{e}%
^{\alpha })=(\ ^{\shortparallel }\mathbf{d}\ ^{\shortparallel }\omega
_{\alpha }-\ ^{\shortparallel }\omega _{\beta }\mathbf{\wedge }_{\star _{N}}%
\mathbf{\ ^{\shortparallel }\Gamma }_{\star \alpha }^{\beta })\otimes
_{\star _{N}}\ ^{\shortparallel }\mathbf{e}^{\alpha }.
\end{equation*}

Finally, we note that the d-operators $\mathbf{\mathbf{\mathbf{\mathbf{\
_{\star }^{\shortparallel }}}}D}$ and $\ _{^{\star }\mathbf{D}%
}^{\shortparallel }\mathbf{d}$ can be generalized in order to include in
N-adapted form the $U\emph{Vec}^{\mathcal{F}}(\mathcal{M},N)$--equivariance
property and/or to define d-connections on d-tensor products. In abstract
geometric form, such constructions are elaborated in sections 4.2 and 4.3 of 
\cite{aschieri17}. Respective N-adapted coefficient formulas can be derived
for explicit actions on N-elongated (co) bases as we computed above.

\subsection{Star generalizations of the torsion, Riemann and Ricci d-tensors}

Considering a star deformed d-connection as a general affine (linear)
connection on corresponding phase spaces and N-adapted tensor and star
products, we can define and compute respective torsion and curvature
d-tensors which are important for elaborating nonholonomic geometric and
physical models. We show how such geometric constructions can be performed
in abstract and N-adapted coefficient forms \cite{blumenhagen16} and
emphasize possible differences when quasi-Hopf d-algebra structures \cite%
{aschieri17} are involved.

\subsubsection{Definition of d-torsions}

\paragraph{Abstract and N-adapted coefficient formulas for star d-torsions: 
\newline
}

A torsion d-tensor, d-torsion, is defined by formula\footnote{%
this is a standard definition in differential geometry by for nonassociative
generalizations contains appropriate insertions of respective $\mathcal{R}$
-matrices and star products into standard formula on $T\mathbf{V},$ when $%
\mathbf{T(X,\mathbf{Y}):=D}_{\mathbf{X}} \mathbf{Y-D}_{\mathbf{Y}}\mathbf{X-
[\mathbf{X,\mathbf{Y}}]}$; we shall use below differential forms for
defining and computing nonassociative star deformed torsion N-adapted
coefficients which is more convenient if quasi-Hopf d-algebras are considered%
}%
\begin{equation}
\mathbf{\mathbf{\mathbf{\mathbf{\ _{\flat }^{\shortparallel }}}T(\mathbf{%
\mathbf{\mathbf{\ ^{\shortparallel }}}}X,\mathbf{\mathbf{\mathbf{\
^{\shortparallel }}}Y})}}:=\mathbf{\mathbf{\mathbf{\ _{\flat
}^{\shortparallel }}}D}_{\ ^{\shortparallel }\mathbf{X}}\ ^{\shortparallel }%
\mathbf{Y-\mathbf{\mathbf{\mathbf{\ _{\flat }^{\shortparallel }}}}D}_{%
\overline{\mathcal{R}}_{N}(\ ^{\shortparallel }\mathbf{Y)}}\overline{%
\mathcal{R}}_{N}(\ ^{\shortparallel }\mathbf{X)-[\mathbf{\mathbf{\mathbf{%
\mathbf{\ ^{\shortparallel }}}}X,\mathbf{\mathbf{\mathbf{\ ^{\shortparallel }%
}}Y}}]}_{\star _{N}},  \notag
\end{equation}%
where we use a convention when, for instance, $\mathbf{\mathbf{\mathbf{%
\mathbf{\mathbf{\ ^{\shortparallel }}}}X}}$ and $\mathbf{\mathbf{\mathbf{\
^{\shortparallel }}}Y}$ are next to each other and not separated by $\
^{\shortparallel }\Gamma .$ Such formulas are consequences of (\ref{aux09a})
when, for instance, some missing terms like $\mathbf{\mathbf{\mathbf{\mathbf{%
\mathbf{\ ^{\shortparallel }}}}X}}\star _{N} \mathbf{\mathbf{\mathbf{\mathbf{%
\ ^{\shortparallel }}}}e}\ \mathbf{\mathbf{\mathbf{\mathbf{\mathbf{\
^{\shortparallel }}}Y=\mathbf{\mathbf{\ ^{\shortparallel }}}}X(\mathbf{%
\mathbf{\mathbf{\ ^{\shortparallel }}}Y)}}}$ are formally added and
substituted.

For simplicity, we elaborate only on N-adapted constructions when 
\begin{eqnarray}
\mathbf{\mathbf{\ _{\flat }^{\shortparallel }T}}^{\gamma }:= &&\
^{\shortparallel }\overrightarrow{\mathbf{\mathbf{\mathbf{D}}}}^{\wedge
_{\star }}\mathbf{\mathbf{\mathbf{\mathbf{\ ^{\shortparallel }}}}e}^{\gamma
}=(\mathbf{\mathbf{\mathbf{\mathbf{\ ^{\shortparallel }}}}e}^{\beta }\wedge
_{\star }\mathbf{\mathbf{\mathbf{\mathbf{\ ^{\shortparallel }}}}e}^{\alpha
})\star _{N}\mathbf{\mathbf{\mathbf{\mathbf{\ _{\flat }^{\shortparallel }}}T}%
}_{\ \beta \alpha }^{\gamma },\mbox{ where }  \notag \\
&&\mathbf{\ _{\flat }^{\shortparallel }\Gamma }_{\beta }^{\gamma }=\mathbf{%
\mathbf{\mathbf{\mathbf{\ _{\flat }^{\shortparallel }}}}\Gamma }_{\ \beta
\alpha }^{\gamma }\mathbf{\mathbf{\mathbf{\mathbf{\ ^{\shortparallel }}}}e}%
^{\alpha }\mbox{
and }\mathbf{\mathbf{\mathbf{\mathbf{\ _{\flat }^{\shortparallel }}}T}}_{\
\beta \alpha }^{\gamma }=\mathbf{\mathbf{\mathbf{\mathbf{\ _{\flat
}^{\shortparallel }}}}\Gamma }_{\ \beta \alpha }^{\gamma }-\mathbf{\mathbf{%
\mathbf{\mathbf{\ _{\flat }^{\shortparallel }}}}\Gamma }_{\ \alpha \beta
}^{\gamma }-\ \mathbf{\mathbf{\mathbf{\mathbf{\ ^{\shortparallel }}}}}%
w_{\alpha \beta }^{\gamma },  \label{dtorsstarform}
\end{eqnarray}%
when the anholonomy coefficients $\ \mathbf{\mathbf{\mathbf{\mathbf{\
^{\shortparallel }}}}}w_{\alpha \beta }^{\gamma }=\{\ \mathbf{\mathbf{%
\mathbf{\mathbf{\ ^{\shortparallel }}}}}C_{ib}^{\ a}=\ \mathbf{\mathbf{%
\mathbf{\mathbf{\ ^{\shortparallel }}}}}e^{a}\ \mathbf{\mathbf{\mathbf{%
\mathbf{\ ^{\shortparallel }}}}}N_{ib},\ \mathbf{\mathbf{\mathbf{\mathbf{\
^{\shortparallel }}}}}C_{jia}=\mathbf{\mathbf{\mathbf{\mathbf{\
^{\shortparallel }}}}e}_{j}\mathbf{\mathbf{\mathbf{\mathbf{\
^{\shortparallel }}}}}N_{ia}-\mathbf{\mathbf{\mathbf{\mathbf{\
^{\shortparallel }}}}e}_{i}\ \mathbf{\mathbf{\mathbf{\mathbf{\
^{\shortparallel }}}}}N_{ja}\}$ are complex phase space analogs being
defined and computed as in formulas (\ref{anhrel}) and (\ref{anhrelcd}).

Explicit computations of the d-torsion coefficients using formulas (\ref%
{aux05a}), (\ref{aux05b}) and (\ref{aux10}) result in such N-adapted
coefficients formulas:%
\begin{eqnarray*}
\ _{\flat }^{\shortparallel }\mathbf{T}(\ ^{\shortparallel }\mathbf{X},\
^{\shortparallel }\mathbf{Y})&:= &{\langle \ ^{\shortparallel }\mathbf{X}%
\otimes _{\star _{N}}\ ^{\shortparallel }\mathbf{Y},\ _{\flat
}^{\shortparallel }\mathbf{T}^{\gamma }\star _{N}\ ^{\shortparallel }\mathbf{%
e}_{\gamma }\rangle }_{\star _{N}}={\langle \ ^{\shortparallel }\mathbf{X}%
\otimes _{\star _{N}}\ ^{\shortparallel }\mathbf{Y},(\ ^{\shortparallel }%
\mathbf{e}^{\beta }\wedge _{\star }\ ^{\shortparallel }\mathbf{e}^{\alpha
})\star _{N}\ _{\flat }^{\shortparallel }\mathbf{T}_{\ \beta \alpha
}^{\gamma }\star _{N}\ ^{\shortparallel }\mathbf{e}_{\gamma }\rangle }%
_{\star _{N}} \\
&=&{\langle \ ^{\shortparallel }\mathbf{X}\otimes _{\star _{N}}\
^{\shortparallel }\mathbf{Y}-\overline{\mathcal{R}}_{N}(\ ^{\shortparallel }%
\mathbf{X})\otimes _{\star _{N}}\overline{\mathcal{R}}_{N}(\
^{\shortparallel }\mathbf{Y}),(\ ^{\shortparallel }\mathbf{e}^{\beta }\wedge
_{\star }\ ^{\shortparallel }\mathbf{e}^{\alpha })\star _{N}\ _{\flat
}^{\shortparallel }\mathbf{T}_{\ \beta \alpha }^{\gamma }\star _{N}\
^{\shortparallel }\mathbf{e}_{\gamma }\rangle }_{\star _{N}}.
\end{eqnarray*}

The N-adapted coefficients of d-torsion come out from formulas%
\begin{equation}
\mathbf{\langle \mathbf{\mathbf{\mathbf{\ _{\flat }^{\shortparallel }}}T(%
\mathbf{\mathbf{\mathbf{\mathbf{\ ^{\shortparallel }}}}e}}}_{\beta }\mathbf{%
\mathbf{,\mathbf{\mathbf{\mathbf{\ ^{\shortparallel }e}}}}}_{\alpha }\mathbf{%
\mathbf{)\mathbf{,\mathbf{\ ^{\shortparallel }}}}e}^{\gamma }\mathbf{\rangle 
}_{\star _{N}}=\mathbf{\mathbf{\mathbf{\mathbf{\ _{\flat }^{\shortparallel }}%
}T}}_{\ \beta \alpha }^{\gamma }=\mathbf{\mathbf{\mathbf{\mathbf{\ _{\flat
}^{\shortparallel }}}}\Gamma }_{\ \beta \alpha }^{\gamma }-\mathbf{\mathbf{%
\mathbf{\mathbf{\ _{\flat }^{\shortparallel }}}}\Gamma }_{\ \alpha \beta
}^{\gamma }-\ \mathbf{\mathbf{\mathbf{\mathbf{\ ^{\shortparallel }}}}}%
w_{\alpha \beta }^{\gamma }  \label{dconstar1}
\end{equation}%
as in (\ref{dtorsstarform}). Such coefficients transform into holonomic ones
considered for coordinate frames in \cite{blumenhagen16}, which in our
notations are written with left labels "$\mathbf{\mathbf{\mathbf{\mathbf{\
^{\shortparallel }}}}}$" for local coordinates on $T_{\shortparallel
}^{\ast}V$ with vanishing anholonomy coefficients and when a N-connection
structure is not distinguished, 
\begin{equation*}
\mathbf{\langle \mathbf{\mathbf{\mathbf{\ _{\flat }^{\shortparallel }}}}}T%
\mathbf{\mathbf{(\mathbf{\mathbf{\mathbf{\ ^{\shortparallel }}}}\partial }}%
_{\beta }\mathbf{\mathbf{,\mathbf{\mathbf{\mathbf{\ ^{\shortparallel }}}}%
\partial }}_{\alpha }\mathbf{\mathbf{)\mathbf{,}}}d\mathbf{\mathbf{\mathbf{%
\mathbf{\ ^{\shortparallel }}}}}u^{\gamma }\mathbf{\rangle }_{\star }=%
\mathbf{\mathbf{\mathbf{\mathbf{\ _{\flat }^{\shortparallel }}}}}T_{\ \beta
\alpha }^{\gamma }=\mathbf{\mathbf{\mathbf{\mathbf{\ _{\flat
}^{\shortparallel }}}}}\Gamma _{\ \beta \alpha }^{\gamma }-\mathbf{\mathbf{%
\mathbf{\mathbf{\ _{\flat }^{\shortparallel }}}}}\Gamma _{\ \alpha \beta
}^{\gamma }.
\end{equation*}%
Introducing in formulas (\ref{dconstar1}) the N-adapted (co) frames $%
^{\shortparallel }\mathbf{e}_{\alpha }=(\ ^{\shortparallel }\mathbf{e}_{i},\
^{\shortparallel }e^{a})$ (\ref{nadapdc}), we obtain a h-v-decomposition for
the coefficients of d-torsion, $\mathbf{\mathbf{\mathbf{\mathbf{\
_{\flat}^{\shortparallel }}}}}\mathcal{T}=\{\mathbf{\mathbf{\mathbf{\mathbf{%
\ _{\flat }^{\shortparallel }}}}T}_{\ \alpha \beta }^{\gamma }=(\ \mathbf{%
\mathbf{\mathbf{\mathbf{\ _{\flat }^{\shortparallel }}}}}T_{\ jk}^{i},\ 
\mathbf{\mathbf{\mathbf{\mathbf{\ _{\flat }^{\shortparallel }}}}}T_{\ j}^{i\
a},\mathbf{\mathbf{\mathbf{\mathbf{\ _{\flat }^{\shortparallel }}}}}T_{aji},%
\mathbf{\mathbf{\mathbf{\mathbf{\ _{\flat }^{\shortparallel }}}}}T_{a\ i}^{\
b},\mathbf{\mathbf{\mathbf{\mathbf{\ _{\flat }^{\shortparallel }}}}}T_{a\
}^{\ bc})\},$ with 
\begin{equation*}
\mathbf{\mathbf{\mathbf{\mathbf{\ _{\flat }^{\shortparallel }}}}}T_{\
jk}^{i}=\mathbf{\mathbf{\mathbf{\mathbf{\ _{\flat }^{\shortparallel }}}}}%
L_{jk}^{i}-\mathbf{\mathbf{\mathbf{\mathbf{\ _{\flat }^{\shortparallel }}}}}%
L_{kj}^{i},\mathbf{\mathbf{\mathbf{\mathbf{\ _{\flat }^{\shortparallel }}}}}%
T_{\ j}^{i\ a}=\mathbf{\mathbf{\mathbf{\mathbf{\ _{\flat }^{\shortparallel }}%
}}}C_{j}^{ia},\ \mathbf{\mathbf{\mathbf{\mathbf{\ _{\flat }^{\shortparallel }%
}}}}T_{aji}=-\ \mathbf{\mathbf{\mathbf{\mathbf{\ ^{\shortparallel }}}}}%
\Omega _{aji},\mathbf{\mathbf{\mathbf{\mathbf{\ _{\flat }^{\shortparallel }}}%
}}T_{c\ j}^{\ a}=\mathbf{\mathbf{\mathbf{\mathbf{\ _{\flat }^{\shortparallel
}}}}}L_{c\ j}^{\ a}-\mathbf{\mathbf{\mathbf{\mathbf{\ ^{\shortparallel }}}}}%
e^{a}(\ ^{\shortmid }N_{cj}),\mathbf{\mathbf{\mathbf{\mathbf{\ _{\flat
}^{\shortparallel }}}}}T_{a\ }^{\ bc}=\mathbf{\mathbf{\mathbf{\mathbf{\
_{\flat }^{\shortparallel }}}}}C_{a}^{\ bc}-\mathbf{\mathbf{\mathbf{\mathbf{%
\ _{\flat }^{\shortparallel }}}}}C_{a}^{\ cb},
\end{equation*}%
where $\ \ _{\flat }^{\shortparallel }\mathbf{D}=\{\ _{\flat
}^{\shortparallel }\widehat{\mathbf{\Gamma }}_{\ \alpha \beta }^{\gamma }=(\
_{\flat }^{\shortparallel }L_{jk}^{i},\ _{\flat }^{\shortparallel }L_{a\
k}^{\ b},\ _{\flat }^{\shortparallel }C_{\ j}^{i\ c},\mathbf{\mathbf{\mathbf{%
\mathbf{\ _{\flat }^{\shortparallel }}}}}C_{a}^{\ bc})\}.$

We extract zero torsion configurations, for instance, for the Levi-Civita
connection, (LC-connection, we shall provide explicit formulas for
respective nonassociative metric structures in next subsection), when $%
\mathbf{\mathbf{\mathbf{\mathbf{\ _{\flat }^{\shortparallel }}}}D\rightarrow 
\mathbf{\mathbf{\mathbf{\ ^{\shortparallel }}}}\nabla ,}$ if we impose the
condition%
\begin{equation}
\ _{\nabla }^{\shortparallel }\mathbf{T}_{\ \beta \alpha }^{\gamma }=0,
\label{lccond}
\end{equation}%
where we put a left label $\nabla $ to such geometric d-objects in order to
emphasize that they are defined by a LC-connection which is a linear
connection but not a d-connection because it does not preserve a
N-connection splitting of type (\ref{ncon}) under parallel transports. In
differential geometry with multiple linear connections, there are used
various types of distortion tensors. For instance, we can consider 
\begin{equation}
\mathbf{\mathbf{\mathbf{\mathbf{\ _{\flat }^{\shortparallel }}}}D=\mathbf{%
\mathbf{\mathbf{\ ^{\shortparallel }}}}\nabla +\mathbf{\mathbf{\mathbf{\
_{\flat }^{\shortparallel }}}}Z,}  \label{distdtens}
\end{equation}%
where $\mathbf{\mathbf{\mathbf{\mathbf{\ _{\flat }^{\shortparallel }}}}Z}$
is a distortion d-tensor even $\mathbf{\mathbf{\mathbf{\mathbf{\
^{\shortparallel }}}}\nabla }$ is not a d-tensor and may be not N-adapted.
In coefficient N-adapted and/or coordinate forms, the coefficients or $%
\mathbf{\mathbf{\mathbf{\mathbf{\ _{\flat }^{\shortparallel }}}}Z}$ are
certain algebraic transforms of coefficients of $\mathbf{\mathbf{\mathbf{%
\mathbf{\ _{\flat }^{\shortparallel }}}T.}}$ Explicit formulas depend on the
type of frames and coordinates we consider. In general form, we write this
through a linear functional $\mathbf{\mathbf{\mathbf{\mathbf{\ _{\flat
}^{\shortparallel }}}}Z}[\mathbf{\mathbf{\mathbf{\mathbf{\ _{\flat
}^{\shortparallel }}}}T}].$

\paragraph{Star d-torsions for d-connections and quasi-Hopf structures : 
\newline
}

Adapting geometric objects both to N-connection and Hopf algebraic
structures result in additional nonholonomy relations. This can be performed
in a well defined geometric form if the concept of d-torsion is introduced
using N-adapted differential forms with coefficients taking values in Hopf
d-algebra, see conventions from subsections \ref{ssdtqh} and \ref{ssdconstar}%
. In such an approach, a star deformed d-torsion $\ _{\star}^{\shortparallel
}\mathcal{T}\in \emph{Vec}_{\star _{N}}\otimes _{\star _{N}} \Omega _{\star
}^{2}$ of a star d-connection $\mathbf{\mathbf{\mathbf{\mathbf{\
^{\shortparallel }}}}D}^{\star }$ (\ref{dconhopf}) is defined as a N-adapted
map with identity 
\begin{equation*}
\mathbf{\langle \mathbf{\mathbf{\mathbf{\mathbf{\ ^{\shortparallel }e}}}}}%
_{\alpha }\otimes _{\star _{N}}\mathbf{\mathbf{\mathbf{\mathbf{\
^{\shortparallel }}}}e}^{\alpha },\ \mathbf{\rangle }_{\star _{N}}:\emph{Vec}%
_{\star _{N}}\rightarrow \emph{Vec}_{\star _{N}}\ 
\end{equation*}%
and expanded to any d-vector field $\mathbf{\ ^{\shortparallel }v=\
^{\shortparallel }\mathbf{e}}_{\alpha }\star _{N}\mathbf{\ ^{\shortparallel
}v}^{\alpha },\mathbf{\ ^{\shortparallel }v}^{\alpha }\in \mathcal{A}%
_{N}^{\star },$ when the associator acts trivially on any basis (co) vector $%
\mathbf{\ }$ $^{\shortparallel }\mathbf{e}_{\alpha }=(\ ^{\shortparallel }%
\mathbf{e}_{i},\ ^{\shortparallel }e^{a})$ (\ref{nadapdc}). Using $\ _{%
\mathbf{D}^{\star }}^{\shortparallel }\mathbf{d}$ $\ $(\ref{aux302}), we
define and compute%
\begin{eqnarray}
\mathbf{\mathbf{\mathbf{\mathbf{\ ^{\shortparallel }}}}}\mathcal{T}^{\star }
&:=&\ _{\mathbf{D}^{\star }}^{\shortparallel }\mathbf{d}\ (\mathbf{\
^{\shortparallel }\mathbf{e}}_{\alpha }\star _{N}\mathbf{\ ^{\shortparallel
}e}^{\alpha })=\mathbf{\mathbf{\mathbf{\mathbf{\ ^{\shortparallel }}}}D}%
^{\star }\mathbf{\ ^{\shortparallel }\mathbf{e}}_{\alpha }{\wedge _{\star
N}\ ^{\shortparallel }\mathbf{e}^{\alpha }=\ ^{\shortparallel }\mathbf{e}}%
_{\alpha }\otimes _{\star _{N}}\left( \mathbf{\ ^{\shortparallel }\Gamma }%
_{\star \beta \gamma }^{\alpha }\star _{N}(\ ^{\shortparallel }\mathbf{e}%
^{\gamma }{\wedge _{\star N}}\ ^{\shortparallel }\mathbf{e}^{\beta })\right)
\notag \\
&:=&\ {^{\shortparallel }\mathbf{e}}_{\alpha }\otimes _{\star _{N}}\mathbf{%
\mathbf{\mathbf{\mathbf{\ _{\star }^{\shortparallel }}}}}\mathcal{T}{%
^{\alpha },\mbox{ where }}  \label{dtorshform} \\
&& \mathbf{\mathbf{\mathbf{\mathbf{\ ^{\shortparallel }}}}}\mathcal{T}{%
_{\star }^{\alpha }}:= \mathbf{\ ^{\shortparallel }\Gamma }_{\star \beta
\gamma }^{\alpha }\star _{N}(\ ^{\shortparallel }\mathbf{e}^{\gamma }{\wedge
_{\star N}}\ ^{\shortparallel }\mathbf{e}^{\beta })=(\mathbf{\
^{\shortparallel }\Gamma }_{\star \beta \gamma }^{\alpha }-\mathbf{\
^{\shortparallel }\Gamma }_{\star \gamma \beta }^{\alpha }+\mathbf{\
^{\shortparallel }w}_{\star \gamma \beta }^{\alpha })\star _{N}(\
^{\shortparallel }\mathbf{e}^{\gamma }\star _{N}\ ^{\shortparallel }\mathbf{e%
}^{\beta }),\mbox{ for }  \notag \\
&& \lbrack \ ^{\shortparallel }\mathbf{e}_{\gamma },\ ^{\shortparallel }%
\mathbf{e}_{\beta }]_{\star _{N}} =\ ^{\shortparallel }\mathbf{e}_{\gamma }{%
\wedge _{\star N}}\ ^{\shortparallel }\mathbf{e}_{\beta }=\ ^{\shortparallel
}\mathbf{e}_{\gamma }\star _{N}\ ^{\shortparallel }\mathbf{e}_{\beta }-\
^{\shortparallel }\mathbf{e}_{\beta }\star _{N}\ ^{\shortparallel }\mathbf{e}%
_{\gamma }=\mathbf{\ ^{\shortparallel }w}_{\star \gamma \beta }^{\alpha
}\star _{N}\ {^{\shortparallel }\mathbf{e}}_{\alpha },  \notag
\end{eqnarray}%
see (\ref{anhrel}) and (\ref{dtorsstarform}). Similar formulas can be
introduced for the dual d-connection $\mathbf{\mathbf{\mathbf{\mathbf{\
_{\star }^{\shortparallel }}}}D}$ (\ref{dualdcon}) with inverse order of
action of d-operators which results in a dual d-torsion $\mathbf{\mathbf{%
\mathbf{\mathbf{\ _{\star }^{\shortparallel }}}}}\mathcal{T}=\{\mathbf{\
_{\star }^{\shortparallel }\Gamma }_{\beta \gamma }^{\alpha }\}.$ In this
work, we omit such nonholonomic geometric constructions which can be
performed in analogous forms.

Additionally, we can consider the star d-torsion as a N-adapted map $\mathbf{%
\mathbf{\mathbf{\mathbf{\ ^{\shortparallel }}}}}\mathcal{T}^{\star }:$ $%
\emph{Vec}_{\star _{N}}\otimes _{\star _{N}}\emph{Vec}_{\star
_{N}}\rightarrow \emph{Vec}_{\star _{N}}$, when 
\begin{equation*}
\mathbf{\mathbf{\mathbf{\mathbf{\ ^{\shortparallel }}}}}\mathcal{T}^{\star }(%
\mathbf{\ ^{\shortparallel }z},\mathbf{\ ^{\shortparallel }v})=\mathbf{%
\langle \mathbf{\mathbf{\mathbf{\ \mathbf{\mathbf{\ ^{\shortparallel }}}}}%
\mathcal{T}}}^{\star },\mathbf{\mathbf{\mathbf{\mathbf{\ }}\
^{\shortparallel }z}}\otimes _{\star _{N}}\mathbf{\mathbf{\ ^{\shortparallel
}v}\rangle }_{\star _{N}}={\ ^{\shortparallel }\mathbf{e}}_{\alpha }\star
_{N}\mathbf{\langle \mathbf{\mathbf{\mathbf{\ \mathbf{\mathbf{\
^{\shortparallel }}}}}\mathcal{T}}}_{\star }^{\alpha },\mathbf{\mathbf{%
\mathbf{\mathbf{\ }}\ ^{\shortparallel }z}}\otimes _{\star _{N}}\mathbf{%
\mathbf{\ ^{\shortparallel }v}\rangle }_{\star _{N}}.
\end{equation*}%
for some d-vectors $\mathbf{\ ^{\shortparallel }z}$ and $\mathbf{\
^{\shortparallel }v.}$ There are satisfied the conditions of right $\mathcal{%
A}_{N}^{\star }$--linearity and star antisymmetry, i.e. $\mathbf{\mathbf{%
\mathbf{\mathbf{\ ^{\shortparallel }}}}}\mathcal{T}^{\star }(\mathbf{\
^{\shortparallel }z},\mathbf{\ ^{\shortparallel }v})=-\mathbf{\mathbf{%
\mathbf{\mathbf{\ ^{\shortparallel }}}}}\mathcal{T}^{\star }(\mathbf{\
_{\intercal }^{\shortparallel }v},\mathbf{\ _{\intercal }^{\shortparallel }z}%
),$ where the \textbf{braiding} d-operator "$\mathbf{\ _{\intercal }^{{}}}$%
", \ equivalently 
\begin{equation}
\tau _{\mathcal{R}}(\mathbf{\ ^{\shortparallel }z\otimes _{\star _{N}}\
^{\shortparallel }v})=(\mathbf{\ _{\intercal }^{\shortparallel }v\otimes
_{\star _{N}}\ _{\intercal }^{\shortparallel }z}),  \label{braidop}
\end{equation}%
results in the properties 
\begin{eqnarray*}
\mathbf{\langle }\ _{\intercal }^{\shortparallel }\mathbf{e}^{\alpha
}\otimes _{\star _{N}}\mathbf{\ _{\intercal }^{\shortparallel }\mathbf{e}}%
^{\beta },\mathbf{\mathbf{\mathbf{\mathbf{\ }}\ ^{\shortparallel }z}}\otimes
_{\star _{N}}\mathbf{\mathbf{\ ^{\shortparallel }v}\rangle }_{\star _{N}} &=&%
\mathbf{\langle }\ ^{\shortparallel }\mathbf{e}^{\beta }\otimes _{\star _{N}}%
\mathbf{\ ^{\shortparallel }\mathbf{e}}^{\alpha },\mathbf{\mathbf{\mathbf{%
\mathbf{\ }}\ _{\intercal }^{\shortparallel }v}}\otimes _{\star _{N}}\mathbf{%
\mathbf{\ _{\intercal }^{\shortparallel }z}\rangle }_{\star _{N}},%
\mbox{ and
} \\
\mathbf{\langle }\ ^{\shortparallel }\mathbf{e}^{\alpha }{\wedge }_{\star
_{N}}\mathbf{\ ^{\shortparallel }\mathbf{e}}^{\beta },\mathbf{\mathbf{%
\mathbf{\mathbf{\ }}\ ^{\shortparallel }z}}\otimes _{\star _{N}}\mathbf{%
\mathbf{\ ^{\shortparallel }v}\rangle }_{\star _{N}} &=&\mathbf{\langle }\
^{\shortparallel }\mathbf{e}^{\alpha }\otimes _{\star _{N}}\mathbf{\
^{\shortparallel }\mathbf{e}}^{\beta },\mathbf{\mathbf{\mathbf{\mathbf{\ }}\
^{\shortparallel }z}}{\wedge }_{\star _{N}}\mathbf{\mathbf{\
^{\shortparallel }v}\rangle }_{\star _{N}}.
\end{eqnarray*}%
This proves the mentioned star antisymmetry but in N-adapted form (in
coordinate frames with re-defined indices, such a proof is given by formulas
(4.31) and (4.32) in \cite{aschieri17}).\footnote{%
See also sections 3.4, 3.5, 3.6, and 4.4 in that work for an abstract tensor
formalism in terms of homomorphic maps, this shows that $\mathbf{\mathbf{%
\mathbf{\mathbf{\ _{\star }^{\shortparallel }}}}}\mathcal{T}$ is a d-tensor
of type $\mathbf{\mathbf{\mathbf{\mathbf{\ _{\star }^{\shortparallel }}}}}%
\mathcal{T}\in \hom _{\star _{N}}(\emph{Vec}_{\star _{N}}{\wedge }_{\star
_{N}}\emph{Vec}_{\star _{N}},\emph{Vec}_{\star _{N}}).$ For N-connection
structures, the abstract and component formulas must be considered both for
h- and (c)v-components.}

Acting with d-operator $\mathbf{\mathbf{\mathbf{\mathbf{\ ^{\shortparallel }}%
}}}\mathcal{T}^{\star }$ (\ref{dtorshform}) on $\ ^{\shortparallel }\mathbf{e%
}_{\alpha }=(\ ^{\shortparallel }\mathbf{e}_{i},\ ^{\shortparallel }e^{a})$ (%
\ref{nadapdc}), we compute the d-torsion coefficients for $\mathbf{\
^{\shortparallel }z=}\ ^{\shortparallel }\mathbf{e}_{\alpha }$ and$\mathbf{\
^{\shortparallel }v=}\ ^{\shortparallel }\mathbf{e}_{\beta }$,%
\begin{equation}
\mathbf{\mathbf{\mathbf{\mathbf{\ ^{\shortparallel }}}}}\mathcal{T}^{\star }(%
\mathbf{\ ^{\shortparallel }e}_{\alpha },\mathbf{\ ^{\shortparallel }e}%
_{\beta })=\mathbf{\ ^{\shortparallel }e}_{\gamma }\star _{N}\mathbf{\langle 
\mathbf{\mathbf{\mathbf{\mathbf{\mathbf{\mathbf{\ ^{\shortparallel }}}}}%
\mathcal{T}}}}_{\star }^{\alpha }\mathbf{,}\ ^{\shortparallel }\mathbf{e}%
_{\alpha }{\wedge }_{\star _{N}}\mathbf{\ ^{\shortparallel }\mathbf{e}}%
_{\beta }\mathbf{\rangle }_{\star _{N}}=\mathbf{\ ^{\shortparallel }e}%
_{\gamma }\star _{N}(\mathbf{\mathbf{\mathbf{\mathbf{\ ^{\shortparallel }}}}%
\Gamma }_{\star \beta \alpha }^{\gamma }-\mathbf{\mathbf{\mathbf{\mathbf{\
^{\shortparallel }}}}\Gamma }_{\star \alpha \beta }^{\gamma }+\ \mathbf{%
\mathbf{\mathbf{\mathbf{\ ^{\shortparallel }}}}}w_{\star \alpha \beta
}^{\gamma })=\mathbf{\ ^{\shortparallel }e}_{\gamma }\star _{N}\mathbf{%
\mathbf{\mathbf{\mathbf{\ ^{\shortparallel }}}}T}_{\star \beta \alpha
}^{\gamma },  \label{aux303}
\end{equation}%
see (\ref{dtorshform}). The h- and c-decomposition of d-torsion is stated in
the form \newline
$\mathbf{\mathbf{\mathbf{\mathbf{\ \ ^{\shortparallel }}}}}\mathcal{T}%
^{\star }=\{\mathbf{\mathbf{\mathbf{\mathbf{\ ^{\shortparallel }}}}T}_{\star
\alpha \beta }^{\gamma }=(\ \mathbf{\mathbf{\mathbf{\mathbf{\
^{\shortparallel }}}}}T_{\star jk}^{i},\ \mathbf{\mathbf{\mathbf{\mathbf{\ \
^{\shortparallel }}}}}T_{\star \ j}^{i\ a},\ \mathbf{\mathbf{\mathbf{\mathbf{%
\ \ ^{\shortparallel }}}}}T_{\star aji},\ \mathbf{\mathbf{\mathbf{\mathbf{\
^{\shortparallel }}}}}T_{\star a\ i}^{\ b},\ \mathbf{\mathbf{\mathbf{\mathbf{%
\ \ ^{\shortparallel }}}}}T_{\star a\ }^{\ bc})\},$ by such N-adapted
coefficients 
\begin{eqnarray}
\mathbf{\mathbf{\mathbf{\mathbf{\ \ ^{\shortparallel }}}}}T_{\star \ jk}^{i}
&=&\mathbf{\mathbf{\mathbf{\mathbf{\ \ ^{\shortparallel }}}}}L_{\star
jk}^{i}-\mathbf{\mathbf{\mathbf{\mathbf{\ \ ^{\shortparallel }}}}}L_{\star
kj}^{i},\mathbf{\mathbf{\mathbf{\mathbf{\ \ ^{\shortparallel }}}}}T_{\star \
j}^{i\ a}=\mathbf{\mathbf{\mathbf{\mathbf{\ ^{\shortparallel }}}}}C_{\star
j}^{ia},\ \mathbf{\mathbf{\mathbf{\mathbf{\ ^{\shortparallel }}}}}T_{\star
aji}=-\ \mathbf{\mathbf{\mathbf{\mathbf{\ \ ^{\shortparallel }}}}}\Omega
_{\star aji},  \notag \\
\mathbf{\mathbf{\mathbf{\mathbf{\ \ ^{\shortparallel }}}}}T_{\star c\ j}^{\
a} &=&\mathbf{\mathbf{\mathbf{\mathbf{\ \ ^{\shortparallel }}}}}L_{\star c\
j}^{\ a}-\mathbf{\mathbf{\mathbf{\mathbf{\ ^{\shortparallel }}}}}e^{a}(%
\mathbf{\mathbf{\mathbf{\mathbf{\ ^{\shortparallel }}}}}N_{\star cj}),%
\mathbf{\mathbf{\mathbf{\mathbf{\ \ ^{\shortparallel }}}}}T_{\star a\ }^{\
bc}=\mathbf{\mathbf{\mathbf{\mathbf{\ ^{\shortparallel }}}}}C_{\star a}^{\
bc}-\mathbf{\mathbf{\mathbf{\mathbf{\ \ ^{\shortparallel }}}}}C_{\star a}^{\
cb},  \label{dtorsnonassoc}
\end{eqnarray}%
where $\ \ ^{\shortparallel }\mathbf{D}^{\star }$ is determined by
h-c-components as in formulas (\ref{irevndecomdc}).

The analogs of zero torsion conditions (\ref{lccond}) and distortion
relations (\ref{distdtens}) are derived respectively, 
\begin{equation*}
\mathbf{\mathbf{\mathbf{\mathbf{\ _{\triangledown }^{\shortparallel }}}}T}%
_{\star \beta \alpha }^{\gamma } = \mathbf{\mathbf{\mathbf{\mathbf{\
_{\triangledown }^{\shortparallel }}}}\Gamma }_{\star \beta \alpha }^{\gamma
}-\mathbf{\mathbf{\mathbf{\mathbf{\ _{\triangledown }^{\shortparallel }}}}%
\Gamma }_{\star \alpha \beta }^{\gamma }+\ \mathbf{\mathbf{\mathbf{\mathbf{\
^{\shortparallel }}}}}w_{\star \alpha \beta }^{\gamma }=0 \mbox{ and } 
\mathbf{\mathbf{\mathbf{\mathbf{\ ^{\shortparallel }}}}D}^{\star } =\mathbf{%
\mathbf{\mathbf{\ ^{\shortparallel }}}\nabla }^{\star }\mathbf{+\mathbf{%
\mathbf{\mathbf{\ ^{\shortparallel }}}}Z}^{\star }.
\end{equation*}%
With respect to a local coordinate basis $\partial _{\tau }$, we can write $%
\mathbf{\mathbf{\mathbf{\mathbf{\ _{\triangledown }^{\shortparallel }}}}T}%
_{\star \beta \alpha }^{\gamma }=\mathbf{\mathbf{\mathbf{\mathbf{\
_{\triangledown }^{\shortparallel }}}}\Gamma }_{\star \beta \alpha }^{\gamma
}-\mathbf{\mathbf{\mathbf{\mathbf{\ _{\triangledown }^{\shortparallel }}}}%
\Gamma }_{\star \alpha \beta }^{\gamma }=0$, which reflects the fact that
such a linear connection $\mathbf{\mathbf{\mathbf{\mathbf{\ ^{\shortparallel}%
}}}\nabla }^{\star }$ is not a d-connection because, in general, it is not
adapted to a N-connection splitting.

Using $\mathbf{^{\shortparallel }z=}\ ^{\shortparallel }\mathbf{e}_{\alpha
}, $ $\mathbf{\ ^{\shortparallel }v=}\ ^{\shortparallel }\mathbf{e}_{\beta }$
and $\mathbf{\mathbf{\mathbf{\mathbf{\ ^{\shortparallel }}}}D}_{\intercal
\alpha }^{\star }=\mathbf{\mathbf{\mathbf{\mathbf{\ ^{\shortparallel }}}}D}_{%
\mathbf{_{\intercal }^{\shortparallel }e}_{\alpha }}^{\star },$ we can
calculate 
\begin{equation*}
\mathbf{\mathbf{\mathbf{\mathbf{\ ^{\shortparallel }}}}}\mathcal{T}^{\star }(%
\mathbf{\ ^{\shortparallel }e}_{\alpha },\mathbf{\ ^{\shortparallel }e}%
_{\beta })=\mathbf{\mathbf{\mathbf{\mathbf{\ ^{\shortparallel }}}}D}_{\beta
}^{\star }\mathbf{\ ^{\shortparallel }e}_{\alpha }-\mathbf{\mathbf{\mathbf{%
\mathbf{\ ^{\shortparallel }}}}D}_{\intercal \alpha }^{\star }\mathbf{\
_{\intercal }^{\shortparallel }e}_{\beta }+[\mathbf{\ ^{\shortparallel }e}%
_{\alpha },\mathbf{\ ^{\shortparallel }e}_{\beta }]_{\star _{N}}=\mathbf{\
^{\shortparallel }e}_{\gamma }\star _{N}(\mathbf{\ ^{\shortparallel }\Gamma }%
_{\star \alpha \beta }^{\gamma }-\mathbf{\ ^{\shortparallel }\Gamma }_{\star
\beta \alpha }^{\gamma }+\mathbf{\ ^{\shortparallel }}w_{\star \alpha \beta
}^{\gamma }).
\end{equation*}%
Such formulas are equivalent to (\ref{aux303}) and provide a proof for the
1st Cartan structure equation which can be written in abstract form for
covariant d-derivatives and general two d-vectors $\mathbf{^{\shortparallel}z%
}$ and$\mathbf{\ ^{\shortparallel }v,}$%
\begin{equation*}
\mathbf{\mathbf{\mathbf{\mathbf{\ ^{\shortparallel }}}}}\mathcal{T}^{\star }(%
\mathbf{\ ^{\shortparallel }z},\mathbf{\ ^{\shortparallel }v})=\ _{\phi
_{1}}^{\shortparallel }\mathbf{D}_{\mathbf{\ }_{\phi _{2}}^{\shortparallel }%
\mathbf{v}}^{\star }\mathbf{\ }_{\phi _{3}}^{\shortparallel }\mathbf{z}-\
_{\phi _{1}}^{\shortparallel }\mathbf{D}_{\mathbf{\ }_{\phi _{2}\intercal
}^{\shortparallel }\mathbf{z}}^{\star }\ _{\phi _{3}\intercal
}^{\shortparallel }\mathbf{v}+[\mathbf{\ ^{\shortparallel }z},\mathbf{\
^{\shortparallel }v}]_{\star _{N}}.
\end{equation*}%
This d-operator is $U\emph{Vec}^{\mathcal{F}}(\mathcal{M},N)$-equivariant
and can be written in terms of N-adapted associative composition $\bullet ,$
pairing and braiding $\tau _{\mathcal{R}}$ (\ref{braidop}),%
\begin{equation*}
\mathbf{\mathbf{\mathbf{\mathbf{\ ^{\shortparallel }}}}}\mathcal{T}^{\star
}(,)=\mathbf{\langle }\ ,\mathbf{\mathbf{\mathbf{\mathbf{\ }}}\rangle }%
_{\star _{N}}\bullet \left( \mathbf{\mathbf{\mathbf{\mathbf{\
^{\shortparallel }}}}D}^{\star }\otimes _{\star _{N}}id\right) -\mathbf{%
\langle }\ ,\mathbf{\mathbf{\mathbf{\mathbf{\ }}}\rangle }_{\star
_{N}}\bullet \left( \mathbf{\mathbf{\mathbf{\mathbf{\ ^{\shortparallel }}}}D}%
^{\star }\otimes _{\star _{N}}id\right) \bullet \tau _{\mathcal{R}%
}+[,]_{\star _{N}},
\end{equation*}%
where $\left( \mathbf{\mathbf{\mathbf{\mathbf{\ ^{\shortparallel }}}}D}%
^{\star }\otimes _{\star _{N}}id\right) (\mathbf{\ ^{\shortparallel
}z\otimes _{\star _{N}}\ ^{\shortparallel }v})=\ _{\overline{\phi }%
_{1}}^{\shortparallel }\mathbf{D}^{\star }(\ _{\overline{\phi }%
_{2}}^{\shortparallel }\mathbf{z)}\otimes _{\star _{N}}\ _{\overline{\phi }%
_{3}}^{\shortparallel }\mathbf{v,}$ $\mathbf{\mathbf{\mathbf{\mathbf{\
^{\shortparallel }}}}}\mathcal{T}^{\star }(\mathbf{\ ^{\shortparallel }z}%
\star _{N}a,\mathbf{\ ^{\shortparallel }v})=\mathbf{\mathbf{\mathbf{\mathbf{%
\ ^{\shortparallel }}}}}\mathcal{T}^{\star }(\mathbf{\ }_{\phi
_{1}}^{\shortparallel }\mathbf{z},_{\phi _{2}}a\star _{N}\ _{\phi
_{2}}^{\shortparallel }\mathbf{v}),$ for $a\in \mathcal{A}_{N}^{\star },$
and $\mathbf{\mathbf{\mathbf{\mathbf{\ ^{\shortparallel }}}}}\mathcal{T}%
^{\star }(\mathbf{\ ^{\shortparallel }z},\mathbf{\ ^{\shortparallel }v})=%
\mathbf{\mathbf{\mathbf{\mathbf{\ ^{\shortparallel }}}}}\mathcal{T}^{\star }(%
\mathbf{\ ^{\shortparallel }z\otimes _{\star _{N}}\ ^{\shortparallel }v}).$

\subsubsection{Riemann (curvature) d-tensors}

We proceed with star generalizations of the Riemann curvature d-tensor for
d-connections. The formulas depend on the type of prescribed nonholonomic
structure and how the geometric constructions are adapted to N-connections
and which conventions are used for left and right covariant derivatives.

\paragraph{Definition of curvature d-tensor for covariant d-derivatives: 
\newline
}

For a d-connection $\ _{\flat }^{\shortparallel }\mathbf{D}$, we define the
curvature $\ _{\flat }^{\shortparallel }\mathbf{R}$ considering that it
matches such geometric conditions:

\begin{enumerate}
\item It is a d-tensor\footnote{%
we chose an inverse sign to that used in \cite{blumenhagen16} in order to
generalize our former works on nonholonomic gravity and
Finsler-Lagrange-Hamilton models} 
\begin{equation}
\mathbf{\mathbf{\mathbf{\mathbf{\ _{\flat }^{\shortparallel }}}R(\mathbf{%
\mathbf{\mathbf{\mathbf{\ ^{\shortparallel }}}}}}X\mathbf{,\mathbf{\mathbf{%
\mathbf{\ ^{\shortparallel }Y}}}}},\mathbf{\mathbf{\mathbf{\mathbf{\mathbf{\
^{\shortparallel }A}}})=(\mathbf{\ _{\flat }^{\shortparallel }}}D}_{\
^{\shortparallel }\mathbf{X}}\bullet \mathbf{\mathbf{\mathbf{\ _{\flat
}^{\shortparallel }}}D}_{\ ^{\shortparallel }\mathbf{Y}}-\mathbf{\mathbf{%
\mathbf{\ _{\flat }^{\shortparallel }}}D}_{\overline{\mathcal{R}}_{N}(\
^{\shortparallel }\mathbf{Y)}}\bullet \mathbf{\mathbf{\mathbf{\ _{\flat
}^{\shortparallel }}}D}_{\overline{\mathcal{R}}_{N}(\ ^{\shortparallel }%
\mathbf{X)}}-\mathbf{\mathbf{\mathbf{\ _{\flat }^{\shortparallel }}}D}_{[\
^{\shortparallel }\mathbf{X,}\ ^{\shortparallel }\mathbf{Y]}_{\star _{N}}})\
^{\shortparallel }\mathbf{A}  \label{starriemdt1}
\end{equation}%
where $\bullet $ extends on $T_{\shortparallel }^{\ast }\mathbf{V}$ and
respective star products the composition (\ref{compcom}) in certain forms
which are compatible to directional covariant derivatives;

\item this is a d-operator of type%
\begin{equation}
\mathbf{\mathbf{\mathbf{\mathbf{\ _{\flat }^{\shortparallel }}}R(\mathbf{%
\mathbf{\mathbf{\mathbf{\ ^{\shortparallel }}}}}}X\mathbf{,\mathbf{\mathbf{%
\mathbf{\ ^{\shortparallel }Y}}}}},\mathbf{\mathbf{\mathbf{\mathbf{\
^{\shortparallel }A}}}):=\langle }\left( \mathbf{\mathbf{\mathbf{\mathbf{%
\mathbf{\ ^{\shortparallel }}}}X\otimes }}_{\star _{N}}\mathbf{\mathbf{%
\mathbf{\mathbf{\mathbf{\ ^{\shortparallel }}}Y}}}\right) \mathbf{\mathbf{%
\otimes }}_{\star _{N}}\mathbf{\mathbf{\mathbf{\mathbf{\mathbf{\
^{\shortparallel }A,}\ _{\flat }^{\shortparallel }}}R}\rangle }_{\star _{N}}=%
\mathbf{\langle }\ _{\phi }\left( \mathbf{\mathbf{\mathbf{\mathbf{\mathbf{\
^{\shortparallel }}}}X\otimes }}_{\star _{N}}\mathbf{\mathbf{\mathbf{\mathbf{%
\mathbf{\ ^{\shortparallel }}}Y}}}\right) ,\ _{\phi }^{\shortparallel }%
\mathbf{\mathbf{\mathbf{\mathbf{\mathbf{A}}}}}^{\alpha }\star _{N}\ \ _{\phi
\flat }^{\shortparallel }\mathbf{\mathbf{R}}_{\ \alpha }^{\beta }\star _{N}%
\mathbf{\mathbf{\mathbf{\mathbf{\ ^{\shortparallel }}}}e}_{\beta }\mathbf{%
\rangle }_{\star _{N}},  \label{starriemdt2}
\end{equation}

\item it is also a 2-form with matrix coefficients which can be also
constructed using the d-connection 1-form $\ ^{\shortparallel }\Gamma =\{\
^{\shortparallel }\Gamma _{\ \beta }^{\gamma }\}$ (\ref{dtorsstarform}), 
\begin{eqnarray}
\mathbf{\mathbf{\mathbf{\mathbf{\ _{\flat }^{\shortparallel }}}R}} &=&d%
\mathbf{\mathbf{\mathbf{\mathbf{\ _{\flat }^{\shortparallel }}}}\Gamma -%
\mathbf{\mathbf{\mathbf{\ _{\flat }^{\shortparallel }}}}\Gamma }\wedge
_{\star }\mathbf{\mathbf{\mathbf{\mathbf{\ _{\flat }^{\shortparallel }}}}%
\Gamma }  \label{starriemdt3} \\
&=&\mathbf{\mathbf{\mathbf{\mathbf{\ ^{\shortparallel }}}}e}^{\alpha }\star
_{N}\ _{\flat }^{\shortparallel }\mathbf{\mathbf{R}}_{\ \alpha }^{\beta
}\star _{N}\mathbf{\mathbf{\mathbf{\mathbf{\ ^{\shortparallel }}}}e}_{\beta
},\mbox{
where }\ _{\flat }^{\shortparallel }\mathbf{\mathbf{R}}_{\ \alpha }^{\beta
}=\ _{\flat }^{\shortparallel }\overrightarrow{\mathbf{\mathbf{\mathbf{D}}}}%
^{\wedge _{\star }}\mathbf{\mathbf{\mathbf{\mathbf{\ _{\flat
}^{\shortparallel }}}}\Gamma }_{\ \alpha }^{\beta }=d\mathbf{\mathbf{\mathbf{%
\mathbf{\ _{\flat }^{\shortparallel }}}}\Gamma }_{\ \alpha }^{\beta }-%
\mathbf{\mathbf{\mathbf{\mathbf{\ _{\flat }^{\shortparallel }}}}\Gamma }_{\
\alpha }^{\gamma }\wedge _{\star }\mathbf{\mathbf{\mathbf{\mathbf{\ _{\flat
}^{\shortparallel }}}}\Gamma }_{\ \gamma }^{\beta },  \notag \\
&&\mbox{ for }\mathbf{\mathbf{\mathbf{\mathbf{\ _{\flat }^{\shortparallel }}}%
}\Gamma }\mathbf{=\mathbf{\mathbf{\mathbf{\ ^{\shortparallel }}}}e}^{\beta }%
\mathbf{\mathbf{\otimes }}_{\star _{N}}\mathbf{\mathbf{\mathbf{\mathbf{\
^{\shortparallel }}}}e}^{\alpha }\star _{N}\mathbf{\mathbf{\mathbf{\mathbf{\
_{\flat }^{\shortparallel }}}}\Gamma }_{\ \beta \alpha }^{\gamma }\star _{N}%
\mathbf{\mathbf{\mathbf{\mathbf{\ ^{\shortparallel }}}}e}_{\gamma }%
\mbox{
and }\mathbf{\mathbf{\mathbf{\mathbf{\ _{\flat }^{\shortparallel }}}}\Gamma }%
_{\ \alpha }^{\beta }:=\mathbf{\mathbf{\mathbf{\mathbf{\ ^{\shortparallel }}}%
}e}^{\gamma }\star _{N}\mathbf{\mathbf{\mathbf{\mathbf{\ _{\flat
}^{\shortparallel }}}}\Gamma }_{\ \alpha \gamma }^{\beta }.  \notag
\end{eqnarray}
\end{enumerate}

Let us explain how the composition $\bullet $ from point 1 can be
constructed explicitly in a form which is compatible with definitions of
star curvature d-tensor 2 and 3. We note that $\mathbf{\mathbf{\mathbf{\
^{\shortparallel }}}D}_{\ ^{\shortparallel }\mathbf{X}}\mathbf{\mathbf{%
\mathbf{\mathbf{\mathbf{\ ^{\shortparallel }Y}}}}}$ encodes two N-adapted
and star involving consecutive d-operators: The first action can be written
from the right as $\ ^{\shortparallel}\overleftarrow{\mathbf{D}}=\mathbf{%
\mathbf{\mathbf{\mathbf{\ ^{\shortparallel }}}}e+\mathbf{\mathbf{\mathbf{\
_{\flat }^{\shortparallel }}}}\Gamma }$ and the second one consists a
contraction with $\mathbf{\mathbf{\mathbf{\mathbf{\mathbf{\ ^{\shortparallel
}X}}}}}$ from the left (we denote it $\mathbf{\mathbf{\mathbf{\mathbf{%
\mathbf{\ }}}}}i_{\mathbf{\mathbf{\mathbf{\mathbf{\mathbf{\ ^{\shortparallel
}X}}}}}}$). We shall apply also an associator $\phi $ in order to construct
brackets of $\mathbf{\mathbf{\mathbf{\mathbf{\mathbf{\ ^{\shortparallel }X}}}%
}}$ and $\ ^{\shortparallel }\mathbf{Y}$. Following such conventions, we
express%
\begin{eqnarray*}
\ _{\flat }^{\shortparallel }\mathbf{D}_{\ ^{\shortparallel }\mathbf{X}}\
^{\shortparallel }\mathbf{Y}:= &&\ (i_{\ ^{\shortparallel }\mathbf{X}})\
^{\shortparallel }\overleftarrow{\mathbf{D}}\ ^{\shortparallel }\mathbf{Y}=\
_{\phi }i_{\ ^{\shortparallel }\mathbf{X}}(\ _{\phi }^{\shortparallel }%
\mathbf{Y}\ _{\phi }^{\shortparallel }\overleftarrow{\mathbf{D}})=\ i_{\
^{\shortparallel }\mathbf{X}}(\ ^{\shortparallel }\mathbf{e}\
^{\shortparallel }\mathbf{Y})+\ _{\phi }i_{\ ^{\shortparallel }\mathbf{X}}(\
_{\phi }^{\shortparallel }\mathbf{Y}\star _{N}\ _{\phi \flat
}^{\shortparallel }\Gamma ) \\
&=&\ ^{\shortparallel }\mathbf{X}\star _{N}(\ ^{\shortparallel }\mathbf{e}\
^{\shortparallel }\mathbf{Y})+\left( \ ^{\shortparallel }\mathbf{X}\otimes
_{\star _{N}}\ ^{\shortparallel }\mathbf{Y}\right) \star _{N}\ _{\flat
}^{\shortparallel }\Gamma
\end{eqnarray*}%
and prescribe that the composition $\bullet $ is a composition of the left
and right acting d-operators of type%
\begin{equation*}
\left( \mathbf{\mathbf{\mathbf{\ _{\flat }^{\shortparallel }}}D}_{\
^{\shortparallel }\mathbf{X}}\bullet \mathbf{\mathbf{\mathbf{\ _{\flat
}^{\shortparallel }}}D}_{\ ^{\shortparallel }\mathbf{Y}}\right) \mathbf{%
\mathbf{\mathbf{\mathbf{\mathbf{\ ^{\shortparallel }A=}}}}}\left( \left( \
^{\shortparallel }i_{\ ^{\shortparallel }\mathbf{X}}\bullet \
^{\shortparallel }i_{\ ^{\shortparallel }\mathbf{Y}}\right) \mathbf{\mathbf{%
\mathbf{\mathbf{\mathbf{\ ^{\shortparallel }A}}}}}\right) \left( \mathbf{%
\mathbf{\mathbf{\ ^{\shortparallel }}}}\overleftarrow{\mathbf{D}}\bullet 
\mathbf{\mathbf{\mathbf{\ ^{\shortparallel }}}}\overleftarrow{\mathbf{D}}%
\right) .
\end{equation*}

To compute coefficient formulas of star Riemann d-tensors of type (\ref%
{starriemdt1}) we have to shift all basis d-vectors in (\ref{starriemdt2})
using formulas (\ref{aux04}). For the operator $\partial $ (\ref{aux02})
this is described in details in Appendix B of \cite{blumenhagen16}. That
proof can be elaborated in a similar abstract geometric and/or in N-adapted
coefficient forms for d-connections.\footnote{%
We do not provide formulas for Bianchi type identities for star torsion and
curvature d-tensors of d-connection $\mathbf{\mathbf{\mathbf{\mathbf{\
^{\shortparallel }D}}}}$ \ (\ref{dcon}) which are considered in holonomic
nonassociative forms in \cite{blumenhagen16} and for commutative
nonholonomic geometries, for instance, in \cite{vacaru18,bubuianu18a}; for
quasi-Hopf d-structures with d-connection \ $\mathbf{\mathbf{\mathbf{\mathbf{%
\ ^{\shortparallel }}}}D}^{\star }$ (\ref{dconhopf}) we derive Bianchi
identities in next subsection.}

\paragraph{N-adapted coefficient formulas of the curvature d-tensor: \newline
}

Here we provide formulas for the curvature d-form (\ref{starriemdt3}) and
some important coefficients in N-adapted bases,%
\begin{equation*}
\mathbf{\mathbf{\mathbf{\mathbf{\ _{\flat }^{\shortparallel }}}R=\mathbf{%
\mathbf{\ ^{\shortparallel }}}}e}^{\gamma }\mathbf{\mathbf{\otimes }}_{\star
_{N}}\mathbf{\mathbf{\mathbf{\mathbf{\ ^{\shortparallel }}}}e}^{\alpha
}\wedge _{\star }\mathbf{\mathbf{\mathbf{\mathbf{\ ^{\shortparallel }}}}e}%
^{\beta }\star _{N}\left( \mathbf{\mathbf{\mathbf{\mathbf{\ ^{\shortparallel
}}}}e}_{\alpha }\mathbf{\mathbf{\mathbf{\mathbf{\ _{\flat }^{\shortparallel }%
}}}\Gamma }_{\ \gamma \beta }^{\nu }-\mathbf{\mathbf{\mathbf{\mathbf{\
_{\flat }^{\shortparallel }}}}\Gamma }_{\ \gamma \beta }^{\mu }\star _{N}%
\mathbf{\mathbf{\mathbf{\mathbf{\ _{\flat }^{\shortparallel }}}}\Gamma }_{\
\mu \beta }^{\nu }-\ _{N}^{\shortparallel }F_{\alpha \quad }^{\quad \tau
\varepsilon }\mathbf{\mathbf{\mathbf{\mathbf{\ ^{\shortparallel }}}}e}_{\tau
}\mathbf{\mathbf{\mathbf{\mathbf{\ _{\flat }^{\shortparallel }}}}\Gamma }_{\
\gamma \beta }^{\mu }\star _{N}\mathbf{\mathbf{\mathbf{\mathbf{\ _{\flat
}^{\shortparallel }}}}\Gamma }_{\ \mu \varepsilon }^{\nu }\right) \star _{N}%
\mathbf{\mathbf{\mathbf{\mathbf{\ ^{\shortparallel }}}}e}_{\nu }
\end{equation*}%
when 
\begin{eqnarray}
\ _{\flat }^{\shortparallel }\mathbf{R}_{\ \tau \alpha \beta }^{\gamma }&:=&
\langle \ _{\flat }^{\shortparallel }\mathbf{R}(\ ^{\shortparallel }\mathbf{%
\ e}_{\beta },\ ^{\shortparallel }\mathbf{e}_{\alpha },\ ^{\shortparallel }%
\mathbf{e}_{\tau }),\mathbf{\mathbf{\mathbf{\mathbf{\ ^{\shortparallel }}}}e}%
^{\gamma }\mathbf{\mathbf{\rangle }}_{\star _{N}}=\mathbf{\mathbf{\mathbf{%
\mathbf{\ ^{\shortparallel }}}}e}_{\beta }\mathbf{\mathbf{\mathbf{\mathbf{\
_{\flat }^{\shortparallel }}}}\Gamma }_{\ \tau \alpha }^{\gamma }-\mathbf{%
\mathbf{\mathbf{\mathbf{\ ^{\shortparallel }}}}e}_{\alpha }\mathbf{\mathbf{%
\mathbf{\mathbf{\ _{\flat }^{\shortparallel }}}}\Gamma }_{\ \tau \beta
}^{\gamma }+\mathbf{\mathbf{\mathbf{\mathbf{\ _{\flat }^{\shortparallel }}}}%
\Gamma }_{\ \tau \alpha }^{\mu }\star _{N}\mathbf{\mathbf{\mathbf{\mathbf{\
_{\flat }^{\shortparallel }}}}\Gamma }_{\ \mu \beta }^{\gamma }-\mathbf{%
\mathbf{\mathbf{\mathbf{\ _{\flat }^{\shortparallel }}}}\Gamma }_{\ \tau
\beta }^{\mu }\star _{N}\mathbf{\mathbf{\mathbf{\mathbf{\ _{\flat
}^{\shortparallel }}}}\Gamma }_{\ \mu \alpha }^{\gamma }  \notag \\
&&+\mathbf{\mathbf{\mathbf{\mathbf{\ _{\flat }^{\shortparallel }}}}\Gamma }%
_{\ \tau \nu }^{\gamma }\star _{N}\mathbf{\mathbf{\mathbf{\mathbf{\
^{\shortparallel }}}}}w_{\ \alpha \beta }^{\nu }+\ _{N}^{\shortparallel
}F_{\alpha \quad }^{\quad \nu \varepsilon }\mathbf{\mathbf{\mathbf{\mathbf{\
^{\shortparallel }}}}e}_{\nu }\mathbf{\mathbf{\mathbf{\mathbf{\ _{\flat
}^{\shortparallel }}}}\Gamma }_{\ \tau \beta }^{\mu }\star _{N}\mathbf{%
\mathbf{\mathbf{\mathbf{\ _{\flat }^{\shortparallel }}}}\Gamma }_{\ \mu
\varepsilon }^{\gamma }-\ _{N}^{\shortparallel }F_{\beta \quad }^{\quad \nu
\varepsilon }\mathbf{\mathbf{\mathbf{\mathbf{\ ^{\shortparallel }}}}e}_{\nu }%
\mathbf{\mathbf{\mathbf{\mathbf{\ _{\flat }^{\shortparallel }}}}\Gamma }_{\
\tau \alpha }^{\mu }\star _{N}\mathbf{\mathbf{\mathbf{\mathbf{\ _{\flat
}^{\shortparallel }}}}\Gamma }_{\ \mu \varepsilon }^{\gamma }.  \label{riem1}
\end{eqnarray}%
In these formulas, the anholonomy coefficients are computed as in (\ref%
{anhrelcd}) but on $T_{\shortparallel }^{\ast }\mathbf{V,}$ when 
\begin{equation*}
\ \mathbf{\mathbf{\mathbf{\mathbf{\ ^{\shortparallel }}}}}w_{\alpha \beta
}^{\gamma }=\{\ \mathbf{\mathbf{\mathbf{\mathbf{\ ^{\shortparallel }}}}}%
w_{ib}^{\ a}=\ \mathbf{\mathbf{\mathbf{\mathbf{\ ^{\shortparallel }}}}}%
e^{a}\ \mathbf{\mathbf{\mathbf{\mathbf{\ ^{\shortparallel }}}}}N_{ib},\ 
\mathbf{\mathbf{\mathbf{\mathbf{\ ^{\shortparallel }}}}}w_{jia}=\ \mathbf{%
\mathbf{\mathbf{\mathbf{\ ^{\shortparallel }}}}e}_{j}\ \mathbf{\mathbf{%
\mathbf{\mathbf{\ ^{\shortparallel }}}}}N_{ia}-\ \mathbf{\mathbf{\mathbf{%
\mathbf{\ ^{\shortparallel }}}}e}_{i}\ \mathbf{\mathbf{\mathbf{\mathbf{\
^{\shortparallel }}}}}N_{ja}\}.
\end{equation*}%
Using N-adapted (co) frames $^{\shortparallel }\mathbf{e}_{\alpha }=(\
^{\shortparallel }\mathbf{e}_{i},\ ^{\shortparallel }e^{a})$ (\ref{nadapdc})
in (\ref{riem1}), we can compute the h-v-components of the Riemannian
d-tensor for a star d-connection $\ _{\flat }^{\shortparallel }\mathbf{D.}$
We provide such a calculus in next subsection, with an additional adapting
to quasi Hopf d-structures which is used for formulating nonassociative
generalizations of the Einstein equations in section \ref{sec4}.

\subsubsection{Star deformed curvature of quasi-Hopf N-adapted structures}

We extend for d-connections and N-adapted frames the definitions and
formulas the star deformed curvature d-tensors. For holonomic basic
structures such constructions are provided in section 4.5 of \cite%
{aschieri17}.

\paragraph{Abstract definition of the curvature d-tensor and N-adapted
differential forms: \newline
}

We define the curvature of a N-adapted covariant derivative $\ _{\mathbf{D}%
^{\star}} ^{\shortparallel }\mathbf{d}$ (\ref{dcov01}) by d-operator%
\begin{equation}
\ ^{\shortparallel }\mathcal{\Re }^{\star }:=\ _{\mathbf{D}^{\star
}}^{\shortparallel }\mathbf{d\bullet }\ _{\mathbf{D}^{\star
}}^{\shortparallel }\mathbf{d:\ }\emph{Vec}_{\star _{N}}\rightarrow \emph{Vec%
}_{\star _{N}}\otimes _{\star _{N}}\Omega _{\star }^{2}.  \label{stardcurvh}
\end{equation}%
This d-operator defines a d-tensor which is $\mathcal{A}_{N}^{\star }$%
--linear and determined by a star d-connection $\ ^{\shortparallel }\mathbf{D%
}^{\star }$ (\ref{dconhopf}). It is also $U\emph{Vec}^{\mathcal{F}}(\mathcal{%
M},N)$-equivariant and can be written in terms of N-adapted associative
composition $\bullet ,$ pairing and braiding $\tau _{\mathcal{R}}$ (\ref%
{braidop}), and associator $\Phi ,$ 
\begin{eqnarray}
\mathbf{\mathbf{\mathbf{\mathbf{\ ^{\shortparallel }}}}}\mathcal{\Re }%
^{\star }:= &&\mathbf{\langle }\ ,\mathbf{\mathbf{\mathbf{\mathbf{\ }}}%
\rangle }_{\star _{N}}\bullet (\mathbf{\mathbf{\mathbf{\mathbf{\
^{\shortparallel }}}}D}^{\star }\otimes _{\star _{N}}id)\bullet (\mathbf{%
\langle }\ ,\mathbf{\mathbf{\mathbf{\mathbf{\ }}}\rangle }_{\star
_{N}}\otimes _{\star _{N}}id)\bullet \Phi _{Vec_{\star _{N}}\otimes _{\star
_{N}}\Omega _{\star }^{1},Vec_{\star _{N}},Vec_{\star _{N}}}^{-1}\bullet (%
\mathbf{\mathbf{\mathbf{\mathbf{\ ^{\shortparallel }}}}D}^{\star }\otimes
_{\star _{N}}id^{\otimes _{\star _{N}}^{2}})  \notag \\
&&\bullet (id^{\otimes _{\star _{N}}^{3}}-id\otimes _{\mathcal{R}}\tau _{%
\mathcal{R}})+\mathbf{\langle }\ ,\mathbf{\mathbf{\mathbf{\mathbf{\ }}}%
\rangle }_{\star _{N}}\bullet \left( \mathbf{\mathbf{\mathbf{\mathbf{\
^{\shortparallel }}}}D}^{\star }\otimes _{\star _{N}}id\right) \bullet
(id\otimes _{\mathcal{R}}[,]_{\star _{N}}).  \label{stardcurvhopf}
\end{eqnarray}

Acting with d-operator $\mathbf{\mathbf{\mathbf{\mathbf{\ ^{\shortparallel }}%
}}}\mathcal{\Re }^{\star }$ (\ref{stardcurvh}) on $^{\shortparallel }\mathbf{%
e}_{\alpha }$ and using respective associators and formulas (\ref{aux300})
and (\ref{aux301}), we compute:%
\begin{eqnarray*}
&&\ ^{\shortparallel }\mathcal{\Re }^{\star }(^{\shortparallel }\mathbf{e}%
_{\alpha })=\ _{\mathbf{D}^{\star }}^{\phi _{1}\shortparallel }\mathbf{%
d\bullet }\ _{\mathbf{D}^{\star }}^{\shortparallel }\mathbf{d}%
(^{\shortparallel }\mathbf{e}_{\alpha })=\ _{\mathbf{D}^{\star
}}^{\shortparallel }\mathbf{d}(\ ^{\shortparallel }\mathbf{D}^{\star }\
^{\shortparallel }\mathbf{e}_{\alpha })=\ _{\mathbf{D}^{\star
}}^{\shortparallel }\mathbf{d}(^{\shortparallel }\mathbf{e}_{\beta }\otimes
_{\star _{N}}\mathbf{\ ^{\shortparallel }\Gamma }_{\star \alpha }^{\beta })
\\
&=&\ _{\overline{\phi }_{1}}^{\shortparallel }\mathbf{D}^{\star }(_{%
\overline{\phi }_{2}}^{\shortparallel }\mathbf{e}_{\beta }){\wedge }_{\star
_{N}}\ _{\overline{\phi }_{3}}^{\shortparallel }\mathbf{\Gamma }_{\star
\alpha }^{\beta }+\ ^{\shortparallel }\mathbf{e}_{\beta }\otimes _{\star
_{N}}\ ^{\shortparallel }\mathbf{d\ ^{\shortparallel }\Gamma }_{\star \alpha
}^{\beta }=(\ ^{\shortparallel }\mathbf{e}_{\gamma }\otimes _{\star _{N}}%
\mathbf{\ ^{\shortparallel }\Gamma }_{\star \beta }^{\gamma }){\wedge }%
_{\star _{N}}\ ^{\shortparallel }\mathbf{\Gamma }_{\star \alpha }^{\beta }+\
^{\shortparallel }\mathbf{e}_{\beta }\otimes _{\star _{N}}\ ^{\shortparallel
}\mathbf{d\ ^{\shortparallel }\Gamma }_{\star \alpha }^{\beta } \\
&=&\ _{\phi _{1}}^{\shortparallel }\mathbf{e}_{\gamma }\otimes _{\star
_{N}}(\ _{\phi _{2}}^{\shortparallel }\mathbf{\Gamma }_{\star \beta
}^{\gamma }{\wedge }_{\star _{N}}\ _{\phi _{3}}^{\shortparallel }\mathbf{%
\Gamma }_{\star \alpha }^{\beta })+\ ^{\shortparallel }\mathbf{e}_{\beta
}\otimes _{\star _{N}}\ ^{\shortparallel }\mathbf{d\ ^{\shortparallel
}\Gamma }_{\star \alpha }^{\beta }=\ ^{\shortparallel }\mathbf{e}_{\gamma
}\otimes _{\star _{N}}(\ ^{\shortparallel }\mathbf{\Gamma }_{\star \beta
}^{\gamma }{\wedge }_{\star _{N}}\ ^{\shortparallel }\mathbf{\Gamma }_{\star
\alpha }^{\beta })+\ ^{\shortparallel }\mathbf{e}_{\beta }\otimes _{\star
_{N}}\ ^{\shortparallel }\mathbf{d\ ^{\shortparallel }\Gamma }_{\star \alpha
}^{\beta } \\
&=&\ ^{\shortparallel }\mathbf{e}_{\gamma }\otimes _{\star _{N}}(\
^{\shortparallel }\mathbf{d\ ^{\shortparallel }\Gamma }_{\star \alpha
}^{\gamma }+^{\shortparallel }\mathbf{\Gamma }_{\star \beta }^{\gamma }{%
\wedge }_{\star _{N}}\ ^{\shortparallel }\mathbf{\Gamma }_{\star \alpha
}^{\beta }):=\ ^{\shortparallel }\mathbf{e}_{\gamma }\otimes _{\star _{N}}%
\mathbf{\mathbf{\mathbf{\mathbf{\ ^{\shortparallel }}}}}\mathcal{R}_{\quad
\alpha }^{\star \gamma },
\end{eqnarray*}%
where we introduce the matrix valued (describing a Hopf d-algebra) star
deformed d-curvature 2-form%
\begin{equation}
\mathbf{\mathbf{\mathbf{\mathbf{\ ^{\shortparallel }}}}}\mathcal{\Re }%
_{\quad \alpha }^{\star \gamma }:=\ ^{\shortparallel }\mathbf{d\
^{\shortparallel }}\Gamma _{\star \alpha }^{\gamma }+^{\shortparallel }%
\mathbf{\Gamma }_{\star \beta }^{\gamma }{\wedge }_{\star _{N}}\
^{\shortparallel }\mathbf{\Gamma }_{\star \alpha }^{\beta }.
\label{strdcurvhf}
\end{equation}%
This formula is a star deformed differential form expression of the
curvature d-operator (\ref{stardcurvhopf}).

\paragraph{Nonassociative Bianchi identities and Hopf-Cartan structure
equations: \newline
}

Taking the exterior derivative of the torsion 2-form $\mathbf{\mathbf{%
\mathbf{\mathbf{\ ^{\shortparallel }}}}}\mathcal{T}{_{\star }^{\alpha }}$ (%
\ref{dtorsstarform}) we prove the 1st Bianchi identity,%
\begin{equation*}
\ ^{\shortparallel }\mathbf{d\mathbf{\mathbf{\mathbf{\ ^{\shortparallel }}}}}%
\mathcal{T}{_{\star }^{\alpha }+}\mathbf{\ ^{\shortparallel }}\Gamma _{\star
\beta }^{\alpha }\mathbf{\mathbf{\mathbf{\mathbf{\ {\wedge }_{\star _{N}}\
^{\shortparallel }}}}}\mathcal{T}{_{\star }^{\beta }=}\mathbf{\mathbf{%
\mathbf{\mathbf{\ ^{\shortparallel }}}}}\mathcal{\Re }_{\quad \beta }^{\star
\alpha },
\end{equation*}%
and \ similar action on $\mathbf{\mathbf{\mathbf{\mathbf{\ ^{\shortparallel }%
}}}}\mathcal{R}_{\quad \alpha }^{\star \gamma }$ (\ref{strdcurvhf}) provides
a proof for the 2d Bianchi identity,%
\begin{eqnarray*}
\ ^{\shortparallel }\mathbf{d\mathbf{\mathbf{\mathbf{\ ^{\shortparallel }}}}}%
\mathcal{\Re }_{\quad \alpha }^{\star \gamma }+\mathbf{\ ^{\shortparallel }}%
\Gamma _{\star \beta }^{\alpha }\mathbf{\mathbf{\mathbf{\mathbf{\ {\wedge }%
_{\star _{N}}\ ^{\shortparallel }}}}}\mathcal{\Re }_{\quad \alpha }^{\star
\beta }-\mathbf{\ ^{\shortparallel }}\mathcal{\Re }_{\star \beta }^{\alpha }%
\mathbf{\mathbf{\mathbf{\mathbf{\ {\wedge }_{\star _{N}}\ ^{\shortparallel }}%
}}}\Gamma _{\quad \alpha }^{\star \beta } &=& \\
\mathbf{\ ^{\shortparallel }}\Gamma _{\star \beta }^{\alpha }\mathbf{\mathbf{%
\mathbf{\mathbf{\ {\wedge }_{\star _{N}}(\mathbf{\mathbf{\mathbf{\mathbf{\
^{\shortparallel }}}}}}}}}\Gamma _{\quad \mu }^{\star \beta }\mathbf{\mathbf{%
\mathbf{\mathbf{\mathbf{\mathbf{\mathbf{\mathbf{{\wedge }_{\star _{N}}\
^{\shortparallel }}}}}}}}}\Gamma _{\quad \alpha }^{\star \mu }\mathbf{%
\mathbf{\mathbf{\mathbf{)-}}}}\ _{\phi _{1}}^{\shortparallel }\Gamma _{\star
\beta }^{\alpha }\mathbf{\mathbf{\mathbf{\mathbf{\ {\wedge }_{\star _{N}}(}}}%
}\ _{\phi _{2}}^{\shortparallel }\Gamma _{\quad \mu }^{\star \beta }\mathbf{%
\mathbf{\mathbf{\mathbf{\mathbf{\mathbf{\mathbf{\mathbf{{\wedge }_{\star
_{N}}\ }}}}}}}}_{\phi _{3}}^{\shortparallel }\Gamma _{\quad \alpha }^{\star
\mu }) &\neq &0,\mbox{ nonassociative}; =0\mbox{ associative}.
\end{eqnarray*}

The d-operator $\ ^{\shortparallel }\mathcal{R}^{\star }$ (\ref%
{stardcurvhopf}) is a fourth rank d-tensor which also satisfies in abstract
form the second Cartan structure equation. We omit such considerations in
this work because they are similar to those in the holonomic form (see
formula (4.55) in \cite{aschieri17}). Nevertheless, we provide an equivalent
N-adapted coefficient version of such an equation which can be obtained from
a geometric calculus of the d-operator $\mathbf{\mathbf{\mathbf{\mathbf{\
^{\shortparallel }}}}}\mathcal{R}^{\star }(\mathbf{\ ^{\shortparallel }z},%
\mathbf{\ ^{\shortparallel }v,\ ^{\shortparallel }q})$ as an action on basis
d-vectors $\mathbf{\ ^{\shortparallel }z=} ^{\shortparallel }\mathbf{e}%
_{\alpha },\mathbf{\ ^{\shortparallel }v=\ ^{\shortparallel }\mathbf{e}}%
_{\beta }$ and $\mathbf{^{\shortparallel }q=\ }^{\shortparallel }\mathbf{e}%
_{\gamma }$ using the braiding d-operator (\ref{braidop}). The corresponding
computations are as follow%
\begin{eqnarray}
\mathbf{\mathbf{\mathbf{\mathbf{\ ^{\shortparallel }}}}D}_{\gamma }^{\star }(%
\mathbf{\mathbf{\mathbf{\mathbf{\ ^{\shortparallel }}}}D}_{\beta }^{\star }%
\mathbf{\ ^{\shortparallel }e}_{\alpha })-\mathbf{\mathbf{\mathbf{\mathbf{\
^{\shortparallel }}}}D}_{\intercal \alpha }^{\star }(\mathbf{\mathbf{\mathbf{%
\mathbf{\ ^{\shortparallel }}}}D}_{\intercal \gamma }^{\star }\mathbf{\
_{\intercal }^{\shortparallel }e}_{\alpha })+\mathbf{\mathbf{\mathbf{\mathbf{%
\ ^{\shortparallel }}}}D}_{[\mathbf{\ ^{\shortparallel }e}_{\gamma },\mathbf{%
\ ^{\shortparallel }e}_{\beta }]_{\star _{N}}}^{\star }\mathbf{\
^{\shortparallel }e}_{\alpha } &=&  \label{2dcartan} \\
\mathbf{\mathbf{\mathbf{\mathbf{\ ^{\shortparallel }}}}D}_{\gamma }^{\star }(%
\mathbf{\mathbf{\mathbf{\mathbf{\ ^{\shortparallel }}}}D}_{\beta }^{\star }%
\mathbf{\ ^{\shortparallel }e}_{\alpha })-\mathbf{\mathbf{\mathbf{\mathbf{\
^{\shortparallel }}}}D}_{\beta }^{\star }(\mathbf{\mathbf{\mathbf{\mathbf{\
^{\shortparallel }}}}D}_{\gamma }^{\star }\mathbf{\ ^{\shortparallel }e}%
_{\alpha })+[\mathbf{\ ^{\shortparallel }e}_{\gamma },\mathbf{\
^{\shortparallel }e}_{\beta }]_{\star _{N}}\mathbf{\ ^{\shortparallel }e}%
_{\alpha } &=&  \notag \\
\mathbf{\mathbf{\mathbf{\mathbf{\ ^{\shortparallel }}}}D}_{\gamma }^{\star
}(\ \mathbf{^{\shortparallel }e}_{\mu }\star _{N}\mathbf{\ ^{\shortparallel
}\Gamma }_{\star \alpha \beta }^{\mu })-\mathbf{\mathbf{\mathbf{\mathbf{\
^{\shortparallel }}}}D}_{\beta }^{\star }(\ \mathbf{^{\shortparallel }e}%
_{\mu }\star _{N}\mathbf{\ ^{\shortparallel }\Gamma }_{\star \alpha \gamma
}^{\mu })+\ \mathbf{^{\shortparallel }e}_{\mu }\star _{N}\mathbf{\
^{\shortparallel }}w_{\star \gamma \beta }^{\mu }\mathbf{\ ^{\shortparallel
}e}_{\alpha } &=&  \notag \\
\mathbf{\mathbf{\mathbf{\mathbf{\ ^{\shortparallel }}}}D}_{\gamma }^{\star
}(\ \mathbf{^{\shortparallel }e}_{\mu }\star _{N}\mathbf{\ ^{\shortparallel
}\Gamma }_{\star \alpha \beta }^{\mu })-\mathbf{\mathbf{\mathbf{\mathbf{\
^{\shortparallel }}}}D}_{\beta }^{\star }(\ \mathbf{^{\shortparallel }e}%
_{\mu }\star _{N}\mathbf{\ ^{\shortparallel }\Gamma }_{\star \alpha \gamma
}^{\mu })+\ \mathbf{^{\shortparallel }e}_{\mu }\star _{N}\mathbf{\
^{\shortparallel }}w_{\star \gamma \beta }^{\mu }\mathbf{\ ^{\shortparallel
}e}_{\alpha } &=&  \notag \\
\langle \ \mathbf{\mathbf{\mathbf{\mathbf{^{\shortparallel }}}}D}^{\star }\ 
\mathbf{^{\shortparallel }e}_{\mu },\mathbf{\ _{\intercal }^{\shortparallel
}e}_{\gamma }\mathbf{\mathbf{\rangle }}_{\star _{N}}\star _{N}\mathbf{\
_{\intercal }^{\shortparallel }\Gamma }_{\star \alpha \beta }^{\mu }-\langle
\ \mathbf{\mathbf{\mathbf{\mathbf{^{\shortparallel }}}}D}^{\star }\ \mathbf{%
^{\shortparallel }e}_{\mu },\mathbf{\ _{\intercal }^{\shortparallel }e}%
_{\beta }\mathbf{\mathbf{\rangle }}_{\star _{N}}\star _{N}\mathbf{\
_{\intercal }^{\shortparallel }\Gamma }_{\star \alpha \gamma }^{\mu }+ && 
\notag \\
\ \mathbf{^{\shortparallel }e}_{\mu }\star _{N}\langle \ ^{\shortparallel }%
\mathbf{d\ ^{\shortparallel }\Gamma }_{\star \alpha \beta }^{\mu },\ \mathbf{%
^{\shortparallel }e}_{\gamma }\mathbf{\mathbf{\rangle }}_{\star _{N}}-\ 
\mathbf{^{\shortparallel }e}_{\mu }\star _{N}\langle \ ^{\shortparallel }%
\mathbf{d\ ^{\shortparallel }\Gamma }_{\star \alpha \gamma }^{\mu },\ 
\mathbf{^{\shortparallel }e}_{\beta }\mathbf{\mathbf{\rangle }}_{\star
_{N}}+\ \mathbf{^{\shortparallel }e}_{\mu }\star _{N}\mathbf{\
^{\shortparallel }}w_{\star \beta \gamma }^{\mu }\mathbf{\ ^{\shortparallel
}e}_{\alpha } &=&\mathbf{\mathbf{\mathbf{\mathbf{\ ^{\shortparallel }}}}}%
\mathcal{\Re }^{\star }(\mathbf{\ ^{\shortparallel }e}_{\alpha },\mathbf{\
^{\shortparallel }e}_{\beta },\mathbf{\ ^{\shortparallel }e}_{\gamma }). 
\notag
\end{eqnarray}%
For holonomic basic structures such constructions are provided in section
4.5 of \cite{aschieri17}.

\paragraph{N-adapted coefficient formulas for the nonassociative Riemann
d-tensor and quasi-Hopf d-algebras: \newline
}

One extracts N-adapted formulas for the star deformed curvature d-tensor of $%
\ \mathbf{\mathbf{\mathbf{\mathbf{^{\shortparallel }}}}D}^{\star }$ stated
in (\ref{2dcartan}) using $\ ^{\shortparallel }\mathcal{\Re }_{\quad \alpha
}^{\star \gamma }$ (\ref{strdcurvhf}), star pairing and with respect to $%
^{\shortparallel }\mathbf{e}_{\alpha }$ (\ref{nadapdc}),%
\begin{eqnarray}
\ ^{\shortparallel }\mathcal{\Re }^{\star } &&(\ ^{\shortparallel }\mathbf{e}%
_{\alpha },\ ^{\shortparallel }\mathbf{e}_{\beta },\ ^{\shortparallel }%
\mathbf{e}_{\gamma })=\langle \ \mathbf{^{\shortparallel }e}_{\mu }\otimes
_{\star _{N}}\mathbf{\mathbf{\mathbf{\mathbf{\ ^{\shortparallel }}}}}%
\mathcal{\Re }_{\quad \alpha }^{\star \mu },\mathbf{\ ^{\shortparallel }e}%
_{\beta }\mathbf{\mathbf{\mathbf{\mathbf{\ {\wedge }_{\star _{N}}}\
^{\shortparallel }e}_{\gamma }\rangle }}_{\star _{N}}  \label{aux15} \\
&=&\ ^{\shortparallel }\mathbf{e}_{\mu }\star _{N}\langle \ ^{\shortparallel
}\mathbf{d\ ^{\shortparallel }}\Gamma _{\star \alpha }^{\mu
}+^{\shortparallel }\mathbf{\Gamma }_{\star \nu }^{\mu }{\wedge }_{\star
_{N}}\ ^{\shortparallel }\mathbf{\Gamma }_{\star \alpha }^{\nu },\mathbf{\
^{\shortparallel }e}_{\beta }\mathbf{\mathbf{\mathbf{\mathbf{\ {\wedge }%
_{\star _{N}}}\ ^{\shortparallel }e}_{\gamma }\rangle }}_{\star _{N}}  \notag
\\
&=&\ \mathbf{^{\shortparallel }e}_{\mu }\star _{N}\langle \ (\ \mathbf{%
^{\shortparallel }e}_{\nu }\mathbf{\ ^{\shortparallel }}\Gamma _{\star
\alpha \varphi }^{\mu }\star _{N}\ \mathbf{^{\shortparallel }e}^{\nu })%
\mathbf{\mathbf{\mathbf{\mathbf{{\wedge }_{\star _{N}}}}}}\ \mathbf{%
^{\shortparallel }e}^{\varphi }+(\ ^{\shortparallel }\mathbf{\Gamma }_{\star
\nu \varphi }^{\mu }\star _{N}\ \mathbf{^{\shortparallel }e}^{\varphi }){%
\wedge }_{\star _{N}}(\ ^{\shortparallel }\mathbf{\Gamma }_{\star \alpha
\lambda }^{\nu }\star _{N}\ \mathbf{^{\shortparallel }e}^{\lambda }),\mathbf{%
\ ^{\shortparallel }e}_{\beta }\mathbf{\mathbf{\mathbf{\mathbf{\ {\wedge }%
_{\star _{N}}}\ ^{\shortparallel }e}_{\gamma }\rangle }}_{\star _{N}}  \notag
\\
&=&\ \mathbf{^{\shortparallel }e}_{\mu }\star _{N}\langle \ \mathbf{%
^{\shortparallel }e}_{\nu }\mathbf{\ ^{\shortparallel }}\Gamma _{\star
\alpha \varphi }^{\mu }\star _{N}(\ \mathbf{^{\shortparallel }e}^{\nu }%
\mathbf{\mathbf{\mathbf{\mathbf{{\wedge }_{\star _{N}}}}}}\ \mathbf{%
^{\shortparallel }e}^{\varphi })+  \notag \\
&&\ ^{\shortparallel }\mathbf{\Gamma }_{\star \nu \varphi }^{\mu }\star
_{N}(\delta _{\ \tau }^{\varphi }\ ^{\shortparallel }\mathbf{\Gamma }_{\star
\alpha \lambda }^{\nu }+i\kappa \mathcal{R}_{\quad \tau }^{\varphi \xi }%
\mathbf{\ ^{\shortparallel }e}_{\xi }\mathbf{\Gamma }_{\star \alpha \lambda
}^{\nu })\star _{N}\ \mathbf{^{\shortparallel }e}^{\tau }\mathbf{\mathbf{%
\mathbf{\mathbf{{\wedge }_{\star _{N}}}}}}\ \mathbf{^{\shortparallel }e}%
^{\lambda },\mathbf{\ ^{\shortparallel }e}_{\beta }\mathbf{\mathbf{\mathbf{%
\mathbf{\ {\wedge }_{\star _{N}}}\ ^{\shortparallel }e}_{\gamma }\rangle }}%
_{\star _{N}}  \notag \\
&=&\ \mathbf{^{\shortparallel }e}_{\mu }\star _{N}\mathbf{\mathbf{\mathbf{%
\mathbf{\ ^{\shortparallel }}}}}\mathcal{\Re }_{\quad \alpha \beta \gamma
}^{\star \mu },  \notag
\end{eqnarray}%
where $\delta _{\ \tau }^{\varphi }$ is the Kronecker delta symbol. In this
formula, we use the N-adapted coefficients of the nonassociative Riemann
d-tensor for the quasi-Hopf d-algebra written in the form%
\begin{eqnarray}
\mathbf{\mathbf{\mathbf{\mathbf{\ ^{\shortparallel }}}}}\mathcal{\Re }%
_{\quad \alpha \beta \gamma }^{\star \mu } &=&\mathbf{\mathbf{\mathbf{%
\mathbf{\ _{1}^{\shortparallel }}}}}\mathcal{\Re }_{\quad \alpha \beta
\gamma }^{\star \mu }+\mathbf{\mathbf{\mathbf{\mathbf{\ _{2}^{\shortparallel
}}}}}\mathcal{\Re }_{\quad \alpha \beta \gamma }^{\star \mu },\mbox{ where }
\label{nadriemhopf} \\
&&\mathbf{\mathbf{\mathbf{\mathbf{\ _{1}^{\shortparallel }}}}}\mathcal{\Re }%
_{\quad \alpha \beta \gamma }^{\star \mu }=\ \mathbf{^{\shortparallel }e}%
_{\gamma }\mathbf{\ ^{\shortparallel }}\Gamma _{\star \alpha \beta }^{\mu
}-\ \mathbf{^{\shortparallel }e}_{\beta }\mathbf{\ ^{\shortparallel }}\Gamma
_{\star \alpha \gamma }^{\mu }+\mathbf{\ ^{\shortparallel }}\Gamma _{\star
\nu \tau }^{\mu }\star _{N}(\delta _{\ \gamma }^{\tau }\mathbf{\
^{\shortparallel }}\Gamma _{\star \alpha \beta }^{\nu }-\delta _{\ \beta
}^{\tau }\mathbf{\ ^{\shortparallel }}\Gamma _{\star \alpha \gamma }^{\nu })+%
\mathbf{\ ^{\shortparallel }}w_{\beta \gamma }^{\tau }\star _{N}\mathbf{\
^{\shortparallel }}\Gamma _{\star \alpha \tau }^{\mu }  \notag \\
&&\mathbf{\mathbf{\mathbf{\mathbf{\ _{2}^{\shortparallel }}}}}\mathcal{\Re }%
_{\quad \alpha \beta \gamma }^{\star \mu }=i\kappa \mathbf{\
^{\shortparallel }}\Gamma _{\star \nu \tau }^{\mu }\star _{N}(\mathcal{R}%
_{\quad \gamma }^{\tau \xi }\ \mathbf{^{\shortparallel }e}_{\xi }\mathbf{\
^{\shortparallel }}\Gamma _{\star \alpha \beta }^{\nu }-\mathcal{R}_{\quad
\beta }^{\tau \xi }\ \mathbf{^{\shortparallel }e}_{\xi }\mathbf{\
^{\shortparallel }}\Gamma _{\star \alpha \gamma }^{\nu }).  \notag
\end{eqnarray}

With respect to $^{\shortparallel }\mathbf{e}_{\alpha }=(\ ^{\shortparallel }%
\mathbf{e}_{i},\ ^{\shortparallel }e^{a})$ (\ref{nadapdc}) and for $\
^{\shortparallel }\mathbf{\Gamma }_{\star \alpha \beta }^{\gamma }=(\ \
^{\shortparallel }L_{\star jk}^{i},\ \ ^{\shortparallel }L_{\star a\ k}^{\
b},\ \ ^{\shortparallel }C_{\star \ j}^{i\ c},\ \ ^{\shortparallel
}C_{\star\ j}^{i\ c})$ (\ref{irevndecomdc}), we find such a
h-c-decomposition: 
\begin{equation}
\ \mathbf{\mathbf{\mathbf{\mathbf{\ _{1}^{\shortparallel }}}}}\mathcal{\Re }%
^{\star }=\mathbf{\{}\ \mathbf{\mathbf{\mathbf{\mathbf{\
_{1}^{\shortparallel }}}}}\mathcal{\Re }_{\quad \alpha \beta \gamma }^{\star
\mu }=(\ \ _{1}^{\shortparallel }R_{\ \star hjk}^{i},\ \
_{1}^{\shortparallel }R_{\star a\ jk}^{\ b},\ \ _{1}^{\shortparallel
}P_{\star \ hj}^{i\ \ \ a},\ \ _{1}^{\shortparallel }P_{\star c\ j}^{\ b\
a},\ \ _{1}^{\shortparallel }S_{\star \ hba}^{i},\ \ _{1}^{\shortparallel
}S_{\star \ bea}^{c})\},\mbox{ where }  \label{nadriemhopf1}
\end{equation}%
\begin{eqnarray}
\ \ _{1}^{\shortparallel }R_{\ \star hjk}^{i} &=&\ \ ^{\shortparallel }%
\mathbf{e}_{k}\ ^{\shortparallel }L_{\star hj}^{i}-\ \ ^{\shortparallel }%
\mathbf{e}_{j}\ ^{\shortparallel }L_{\star hk}^{i}+\ \ ^{\shortparallel
}L_{\star hj}^{m}\star _{N}\quad ^{\shortparallel }L_{\star mk}^{i}-\ \
^{\shortparallel }L_{\star hk}^{m}\star _{N}\ ^{\shortparallel }L_{\star
mj}^{i}-\ ^{\shortparallel }C_{\star \ h}^{i\ a}\ \star _{N}\
^{\shortparallel }\Omega _{\star akj},  \notag \\
\ _{1}^{\shortparallel }R_{\star a\ jk}^{\ b} &=&\ ^{\shortparallel }\mathbf{%
e}_{k}\ ^{\shortparallel }\acute{L}_{\star a\ j}^{\ b}-\ ^{\shortparallel }%
\mathbf{e}_{j}\ ^{\shortparallel }\acute{L}_{\star a\ k}^{\ b}+\
^{\shortparallel }\acute{L}_{\star c\ j}^{\ b}\star _{N}\ ^{\shortparallel }%
\acute{L}_{\star a\ k}^{\ c}-\ ^{\shortparallel }\acute{L}_{\star c\ k}^{\
b}\star _{N}\ ^{\shortparallel }\acute{L}_{\star a\ j}^{\ c}-\
^{\shortparallel }C_{\star a\ }^{\ bc}\ \star _{N}^{\shortparallel }\Omega
_{\star ckj},  \notag \\
\ \ _{1}^{\shortparallel }P_{\star \ hj}^{i\ \ \ a} &=&\ ^{\shortparallel
}e^{a}\ \ ^{\shortparallel }L_{\star \ jk}^{i}-\ \ ^{\shortparallel
}D_{k}^{\star }\ \star _{N}\ ^{\shortparallel }\acute{C}_{\star \ j}^{i\
a}+\ ^{\shortparallel }\acute{C}_{\star \ j}^{i\ b}\star _{N}\
^{\shortparallel }T_{\star bk}^{\ \ \ a},\   \notag \\
\ _{1}^{\shortparallel }P_{\star c\ k}^{\ b\ a} &=&\ ^{\shortparallel
}e^{a}\ ^{\shortparallel }\acute{L}_{\star c\ k}^{\ b}-\ ^{\shortparallel
}D_{k}^{\star }\ \star _{N}\ ^{\shortparallel }C_{\star c\ }^{\ ba}+\
^{\shortparallel }C_{\star \ bd}^{c}\ \star _{N}\ ^{\shortparallel }T_{\star
\ ka}^{c},  \notag \\
\ _{1}^{\shortparallel }S_{\star \ j}^{i\quad bc} &=&\ ^{\shortparallel
}e^{c}\ ^{\shortparallel }\acute{C}_{\star \ j}^{i\ b}-\ \ ^{\shortparallel
}e^{b}\ ^{\shortparallel }\acute{C}_{\star \ j}^{i\quad c}+\ \
^{\shortparallel }\acute{C}_{\star \ h}^{j\ b}\ \star _{N}\ ^{\shortparallel
}\acute{C}_{\star \ j}^{i\ c}-\ \ ^{\shortparallel }\acute{C}_{\star \
h}^{j\quad c}\star _{N}\ \ ^{\shortparallel }\acute{C}_{\star \ j}^{i\quad
b},  \notag \\
\ \ _{1}^{\shortparallel }S_{\star a\ }^{\quad bcd} &=&\ ^{\shortparallel
}e^{d}\ \ ^{\shortparallel }C_{\star a\ }^{\quad bc}-\ \ ^{\shortparallel
}e^{c}\ \ ^{\shortparallel }C_{\star a\ }^{\quad bd}+\ \ ^{\shortparallel
}C_{\star e\ }^{\quad bc}\ \star _{N}\ ^{\shortparallel }C_{\star a\
}^{\quad ed}-\ \ ^{\shortparallel }C_{\star e}^{\quad bd}\ \star _{N}\
^{\shortparallel }C_{\star a}^{\quad ec}.  \label{nadriemhopf1coef}
\end{eqnarray}%
In these formulas, there are involved nonassociative d-torsion coefficients (%
\ref{dtorsnonassoc}) and the nonholonomic distributions and respective
N-adapted frames considered in (\ref{nadriemhopf}). The coefficients (\ref%
{nadriemhopf1coef}) transform into respective N-adapted coefficients of
commutative Riemann d-tensor, see details in \cite{vacaru18,bubuianu18a}.

In a similar form, introducing N-adapted coefficients (\ref{nadapdc}) and (%
\ref{irevndecomdc}), we can compute h-v-decompositions of $\mathbf{\mathbf{%
\mathbf{\mathbf{\ _{2}^{\shortparallel }}}}}\mathcal{\Re }_{\quad \alpha
\beta \gamma }^{\star \mu }.$ We do not provide such formulas because they
are not used in this and related series of works on nonassociative geometry
and gravity.

Finally, we note that the nonassociative Riemann d-tensor (\ref{nadriemhopf}%
) is constructed for an arbitrary star deformed d-connection $\
^{\shortparallel }\mathbf{D}^{\star }$ (\ref{dconhopf}). This defines a
model of nonassociative geometry of affine (linear) connections adapted to
quasi-Hopf structures with N-connection.

\subsubsection{Star deformed Ricci d-tensors without and with quasi-Hopf
structure}

For elaborated nonassociative nonholonomic geometric models, the Ricc tensor
is the trace of respective Riemann tensor. In N-adapted form, we work with
d-tensors. In general, such a second rank Ricci d-tensor is not symmetric.
It is a d-object which be adapted both to N-connection and quasi-Hopf
structures.

\paragraph{Abstract definitions and N-adapted coefficients for star deformed
Ricci d-tensors: \newline
}

Using the curvature d-tensor (\ref{starriemdt1}), we define the star
deformed Ricci d-tensor 
\begin{equation*}
\mathbf{\mathbf{\mathbf{\mathbf{\ _{\flat }^{\shortparallel }}}}}\mathcal{R}%
ic:=\langle \mathbf{\mathbf{\mathbf{\mathbf{\mathbf{\mathbf{\ _{\flat
}^{\shortparallel }}}R(}}}}\ ^{\shortparallel }\mathbf{\mathbf{\mathbf{X%
\mathbf{,\mathbf{\mathbf{\mathbf{\ ^{\shortparallel }Y}}}}},}}\
^{\shortparallel }\mathbf{e}_{\gamma }\mathbf{\mathbf{\mathbf{\mathbf{),%
\mathbf{\ ^{\shortparallel }e}}}}}^{\gamma }\mathbf{\mathbf{\rangle }}%
_{\star _{N}}.
\end{equation*}%
For N-adapted bases with $\mathbf{\mathbf{\mathbf{\mathbf{\mathbf{\mathbf{%
\mathbf{\mathbf{\ ^{\shortparallel }}}}}}X=\mathbf{\mathbf{\
^{\shortparallel }e}}}}}_{\alpha }$ and$\mathbf{\mathbf{\mathbf{\mathbf{%
\mathbf{\mathbf{\mathbf{\ ^{\shortparallel }Y=}}\ ^{\shortparallel }e}}}}}%
_{\beta }$, this mean that we contract the first and forth indices when 
\begin{equation*}
\mathbf{\mathbf{\mathbf{\mathbf{\ _{\flat }^{\shortparallel }}}}}\mathcal{R}%
ic=\{\mathbf{\mathbf{\mathbf{\mathbf{\ _{\flat }^{\shortparallel }}}}}%
\mathcal{R}_{\alpha \beta }:=\mathbf{\mathbf{\mathbf{\mathbf{\ _{\flat
}^{\shortparallel }}}R}}_{\ \alpha \beta \gamma }^{\gamma }\},\mbox{ for }%
\mathbf{\mathbf{\mathbf{\mathbf{\ _{\flat }^{\shortparallel }}}}}\mathcal{R}%
_{\alpha \beta }\neq \mathbf{\mathbf{\mathbf{\mathbf{\ _{\flat
}^{\shortparallel }}}}}\mathcal{R}_{\beta \alpha }.
\end{equation*}%
We compute the N-adapted coefficients $\mathbf{\mathbf{\mathbf{\mathbf{\
_{\flat }^{\shortparallel }}}}}\mathcal{R}_{\alpha \beta }$ is explicit form
using formulas (\ref{riem1}), 
\begin{eqnarray}
\mathbf{\mathbf{\mathbf{\mathbf{\ _{\flat }^{\shortparallel }}}}}\mathcal{R}%
_{\ \tau \alpha }&:= &\mathbf{\mathbf{\mathbf{\mathbf{\ ^{\shortparallel }}}}%
e}_{\gamma }\mathbf{\mathbf{\mathbf{\mathbf{\ _{\flat }^{\shortparallel }}}}%
\Gamma }_{\ \tau \alpha }^{\gamma }-\mathbf{\mathbf{\mathbf{\mathbf{\
^{\shortparallel }}}}e}_{\alpha }\mathbf{\mathbf{\mathbf{\mathbf{\ _{\flat
}^{\shortparallel }}}}\Gamma }_{\ \tau \gamma }^{\gamma }+\mathbf{\mathbf{%
\mathbf{\mathbf{\ _{\flat }^{\shortparallel }}}}\Gamma }_{\ \tau \alpha
}^{\mu }\star _{N}\mathbf{\mathbf{\mathbf{\mathbf{\ _{\flat
}^{\shortparallel }}}}\Gamma }_{\ \mu \gamma }^{\gamma }-\mathbf{\mathbf{%
\mathbf{\mathbf{\ _{\flat }^{\shortparallel }}}}\Gamma }_{\ \tau \gamma
}^{\mu }\star _{N}\mathbf{\mathbf{\mathbf{\mathbf{\ _{\flat
}^{\shortparallel }}}}\Gamma }_{\ \mu \alpha }^{\gamma }+\mathbf{\mathbf{%
\mathbf{\mathbf{\ _{\flat }^{\shortparallel }}}}\Gamma }_{\ \tau \nu
}^{\gamma }\star _{N}\mathbf{\mathbf{\mathbf{\mathbf{\ ^{\shortparallel }}}}}%
w_{\ \alpha \gamma }^{\nu }  \label{ricci01a} \\
&&+\ _{N}^{\shortparallel }F_{\alpha \quad }^{\quad \nu \varepsilon }\mathbf{%
\mathbf{\mathbf{\mathbf{\ ^{\shortparallel }}}}e}_{\nu }\mathbf{\mathbf{%
\mathbf{\mathbf{\ _{\flat }^{\shortparallel }}}}\Gamma }_{\ \tau \gamma
}^{\mu }\star _{N}\mathbf{\mathbf{\mathbf{\mathbf{\ _{\flat
}^{\shortparallel }}}}\Gamma }_{\ \mu \varepsilon }^{\gamma }-\
_{N}^{\shortparallel }F_{\gamma \quad }^{\quad \nu \varepsilon }\mathbf{%
\mathbf{\mathbf{\mathbf{\ ^{\shortparallel }}}}e}_{\nu }\mathbf{\mathbf{%
\mathbf{\mathbf{\ _{\flat }^{\shortparallel }}}}\Gamma }_{\ \tau \alpha
}^{\mu }\star _{N}\mathbf{\mathbf{\mathbf{\mathbf{\ _{\flat
}^{\shortparallel }}}}\Gamma }_{\ \mu \varepsilon }^{\gamma }.  \notag
\end{eqnarray}

Considering a h-c-splitting for $\mathbf{\mathbf{\mathbf{\mathbf{\
^{\shortparallel }}}}e}_{\gamma }$ and$\mathbf{\mathbf{\mathbf{\mathbf{\
_{\flat }^{\shortparallel }}}}\Gamma }_{\ \tau \alpha }^{\gamma },$ we find
such a N-adapted decomposition of (\ref{ricci01a})%
\begin{equation}
\ _{\flat }^{\shortparallel }\mathcal{R}_{\tau \alpha}:= \{\ _{\flat
}^{\shortparallel }R_{\ hj}=\ _{\flat}^{\shortparallel }R_{\ hji}^{i}, \
_{\flat }^{\shortparallel }P_{j}^{a}=-\ _{\flat }^{\shortparallel }R_{\
ji}^{i\quad a},\ _{\flat }^{\shortparallel }P_{k}^{b}=\ _{\flat
}^{\shortparallel }R_{c\ k}^{\ b\ c},\ _{\flat }^{\shortparallel }S^{bc}= \
_{\flat}^{\shortparallel }S_{a\ }^{\ bca}\},  \label{ricci01b}
\end{equation}%
see similar associative and commutative formulas in (A1) of references \cite%
{vacaru18,bubuianu18a}. For star deformed nonholonomic configurations, we
have to identify and summarize on respective h- and c-indices, use the star
product $\star _{N}$ and consider additional contributions of terms
proportional to $\ _{N}^{\shortparallel }F_{\gamma}^{\quad \nu\varepsilon }.$
We omit such details in this work and consider below a similar N-adapted
calculus with quasi Hopf d-objects which allow us to compute the
coefficients of star deformed Ricci d-tensors.

\paragraph{The star deformed Ricci d-tensor for quasi-Hopf N-adapted
structures: \newline
}

The Ricci d-tensor for nonholonomic and noncommutative models was studied in
our previous works \cite{vacaru01,vacaru03,vacaru09a,vacaru16}, see also
parts II and III of monograph \cite{vacaru05a}. In this subsection, we
provide definitions and computations extending the nonholonomic geometric
approach for nonassociative star deformations of the Ricci tensor on
quasi-Hopf structures generalizing the constructions from and references
therein. Using star curvature d-operator (\ref{stardcurvhopf}), we define 
\begin{equation*}
\mathbf{\mathbf{\mathbf{\mathbf{\ ^{\shortparallel }}}}}\mathcal{\Re }%
ic^{\star }(\mathbf{\ ^{\shortparallel }z},\mathbf{\ ^{\shortparallel }v}):=%
\mathbf{\langle \mathbf{\mathbf{\mathbf{\ ^{\shortparallel }}}}}\mathcal{\Re 
}^{\star }(\mathbf{\ ^{\shortparallel }z},\mathbf{\ ^{\shortparallel }v,}\
^{\shortparallel }\mathbf{e}_{\alpha })\ ,\mathbf{\mathbf{\ ^{\shortparallel
}\mathbf{e}}^{\alpha }\rangle }_{\star _{N}},
\end{equation*}%
for any d-vectors $\mathbf{\ ^{\shortparallel }z},\mathbf{\ ^{\shortparallel
}v\in }$ $\emph{Vec}_{\star _{N}}$ and N-adapted (co) frames $\
^{\shortparallel }\mathbf{e}_{\alpha }$ and $\mathbf{\ ^{\shortparallel }%
\mathbf{e}}^{\beta }$, see (\ref{nadapdc}). This states that we can contract
the first and forth indices in a N-adapted decomposition of a star Riemann
d-tensor in order to compute the coefficients of a respective star Ricci
d-tensor.

The N-adapted coefficients of the star Ricci d-tensor are computed following
formulas 
\begin{eqnarray}
\mathbf{\mathbf{\mathbf{\mathbf{\ ^{\shortparallel }}}}}\mathcal{\Re }%
ic^{\star } &=&\mathbf{\mathbf{\mathbf{\mathbf{\ ^{\shortparallel }R}}}}%
ic_{\alpha \beta }^{\star }\star _{N}(\ \mathbf{^{\shortparallel }e}^{\alpha
}\otimes _{\star _{N}}\ \mathbf{^{\shortparallel }e}^{\beta }),\mbox{ where }
\label{driccina} \\
&&\mathbf{\mathbf{\mathbf{\mathbf{\ ^{\shortparallel }R}}}}ic_{\alpha \beta
}^{\star }:=\mathbf{\mathbf{\mathbf{\mathbf{\ ^{\shortparallel }}}}}\mathcal{%
\Re }ic^{\star }(\mathbf{\ }\ ^{\shortparallel }\mathbf{e}_{\alpha },\
^{\shortparallel }\mathbf{e}_{\beta })=\mathbf{\langle }\ \mathbf{\mathbf{%
\mathbf{\mathbf{\ ^{\shortparallel }R}}}}ic_{\mu \nu }^{\star }\star _{N}(\ 
\mathbf{^{\shortparallel }e}^{\mu }\otimes _{\star _{N}}\ \mathbf{%
^{\shortparallel }e}^{\nu }),\mathbf{\mathbf{\ }\ ^{\shortparallel }\mathbf{e%
}}_{\alpha }\mathbf{\otimes _{\star _{N}}\ ^{\shortparallel }\mathbf{e}}%
_{\beta }\mathbf{\rangle }_{\star _{N}}.  \notag
\end{eqnarray}%
For $\ ^{\shortparallel }\mathcal{\Re }^{\star }(\ ^{\shortparallel }\mathbf{%
e}_{\alpha },\ ^{\shortparallel }\mathbf{e}_{\beta },\ ^{\shortparallel }%
\mathbf{e}_{\gamma })$ (\ref{aux15}) and using the property that the
associator acts trivially on basis d-vectors and N-adapted forms (and
contracting respective indices in (\ref{nadriemhopf})), we calculate
explicitly 
\begin{eqnarray*}
\mathbf{\mathbf{\mathbf{\mathbf{\ ^{\shortparallel }R}}}}ic_{\alpha \beta
}^{\star } &=&\mathbf{\mathbf{\mathbf{\mathbf{\ ^{\shortparallel }}}}}%
\mathcal{\Re }_{\quad \alpha \beta \mu }^{\star \mu }=\mathbf{\mathbf{%
\mathbf{\mathbf{\ _{1}^{\shortparallel }R}}}}ic_{\alpha \beta }^{\star }+%
\mathbf{\mathbf{\mathbf{\mathbf{\ _{2}^{\shortparallel }R}}}}ic_{\alpha
\beta }^{\star },\mbox{ for } \\
&&\mathbf{\mathbf{\mathbf{\mathbf{\ _{1}^{\shortparallel }R}}}}ic_{\alpha
\beta }^{\star }=\mathbf{\mathbf{\mathbf{\mathbf{\ _{1}^{\shortparallel }}}}}%
\mathcal{\Re }_{\quad \alpha \beta \mu }^{\star \mu }\mbox{ and }\mathbf{%
\mathbf{\mathbf{\mathbf{\ _{2}^{\shortparallel }R}}}}ic_{\alpha \beta
}^{\star }=\mathbf{\mathbf{\mathbf{\mathbf{\ _{2}^{\shortparallel }}}}}%
\mathcal{\Re }_{\quad \alpha \beta \mu }^{\star \mu }.
\end{eqnarray*}%
The h- and c-decompositions are computed by introducing in these formulas $\
^{\shortparallel }\mathbf{e}_{\alpha }=(\ ^{\shortparallel }\mathbf{e}_{i},\
^{\shortparallel }e^{a})$ (\ref{nadapdc}) and $\ ^{\shortparallel }\mathbf{%
\Gamma }_{\star \alpha \beta }^{\gamma }=(\ \ ^{\shortparallel }L_{\star
jk}^{i},\ \ ^{\shortparallel }L_{\star a\ k}^{\ b},\ \ ^{\shortparallel
}C_{\star \ j}^{i\ c},\ \ ^{\shortparallel }C_{\star \ j}^{i\ c})$ (\ref%
{irevndecomdc}) as in (\ref{nadriemhopf1}). Such computations are similar to
(\ref{ricci01b}) and associative and commutative formulas in (A1) of
references \cite{vacaru18,bubuianu18a}). We express the result as%
\begin{equation}
\ \mathbf{\mathbf{\mathbf{\mathbf{^{\shortparallel }R}}}}ic_{\alpha \beta
}^{\star }=\{ \ ^{\shortparallel }R_{\ \star hj}=\ \ ^{\shortparallel }%
\mathcal{\Re }_{\ \star hji}^{i},\ \ ^{\shortparallel }P_{\star j}^{\ a}=-\
^{\shortparallel }\mathcal{\Re }_{\star \ ji}^{i\quad a}\ ,\
^{\shortparallel }P_{\star \ k}^{b\quad }=\ ^{\shortparallel }\mathcal{\Re }%
_{\star c\ k}^{\ b\ c},\ ^{\shortparallel }S_{\star \ }^{bc\quad }=\
^{\shortparallel }\mathcal{\Re }_{\star a\ }^{\ bca}\ \}.  \label{driccinahc}
\end{equation}%
It is possible to decompose this d-tensor in powers of $i\kappa \mathcal{R}%
_{\quad \gamma }^{\alpha \beta },$ 
\begin{equation*}
\ \mathbf{\mathbf{\mathbf{\mathbf{^{\shortparallel }R}}}}ic_{\alpha \beta
}^{\star }=(i\kappa )^{0}\ _{[0]}^{\shortparallel }\mathbf{\mathbf{\mathbf{%
\mathbf{R}}}}ic_{\alpha \beta }^{\star }+(i\kappa )^{1}\
_{[1]}^{\shortparallel }\mathbf{\mathbf{\mathbf{\mathbf{R}}}}ic_{\alpha
\beta }^{\star }-\kappa ^{2}\ _{[2]}^{\shortparallel }\mathbf{\mathbf{%
\mathbf{\mathbf{R}}}}ic_{\alpha \beta }^{\star },
\end{equation*}%
where the term $\ _{[0]}^{\shortparallel }\mathbf{\mathbf{\mathbf{\mathbf{R}}%
}}ic_{\alpha \beta }^{\star }$ contains certain nonassociative and
noncommutative contributions from star products of type (\ref{starpn}) or (%
\ref{starplie}) and their generalizations for quasi-Hopf structures.
Formulas (4.83) in \cite{aschieri17} provide a holonomic version of such a
decomposition on powers of $\kappa .$\footnote{%
From a formal geometric point of view, we can generalize those formulas
considering that $\partial \rightarrow \ ^{\shortparallel }\mathbf{e}$ and
the nontrivial N-connection structure results in non-zero anholonomy
coefficients $\ ^{\shortparallel }w_{\ \alpha \gamma }^{\nu }$. In our
works, we consider nonholonomic structures which allows us to apply the
AFCDM. This method does not work in a straightforward form for the
nonassociative models elaborated in \cite{blumenhagen16,aschieri17} because
the gravitational equations formulated in coordinate bases do not decouple
in a general form.} We can chose such nonholonomic distributions on a curved
nonholonomic phase space $\mathcal{M}$ when $\ _{[0]}^{\shortparallel }%
\mathbf{\mathbf{\mathbf{\mathbf{R}}}}ic_{\alpha \beta \mid \hbar \rightarrow
0,\kappa \rightarrow 0}^{\star }=\mathbf{R}ic_{\alpha \beta }$ is the Ricci
d-tensor for a real nonholonomic cotangent Lorentz bundle $T^{\ast }\mathbf{%
V.}$ For nontrivial $_{[1]}^{\shortparallel}\mathbf{\mathbf{\mathbf{\mathbf{R%
}}}}ic_{\alpha \beta }^{\star }$ and $\ _{[2]}^{\shortparallel}\mathbf{R}%
ic_{\alpha \beta }^{\star }$, a nonassociative and/or noncommutative $%
\mathcal{M}$ is a complex cotangent manifold with geometric d-objects
determined by string contributions and adapted both to the N-connection and
quasi-Hopf algebra structure. A d-connection $\ ^{\shortparallel }\mathbf{%
\Gamma }_{\star \alpha \beta }^{\gamma }$ is, in general, with a nonzero
torsion. We can extract various types of real or complex configurations by
imposing self-consistent nonholonomic constraints, or to consider limits of
small parameters $\hbar \rightarrow 0$ and/or $\kappa \rightarrow 0$ for
certain well-defined smooth conditions.

\subsection{(Non) symmetric star deformed d-metrics and quasi-Hopf structures%
}

In (pseudo) Riemannian geometry and gravity theories constructed with
(modified) Lagrangians on a Lorentz manifold $V$, the metric $g=\{g_{\alpha
\beta }\}\in $ $T^{\ast }V\otimes T^{\ast }V$ is considered as a fundamental
geometric object defining respectively the gravitational field. For a
nontrivial N-connection structure $\mathbf{g}$ on $V,$ such a metric field
can be transformed into a d-metric, i.e. into a symmetric metric d-tensor
with conventional h- and v-splitting, $\mathbf{g}=(h\mathbf{g},v\mathbf{g}%
)=\{g_{\alpha \beta }=(g_{ij},g_{ab})\},$ when the coefficients are computed
with respect to a N-adapted dual basis $\mathbf{e}^{\alpha }\in T^{\ast }%
\mathbf{TV}$ (\ref{nadap}). To elaborate geometric models on real and/or
complex phase spaces, respectively, on $\mathcal{M}=\mathbf{T}_{\shortmid
}^{\ast }\mathbf{V}$ and/or $\mathcal{M}=\mathbf{T}_{\shortparallel }^{\ast }%
\mathbf{V,}$ we follow the Convention 2 and consider d-tensors $\
^{\shortmid }\mathbf{g}=(h\ ^{\shortmid }\mathbf{g},c\ ^{\shortmid }\mathbf{g%
})=\{\ ^{\shortmid }\mathbf{g}_{\alpha \beta }=(\mathbf{g}_{ij},\
^{\shortmid }\mathbf{g}^{ab})\},$ when the coefficients are computed respect
to tensor products of $\ ^{\shortmid }\mathbf{e}^{\alpha }\in T^{\ast }%
\mathbf{T}_{\shortmid }^{\ast }\mathbf{V}$ (\ref{nadapd}) and $\
^{\shortparallel }\mathbf{g}=(h\ ^{\shortparallel }\mathbf{g},c\
^{\shortparallel }\mathbf{g})=\{\ ^{\shortparallel }\mathbf{g}_{\alpha
\beta}=(\mathbf{g}_{ij},\ ^{\shortparallel }\mathbf{g}^{ab})\},$ with
coefficients computed with respect to tensor products of $\ ^{\shortparallel
}\mathbf{e}^{\alpha }\in T^{\ast }\mathbf{T}_{\shortparallel }^{\ast }%
\mathbf{V}$ (\ref{nadapdc}). Nonassociative star deformations of such
commutative metrics can be elaborated in a N-adapted form generalizing all
constructions from \cite{blumenhagen16,aschieri17}. The main issue of such
nonassociative generalizations is that star deformations, for instance, on
tensor products $T\mathbf{T}_{\shortparallel }^{\ast }\mathbf{V}\otimes
_{\star N}T\mathbf{T}_{\shortparallel }^{\ast }\mathbf{V}$ result both in 
\begin{eqnarray}
\mbox{ symmetric },\ _{\star }^{\shortparallel }\mathbf{g} &=&(h\ _{\star
}^{\shortparallel }\mathbf{g},c\ _{\star }^{\shortparallel }\mathbf{g})=\{\
_{\star }^{\shortparallel }\mathbf{g}_{\alpha \beta }=\
_{\star}^{\shortparallel }\mathbf{g}_{\beta \alpha }=(\ _{\star
}^{\shortparallel }\mathbf{g}_{ij}=\ _{\star }^{\shortparallel }\mathbf{g}%
_{ji},\ _{\star }^{\shortparallel }\mathbf{g}^{ab}=\ \ _{\star
}^{\shortparallel }\mathbf{g}^{ba})\},\mbox{ and }  \label{dmss} \\
\mbox{ nonsymmetric },\ _{\star }^{\shortparallel }\mathfrak{g} &=&(h\
_{\star }^{\shortparallel }\mathfrak{g},c\ _{\star }^{\shortparallel }%
\mathfrak{g})=\{\ _{\star }^{\shortparallel }\mathfrak{g}_{\alpha \beta }=(\
_{\star }^{\shortparallel }\mathfrak{g}_{ij}\neq \ _{\star }^{\shortparallel
}\mathfrak{g}_{ji},\ \ _{\star }^{\shortparallel }\mathfrak{g}^{ab}\neq \
_{\star }^{\shortparallel }\mathfrak{g}^{ba})\neq \ _{\star
}^{\shortparallel }\mathfrak{g}_{\beta \alpha }\},  \notag
\end{eqnarray}%
d-metric structures. In coordinate bases, when $\ ^{\shortparallel }\mathbf{e%
}^{\alpha }\rightarrow \ ^{\shortparallel }e^{\alpha }=d\
^{\shortparallel}u^{\alpha }\in T^{\ast }T_{\shortparallel }^{\ast }V$, such
star deformed metric structures are denoted respectively in generic
off-diagonal 
\begin{equation}
\mbox{ symmetric },\ _{\star }^{\shortparallel }g=\{\ _{\star
}^{\shortparallel }g_{\alpha \beta }=\ _{\star }^{\shortparallel }g_{\beta
\alpha }\},\mbox{ and nonsymmetric },\ _{\star }^{\shortparallel }\mathsf{G}%
=\{\ _{\star }^{\shortparallel }\mathsf{G}_{\alpha \beta }\neq \ _{\star
}^{\shortparallel }\mathsf{G}_{\beta \alpha }\},  \label{offdns}
\end{equation}%
metric structures. For a fixed N-connection and respective coordinate
structures, we have $\ _{\star }^{\shortparallel }\mathbf{g}=\
_{\star}^{\shortparallel }g$ and, respectively, $\ _{\star }^{\shortparallel
}\mathfrak{g}=\ _{\star }^{\shortparallel }\mathsf{G,}$ but with different
block $(4\times 4)+(4\times 4)$ and $8\times 8$ matrices of coefficients
when N-adapted, or local coordinate (co) frame decompositions are
considered. In this and next sections, we shall follow the approach with
quasi-Hopf d-algebras because allows to elaborate on a N-adapted
diffeomorphisms and star deformations which allows more straightforward
quantum generalizations. For instance, we use the symbol $\mathsf{G}$ for
nonsymmetric metrics as in Section 5 of \cite{aschieri17} reconsidering the
formulas in N-adapted form and with boldface symbols and labels $N,\
_{\star}^{\shortparallel },$ etc. In a similar form, we can preform a
nonholonomic generalization of the formulas from Section 6 of \cite%
{blumenhagen16} using a corresponding abstract geometric calculus and
re-definition of nonholonomic structures as we distinguished in previous
sections. We omit in this and next sections dubbing of nonassociative
constructions for different nonholonomic distributions determining different
parameterizations of symmetric and nonsymmetric metric structures.

The goal of this subsection is to study main properties of (non) symmetric
metric and related linear connection structures which are generated by
N-adapted star deformations to quasi-Hopf structures on phase spaces $%
\mathcal{M}=\mathbf{T}_{\shortparallel }^{\ast }\mathbf{V.}$

\subsubsection{The geometry of phase space commutative metrics and canonical
d-connections}

A d--metric $\ ^{\shortparallel }\mathbf{g}$ on $\mathbf{T}%
_{\shortparallel}^{\ast }\mathbf{V}$ is a symmetric d-tensor 
\begin{equation}
\ ^{\shortparallel }\mathbf{g}=\ ^{\shortparallel }\mathbf{g}_{\alpha \beta
}(x,^{\shortparallel }p)\ ^{\shortparallel }\mathbf{e}^{\alpha }\mathbf{%
\otimes \ ^{\shortparallel }e}^{\beta }=\ ^{\shortparallel }g_{\underline{%
\alpha }\underline{\beta }}(x,\ ^{\shortparallel }p)d\ ^{\shortparallel }u^{%
\underline{\alpha }}\mathbf{\otimes }d\ ^{\shortparallel }u^{\underline{%
\beta }},  \label{commetr}
\end{equation}%
which can be parameterized via frame transforms, $\ ^{\shortparallel }%
\mathbf{g}_{\alpha \beta }=\ \ ^{\shortparallel }e_{\ \alpha }^{\underline{%
\alpha }}\ \ ^{\shortparallel }e_{\ \beta }^{\underline{\beta }}\ \
^{\shortparallel }g_{\underline{\alpha }\underline{\beta }}$ in off-diagonal
forms, for instance, with respect to local coordinate dual basis $d\
^{\shortparallel }u^{\underline{\alpha }},$ 
\begin{equation}
\ \ ^{\shortparallel }g_{\underline{\alpha }\underline{\beta }}=\left[ 
\begin{array}{cc}
\ \ ^{\shortparallel }g_{ij}(x)+\ ^{\shortparallel }g^{ab}(x,\
^{\shortparallel }p)\ ^{\shortparallel }N_{ia}(x,^{\shortparallel }p)\ \
^{\shortparallel }N_{jb}(x,\ ^{\shortparallel }p) & \ \ ^{\shortparallel
}g^{ae}\ \ ^{\shortparallel }N_{je}(x,\ ^{\shortparallel }p) \\ 
\ \ ^{\shortparallel }g^{be}\ \ ^{\shortparallel }N_{ie}(x,\
^{\shortparallel }p) & \ \ ^{\shortparallel }g^{ab}(x,\ ^{\shortparallel
}p)\ 
\end{array}%
\right]  \label{offd}
\end{equation}

Using a d-metric $\ ^{\shortparallel }\mathbf{g,}$ we can work in equivalent
form with two different linear connections: 
\begin{equation}
(\ ^{\shortparallel }\mathbf{g,\ ^{\shortparallel }N})\rightarrow \left\{ 
\begin{array}{cc}
\ ^{\shortparallel }\mathbf{\nabla :} & \ ^{\shortparallel }\mathbf{\nabla \
^{\shortparallel }g}=0;\ _{\nabla }^{\shortparallel }\mathcal{T}=0,%
\mbox{\
for  the LC--connection } \\ 
\ ^{\shortparallel }\widehat{\mathbf{D}}: & \ ^{\shortparallel }\widehat{%
\mathbf{D}}\ ^{\shortparallel }\mathbf{g}=0;\ h\ ^{\shortparallel }\widehat{%
\mathcal{T}}=0,c\ ^{\shortparallel }\widehat{\mathcal{T}}=0,hc\
^{\shortparallel }\widehat{\mathcal{T}}\neq 0,\mbox{ for the canonical
d--connection  },%
\end{array}%
\right.  \label{twocon}
\end{equation}%
where $\ ^{\shortparallel }\widehat{\mathbf{D}}=(h\ ^{\shortparallel }%
\widehat{\mathbf{D}},c\ ^{\shortparallel }\widehat{\mathbf{D}}),$ with
covertical splitting, is a d-connection adapted to a N--connection structure 
$\ ^{\shortparallel }\mathbf{N.}$ The LC-connection $\ ^{\shortparallel }%
\mathbf{\nabla }$ is not a d-connection because it does not preserve the
N-connection splitting under parallel transports. For such linear
connections, we can define and compute in standard forms respective
torsions, $\ _{\nabla }^{\shortparallel }\mathcal{T}=0$ and $\
^{\shortparallel }\widehat{\mathcal{T}},$ and curvatures, $\
_{\nabla}^{\shortparallel }\mathcal{R}=\{\ \ _{\nabla }^{\shortparallel
}R_{\ \beta \gamma \delta }^{\alpha }\}$ and $\ ^{\shortparallel }\widehat{%
\mathcal{R}}=\{\ ^{\shortparallel }\widehat{\mathbf{R}}_{\ \beta \gamma
\delta }^{\alpha}\}$, which can be defined and computed in coordinate free
and/or coefficient forms with respect to arbitrary, coordinate, and/or
N-adapted frames. Even $\ ^{\shortparallel }\mathbf{\nabla }$ is not a
d-connection, there is a canonical distortion relation which is N-adapted, 
\begin{equation}
\ ^{\shortparallel }\widehat{\mathbf{D}}=\ ^{\shortparallel }\nabla +\
^{\shortparallel }\widehat{\mathbf{Z}}.  \label{candistr}
\end{equation}%
The distortion d-tensor, $\ ^{\shortparallel }\widehat{\mathbf{Z}}=\{\
^{\shortparallel }\widehat{\mathbf{Z}}_{\ \beta \gamma }^{\alpha }[\
^{\shortparallel }\widehat{\mathbf{T}}_{\ \beta \gamma }^{\alpha }]\},$ is
an algebraic combination of the coefficients of the corresponding torsion
d-tensor $\ ^{\shortparallel }\widehat{\mathcal{T}}=\{\ ^{\shortparallel }%
\widehat{\mathbf{T}}_{\ \beta \gamma }^{\alpha }\}$ of $\ ^{\shortparallel } 
\widehat{\mathbf{D}}.$ This distortion is determined by the N-connection
coefficients and anholonomy coefficients $\ ^{\shortparallel }w_{\alpha
\beta }^{\gamma },$ see similar formulas (\ref{anhrelcd}).

For such linear connections determined by a metric structure, we can define
and compute respective Ricci tensors. For instance, the canonical Ricci
Ricci d-tensor is $\ ^{\shortparallel }\widehat{\mathcal{R}}ic=\{\
^{\shortparallel} \widehat{\mathbf{R}}_{\ \beta \gamma }:=\ ^{\shortparallel
}\widehat{\mathbf{R}}_{\ \alpha \beta \gamma }^{\gamma }\}$ and, for the
LC-connection, $\ ^{\shortparallel }Ric=\{\ ^{\shortparallel }R_{\ \beta
\gamma }:=\ ^{\shortparallel }R_{\ \alpha \beta \gamma }^{\gamma }\}$ (we
may omit the label $\nabla $ writing formulas in not "boldface" forms if
such formulas do not result in ambiguities). The canonical d-tensor $\
^{\shortparallel }\widehat{\mathcal{R}}ic$ is characterized by $h$ -$c$
N-adapted coefficients, 
\begin{equation}
\ ^{\shortparallel }\widehat{\mathbf{R}}_{\alpha \beta }=\{\
^{\shortparallel }\widehat{R}_{ij}:=\ ^{\shortparallel }\widehat{R}_{\
ijk}^{k},\ \ ^{\shortparallel }\widehat{R}_{i}^{\ a}:=-\ ^{\shortparallel }%
\widehat{R}_{\ ik}^{k\ a},\ \ ^{\shortparallel }\widehat{R}_{\ i}^{a}:=\
^{\shortparallel }\widehat{R}_{b\ i}^{a\quad b},\ \ ^{\shortparallel }%
\widehat{R}^{ab}:=\ ^{\shortparallel }\widehat{R}_{c}^{\quad abc}\}.
\label{candricci}
\end{equation}%
Respectively, there are two different scalar curvatures, $\ ^{\shortparallel
}\widehat{\mathbf{R}}:=\ ^{\shortparallel }\mathbf{g}^{\alpha \beta }\
^{\shortparallel }\widehat{\mathbf{R}}_{\alpha \beta }=\ ^{\shortparallel
}g^{ij}\ ^{\shortparallel }\widehat{R}_{ij}+\ ^{\shortparallel }g_{ab}\
^{\shortparallel }\widehat{R}^{ab}$ and $\ ^{\shortparallel }R:=\
^{\shortparallel }\mathbf{g}^{\alpha \beta }\ ^{\shortparallel }R_{\alpha
\beta }.$

We conclude that we can construct a standard (pseudo) Riemannian commutative
geometry and/or a nonholonomic model with nonholonomic induced torsion by
the N-connection structure on $\mathbf{T}_{\shortparallel }^{\ast }\mathbf{V.%
}$ This can be described equivalently by two different geometric data $%
\left(\ ^{\shortparallel }\mathbf{g,\ ^{\shortparallel }\nabla }\right)$ and 
$(\ ^{\shortparallel }\mathbf{g,\ ^{\shortparallel }N,}\ ^{\shortparallel}%
\widehat{\mathbf{D}}).$ Using the canonical distortion relation (\ref%
{candistr}), we can compute respective distortions of geometric objects and
certain important geometric/ physical equations, for instance, of curvature
and Ricci tensors, 
\begin{equation}
\ ^{\shortparallel }\widehat{\mathcal{R}}=\ \ _{\nabla }^{\shortparallel }%
\mathcal{R+}\ \ \ _{\nabla }^{\shortparallel }\widehat{\mathcal{Z}}%
\mbox{
and }\ ^{\shortparallel }\widehat{\mathcal{R}}ic=\ \ _{\nabla
}^{\shortparallel }Ric+\ \ _{\nabla }^{\shortparallel }\widehat{\mathcal{Z}}%
ic,  \label{candriccidist}
\end{equation}%
with corresponding distortion tensors $\ _{\nabla }^{\shortparallel} 
\widehat{\mathcal{Z}}$ and $\ \ _{\nabla }^{\shortparallel }\widehat{%
\mathcal{Z}}ic.$

We provide here the N-adapted coefficients of canonical d-connection $%
\mathbf{T}_{\shortparallel }^{\ast }\mathbf{V}$ (see similar proofs for $%
\mathbf{T}_{\shortmid }^{\ast }\mathbf{V}$ in \cite{vacaru18,bubuianu18a}):%
\begin{eqnarray}
T\mathbf{T}_{\shortparallel }^{\ast }\mathbf{V},\ ^{\shortparallel }\widehat{%
\mathbf{D}} &=&\{\ ^{\shortparallel }\widehat{\mathbf{\Gamma }}_{\ \alpha
\beta }^{\gamma }=(\ \ ^{\shortparallel }\widehat{L}_{jk}^{i},\ \
^{\shortparallel }\widehat{L}_{a\ k}^{\ b},\ \ ^{\shortparallel }\widehat{C}%
_{\ j}^{i\ c},\ \ ^{\shortparallel }\widehat{C}_{\ j}^{i\ c})\},\mbox{ for }%
\lbrack \ \ ^{\shortparallel }\mathbf{g}_{\alpha \beta }=(\ \
^{\shortparallel }g_{jr},\ \ ^{\shortparallel }g^{ab})\mathbf{,\ \
^{\shortparallel }N=\{}\ \ ^{\shortparallel }N_{ai}\mathbf{\}]},  \notag \\
\ \ ^{\shortparallel }\widehat{L}_{jk}^{i} &=&\frac{1}{2}\ \
^{\shortparallel }g^{ir}(\ \ ^{\shortparallel }\mathbf{e}_{k}\ \
^{\shortparallel }g_{jr}+\ \ ^{\shortparallel }\mathbf{e}_{j}\ \
^{\shortparallel }g_{kr}-\ \ ^{\shortparallel }\mathbf{e}_{r}\ \
^{\shortparallel }g_{jk}),\   \notag \\
\ \ ^{\shortparallel }\widehat{L}_{a\ k}^{\ b} &=&\ \ ^{\shortparallel
}e^{b}(\ \ ^{\shortparallel }N_{ak})+\frac{1}{2}\ \ ^{\shortparallel
}g_{ac}(\ \ ^{\shortparallel }e_{k}\ \ ^{\shortparallel }g^{bc}-\ \
^{\shortparallel }g^{dc}\ \ \ ^{\shortparallel }e^{b}\ \ ^{\shortparallel
}N_{dk}-\ \ ^{\shortparallel }g^{db}\ \ \ ^{\shortparallel }e^{c}\ \
^{\shortparallel }N_{dk}),  \notag \\
\ \ ^{\shortparallel }\widehat{C}_{\ j}^{i\ c} &=&\frac{1}{2}\ \
^{\shortparallel }g^{ik}\ \ ^{\shortparallel }e^{c}\ \ ^{\shortparallel
}g_{jk},\ \ \ ^{\shortparallel }\widehat{C}_{\ a}^{b\ c}=\frac{1}{2}\ \
^{\shortparallel }g_{ad}(\ \ ^{\shortparallel }e^{c}\ \ ^{\shortparallel
}g^{bd}+\ \ ^{\shortparallel }e^{b}\ \ ^{\shortparallel }g^{cd}-\ \
^{\shortparallel }e^{d}\ \ ^{\shortparallel }g^{bc}).  \label{canhc}
\end{eqnarray}%
By straightforward computations with such coefficients, we can check that
there are satisfied all necessary conditions from the definition of the
canonical d-connection in (\ref{twocon}). Introducing (\ref{canhc}) in
formulas (\ref{candricci}) and (\ref{candriccidist}), we can compute
respective N-adapted and h-/c-components,for the canonical Ricci d-tensor,
corresponding Ricci scalars and distortions.

In our partner works, we prove that using nonassociative extensions of the
canonical d-connection, for certain well defined nonholonomic
configurations, various types of gravitational and matter field equations
rewritten in nonholonomic variables $(\ ^{\shortparallel }\mathbf{g,\
^{\shortparallel }N},\ ^{\shortparallel }\widehat{\mathbf{D}})$ can be
decoupled and integrated in some general forms.\footnote{%
We emphasize that it is not possible to decouple the gravitational and filed
equations, and geometric flow equations, for generic off-diagonal metrics if
we work from the very beginning with the data $\left(\ ^{\shortparallel }%
\mathbf{g,\ ^{\shortparallel }\nabla }\right) $. Following our works (see
also references therein), we apply the strategy that having constructed
certain classes of generalized solutions (following the AFCDM), necessary
type LC-configurations can be extracted if the condition $\ ^{\shortparallel}%
\widehat{\mathcal{T}}=0$ is imposed at the end.} This motivates our approach
with two linear connections structures (\ref{twocon}). In similar forms, we
can consider another type d-connection structure $^{\shortparallel}\mathbf{D}
$ (for a general d-connection, we omit the "hat" symbol) and compute
respective d-connection N-adapted coefficients, d-curvature etc. If $\
^{\shortparallel }\mathbf{D}\ ^{\shortparallel }\mathbf{g}=\
^{\shortparallel }\mathbf{Q\neq 0,}$ (we note that by definition $\
^{\shortparallel }\widehat{\mathbf{Q}}=0,$ i.e. the canonical d-connection
is metric compatible) we have to work with a so-called nonmetricity d-tensor 
$\ ^{\shortparallel }\mathbf{Q}$ adapted to a N-connection structure as in
Parts II and III of \cite{vacaru05a} for nonholonomic metric-affine and
generalized Finsler-Lagrange-Hamilton spaces in commutative and
noncommutative geometries. Formulas from this subsection can be similarly
introduced for real tangent and contangent Lorentz bundles, $\mathbf{TV}$
and $\mathbf{T}_{\shortmid }^{\ast }\mathbf{V}$, with corresponding
canonical geometric data $(\mathbf{g,N},\mathbf{e,\partial ,}\widehat{%
\mathbf{D}}=\nabla +\widehat{\mathbf{Z}})$ and $(\ ^{\shortmid }\mathbf{g,\
^{\shortmid }N},\mathbf{\ ^{\shortmid }e,\ ^{\shortmid }\partial ,}\
^{\shortmid }\widehat{\mathbf{D}}=\ ^{\shortmid }\nabla +\ ^{\shortmid }%
\widehat{\mathbf{Z}}).$

We note that the star deformation of geometric and physical models following
Convention 1 is based on extensions%
\begin{equation*}
(\mathbf{\ ^{\shortparallel }}g\mathbf{,\ ^{\shortparallel }\partial }
_{_{\alpha }}\mathbf{,\ ^{\shortparallel }}\nabla )\rightarrow (\star \ , 
\mathcal{A}^{\star },\ _{\star }^{\shortparallel }g,\ _{\star
}^{\shortparallel }\mathsf{G},\ \ ^{\shortparallel }\partial _{\alpha }%
\mathbf{,\mathbf{\mathbf{\mathbf{\ ^{\shortparallel }}}}}D^{\star }=\mathbf{%
\ ^{\shortparallel }}\nabla ^{\star }+\mathbf{\ ^{\shortparallel }}Z^{\star
})
\end{equation*}%
as in \cite{blumenhagen16,aschieri17}. On a target star deformed phase
space, we can elaborate on nonassociative geometric models when the
geometric objects are described by coefficients and equations defined with
respect to $d\ ^{\shortparallel }u^{\alpha }$ and$\ ^{\shortparallel
}\partial _{\alpha }$ when there are possible different models of
nonassociative covariant calculus. In general, the AFCDM can not applied for
physically important systems of nonlinear PDEs encoding of geometric and
physical objects.

In this and next sections, for prescriptions of Convention 2 on N-adapted
star deformations, we work with 
\begin{equation*}
(\ \ ^{\shortparallel }\mathbf{g,\ \ ^{\shortparallel }N},\mathbf{\ \
^{\shortparallel }e}_{\alpha }\mathbf{,}\ ^{\shortparallel }\widehat{\mathbf{%
D}}=\ \ ^{\shortparallel }\nabla +\ \ ^{\shortparallel }\widehat{\mathbf{Z}}%
)\rightarrow (\star _{N},\ \ \mathcal{A}_{N}^{\star },\ _{\star
}^{\shortparallel }\mathbf{g,\ _{\star }^{\shortparallel }\mathfrak{g,}\ \
^{\shortparallel }N},\mathbf{\ \ ^{\shortparallel }e}_{\alpha }\mathbf{,\ 
\mathbf{\mathbf{\mathbf{\ ^{\shortparallel }}}}D}^{\star }=\ \
^{\shortparallel }\nabla ^{\star }+\ \ ^{\shortparallel }\widehat{\mathbf{Z}}%
^{\star }).
\end{equation*}%
In such cases, we transform nonholonomic curved tangent Lorentz bundles into
nonassociative nonholonomic phase spaces with complex momentum like
coordinates.

\subsubsection{Nonassociative Levi Civita connections and canonical
d-connections}

Let us study metric aspects of nonassociative nonholonomic differential
geometry and how to construct star deformations of the LC-connections and
canonical d-connections on phase spaces with N-adapted quasi-Hopf
structures. A star metric symmetric d-tensor is taken in the form $\ _{\star
}^{\shortparallel }\mathbf{g}$ (\ref{dmss}). It can be represented as 
\begin{equation*}
\ _{\star }^{\shortparallel }\mathbf{g=\ }_{\star }^{\shortparallel }\mathbf{%
g}_{\alpha \beta }\star _{N}(\ ^{\shortparallel }\mathbf{e}^{\alpha }\otimes
_{\star N}\ ^{\shortparallel }\mathbf{e}^{\beta })\in \Omega _{\star
}^{1}\otimes _{\star N}\Omega _{\star }^{1}
\end{equation*}%
with real-valued N-adapted coefficients $\ _{\star }^{\shortparallel }%
\mathbf{g(\ ^{\shortparallel }\mathbf{e}}_{\alpha }\mathbf{,\
^{\shortparallel }\mathbf{e}}_{\beta }\mathbf{)=}$ $\ _{\star
}^{\shortparallel }\mathbf{g}_{\alpha \beta }=\ _{\star }^{\shortparallel} 
\mathbf{g}_{\beta \alpha }$ $\in \mathcal{A}_{N}^{\star }$ because of the
property $\ _{\star }^{\shortparallel }\mathbf{g(\ ^{\shortparallel }z,\
^{\shortparallel }v)=}\ _{\star }^{\shortparallel }\mathbf{g}(\mathbf{\
_{\intercal }^{\shortparallel }v},\mathbf{\ _{\intercal }^{\shortparallel }z}%
)$ for all $\mathbf{\ ^{\shortparallel }z,\ ^{\shortparallel }v\in }\emph{Vec%
}_{\star _{N}}$ when the braiding d-operator "$\mathbf{\ _{\intercal }^{{}}}$
"\thinspace\ is defined by (\ref{braidop}). We state the d-metric
compatibility with a star d-connection $\mathbf{\mathbf{\mathbf{\mathbf{\
^{\shortparallel }}}}D}^{\star }$ (\ref{dconhopf}) by the condition 
\begin{equation}
\mathbf{\mathbf{\mathbf{\mathbf{\ ^{\shortparallel }}}}D}^{\star }\ _{\star
}^{\shortparallel }\mathbf{g}=0.  \label{mcompnas}
\end{equation}

\paragraph{Nonsymmetric metrics and d-metrics and their inverses: \newline
}

For nonassociative phase space, we introduce a nonsymmetric metric (\ref%
{offdns}) in generic off-diagonal form with respect to a local coordinate
base,%
\begin{equation}
\ _{\star }^{\shortparallel }\mathsf{G}_{\alpha \beta }=\ _{\star
}^{\shortparallel }g_{\alpha \beta }-i\kappa \mathcal{R}_{\quad \alpha
}^{\tau \xi }\ \mathbf{^{\shortparallel }}\partial _{\xi }\ _{\star
}^{\shortparallel }g_{\beta \tau }.  \label{offdns1}
\end{equation}%
In N-adapted form with respect to $\ \mathbf{^{\shortparallel }e}_{\xi },$
such a nonsymmetric d-metric structure (\ref{dmss}) is parameterized 
\begin{equation}
\ _{\star }^{\shortparallel }\mathfrak{g}_{\alpha \beta }=\ _{\star
}^{\shortparallel }\mathbf{g}_{\alpha \beta }-i\kappa \overline{\mathcal{R}}%
_{\quad \alpha }^{\tau \xi }\ \mathbf{^{\shortparallel }e}_{\xi }\ _{\star
}^{\shortparallel }\mathbf{g}_{\beta \tau },  \label{dmss1}
\end{equation}%
where $\overline{\mathcal{R}}_{\quad \alpha }^{\tau \xi }$ and $\mathcal{R}%
_{\quad \alpha }^{\tau \xi }$ are related by a star nonsymmetric
generalization of (\ref{commetr}) with respective frame transforms, 
\begin{equation*}
\ _{\star }^{\shortparallel }\mathfrak{g} =\ _{\star}^{\shortparallel } 
\mathfrak{g}_{\alpha \beta }\star _{N}(\ ^{\shortparallel }\mathbf{e}%
^{\alpha }\otimes _{\star N}\ ^{\shortparallel }\mathbf{e}^{\beta })=\
_{\star }^{\shortparallel }\mathsf{G}_{\alpha \beta }\star (\ d\
^{\shortparallel }u^{\alpha }\otimes _{\star }d\ ^{\shortparallel }u^{\beta
}),
\end{equation*}%
where $\ _{\star }^{\shortparallel }\mathfrak{g}_{\alpha \beta} \neq \
_{\star }^{\shortparallel }\mathfrak{g}_{\beta \alpha }$ and $\ _{\star
}^{\shortparallel }\mathsf{G}_{\alpha \beta }\neq \ _{\star}^{\shortparallel
}\mathsf{G}_{\beta \alpha }.$ Such nonsymmetric d-metric and metric
structures are nondegenerate and can be sought as R-flux corrected
"effective metrics" from string theory. For quasi-Hopf configurations, we
should consider the formalism of constructing inversions of matrices in $%
\mathcal{A}^{\star },$ see details in section 5.2 of \cite{aschieri17}. We
can compute the inverse matrix $\ ^{\shortparallel }\overline{\mathsf{G}}%
^{-1}=\{\ _{\star }^{\shortparallel }\mathsf{G}^{\alpha \beta }\}$ of the
matrix $\ ^{\shortparallel }\overline{\mathsf{G}}=\{\ _{\star
}^{\shortparallel }\mathsf{G}_{\alpha \beta }\}$ as a solution of algebraic
equations $\ _{\star }^{\shortparallel }\mathsf{G}^{\alpha \beta }\cdot \
_{\star }^{\shortparallel }\mathsf{G}_{\beta \gamma }=\ _{\star
}^{\shortparallel }\mathsf{G}_{\gamma \beta }\cdot $ $\ _{\star
}^{\shortparallel }\mathsf{G}^{\beta \alpha }=\delta _{\beta }^{\alpha }.$
In compact symbolic matrix form, we can write 
\begin{equation*}
\ _{\star }^{\shortparallel }\overline{\mathsf{G}}^{-1}=(\mathbb{I}+i\kappa
\ ^{\shortparallel }\overline{g}^{-1}\mathcal{R\partial }\ ^{\shortparallel }%
\overline{g})^{-1}\ ^{\shortparallel }\overline{g}^{-1},
\end{equation*}%
where $\mathbb{I}$ is the $4\times 4$ unity matrix, $\mathcal{R\partial }\
^{\shortparallel }\overline{g}=\mathcal{R}_{\quad \alpha }^{\tau \xi }\ \ 
\mathbf{^{\shortparallel }}\partial _{\xi }\ ^{\shortparallel }g_{\beta \tau
},$ the symmetric matrix $\ ^{\shortparallel }\overline{g}^{-1}=\{\
^{\shortparallel }g^{\beta \tau }\}$ is the inverse of symmetric $\
^{\shortparallel }\overline{g}=\{\ ^{\shortparallel }\mathbf{g}_{\beta \tau
}\}.$ Such matrix representations of formulas are understood as geometric
series, when, for instance, 
\begin{equation}
\ _{\star }^{\shortparallel }\mathsf{G}^{\alpha \beta }=\ ^{\shortparallel
}g^{\alpha \beta }-i\kappa \ ^{\shortparallel }g^{\alpha \tau }\mathcal{R}%
_{\quad \tau }^{\mu \nu }(\partial _{\mu }\ ^{\shortparallel }g_{\nu
\varepsilon })\ ^{\shortparallel }g^{\varepsilon \beta }+O(\kappa ^{2}).
\label{offdns1inv}
\end{equation}%
For any star like product $\ast ,$ we can compute $\ _{\ast}^{\shortparallel
}\overline{\mathsf{G}}^{-1}=\{\ _{\ast }^{\shortparallel }\mathsf{G}^{\alpha
\beta }\}$ as the $\ast $-inverse of the matrix $\ _{\ast}^{\shortparallel }%
\overline{\mathsf{G}}=\{\ _{\ast }^{\shortparallel }\mathsf{G}_{\alpha \beta
}\}$ as a solution of star like equation $\ _{\ast }^{\shortparallel }%
\mathsf{G}^{\alpha \beta }\ast \ _{\ast }^{\shortparallel }\mathsf{G}_{\beta
\gamma }=\ _{\ast }^{\shortparallel }\mathsf{G}_{\gamma \beta }\ast $ $\
_{\ast }^{\shortparallel }\mathsf{G}^{\beta \alpha }=\delta _{\gamma
}^{\alpha },$ where $\cdot \rightarrow \ast $ for a star associated to $%
\hbar $ and which can be different from the $\star $-product. Such equations
define, for instance, power series on $\hbar ,$%
\begin{equation}
\ _{\ast }^{\shortparallel }\mathsf{G}^{\alpha \beta }=\ _{\star
}^{\shortparallel }\mathsf{G}^{\alpha \beta }+\sum_{m=1}^{\infty }\ _{\star
}^{\shortparallel }\mathsf{G}^{\alpha \gamma }\ast (\mathbb{I-}\ _{\star
}^{\shortparallel }\overline{\mathsf{G}}\ast \ _{\star }^{\shortparallel }%
\overline{\mathsf{G}}^{-1})_{\quad \gamma }^{\ast m\ \beta }.  \label{aux34}
\end{equation}

Similar formulas can be written in block $(2\times 2)\times (2\times 2)$
forms for matrices constructed from N-adapted coefficients of a nonsymmetric
star d-metric $\ _{\star }^{\shortparallel }\mathfrak{g}_{\alpha \beta }$
and, respectively, of a symmetric star d-metric $\ _{\star}^{\shortparallel }%
\mathbf{g}_{\alpha \beta }.$

\paragraph{Star deformed LC-connection and canonical d-connection: \newline
}

We define star deformations of the two linear connections structure (\ref%
{twocon}) using a star d-metric $\ _{\star }^{\shortparallel }\mathbf{g}$
and work in equivalent forms with respective two different linear
connections: 
\begin{equation}
(\ _{\star }^{\shortparallel }\mathbf{g,\ ^{\shortparallel }N})\rightarrow
\left\{ 
\begin{array}{cc}
\ _{\star }^{\shortparallel }\mathbf{\nabla :} & 
\begin{array}{c}
\ _{\star }^{\shortparallel }\mathbf{\nabla \ \ _{\star }^{\shortparallel }g}%
=0;\ _{\nabla }^{\shortparallel }\mathcal{T}^{\star }=0,%
\mbox{\
for  the star  LC--connection }; \\ 
\end{array}
\\ 
\mathbf{\mathbf{\mathbf{\mathbf{\ ^{\shortparallel }}}}}\widehat{\mathbf{D}}%
^{\star }: & 
\begin{array}{c}
\mathbf{\mathbf{\mathbf{\mathbf{\ ^{\shortparallel }}}}}\widehat{\mathbf{D}}%
^{\star }\ \mathbf{\ _{\star }^{\shortparallel }g}=0;\ h\ ^{\shortparallel }%
\widehat{\mathcal{T}}^{\star }=\{\mathbf{\mathbf{\mathbf{\mathbf{\ \
^{\shortparallel }}}}}\widehat{T}_{\star \ jk}^{i}\}=0,c\ ^{\shortparallel }%
\widehat{\mathcal{T}}^{\star }=\{\mathbf{\mathbf{\mathbf{\mathbf{\ \
^{\shortparallel }}}}}\widehat{T}_{\star a\ }^{\ bc}\}=0,hc\
^{\shortparallel }\widehat{\mathcal{T}}^{\star }\neq 0, \\ 
\mbox{ for the star canonical d--connection  }.%
\end{array}%
\end{array}%
\right.  \label{twoconnonas}
\end{equation}%
In these formulas, the nonassociative metric compatibility is given
respectively as in (\ref{mcompnas}) and the star d-torsion components are
parameterized as in formulas (\ref{dtorsnonassoc}), when 
\begin{equation*}
\ ^{\shortparallel }\widehat{\mathcal{T}}^{\star }=\{\mathbf{\mathbf{\mathbf{%
\mathbf{\ ^{\shortparallel }}}}}\widehat{\mathbf{T}}_{\star \alpha \beta
}^{\gamma }=(\ \mathbf{\mathbf{\mathbf{\mathbf{\ ^{\shortparallel }}}}}%
\widehat{T}_{\star jk}^{i}=0,\ \mathbf{\mathbf{\mathbf{\mathbf{\ \
^{\shortparallel }}}}}\widehat{T}_{\star \ j}^{i\ a},\ \mathbf{\mathbf{%
\mathbf{\mathbf{\ \ ^{\shortparallel }}}}}\widehat{T}_{\star aji},\ \mathbf{%
\mathbf{\mathbf{\mathbf{\ ^{\shortparallel }}}}}\widehat{T}_{\star a\ i}^{\
b},\ \mathbf{\mathbf{\mathbf{\mathbf{\ \ ^{\shortparallel }}}}}\widehat{T}%
_{\star a\ }^{\ bc}=0)\},
\end{equation*}%
by such N-adapted coefficients with such imposed nonholonomic conditions 
\begin{eqnarray}
\mathbf{\mathbf{\mathbf{\mathbf{\ \ ^{\shortparallel }}}}}\widehat{T}_{\star
\ jk}^{i} &=&\mathbf{\mathbf{\mathbf{\mathbf{\ \ ^{\shortparallel }}}}}%
\widehat{L}_{\star jk}^{i}-\mathbf{\mathbf{\mathbf{\mathbf{\ \
^{\shortparallel }}}}}\widehat{L}_{\star kj}^{i}=0,\mathbf{\mathbf{\mathbf{%
\mathbf{\ \ ^{\shortparallel }}}}}\widehat{T}_{\star \ j}^{i\ a}=\mathbf{%
\mathbf{\mathbf{\mathbf{\ ^{\shortparallel }}}}}\widehat{C}_{\star j}^{ia},\ 
\mathbf{\mathbf{\mathbf{\mathbf{\ ^{\shortparallel }}}}}\widehat{T}_{\star
aji}=-\ \mathbf{\mathbf{\mathbf{\mathbf{\ \ ^{\shortparallel }}}}}\Omega
_{\star aji},  \notag \\
\mathbf{\mathbf{\mathbf{\mathbf{\ \ ^{\shortparallel }}}}}\widehat{T}_{\star
c\ j}^{\ a} &=&\mathbf{\mathbf{\mathbf{\mathbf{\ \ ^{\shortparallel }}}}}%
\widehat{L}_{\star c\ j}^{\ a}-\mathbf{\mathbf{\mathbf{\mathbf{\
^{\shortparallel }}}}}e^{a}(\mathbf{\mathbf{\mathbf{\mathbf{\
^{\shortparallel }}}}}N_{\star cj}),\mathbf{\mathbf{\mathbf{\mathbf{\ \
^{\shortparallel }}}}}\widehat{T}_{\star a\ }^{\ bc}=\mathbf{\mathbf{\mathbf{%
\mathbf{\ ^{\shortparallel }}}}}\widehat{C}_{\star a}^{\ bc}-\mathbf{\mathbf{%
\mathbf{\mathbf{\ \ ^{\shortparallel }}}}}\widehat{C}_{\star a}^{\ cb}=0.
\label{candtorsnonas}
\end{eqnarray}%
We can consider a set of N-adapted Christoffel symbols 
\begin{eqnarray*}
&&\ _{[0]}^{\shortparallel }\widehat{\mathbf{\Gamma }}_{\star \gamma \alpha
\beta }=\ _{\star }^{\shortparallel }\mathbf{g}_{\gamma \tau }\
_{[0]}^{\shortparallel }\widehat{\mathbf{\Gamma }}_{\star \beta \alpha
}^{\tau }= \\
&&(\ _{[0]}^{\shortparallel }\widehat{L}_{\star ijk}=\ _{\star
}^{\shortparallel }g_{im}\ \ _{[0]}^{\shortparallel }\widehat{L}_{\star
jk}^{m},\ _{[0]}^{\shortparallel }\widehat{L}_{\star \ k}^{\ ab}=\ _{\star
}^{\shortparallel }g^{ac}\ \ _{[0]}^{\shortparallel }\widehat{L}_{\star c\
k}^{\ b},\ _{[0]}^{\shortparallel }\widehat{C}_{\star ij}^{\quad c}=\
_{\star }^{\shortparallel }g_{im}\ _{[0]}^{\shortparallel }\widehat{C}%
_{\star \ j}^{m\ c},\ _{[0]}^{\shortparallel }\widehat{C}_{\star }^{bec}=\
_{\star }^{\shortparallel }g^{ae}\ _{[0]}^{\shortparallel }\widehat{C}%
_{\star \ e}^{b\ c}),
\end{eqnarray*}%
where h-c-components are similar to (\ref{canhc}) with $^{\shortparallel }%
\mathbf{e}_{\alpha }=(\ ^{\shortparallel }\mathbf{e}_{i},\ ^{\shortparallel
}e^{a})$ (\ref{nadapdc}), and$\mathbf{\ \ ^{\shortparallel }N=\{}\
^{\shortparallel }N_{ai}\mathbf{\}}$ but with star labels for $\ _{\star
}^{\shortparallel }\mathbf{g}_{\alpha \beta }=(\ _{\star}^{\shortparallel
}g_{jr},\ _{\star }^{\shortparallel }g^{ab})$ (\ref{dmss}), \ 
\begin{eqnarray}
\ \ _{[0]}^{\shortparallel }\widehat{L}_{\star ijk} &=&\frac{1}{2}\ (\
^{\shortparallel }\mathbf{e}_{k}\ \ _{\star }^{\shortparallel }g_{ji}+\
^{\shortparallel }\mathbf{e}_{j}\ _{\star }^{\shortparallel }g_{ki}-\
^{\shortparallel }\mathbf{e}_{i}\ _{\star }^{\shortparallel }g_{jk}),\ 
\label{0canconnonas} \\
\ \ _{[0]}^{\shortparallel }\widehat{L}_{\star \ k}^{\ ab} &=&\ _{\star
}^{\shortparallel }g^{ac}\ ^{\shortparallel }e^{b}(\ ^{\shortparallel
}N_{ck})+\frac{1}{2}\ (\ ^{\shortparallel }e_{k}\ \ _{\star
}^{\shortparallel }g^{ba}-\ \ \ _{\star }^{\shortparallel }g^{da}\
^{\shortparallel }e^{b}\ ^{\shortparallel }N_{dk}-\ \ _{\star
}^{\shortparallel }g^{db}\ \ ^{\shortparallel }e^{a}\ ^{\shortparallel
}N_{dk}),  \notag \\
\ _{[0]}^{\shortparallel }\widehat{C}_{\star ij}^{\quad c} &=&\frac{1}{2}\ \
^{\shortparallel }e^{c}\ \ _{\star }^{\shortparallel }g_{ij},\ \
_{[0]}^{\shortparallel }\widehat{C}_{\star }^{bec}=\frac{1}{2}\ (\
^{\shortparallel }e^{c}\ \ _{\star }^{\shortparallel }g^{be}+\
^{\shortparallel }e^{b}\ \ _{\star }^{\shortparallel }g^{ce}-\
^{\shortparallel }e^{e}\ \ \ _{\star }^{\shortparallel }g^{bc}).  \notag
\end{eqnarray}%
It is possible to introduce 
\begin{equation}
\ _{[1]}^{\shortparallel }\widehat{\mathbf{\Gamma }}_{\star \alpha \beta \mu
}=\frac{1}{2}\overline{\mathcal{R}}_{\quad \mu }^{\xi \tau }\ (\ \mathbf{%
^{\shortparallel }e}_{\xi }\ \mathbf{^{\shortparallel }e}_{\alpha }\ _{\star
}^{\shortparallel }\mathbf{g}_{\beta \tau }+\ \mathbf{^{\shortparallel }e}%
_{\xi }\mathbf{^{\shortparallel }e}_{\beta }\ _{\star }^{\shortparallel }%
\mathbf{g}_{\alpha \tau })  \label{aux311}
\end{equation}%
and define the N-adapted coefficients of the star canonical d-connection $%
\mathbf{\mathbf{\mathbf{\mathbf{\ ^{\shortparallel }}}}}\widehat{\mathbf{D}}%
^{\star }=\{\ ^{\shortparallel }\widehat{\mathbf{\Gamma }}_{\star \alpha
\beta }^{\gamma }\},$ where 
\begin{equation}
\ ^{\shortparallel }\widehat{\mathbf{\Gamma }}_{\star \alpha \beta }^{\gamma
}=\ _{[0]}^{\shortparallel }\widehat{\mathbf{\Gamma }}_{\star \beta \alpha
}^{\tau }+i\kappa \ _{[1]}^{\shortparallel }\widehat{\mathbf{\Gamma }}%
_{\star \beta \alpha }^{\tau }=(\ ^{\shortparallel }\widehat{L}_{\star
jk}^{i},\ ^{\shortparallel }\widehat{L}_{\star a\ k}^{\ b},\
^{\shortparallel }\widehat{C}_{\star \ j}^{i\ c},\ ^{\shortparallel }%
\widehat{C}_{\star ab}^{\quad c})  \label{aux311a}
\end{equation}%
are parameterized in the form (\ref{irevndecomdc}) and (for the nonsymmetric
d-metric $\ _{\star }^{\shortparallel }\mathfrak{g}_{\alpha \beta }$ (\ref%
{dmss1})) subjected to the condition%
\begin{equation}
\ \mathbf{^{\shortparallel }e}^{\tau }\star _{N}\ _{\star }^{\shortparallel }%
\mathfrak{g}_{\tau \gamma }\star _{N}\ ^{\shortparallel }\widehat{\mathbf{%
\Gamma }}_{\star \alpha \beta }^{\gamma }=\ \mathbf{^{\shortparallel }e}%
^{\tau }\star _{N}(\ _{[0]}^{\shortparallel }\widehat{\mathbf{\Gamma }}%
_{\star \tau \alpha \beta }+i\kappa \ _{[1]}^{\shortparallel }\widehat{%
\mathbf{\Gamma }}_{\star \tau \alpha \beta }).  \label{eqnasdmdc}
\end{equation}%
In these formulas, the h-c components of $\ _{[0]}^{\shortparallel }\widehat{%
\mathbf{\Gamma }}_{\star \tau \alpha \beta }$ are determined by (\ref%
{0canconnonas}) and $\ _{[1]}^{\shortparallel }\widehat{\mathbf{\Gamma }}%
_{\star \alpha \beta \mu }$(\ref{aux311}) can be computed with h- and
c-compositions of $\ _{\star }^{\shortparallel }\mathbf{g}_{\beta \tau }.$

In a local coordinate basis $\ \mathbf{^{\shortparallel }\partial }_{\tau }$%
, the equations (\ref{eqnasdmdc}) for the LC-connection$\ ^{\shortparallel } 
\mathbf{\nabla =\{\mathbf{\ ^{\shortparallel }}}\Gamma _{\star \beta \alpha
}^{\gamma }\mathbf{\}}$ can be written in the form 
\begin{equation*}
\ _{\star }^{\shortparallel }\mathsf{G}_{\gamma \nu }\star \mathbf{\mathbf{\
^{\shortparallel }}}\Gamma _{\star \alpha \beta }^{\nu }=\frac{1}{2}\left( \ 
\mathbf{^{\shortparallel }\partial }_{\beta }\ _{\star }^{\shortparallel
}g_{\alpha \gamma }+\ \mathbf{^{\shortparallel }\partial }_{\alpha }\
_{\star }^{\shortparallel }g_{\beta \gamma }-\ \mathbf{^{\shortparallel
}\partial }_{\gamma }\ _{\star }^{\shortparallel }g_{\alpha \beta }+i\kappa 
\mathcal{R}_{\ \gamma }^{\xi \tau }\ (\ \mathbf{^{\shortparallel }}\partial
_{\xi }\ \mathbf{^{\shortparallel }}\partial _{\beta }\ _{\star
}^{\shortparallel }g_{\alpha \tau }+\ \mathbf{^{\shortparallel }}\partial
_{\xi }\ \mathbf{^{\shortparallel }}\partial _{\alpha }\ _{\star
}^{\shortparallel }g_{\beta \tau })\right) ,
\end{equation*}%
where the off-diagonal nonsymmetric metric $\ _{\star }^{\shortparallel }%
\mathsf{G}_{\alpha \beta }$ is computed as in (\ref{offdns1}). Such a
formula is equivalent to (5.13) in \cite{aschieri17} and allow to extract
the coefficients $\mathbf{\mathbf{\ ^{\shortparallel }}}\Gamma _{\star\alpha
\beta }^{\nu }$ with respect to local coordinate dual frames. Changing $\ 
\mathbf{^{\shortparallel }}\partial _{\alpha }\rightarrow \ \mathbf{%
^{\shortparallel }e}_{\alpha },$ we can compute N-adapted coefficients of $\
^{\shortparallel }\mathbf{\nabla }$ and nonassociative star deformation of
the canonical distortion relation (\ref{candistr}), 
\begin{equation}
\ ^{\shortparallel }\widehat{\mathbf{D}}^{\star }=\ ^{\shortparallel }\nabla
^{\star }+\ _{\star }^{\shortparallel }\widehat{\mathbf{Z}}.
\label{candistrnas}
\end{equation}%
The distortion d-tensor, $\ ^{\shortparallel }\widehat{\mathbf{Z}}=\{\
^{\shortparallel }\widehat{\mathbf{Z}}_{\ \beta \gamma }^{\alpha }[\
^{\shortparallel }\widehat{\mathbf{T}}_{\ \beta \gamma }^{\alpha }]\},$ is
an algebraic combination of the coefficients of the nonassociative canonical
d-tensor $\ ^{\shortparallel }\widehat{\mathcal{T}}=\{\ ^{\shortparallel }%
\widehat{\mathbf{T}}_{\ \beta \gamma }^{\alpha }\}$ with h- c-components (%
\ref{candtorsnonas}) computed using formulas (\ref{0canconnonas}).

We conclude that we can describe any model of nonassociative phase geometry
determined by a nonsymmetric d-metric structure $\ _{\star }^{\shortparallel}%
\mathfrak{g}_{\alpha \beta }=\ _{\star }^{\shortparallel }\mathbf{g}_{\alpha
\beta }-i\kappa \overline{\mathcal{R}}_{\quad \alpha }^{\tau \xi }\
^{\shortparallel } \mathbf{e}_{\xi }\ _{\star }^{\shortparallel }\mathbf{g}%
_{\beta \tau }$ (for any prescribed N-connection structure and R-flux, we
can consider only the symmetric d-metric $\ _{\star }^{\shortparallel }%
\mathbf{g}_{\alpha \beta }$) can be described equivalently in terms of the
star LC-connection $\ ^{\shortparallel }\nabla ^{\star }$ and/or the
canonical d-connection $\ ^{\shortparallel }\widehat{\mathbf{D}}^{\star }.$
For any nonholonomic configuration determined by geometric data $(\
_{\star}^{\shortparallel }\mathfrak{g}_{\alpha \beta },\ ^{\shortparallel }%
\widehat{\mathbf{D}}^{\star }),$ we can extract nonassociative
LC-configurations imposing zero torsion conditions%
\begin{equation}
\ _{\star }^{\shortparallel }\widehat{\mathbf{Z}}=0,%
\mbox{ which is
equivalent to }\ ^{\shortparallel }\widehat{\mathbf{D}}_{\mid \
^{\shortparallel }\widehat{\mathbf{T}}=0}^{\star }=\ ^{\shortparallel
}\nabla ^{\star }.  \label{lccondnonass}
\end{equation}%
Even such conditions are satisfied for certain geometric models, the
anholonomy coefficients $\ ^{\shortparallel }w_{\alpha \beta }^{\gamma }$
may be not zero, see similar formulas (\ref{anhrelcd}), which means that we
work with general nonholonomic frames and involved (non) symmetric metrics
are generic off-diagonal (i.e. can not be diagonalized by coordinate
transforms on a finite spacetime or phase space region) if we re-write them
with respect to local coordinate frames.

\paragraph{Decompositions on small parameters of nonassociative d-metric/
-connection coefficients: \newline
}

For nonassociative geometry with quasi-Hopf structures, such coordinate
frame formulas for the (non) symmetric metrics and LC-connections with small
parameters $\hbar $ and $\kappa $ are provided in section 5.3 of \cite%
{aschieri17}. All nonassociative coefficient formulas from that work can be
re-written in N-adapted for two linear connection structures (\ref%
{twoconnonas}) using such direct or inverse nonholonomic transforms and
canonical distortions (\ref{candistrnas}) of geometric data: 
\begin{equation}
(\ \mathbf{^{\shortparallel }\partial }_{\alpha },\ _{\star
}^{\shortparallel }\mathsf{G}_{\gamma \nu },\ ^{\shortparallel }\nabla
^{\star })\longleftrightarrow (\ \mathbf{^{\shortparallel }e}_{\alpha },\
^{\shortparallel }w_{\alpha \beta }^{\gamma },\ _{\star }^{\shortparallel }%
\mathfrak{g}_{\alpha \beta },\ ^{\shortparallel }\widehat{\mathbf{D}}^{\star
}=\ ^{\shortparallel }\nabla ^{\star }+\ _{\star }^{\shortparallel }\widehat{%
\mathbf{Z}}),  \label{conv2a}
\end{equation}%
where certain new terms of the d-curvature and d-torsion, and Ricci
d-tensor, additionally to those with $\ \mathbf{^{\shortparallel }\partial }%
_{\alpha }$ (which are holonomic) appear for frame transforms to
nonholonomic N-adapted frames$\ \mathbf{^{\shortparallel }e}_{\alpha }$ and
their nontrivial anholonomy coefficeints $\ ^{\shortparallel }w_{\alpha
\beta}^{\gamma }.$ We provide here certain important formulas which show the
nonassociative geometric formalism is different for holonomic and
nonholonomic structures.

The formulas (\ref{offdns1}) and (\ref{offdns1inv}) for the nonsymmetric
off-diagonal metric and it inverse metric can be written respectively in the
form 
\begin{eqnarray}
\ _{\star }^{\shortparallel }\mathsf{G}_{\alpha \beta } &=&\ _{\star
}^{\shortparallel }g_{\alpha \beta }-i\kappa \mathcal{R}_{\quad \alpha
}^{\tau \xi }\ \ \mathbf{^{\shortparallel }}\partial _{\xi }\ _{\star
}^{\shortparallel }g_{\beta \tau }:=\ _{\star }^{\shortparallel }\mathsf{G}%
_{\alpha \beta }^{[0]}+\ _{\star }^{\shortparallel }\mathsf{G}_{\alpha \beta
}^{[1]}(\kappa )\mbox{ and }  \label{aux35} \\
\ _{\star }^{\shortparallel }\mathsf{G}^{\alpha \beta } &=&\
^{\shortparallel }g^{\alpha \beta }-i\kappa \ ^{\shortparallel }g^{\alpha
\tau }\mathcal{R}_{\quad \tau }^{\mu \nu }(\ \mathbf{^{\shortparallel }}%
\partial _{\mu }\ ^{\shortparallel }g_{\nu \varepsilon })\ ^{\shortparallel
}g^{\varepsilon \beta }+O(\kappa ^{2}):=\ _{\star }^{\shortparallel }\mathsf{%
G}_{[0]}^{\alpha \beta }+\ _{\star }^{\shortparallel }\mathsf{G}%
_{[1]}^{\alpha \beta }(\kappa )+O(\kappa ^{2}),  \notag \\
\ _{\ast }^{\shortparallel }\mathsf{G}^{\alpha \beta } &=&2\ _{\star
}^{\shortparallel }\mathsf{G}^{\alpha \beta }-\ _{\star }^{\shortparallel }%
\mathsf{G}^{\alpha \gamma }\ast \ _{\star }^{\shortparallel }\mathsf{G}%
_{\gamma \tau }\ast \ _{\star }^{\shortparallel }\mathsf{G}^{\tau \beta
}+O(\hbar ^{2}),\mbox{ in this case, we follow the rules (\ref{aux34})}. 
\notag
\end{eqnarray}%
This allow to derive formulas which are parametric on $\hbar $ and $\kappa $
and can be written in generic off-diagonal and/or N-adapted forms. We
introduce 
\begin{equation}
\ _{\star }^{\shortparallel }\mathsf{W}_{\gamma \alpha \beta }=\ _{\star
}^{\shortparallel }g_{\gamma \mu }\mathbf{\mathbf{\ ^{\shortparallel }}}%
\Gamma _{\star \alpha \beta }^{\nu }+i\kappa \mathcal{R}_{\quad \alpha
}^{\tau \xi }\ \ \mathbf{^{\shortparallel }}\partial _{\xi }\ _{\star
}^{\shortparallel }g_{\beta \tau }:=\ _{\star }^{\shortparallel }\mathsf{W}%
_{\gamma \alpha \beta }^{[0]}+\ _{\star }^{\shortparallel }\mathsf{W}%
_{\gamma \alpha \beta }^{[1]}(\kappa ),  \label{aux36}
\end{equation}%
where $\mathbf{\mathbf{\ ^{\shortparallel }}}\Gamma _{\star \alpha \beta
\mid \hbar ,\kappa =0}^{\nu }=\ ^{\shortparallel }\Gamma _{\alpha \beta
}^{\nu }$ can be considered as the coefficients of the commutative
LC-connection $\ ^{\shortparallel }\nabla $ in (\ref{twocon}). Such formulas
are written with respect to coordinate basis but we can also consider
arbitrary frame transforms. For nontrivial $\hbar $ and $\kappa ,$ we work
with a nonassociative $\ ^{\shortparallel }\nabla ^{\star }$ from (\ref%
{twoconnonas}).

The techniques of decomposition on small parameters used in (\ref{aux35})
and (\ref{aux36}) can be extended in a similar to canonical data $(\
_{\star}^{\shortparallel }\mathfrak{g}_{\alpha \beta },\ ^{\shortparallel }%
\widehat{\mathbf{D}}^{\star })$ following Convention 2 and rule (\ref{conv2a}%
), working in N-adapted (co) bases. Up to first order on $\hbar $ and first
order on $\kappa ,$ we express conventionally 
\begin{eqnarray}
\ _{\star }^{\shortparallel }\mathfrak{g}_{\alpha \beta } &=&\ _{\star
}^{\shortparallel }\mathbf{g}_{\alpha \beta }-i\kappa \overline{\mathcal{R}}%
_{\quad \alpha }^{\tau \xi }\ \mathbf{^{\shortparallel }e}_{\xi }\ _{\star
}^{\shortparallel }\mathbf{g}_{\beta \tau }:=\ _{\star }^{\shortparallel }%
\mathfrak{g}_{\alpha \beta }^{[0]}+\ _{\star }^{\shortparallel }\mathfrak{g}%
_{\alpha \beta }^{[1]}(\kappa )\mbox{ and }  \notag \\
\ _{\star }^{\shortparallel }\mathfrak{g}^{\alpha \beta } &=&\ \ _{\star
}^{\shortparallel }\mathbf{g}^{\alpha \beta }-i\kappa \ \ _{\star
}^{\shortparallel }\mathbf{g}^{\alpha \tau }\overline{\mathcal{R}}_{\quad
\tau }^{\mu \nu }(\ \mathbf{^{\shortparallel }e}_{\mu }\ \ _{\star
}^{\shortparallel }\mathbf{g}_{\nu \varepsilon })\ \ _{\star
}^{\shortparallel }\mathbf{g}^{\varepsilon \beta }+O(\kappa ^{2})\ :=\
_{\star }^{\shortparallel }\mathfrak{g}_{[0]}^{\alpha \beta }+\ _{\star
}^{\shortparallel }\mathfrak{g}_{[1]}^{\alpha \beta }(\kappa )+O(\kappa
^{2}),  \notag \\
\ _{\ast }^{\shortparallel }\mathfrak{g}^{\alpha \beta } &=&2\ _{\star
}^{\shortparallel }\mathfrak{g}^{\alpha \beta }-\ _{\star }^{\shortparallel }%
\mathfrak{g}^{\alpha \gamma }\ast \ _{\star }^{\shortparallel }\mathfrak{g}%
_{\gamma \tau }\ast \ _{\star }^{\shortparallel }\mathfrak{g}^{\tau \beta
}+O(\hbar ^{2}),  \label{aux37}
\end{eqnarray}%
with possible h- and c-splitting of the star d-metric $\
_{\star}^{\shortparallel }\mathfrak{g}_{\gamma \tau }$ (\ref{dmss1}) by the
N-connection structure.

We introduce also 
\begin{equation}
\ _{\star }^{\shortparallel }\widehat{\mathsf{W}}_{\gamma \alpha \beta }=\
_{\star }^{\shortparallel }\mathbf{g}_{\gamma \mu }\mathbf{\mathbf{\
^{\shortparallel }}}\widehat{\Gamma }_{\star \alpha \beta }^{\nu }+i\kappa 
\overline{\mathcal{R}}_{\quad \alpha }^{\tau \xi }\ \mathbf{^{\shortparallel
}e}_{\xi }\ _{\star }^{\shortparallel }\mathbf{g}_{\beta \tau }:=\ _{\star
}^{\shortparallel }\widehat{\mathsf{W}}_{\gamma \alpha \beta }^{[0]}+\
_{\star }^{\shortparallel }\widehat{\mathsf{W}}_{\gamma \alpha \beta
}^{[1]}(\kappa ),  \label{aux38}
\end{equation}%
which are related via star canonical distortion formulas (\ref{candistrnas})
to similar d-objects (\ref{aux36}). It should be noted that $\
_{\star}^{\shortparallel }\widehat{\mathsf{W}}_{\gamma \alpha \beta }$
transforms into $\ _{\star }^{\shortparallel }\mathsf{W}_{\gamma \alpha
\beta }$ if there are imposed the zero torsion conditions (\ref{lccondnonass}%
).

We can the formalism of Section 5.2 of \cite{aschieri17} on decomposition of
the coefficients of metrics and linear connections with on parameters $\hbar$
and $\kappa $ in N-adapted form for the star d-metrics and canonical
d-connections. The coordinate formulas for the LC-connections presented in
that work (formulas (5.55)-(5.65)) can be reconsidered in N-adapted bases
and than distorted following (\ref{candtorsnonas}). Here we present only the
abstract geometric decomposition of the mentioned quantum and string
parameters but omit cumbersome explicit coefficient formulas (they will be
not used for constructing solutions and quantization of nonassociative
theories in our future works) parameterizations. For the nonsymmetric
d-metric, we write(\ref{aux37}) in the form 
\begin{equation}
\ _{\ast }^{\shortparallel }\mathfrak{g}_{[0]}^{\alpha \beta }=\ _{\star
}^{\shortparallel }\mathfrak{g}_{[00]}^{\alpha \beta }+\ _{\star
}^{\shortparallel }\mathfrak{g}_{[01]}^{\alpha \beta }(\hbar )+O(\hbar
^{2},\kappa ^{2})\mbox{ and }\ _{\ast }^{\shortparallel }\mathfrak{g}%
_{[1]}^{\alpha \beta }=\ _{\star }^{\shortparallel }\mathfrak{g}%
_{[10]}^{\alpha \beta }(\kappa )+\ _{\star }^{\shortparallel }\mathfrak{g}%
_{[11]}^{\alpha \beta }(\hbar \kappa )+O(\hbar ^{2},\kappa ^{2}),
\label{aux38a}
\end{equation}%
\begin{eqnarray*}
\mbox{ where }\ _{\star }^{\shortparallel }\mathfrak{g}_{[00]}^{\alpha \beta
} &=&\ ^{\shortparallel }\mathbf{g}^{\alpha \beta }, \\
\ _{\star }^{\shortparallel }\mathfrak{g}_{[01]}^{\alpha \beta } &=&\frac{%
i\hbar }{2}(^{\shortparallel }\partial ^{n+i}\ ^{\shortparallel }\mathbf{g}%
^{\alpha \gamma }\ \ ^{\shortparallel }\mathbf{e}_{i}\ ^{\shortparallel }%
\mathbf{g}_{\gamma \tau }-\ ^{\shortparallel }\mathbf{e}_{i}\
^{\shortparallel }\mathbf{g}^{\alpha \gamma }\ ^{\shortparallel }\partial
^{n+i}\ ^{\shortparallel }\mathbf{g}_{\gamma \tau })\ ^{\shortparallel }%
\mathbf{g}^{\tau \beta },\mbox{ for }\ ^{\shortparallel }\partial ^{n+i}=%
\frac{\partial }{\partial \ ^{\shortparallel }p_{n+i}}, \\
\ _{\star }^{\shortparallel }\mathfrak{g}_{[10]}^{\alpha \beta } &=&i\kappa 
\overline{\mathcal{R}}_{\quad \gamma }^{\tau \xi }\ ^{\shortparallel }%
\mathbf{g}^{\alpha \gamma }(\ \mathbf{^{\shortparallel }e}_{\xi }\
^{\shortparallel }\mathbf{g}_{\mu \tau })\ ^{\shortparallel }\mathbf{g}^{\mu
\beta },
\end{eqnarray*}%
\begin{eqnarray*}
\ _{\star }^{\shortparallel }\mathfrak{g}_{[11]}^{\alpha \beta } &=&\frac{%
\hbar \kappa }{2}\overline{\mathcal{R}}_{\quad \gamma }^{\tau \xi }\{(\ 
\mathbf{^{\shortparallel }e}_{\xi }\ ^{\shortparallel }\mathbf{g}_{\mu \tau
})[\ ^{\shortparallel }\mathbf{g}^{\nu \gamma }\ ^{\shortparallel }\mathbf{g}%
^{\mu \beta }(\ \mathbf{^{\shortparallel }e}_{i}\ ^{\shortparallel }\mathbf{g%
}^{\alpha \lambda }\ ^{\shortparallel }\partial ^{n+i}\ ^{\shortparallel }%
\mathbf{g}_{\lambda \nu }-\ ^{\shortparallel }\partial ^{n+i}\
^{\shortparallel }\mathbf{g}^{\alpha \lambda }\ \ \mathbf{^{\shortparallel }e%
}_{i}\ ^{\shortparallel }\mathbf{g}_{\lambda \nu }) \\
&&+\ ^{\shortparallel }\mathbf{g}^{\alpha \gamma }\ ^{\shortparallel }%
\mathbf{g}^{\nu \beta }(\ \mathbf{^{\shortparallel }e}_{i}\ ^{\shortparallel
}\mathbf{g}^{\mu \lambda }\ ^{\shortparallel }\partial ^{n+i}\
^{\shortparallel }\mathbf{g}_{\lambda \nu }-\ ^{\shortparallel }\partial
^{n+i}\ ^{\shortparallel }\mathbf{g}^{\mu \lambda }\ \ \mathbf{%
^{\shortparallel }e}_{i}\ ^{\shortparallel }\mathbf{g}_{\lambda \nu }) \\
&&+(\ \mathbf{^{\shortparallel }e}_{i}\ ^{\shortparallel }\mathbf{g}^{\alpha
\gamma }\ ^{\shortparallel }\partial ^{n+i}\ ^{\shortparallel }\mathbf{g}%
^{\mu \beta }-\ ^{\shortparallel }\partial ^{n+i}\ ^{\shortparallel }\mathbf{%
g}^{\alpha \gamma }\ \ \mathbf{^{\shortparallel }e}_{i}\ ^{\shortparallel }%
\mathbf{g}^{\mu \beta })] \\
&&-(\ \ \mathbf{^{\shortparallel }e}_{i}\ \mathbf{^{\shortparallel }e}_{\xi
}\ ^{\shortparallel }\mathbf{g}_{\nu \tau })(\ \ ^{\shortparallel }\mathbf{g}%
^{\alpha \gamma }\ ^{\shortparallel }\partial ^{n+i}\ ^{\shortparallel }%
\mathbf{g}^{\nu \beta }-\ ^{\shortparallel }\mathbf{g}^{\nu \beta }\
^{\shortparallel }\partial ^{n+i}\ ^{\shortparallel }\mathbf{g}^{\alpha
\gamma }) \\
&&+(\ \ ^{\shortparallel }\partial ^{n+i}\ \mathbf{^{\shortparallel }e}_{\xi
}\ ^{\shortparallel }\mathbf{g}_{\nu \tau })(\ \ ^{\shortparallel }\mathbf{g}%
^{\alpha \gamma }\ \ \mathbf{^{\shortparallel }e}_{i}\ ^{\shortparallel }%
\mathbf{g}^{\nu \beta }-\ ^{\shortparallel }\mathbf{g}^{\nu \beta }\ \ 
\mathbf{^{\shortparallel }e}_{i}\ ^{\shortparallel }\mathbf{g}^{\alpha
\gamma })\}+O(\hbar ^{2},\kappa ^{2}).
\end{eqnarray*}%
We can organize the nonholonomic frame structure (\ref{nhframtr}) in such a
form when for $\ \mathbf{^{\shortparallel }e}_{\alpha }\rightarrow $ $\ 
\mathbf{^{\shortparallel }\partial }_{\alpha }$ (and, inversely, following
the Convention 2, above formulas transform in coordinate basis analogs of
respective formulas (5.59) - (5.61) from \cite{aschieri17}.

Using these formulas and (\ref{aux38}) for $\ _{\star }^{\shortparallel }%
\mathfrak{g}_{\alpha \beta }=\ _{\star }^{\shortparallel }\mathfrak{g}%
_{\alpha \beta }^{[0]}+\ _{\star }^{\shortparallel }\mathfrak{g}_{\alpha
\beta }^{[1]},$ $\ _{\ast }^{\shortparallel }\mathfrak{g}^{\alpha \beta }=\
_{\ast }^{\shortparallel }\mathfrak{g}_{[0]}^{\alpha \beta }+\ _{\ast
}^{\shortparallel }\mathfrak{g}_{[1]}^{\alpha \beta }$ and $\
^{\shortparallel }\widehat{\mathbf{\Gamma }}_{\star \alpha \beta }^{\gamma }$
(\ref{eqnasdmdc}), we express%
\begin{eqnarray}
\ _{[0]}^{\shortparallel }\widehat{\Gamma }_{\ast \alpha \beta }^{\nu } &=&\
_{\ast }^{\shortparallel }\mathfrak{g}_{[0]}^{\nu \gamma }\ast \ _{\ast
}^{\shortparallel }\widehat{\mathsf{W}}_{\gamma \alpha \beta }^{[0]}:=\
_{[00]}^{\shortparallel }\widehat{\Gamma }_{\star \alpha \beta }^{\nu }+\
_{[01]}^{\shortparallel }\widehat{\Gamma }_{\star \alpha \beta }^{\nu
}(\hbar )+O(\hbar ^{2}),  \label{aux39} \\
\ _{[1]}^{\shortparallel }\widehat{\Gamma }_{\ast \alpha \beta }^{\nu } &=&\
_{\ast }^{\shortparallel }\mathfrak{g}_{[0]}^{\nu \gamma }\ast \ _{\ast
}^{\shortparallel }\widehat{\mathsf{W}}_{\gamma \alpha \beta }^{[1]}+\
_{\ast }^{\shortparallel }\mathfrak{g}_{[1]}^{\nu \gamma }\ast \ _{\ast
}^{\shortparallel }\widehat{\mathsf{W}}_{\gamma \alpha \beta }^{[0]}:=\
_{[10]}^{\shortparallel }\widehat{\Gamma }_{\star \alpha \beta }^{\nu
}(\kappa )+\ _{[11]}^{\shortparallel }\widehat{\Gamma }_{\star \alpha \beta
}^{\nu }(\hbar \kappa )+O(\hbar ^{2}),  \notag
\end{eqnarray}%
where%
\begin{eqnarray}
\ _{[00]}^{\shortparallel }\widehat{\Gamma }_{\star \alpha \beta }^{\nu }
&=&\ ^{\shortparallel }\widehat{\Gamma }_{\ \alpha \beta }^{\nu },\
_{[01]}^{\shortparallel }\widehat{\Gamma }_{\star \alpha \beta }^{\nu }=-%
\frac{i\hbar }{2}\ ^{\shortparallel }\mathbf{g}^{\nu \gamma }(\ \mathbf{%
^{\shortparallel }e}_{i}\ ^{\shortparallel }\mathbf{g}_{\gamma \tau }\
^{\shortparallel }\partial ^{n+i}\ ^{\shortparallel }\widehat{\Gamma }_{\
\alpha \beta }^{\tau }\ -\ ^{\shortparallel }\partial ^{n+i}\ \
^{\shortparallel }\mathbf{g}_{\gamma \tau }\ \ \mathbf{^{\shortparallel }e}%
_{i}\ ^{\shortparallel }\widehat{\Gamma }_{\ \alpha \beta }^{\tau }),
\label{aux51} \\
\ _{[10]}^{\shortparallel }\widehat{\Gamma }_{\star \alpha \beta }^{\nu }
&=&i\kappa \overline{\mathcal{R}}_{\quad \quad \quad }^{n+i\ n+j\ n+k}\ (\ \
^{\shortparallel }\mathbf{g}_{~n+k}^{\nu }\quad ^{\shortparallel }\mathbf{g}%
_{jm}(\ \mathbf{^{\shortparallel }e}_{i}\ ^{\shortparallel }\widehat{\Gamma }%
_{\ \alpha \beta }^{m})-\ ^{\shortparallel }\mathbf{g}_{\ }^{\nu \mu }\ \ 
\mathbf{^{\shortparallel }}p_{n+j}(\ \ \mathbf{^{\shortparallel }e}_{k}\
^{\shortparallel }\mathbf{g}_{\mu \tau })(\ \mathbf{^{\shortparallel }e}%
_{i}\ ^{\shortparallel }\widehat{\Gamma }_{\ \alpha \beta }^{\tau })), 
\notag
\end{eqnarray}%
\begin{eqnarray}
\ _{[11]}^{\shortparallel }\widehat{\Gamma }_{\star \alpha \beta }^{\nu }
&=&i\frac{\kappa \hbar }{2}\overline{\mathcal{R}}_{\quad \quad \quad }^{n+i\
n+j\ n+k}\{-(\ \mathbf{^{\shortparallel }e}_{l}\ ^{\shortparallel }\mathbf{g}%
_{~n+k}^{\nu })\ ^{\shortparallel }\partial ^{n+l}[\ \mathbf{%
^{\shortparallel }e}_{i}(~^{\shortparallel }\mathbf{g}_{jm}\
^{\shortparallel }\widehat{\Gamma }_{\ \alpha \beta }^{m})]  \notag \\
&&+(\ ^{\shortparallel }\partial ^{n+l}\ ^{\shortparallel }\mathbf{g}%
_{~~n+k}^{\nu })[~\mathbf{^{\shortparallel }e}_{l}(\ \ \mathbf{%
^{\shortparallel }e}_{i}~^{\shortparallel }\mathbf{g}_{jm}\ ^{\shortparallel
}\widehat{\Gamma }_{\ \alpha \beta }^{m})]  \notag \\
&&+[(\ \mathbf{^{\shortparallel }e}_{l}~^{\shortparallel }\mathbf{g}%
_{~}^{\nu \gamma })(\ ^{\shortparallel }\partial ^{n+l}\ ^{\shortparallel }%
\mathbf{g}_{\gamma \tau })-(\ ^{\shortparallel }\partial
^{n+l}~^{\shortparallel }\mathbf{g}_{~}^{\nu \gamma })(\ \mathbf{%
^{\shortparallel }e}_{l}\ ^{\shortparallel }\mathbf{g}_{\gamma \tau })]\
^{\shortparallel }\mathbf{g}_{~n+k}^{\tau }(\ \ \mathbf{^{\shortparallel }e}%
_{i}~^{\shortparallel }\mathbf{g}_{jm}\ ^{\shortparallel }\widehat{\Gamma }%
_{\ \alpha \beta }^{m})  \notag \\
&&+\ \mathbf{^{\shortparallel }e}_{l}[\ ^{\shortparallel }\mathbf{g}%
_{~n+k}^{\nu }~^{\shortparallel }\mathbf{g}_{~}^{qm}(\ \mathbf{%
^{\shortparallel }e}_{i}\ ^{\shortparallel }\mathbf{g}_{qj})]\
^{\shortparallel }\partial ^{n+l}(\ ~^{\shortparallel }\mathbf{g}_{mo}\
^{\shortparallel }\widehat{\Gamma }_{\ \alpha \beta }^{o})  \notag \\
&&-\ ^{\shortparallel }\partial ^{n+l}[(\ ^{\shortparallel }\mathbf{g}%
_{~n+k}^{\nu }~^{\shortparallel }\mathbf{g}_{~}^{qm}(\ \mathbf{%
^{\shortparallel }e}_{i}\ ^{\shortparallel }\mathbf{g}_{qj})]\ \mathbf{%
^{\shortparallel }e}_{l}(\ ~^{\shortparallel }\mathbf{g}_{mo}\
^{\shortparallel }\widehat{\Gamma }_{\ \alpha \beta }^{o})  \notag \\
&&-(\ \mathbf{^{\shortparallel }e}_{i}\ ^{\shortparallel }\mathbf{g}_{qj})[\
^{\shortparallel }\mathbf{g}_{~n+k}^{\tau }\left( (\ \mathbf{%
^{\shortparallel }e}_{l}~^{\shortparallel }\mathbf{g}_{~}^{\nu \gamma })(\
^{\shortparallel }\partial ^{n+l}\ ^{\shortparallel }\mathbf{g}_{\gamma \tau
})-(\ ^{\shortparallel }\partial ^{n+l}~^{\shortparallel }\mathbf{g}%
_{~}^{\nu \gamma })(\ \mathbf{^{\shortparallel }e}_{l}\ ^{\shortparallel }%
\mathbf{g}_{\gamma \tau })\right) \ ^{\shortparallel }\widehat{\Gamma }_{\
\alpha \beta }^{q}  \notag \\
&&+\ ^{\shortparallel }\mathbf{g}_{~n+k}^{\nu }\left( (\ \mathbf{%
^{\shortparallel }e}_{l}~^{\shortparallel }\mathbf{g}_{~}^{\nu \gamma })(\
^{\shortparallel }\partial ^{n+k}\ ^{\shortparallel }\mathbf{g}_{\gamma \tau
})-(\ ^{\shortparallel }\partial ^{n+k}~^{\shortparallel }\mathbf{g}%
_{~}^{\nu \gamma })(\ \mathbf{^{\shortparallel }e}_{l}\ ^{\shortparallel }%
\mathbf{g}_{\gamma \tau })\right) \ ^{\shortparallel }\widehat{\Gamma }_{\
\alpha \beta }^{\tau }  \notag \\
&&+\left( \ (~\mathbf{^{\shortparallel }e}_{l}\ ^{\shortparallel }\mathbf{g}%
_{~n+k}^{\nu })(\ ^{\shortparallel }\partial ^{n+l}~^{\shortparallel }%
\mathbf{g}_{~}^{\mu \gamma })-(\ ^{\shortparallel }\partial ^{n+k}~\
^{\shortparallel }\mathbf{g}_{~n+k}^{\nu })(~\mathbf{^{\shortparallel }e}%
_{l}\ ^{\shortparallel }\mathbf{g}_{~}^{\mu \gamma })\ \right)
~^{\shortparallel }\mathbf{g}_{\gamma \tau }\ ^{\shortparallel }\widehat{%
\Gamma }_{\ \alpha \beta }^{\tau }]  \notag
\end{eqnarray}%
\begin{eqnarray*}
&&+(~\mathbf{^{\shortparallel }e}_{l}~\mathbf{^{\shortparallel }e}_{i}\
^{\shortparallel }\mathbf{g}_{qj})[(\ ^{\shortparallel }\mathbf{g}%
_{~n+k}^{\nu })(\ ^{\shortparallel }\partial ^{n+l}~^{\shortparallel }%
\mathbf{g}_{~}^{qs})\ ^{\shortparallel }\mathbf{g}_{sp}\ ^{\shortparallel }%
\widehat{\Gamma }_{\ \alpha \beta }^{p}-(\ ^{\shortparallel }\partial
^{n+l}~\ ^{\shortparallel }\mathbf{g}_{~n+k}^{\nu })\ ^{\shortparallel }%
\widehat{\Gamma }_{\ \alpha \beta }^{q}] \\
&&-(\ ^{\shortparallel }\partial ^{n+l}~\mathbf{^{\shortparallel }e}_{i}\
^{\shortparallel }\mathbf{g}_{qj})[(\ ^{\shortparallel }\mathbf{g}%
_{~n+k}^{\nu })(~\mathbf{^{\shortparallel }e}_{l}\ ~^{\shortparallel }%
\mathbf{g}_{~}^{qs})\ ^{\shortparallel }\mathbf{g}_{sp}\ ^{\shortparallel }%
\widehat{\Gamma }_{\ \alpha \beta }^{p}-(~\mathbf{^{\shortparallel }e}_{l}~\
^{\shortparallel }\mathbf{g}_{~n+k}^{\nu })\ ^{\shortparallel }\widehat{%
\Gamma }_{\ \alpha \beta }^{q}] \\
&&+p_{n+j}[~\mathbf{^{\shortparallel }e}_{l}\left( ~^{\shortparallel }%
\mathbf{g}_{~}^{\nu \tau }(~\mathbf{^{\shortparallel }e}_{k}~^{%
\shortparallel }\mathbf{g}_{\tau \mu })\right) \ ^{\shortparallel }\partial
^{n+l}(~\mathbf{^{\shortparallel }e}_{i}\ \ ^{\shortparallel }\widehat{%
\Gamma }_{\ \alpha \beta }^{\mu })-\ ^{\shortparallel }\partial ^{n+l}\left(
~^{\shortparallel }\mathbf{g}_{~}^{\nu \tau }(~\mathbf{^{\shortparallel }e}%
_{k}~^{\shortparallel }\mathbf{g}_{\tau \mu })\right) ~\mathbf{%
^{\shortparallel }e}_{l}(~\mathbf{^{\shortparallel }e}_{i}\ \
^{\shortparallel }\widehat{\Gamma }_{\ \alpha \beta }^{\mu })] \\
&&+p_{n+j}~^{\shortparallel }\mathbf{g}_{~}^{\nu \tau }(~\mathbf{%
^{\shortparallel }e}_{k}~^{\shortparallel }\mathbf{g}_{\tau \mu })~\mathbf{%
^{\shortparallel }e}_{i}\ [\left( (~\mathbf{^{\shortparallel }e}%
_{l}~^{\shortparallel }\mathbf{g}_{~}^{\mu \gamma })\ ^{\shortparallel
}\partial ^{n+l}(~^{\shortparallel }\mathbf{g}_{\gamma \tau }\
^{\shortparallel }\widehat{\Gamma }_{\ \alpha \beta }^{\tau })-(\
^{\shortparallel }\partial ^{n+l}~^{\shortparallel }\mathbf{g}_{~}^{\mu
\gamma })~\mathbf{^{\shortparallel }e}_{l}(~^{\shortparallel }\mathbf{g}%
_{\gamma \tau }\ ^{\shortparallel }\widehat{\Gamma }_{\ \alpha \beta }^{\tau
})\right) \\
&&-\left( (~\mathbf{^{\shortparallel }e}_{l}~^{\shortparallel }\mathbf{g}%
_{~}^{\mu \gamma })(\ ^{\shortparallel }\partial ^{n+l}~^{\shortparallel }%
\mathbf{g}_{\gamma \tau })-(\ ^{\shortparallel }\partial
^{n+l}~^{\shortparallel }\mathbf{g}_{~}^{\mu \gamma })(~\mathbf{%
^{\shortparallel }e}_{l}~^{\shortparallel }\mathbf{g}_{\gamma \tau })\right)
\ ^{\shortparallel }\widehat{\Gamma }_{\ \alpha \beta }^{\tau }] \\
&&+p_{n+j}\left( (~\mathbf{^{\shortparallel }e}_{l}~^{\shortparallel }%
\mathbf{g}_{~}^{\nu \tau })(\ ^{\shortparallel }\partial ^{n+l}~~\mathbf{%
^{\shortparallel }e}_{k}~^{\shortparallel }\mathbf{g}_{\tau \mu })-(\
^{\shortparallel }\partial ^{n+l}~^{\shortparallel }\mathbf{g}_{~}^{\nu \tau
})(~~\mathbf{^{\shortparallel }e}_{l}~\mathbf{^{\shortparallel }e}%
_{k}~^{\shortparallel }\mathbf{g}_{\tau \mu })\right) (~~\mathbf{%
^{\shortparallel }e}_{i}\ ^{\shortparallel }\widehat{\Gamma }_{\ \alpha
\beta }^{\mu }) \\
&&-p_{n+j}\left( (~\mathbf{^{\shortparallel }e}_{l}~^{\shortparallel }%
\mathbf{g}_{~}^{\nu \tau })(\ ^{\shortparallel }\partial ^{n+l}~~\mathbf{%
^{\shortparallel }e}_{k}~^{\shortparallel }\mathbf{g}_{\tau \varepsilon
})-(\ ^{\shortparallel }\partial ^{n+l}~^{\shortparallel }\mathbf{g}%
_{~}^{\nu \tau })(~~\mathbf{^{\shortparallel }e}_{l}~^{\shortparallel }%
\mathbf{g}_{\tau \varepsilon })\right) ~^{\shortparallel }\mathbf{g}%
_{~}^{\varepsilon \lambda }(~\mathbf{^{\shortparallel }e}_{k}~^{%
\shortparallel }\mathbf{g}_{\lambda \mu })(~\mathbf{^{\shortparallel }e}%
_{i}\ ^{\shortparallel }\widehat{\Gamma }_{\ \alpha \beta }^{\mu }) \\
&&+(~\mathbf{^{\shortparallel }e}_{l}~^{\shortparallel }\mathbf{g}_{~}^{\nu
\tau })(~\mathbf{^{\shortparallel }e}_{j}~^{\shortparallel }\mathbf{g}_{\tau
\mu })(~\mathbf{^{\shortparallel }e}_{k}\ ^{\shortparallel }\widehat{\Gamma }%
_{\ \alpha \beta }^{\mu })\},
\end{eqnarray*}%
where, for instance, $\ $the respective nontrivial c-component of a d-metric
is computed in the form$\ ^{\shortparallel }\mathbf{g}_{ n+k}^{\nu }=(\
^{\shortparallel }g_{~b}^{i}=0,\ ^{\shortparallel }g_{ab}),\,$ for $b=n+k.$
Above formulas are for the canonical d-connection $\ ^{\shortparallel }%
\widehat{\mathbf{D}}^{\star }$ (\ref{candistrnas}). They can be transformed
into formulas for the nonassociative LC-connection for zero distortion
d-tensors when $\ ^{\shortparallel }\widehat{\mathbf{D}}_{\mid \
^{\shortparallel }\widehat{\mathbf{T}}=0}^{\star }=\ ^{\shortparallel
}\nabla ^{\star },$ see (\ref{lccondnonass}). If necessary, we state the
nonholonomic frame structure (\ref{nhframtr}) in such a form when for $\ 
\mathbf{^{\shortparallel }e}_{\alpha }\rightarrow $ $\ \mathbf{\
^{\shortparallel }\partial }_{\alpha }$ (and, inversely, following the
Convention 2). This results in coordinate basis formulas (5.62) - (5.65)
from \cite{aschieri17}.

All components can be subjected to h- and c-decompositions with respect to $%
\ ^{\shortparallel }\mathbf{e}^{\alpha }$ and $\ ^{\shortparallel }\mathbf{e}%
_{\alpha }$ (\ref{nadapdc}), when $\ _{[00]}^{\shortparallel } \widehat{%
\Gamma }_{\ast \alpha \beta }^{\nu }=\ ^{\shortparallel }\widehat{\Gamma }%
_{\alpha \beta }^{\nu }$ from (\ref{twocon}) with associative canonical
d-connection h- and c-coefficients (\ref{canhc}). We omit such h- and
c-terms for other components in (\ref{aux39}).

\paragraph{Nonassociative canonical Riemann and Ricci d-tensors for
quasi-Hopf structures: \newline
}

In "hat" nonholonomic variables, the formula (\ref{nadriemhopf}) defines the
N-adapted coefficients of the nonassociative canonical Riemann d-tensor, 
\begin{eqnarray}
\mathbf{\mathbf{\mathbf{\mathbf{\ ^{\shortparallel }}}}}\widehat{\mathcal{%
\Re }}_{\quad \alpha \beta \gamma }^{\star \mu } &=&\mathbf{\mathbf{\mathbf{%
\mathbf{\ _{1}^{\shortparallel }}}}}\widehat{\mathcal{\Re }}_{\quad \alpha
\beta \gamma }^{\star \mu }+\mathbf{\mathbf{\mathbf{\mathbf{\
_{2}^{\shortparallel }}}}}\widehat{\mathcal{\Re }}_{\quad \alpha \beta
\gamma }^{\star \mu },\mbox{ where }  \label{nadriemhopfcan} \\
&&\mathbf{\mathbf{\mathbf{\mathbf{\ _{1}^{\shortparallel }}}}}\widehat{%
\mathcal{\Re }}_{\quad \alpha \beta \gamma }^{\star \mu }=\ \mathbf{%
^{\shortparallel }e}_{\gamma }\mathbf{\ ^{\shortparallel }}\widehat{\Gamma }%
_{\star \alpha \beta }^{\mu }-\ \mathbf{^{\shortparallel }e}_{\beta }\mathbf{%
\ ^{\shortparallel }}\widehat{\Gamma }_{\star \alpha \gamma }^{\mu }+\mathbf{%
\ ^{\shortparallel }}\widehat{\Gamma }_{\star \nu \tau }^{\mu }\star
_{N}(\delta _{\ \gamma }^{\tau }\mathbf{\ ^{\shortparallel }}\widehat{\Gamma 
}_{\star \alpha \beta }^{\nu }-\delta _{\ \beta }^{\tau }\mathbf{\
^{\shortparallel }}\widehat{\Gamma }_{\star \alpha \gamma }^{\nu })+\mathbf{%
\ ^{\shortparallel }}w_{\beta \gamma }^{\tau }\star _{N}\mathbf{\
^{\shortparallel }}\widehat{\Gamma }_{\star \alpha \tau }^{\mu }  \notag \\
&& \mbox{ and } \ _{2}^{\shortparallel }\widehat{\mathcal{\Re }}_{\quad
\alpha \beta \gamma }^{\star \mu }=i\kappa \ ^{\shortparallel }\widehat{%
\Gamma }_{\star \nu \tau }^{\mu }\star _{N}(\mathcal{R}_{\quad \gamma
}^{\tau \xi }\ \mathbf{^{\shortparallel }e}_{\xi }\mathbf{\ ^{\shortparallel
}}\widehat{\Gamma }_{\star \alpha \beta }^{\nu }-\mathcal{R}_{\quad \beta
}^{\tau \xi }\ \mathbf{^{\shortparallel }e}_{\xi }\mathbf{\ ^{\shortparallel
}}\widehat{\Gamma }_{\star \alpha \gamma }^{\nu}).  \notag
\end{eqnarray}%
Using (\ref{aux311a}), we write 
\begin{equation*}
\ ^{\shortparallel }\widehat{\mathbf{\Gamma }}_{\star \alpha \beta }^{\gamma
}=\ _{[0]}^{\shortparallel }\widehat{\mathbf{\Gamma }}_{\star \beta \alpha
}^{\tau }+i\kappa \ _{[1]}^{\shortparallel }\widehat{\mathbf{\Gamma }}%
_{\star \beta \alpha }^{\tau }=\ _{[00]}^{\shortparallel }\widehat{\Gamma }%
_{\ast \alpha \beta }^{\nu }+\ _{[01]}^{\shortparallel }\widehat{\Gamma }%
_{\ast \alpha \beta }^{\nu }(\hbar )+\ _{[10]}^{\shortparallel }\widehat{%
\Gamma }_{\ast \alpha \beta }^{\nu }(\kappa )+\ _{[11]}^{\shortparallel }%
\widehat{\Gamma }_{\ast \alpha \beta }^{\nu }(\hbar \kappa )+O(\hbar
^{2},\kappa ^{2},...)
\end{equation*}%
and decompose (\ref{nadriemhopfcan}) as%
\begin{equation*}
\mathbf{\mathbf{\mathbf{\mathbf{\ ^{\shortparallel }}}}}\widehat{\mathcal{%
\Re }}_{\quad \alpha \beta \gamma }^{\star \mu }=\mathbf{\mathbf{\mathbf{%
\mathbf{\ }}}}\ _{[00]}^{\shortparallel }\widehat{\mathcal{\Re }}_{\quad
\alpha \beta \gamma }^{\star \mu }+\mathbf{\mathbf{\mathbf{\mathbf{\ }}}}\
_{[01]}^{\shortparallel }\widehat{\mathcal{\Re }}_{\quad \alpha \beta \gamma
}^{\star \mu }(\hbar )+\mathbf{\mathbf{\mathbf{\mathbf{\ }}}}\
_{[10]}^{\shortparallel }\widehat{\mathcal{\Re }}_{\quad \alpha \beta \gamma
}^{\star \mu }(\kappa )+\mathbf{\mathbf{\mathbf{\mathbf{\ }}}}\
_{[11]}^{\shortparallel }\widehat{\mathcal{\Re }}_{\quad \alpha \beta \gamma
}^{\star \mu }(\hbar \kappa )+O(\hbar ^{2},\kappa ^{2},...).
\end{equation*}%
We can fix such nonholonomic distributions on $\mathcal{M}= \mathbf{T}
_{\shortparallel }^{\ast }\mathbf{V}$ when $\mathbf{\mathbf{\mathbf{\mathbf{%
\ }}}}\ _{[00]}^{\shortparallel }\widehat{\mathcal{\Re }}_{\quad \alpha
\beta \gamma }^{\star \mu }$ is a "hat" variant of (\ref{nadriemhopf}) with
h- and c-coefficients (\ref{nadriemhopf1coef}). The explicit formulas for
the terms up till order $\hbar ,\kappa $ and $\hbar \kappa $ and their h-
and c-decompositions are quite cumbersome and we omit such technical results
in this paper.

Contracting on the fist and forth indices above formulas for $\
^{\shortparallel }\widehat{\Gamma }_{\star \alpha \gamma }^{\mu }$ and
respective h- and c-decompositions we define the nonassociative canonical
Ricci d-tensor as a "hat" variant of (\ref{driccina}), 
\begin{eqnarray}
\mathbf{\mathbf{\mathbf{\mathbf{\ ^{\shortparallel }}}}}\widehat{\mathcal{%
\Re }}ic^{\star } &=&\mathbf{\mathbf{\mathbf{\mathbf{\ ^{\shortparallel }}}}}%
\widehat{\mathbf{\mathbf{\mathbf{\mathbf{R}}}}}ic_{\alpha \beta }^{\star
}\star _{N}(\ \mathbf{^{\shortparallel }e}^{\alpha }\otimes _{\star _{N}}\ 
\mathbf{^{\shortparallel }e}^{\beta }),\mbox{ where }
\label{driccicanonstar} \\
&&\mathbf{\mathbf{\mathbf{\mathbf{\ ^{\shortparallel }}}}}\widehat{\mathbf{%
\mathbf{\mathbf{\mathbf{R}}}}}ic_{\alpha \beta }^{\star }:=\mathbf{\mathbf{%
\mathbf{\mathbf{\ ^{\shortparallel }}}}}\widehat{\mathcal{\Re }}ic^{\star }(%
\mathbf{\ }\ ^{\shortparallel }\mathbf{e}_{\alpha },\ ^{\shortparallel }%
\mathbf{e}_{\beta })=\mathbf{\langle }\ \mathbf{\mathbf{\mathbf{\mathbf{\
^{\shortparallel }}}}}\widehat{\mathbf{\mathbf{\mathbf{\mathbf{R}}}}}ic_{\mu
\nu }^{\star }\star _{N}(\ \mathbf{^{\shortparallel }e}^{\mu }\otimes
_{\star _{N}}\ \mathbf{^{\shortparallel }e}^{\nu }),\mathbf{\mathbf{\ }\
^{\shortparallel }\mathbf{e}}_{\alpha }\mathbf{\otimes _{\star _{N}}\
^{\shortparallel }\mathbf{e}}_{\beta }\mathbf{\rangle }_{\star _{N}}.  \notag
\end{eqnarray}%
The respective parametric decompositions are of type 
\begin{eqnarray}
\mathbf{\mathbf{\mathbf{\mathbf{\ ^{\shortparallel }}}}}\widehat{\mathbf{%
\mathbf{\mathbf{\mathbf{R}}}}}ic_{\alpha \beta }^{\star } &=&\mathbf{\mathbf{%
\mathbf{\mathbf{\ ^{\shortparallel }}}}}\widehat{\mathcal{\Re }}_{\quad
\alpha \beta \mu }^{\star \mu }=\mathbf{\mathbf{\mathbf{\mathbf{\ \mathbf{%
\mathbf{\mathbf{\mathbf{\ }}}}}}}}\ _{[00]}^{\shortparallel }\widehat{%
\mathbf{\mathbf{\mathbf{\mathbf{R}}}}}ic_{\alpha \beta }^{\star }+\mathbf{%
\mathbf{\mathbf{\mathbf{\ \ }}}}_{[01]}^{\shortparallel }\widehat{\mathbf{%
\mathbf{\mathbf{\mathbf{R}}}}}ic_{\alpha \beta }^{\star }(\hbar )+\mathbf{%
\mathbf{\mathbf{\mathbf{\ }}}}_{[10]}^{\shortparallel }\widehat{\mathbf{%
\mathbf{\mathbf{\mathbf{R}}}}}ic_{\alpha \beta }^{\star }(\kappa )+\mathbf{%
\mathbf{\mathbf{\mathbf{\ }}}}_{[11]}^{\shortparallel }\widehat{\mathbf{%
\mathbf{\mathbf{\mathbf{R}}}}}ic_{\alpha \beta }^{\star }(\hbar \kappa
)+O(\hbar ^{2},\kappa ^{2},...),  \notag \\
&\mbox{where}&\ _{[00]}^{\shortparallel }\widehat{\mathbf{R}}ic_{\alpha
\beta }^{\star }=\ _{[00]}^{\shortparallel }\widehat{\mathcal{\Re }}_{\quad
\alpha \beta \mu }^{\star \mu }\mathbf{\mathbf{\mathbf{\mathbf{\ ,}}}}\
_{[01]}^{\shortparallel }\mathbf{\mathbf{\mathbf{\mathbf{\widehat{\mathbf{%
\mathbf{\mathbf{\mathbf{R}}}}}}}}}ic_{\alpha \beta }^{\star }=\
_{[01]}^{\shortparallel }\mathbf{\mathbf{\mathbf{\mathbf{\widehat{\mathcal{%
\Re }}}}}}_{\quad \alpha \beta \mu }^{\star \mu },  \label{driccicanonstar1}
\\
&&\ _{[10]}^{\shortparallel }\mathbf{\mathbf{\mathbf{\mathbf{\widehat{%
\mathbf{\mathbf{\mathbf{\mathbf{R}}}}}}}}}ic_{\alpha \beta }^{\star }=\
_{[10]}^{\shortparallel }\mathbf{\mathbf{\mathbf{\mathbf{\widehat{\mathcal{%
\Re }}}}}}_{\quad \alpha \beta \mu }^{\star \mu },\ _{[11]}^{\shortparallel }%
\widehat{\mathbf{\mathbf{\mathbf{\mathbf{R}}}}}ic_{\alpha \beta }^{\star }=\
_{[11]}^{\shortparallel }\mathbf{\mathbf{\mathbf{\mathbf{\widehat{\mathcal{%
\Re }}}}}}_{\quad \alpha \beta \mu }^{\star \mu }.  \notag
\end{eqnarray}%
We present here the explicit N-adapted formulas for such coefficients: 
\begin{eqnarray}
\ _{[00]}^{\shortparallel }\widehat{\mathbf{R}}ic_{\beta \gamma }^{\star }
&=&\ ^{\shortparallel }\widehat{\mathbf{R}}ic_{\beta \gamma }=  \label{ric50}
\\
\ ^{\shortparallel }\widehat{\mathbf{R}}_{\beta \gamma } &=&\mathbf{\mathbf{%
\mathbf{\mathbf{\ ^{\shortparallel }}}}e}_{\alpha }\mathbf{\mathbf{\mathbf{%
\mathbf{\ ^{\shortparallel }}}}}\widehat{\mathbf{\Gamma }}_{\ \beta \gamma
}^{\alpha }-\mathbf{\mathbf{\mathbf{\mathbf{\ ^{\shortparallel }}}}e}_{\beta
}\mathbf{\mathbf{\mathbf{\mathbf{\ ^{\shortparallel }}}}}\widehat{\mathbf{%
\Gamma }}_{\ \gamma \alpha }^{\alpha }+\mathbf{\mathbf{\mathbf{\mathbf{\
^{\shortparallel }}}}}\widehat{\mathbf{\Gamma }}_{\ \beta \gamma }^{\mu }%
\mathbf{\mathbf{\mathbf{\mathbf{\ ^{\shortparallel }}}}}\widehat{\mathbf{%
\Gamma }}_{\ \mu \alpha }^{\alpha }-\mathbf{\mathbf{\mathbf{\mathbf{\
^{\shortparallel }}}}}\widehat{\mathbf{\Gamma }}_{\ \beta \alpha }^{\mu }%
\mathbf{\mathbf{\mathbf{\mathbf{\ ^{\shortparallel }}}}}\widehat{\mathbf{%
\Gamma }}_{\ \mu \gamma }^{\alpha }+\mathbf{\mathbf{\mathbf{\mathbf{\
^{\shortparallel }}}}}w_{\ \gamma \alpha }^{\mu }\mathbf{\mathbf{\mathbf{%
\mathbf{\ ^{\shortparallel }}}}}\widehat{\mathbf{\Gamma }}_{\ \beta \mu
}^{\alpha },  \notag
\end{eqnarray}%
see (\ref{candricci}) on further h-c-decompositions.

The order $\hbar $ contribution to star deformations of the Ricci d-tensor
for the canonical d-connection is 
\begin{eqnarray}
\mathbf{\mathbf{\mathbf{\mathbf{\ \ }}}}_{[01]}^{\shortparallel }\widehat{%
\mathbf{\mathbf{\mathbf{\mathbf{R}}}}}ic_{\beta \gamma }^{\star } &=&\mathbf{%
\mathbf{\mathbf{\mathbf{\ ^{\shortparallel }}}}e}_{\alpha }\mathbf{\mathbf{%
\mathbf{\mathbf{\ }}}}_{[01]}^{\shortparallel }\widehat{\mathbf{\Gamma }}_{\
\beta \gamma }^{\alpha }-\mathbf{\mathbf{\mathbf{\mathbf{\ ^{\shortparallel }%
}}}e}_{\beta }\ _{[01]}^{\shortparallel }\widehat{\mathbf{\Gamma }}_{\
\gamma \alpha }^{\alpha }+\mathbf{\mathbf{\mathbf{\mathbf{\ }}}}%
_{[01]}^{\shortparallel }\widehat{\mathbf{\Gamma }}_{\ \beta \gamma }^{\mu }%
\mathbf{\mathbf{\mathbf{\mathbf{\ ^{\shortparallel }}}}}\widehat{\mathbf{%
\Gamma }}_{\ \mu \alpha }^{\alpha }+\mathbf{\mathbf{\mathbf{\mathbf{\
^{\shortparallel }}}}}\widehat{\mathbf{\Gamma }}_{\ \beta \gamma }^{\mu }%
\mathbf{\mathbf{\mathbf{\mathbf{\ \ }}}}_{[01]}^{\shortparallel }\widehat{%
\mathbf{\Gamma }}_{\ \mu \alpha }^{\alpha }  \notag \\
&&-\mathbf{\mathbf{\mathbf{\mathbf{\ \ }}}}_{[01]}^{\shortparallel }\widehat{%
\mathbf{\Gamma }}_{\ \beta \alpha }^{\mu }\mathbf{\mathbf{\mathbf{\mathbf{\
^{\shortparallel }}}}}\widehat{\mathbf{\Gamma }}_{\ \mu \gamma }^{\alpha }-%
\mathbf{\mathbf{\mathbf{\mathbf{\ ^{\shortparallel }}}}}\widehat{\mathbf{%
\Gamma }}_{\ \beta \alpha }^{\mu }\mathbf{\mathbf{\mathbf{\mathbf{\ }}}}%
_{[01]}^{\shortparallel }\widehat{\mathbf{\Gamma }}_{\ \mu \gamma }^{\alpha
}+\mathbf{\mathbf{\mathbf{\mathbf{\ ^{\shortparallel }}}}}w_{\ \gamma \alpha
}^{\mu }\mathbf{\mathbf{\mathbf{\mathbf{\ }}}}_{[01]}^{\shortparallel }%
\widehat{\mathbf{\Gamma }}_{\ \beta \mu }^{\alpha }  \label{ric51} \\
&&+\frac{i\hbar }{2}[(\ \mathbf{^{\shortparallel }e}_{l}~\mathbf{\mathbf{%
\mathbf{\mathbf{^{\shortparallel }}}}}\widehat{\mathbf{\Gamma }}_{\ \mu
\alpha }^{\alpha })(\ ^{\shortparallel }\partial ^{n+l}\ \mathbf{\mathbf{%
\mathbf{\mathbf{\ ^{\shortparallel }}}}}\widehat{\mathbf{\Gamma }}_{\ \beta
\gamma }^{\mu })-(\ ^{\shortparallel }\partial ^{n+l}\ \ ~\mathbf{\mathbf{%
\mathbf{\mathbf{^{\shortparallel }}}}}\widehat{\mathbf{\Gamma }}_{\ \mu
\alpha }^{\alpha })(~\mathbf{^{\shortparallel }e}_{l}\mathbf{\mathbf{\mathbf{%
\mathbf{\ ^{\shortparallel }}}}}\widehat{\mathbf{\Gamma }}_{\ \beta \gamma
}^{\mu })]  \notag \\
&&-\frac{i\hbar }{2}[(\ \mathbf{^{\shortparallel }e}_{l}~\mathbf{\mathbf{%
\mathbf{\mathbf{\ ^{\shortparallel }}}}}\widehat{\mathbf{\Gamma }}_{\ \mu
\gamma }^{\alpha })(\ ^{\shortparallel }\partial ^{n+l}\ \mathbf{\mathbf{%
\mathbf{\mathbf{\ ^{\shortparallel }}}}}\widehat{\mathbf{\Gamma }}_{\ \beta
\alpha }^{\mu })-(\ ^{\shortparallel }\partial ^{n+l}\ \ \mathbf{\mathbf{%
\mathbf{\mathbf{\ ^{\shortparallel }}}}}\widehat{\mathbf{\Gamma }}_{\ \mu
\gamma }^{\alpha })(~\mathbf{^{\shortparallel }e}_{l}\mathbf{\mathbf{\mathbf{%
\mathbf{\ \ ^{\shortparallel }}}}}\widehat{\mathbf{\Gamma }}_{\ \beta \alpha
}^{\mu })].  \notag
\end{eqnarray}%
Inserting formulas for coefficients of type $\ _{[01]}^{\shortparallel} 
\widehat{\Gamma }_{\star \alpha \beta }^{\nu }$ (\ref{aux51}), we can
complete computation of such terms (we omit such cumbersome formulas with
h-c-splitting, and similar ones with labels [1,0], [1,1] etc., because they
will be not used in this and partner works).

Using respectively $\ _{[00]}^{\shortparallel }\widehat{\Gamma }_{\star
\alpha \beta }^{\nu }=$ $\ ^{\shortparallel }\widehat{\Gamma }_{~\alpha
\beta }^{\nu }$ and $\ _{[10]}^{\shortparallel }\widehat{\Gamma }_{\star
\alpha \beta }^{\nu }$ from (\ref{aux39}) and (\ref{aux51}), we compute the
order $\kappa $ of the nonassociative Ricci d-tensor 
\begin{eqnarray}
\mathbf{\mathbf{\mathbf{\mathbf{\ \ }}}}_{[10]}^{\shortparallel }\widehat{%
\mathbf{\mathbf{\mathbf{\mathbf{R}}}}}ic_{\beta \gamma }^{\star } &=&\mathbf{%
\mathbf{\mathbf{\mathbf{\ ^{\shortparallel }}}}e}_{\alpha }\mathbf{\mathbf{%
\mathbf{\mathbf{\ }}}}_{[10]}^{\shortparallel }\widehat{\mathbf{\Gamma }}_{\
\beta \gamma }^{\alpha }-\mathbf{\mathbf{\mathbf{\mathbf{\ ^{\shortparallel }%
}}}e}_{\beta }\ _{[10]}^{\shortparallel }\widehat{\mathbf{\Gamma }}_{\
\gamma \alpha }^{\alpha }+\mathbf{\mathbf{\mathbf{\mathbf{\ }}}}%
_{[10]}^{\shortparallel }\widehat{\mathbf{\Gamma }}_{\ \beta \gamma }^{\mu }%
\mathbf{\mathbf{\mathbf{\mathbf{\ ^{\shortparallel }}}}}\widehat{\mathbf{%
\Gamma }}_{\ \mu \alpha }^{\alpha }+\mathbf{\mathbf{\mathbf{\mathbf{\
^{\shortparallel }}}}}\widehat{\mathbf{\Gamma }}_{\ \beta \gamma }^{\mu }%
\mathbf{\mathbf{\mathbf{\mathbf{\ \ }}}}_{[10]}^{\shortparallel }\widehat{%
\mathbf{\Gamma }}_{\ \mu \alpha }^{\alpha }  \notag \\
&&-\mathbf{\mathbf{\mathbf{\mathbf{\ \ }}}}_{[10]}^{\shortparallel }\widehat{%
\mathbf{\Gamma }}_{\ \beta \alpha }^{\mu }\mathbf{\mathbf{\mathbf{\mathbf{\
^{\shortparallel }}}}}\widehat{\mathbf{\Gamma }}_{\ \mu \gamma }^{\alpha }-%
\mathbf{\mathbf{\mathbf{\mathbf{\ ^{\shortparallel }}}}}\widehat{\mathbf{%
\Gamma }}_{\ \beta \alpha }^{\mu }\mathbf{\mathbf{\mathbf{\mathbf{\ }}}}%
_{[10]}^{\shortparallel }\widehat{\mathbf{\Gamma }}_{\ \mu \gamma }^{\alpha
}+\mathbf{\mathbf{\mathbf{\mathbf{\ ^{\shortparallel }}}}}w_{\ \gamma \alpha
}^{\mu }\mathbf{\mathbf{\mathbf{\mathbf{\ }}}}_{[10]}^{\shortparallel }%
\widehat{\mathbf{\Gamma }}_{\ \beta \mu }^{\alpha }  \label{ric52} \\
&&+i\kappa \overline{\mathcal{R}}_{\quad \quad \quad }^{n+i\ b\ n+k}\mathbf{%
\mathbf{\mathbf{\mathbf{\ }}}}\ ^{\shortparallel }p_{b}[(\mathbf{\mathbf{%
\mathbf{\mathbf{\ ^{\shortparallel }}}}e}_{k}\mathbf{\mathbf{\mathbf{\mathbf{%
\ ^{\shortparallel }}}}}\widehat{\mathbf{\Gamma }}_{\ \mu \alpha }^{\alpha
})(\mathbf{\mathbf{\mathbf{\mathbf{\ ^{\shortparallel }}}}e}_{i}\mathbf{%
\mathbf{\mathbf{\mathbf{\ ^{\shortparallel }}}}}\widehat{\mathbf{\Gamma }}%
_{\ \beta \gamma }^{\mu })-(\mathbf{\mathbf{\mathbf{\mathbf{\
^{\shortparallel }}}}e}_{k}\mathbf{\mathbf{\mathbf{\mathbf{\
^{\shortparallel }}}}}\widehat{\mathbf{\Gamma }}_{\ \mu \gamma }^{\alpha })(%
\mathbf{\mathbf{\mathbf{\mathbf{\ ^{\shortparallel }}}}e}_{i}\mathbf{\mathbf{%
\mathbf{\mathbf{\ ^{\shortparallel }}}}}\widehat{\mathbf{\Gamma }}_{\ \beta
\alpha }^{\mu })]  \notag \\
&&-i\kappa \overline{\mathcal{R}}_{\quad \quad \quad }^{n+i\ n+j\ n+k}\delta
_{\gamma k}\mathbf{\mathbf{\mathbf{\mathbf{\ ^{\shortparallel }}}}}\widehat{%
\mathbf{\Gamma }}_{\ \mu i}^{\alpha }(\mathbf{\mathbf{\mathbf{\mathbf{\
^{\shortparallel }}}}e}_{j}\mathbf{\mathbf{\mathbf{\mathbf{\
^{\shortparallel }}}}}\widehat{\mathbf{\Gamma }}_{\ \beta \alpha }^{\mu }) 
\notag \\
&&+i\kappa \overline{\mathcal{R}}_{\quad \quad \quad }^{n+i\ n+j\ a}\delta
_{a}^{\gamma }[\mathbf{\mathbf{\mathbf{\mathbf{\ ^{\shortparallel }}}}e}_{j}%
\mathbf{\mathbf{\mathbf{\mathbf{\ ^{\shortparallel }}}}e}_{\gamma }\widehat{%
\mathbf{\Gamma }}_{\ \beta i}^{\alpha }-(\mathbf{\mathbf{\mathbf{\mathbf{\
^{\shortparallel }}}}e}_{j}\mathbf{\mathbf{\mathbf{\mathbf{\
^{\shortparallel }}}}}\widehat{\mathbf{\Gamma }}_{\ \mu i}^{\alpha })\mathbf{%
\mathbf{\mathbf{\mathbf{\ ^{\shortparallel }}}}}\widehat{\mathbf{\Gamma }}%
_{\ \beta \gamma }^{\mu }+(\mathbf{\mathbf{\mathbf{\mathbf{\
^{\shortparallel }}}}e}_{j}\mathbf{\mathbf{\mathbf{\mathbf{\
^{\shortparallel }}}}}\widehat{\mathbf{\Gamma }}_{\ \mu \gamma }^{\alpha })%
\mathbf{\mathbf{\mathbf{\mathbf{\ ^{\shortparallel }}}}}\widehat{\mathbf{%
\Gamma }}_{\ \beta i}^{\mu }+~\mathbf{\mathbf{\mathbf{\mathbf{%
^{\shortparallel }}}}}\widehat{\mathbf{\Gamma }}_{\ \mu \gamma }^{\alpha }(%
\mathbf{\mathbf{\mathbf{\mathbf{\ ^{\shortparallel }}}}e}_{j}\mathbf{\mathbf{%
\mathbf{\mathbf{\ \ ^{\shortparallel }}}}}\widehat{\mathbf{\Gamma }}_{\
\beta i}^{\mu })],  \notag
\end{eqnarray}%
where, for instance, $\delta _{\gamma k}$ and $\delta _{a}^{\gamma }$ are
Kronecker symbols. Terms (\ref{ric51}) and (\ref{ric52}) for the star
deformed Ricci d-tensor are imaginary and they preserve such a property for
zero torsion distortions to LC-configurations, for instance, for holonomic
geometric models similar to that elaborated in \cite{aschieri17}. Without
additional assumptions on the type of physical models we elaborate on phase
space and for projections on base spacetime configurations, we can not
speculate on possible gravitational and matter field effects of R-fluxes. If
we consider on total phase space nonholonomic Finsler-Lagrange-Hamilton
variables which can be transformed equivalently in almost symplectic
structures, we can develop a direction with real phase space constructions
when the complex unity $i$ and respective complex terms are substituted in
low energy limits by geometric N-adapted objects with almost complex
structure and respective almost K\"{a}hler geometric models. For instance,
such nonholonomic models were studied in \cite{vacaru08a} (for almost
complex/ symplectic constructions both with symmetric and nonsymmetric
metrics). A series of works \cite{vacaru07a,vacaru08b,vacaru08a,vacaru07,
vacaru07b,vacaru16} were devoted to classical commutative and noncommutative
gravity and geometric flow theories with almost K\"{a}hler structure and
their deformation and A-brane quantization. Such approaches can be
generalized self-consistently for R-flux star deformations as we prove in
other partner works. For certain classes of models, $[01]$ and $[10]$
contributions can be considered as real but almost complex ones (mimicking
complex structures) or with complex imaginary terms.

We can transform the coordinate formulas (5.78) - (5.79) from \cite%
{aschieri17}, for the LC-connection, into respective formulas (\ref{ric51})
and (\ref{ric52}) for the canonical d-connection following such rules (for
our conventions on indices and N-adapted values on nonholonomic phase spaces
stated by Convention 2): 1) we change $\ \ \mathbf{^{\shortparallel
}\partial }_{\alpha }\rightarrow $\ $\ \mathbf{^{\shortparallel }e}_{\alpha
} $ and 2)\ $\ [\ \mathbf{^{\shortparallel }\partial }_{\alpha },\ \ \mathbf{%
^{\shortparallel }\partial }_{\beta }]\rightarrow $ $[\ \mathbf{%
^{\shortparallel }e}_{\alpha },\ \mathbf{^{\shortparallel }e}_{\beta
}]+w_{\alpha \beta }^{\gamma }\ \mathbf{^{\shortparallel }e}_{\gamma };$ 3) $%
\mathbf{\mathbf{\mathbf{\mathbf{\ ^{\shortparallel }}}}}\Gamma _{\ \beta
\gamma }^{\mu }\rightarrow \mathbf{\mathbf{\mathbf{\mathbf{\
^{\shortparallel }}}}}\widehat{\mathbf{\Gamma }}_{\ \beta \gamma }^{\mu },$
i.e.$\mathbf{\mathbf{\mathbf{\mathbf{\ ^{\shortparallel }}}}}\nabla
\rightarrow \mathbf{\mathbf{\mathbf{\mathbf{\ ^{\shortparallel }}}}}\widehat{%
\mathbf{D}};$ 4) $\mathbf{\mathbf{\mathbf{\mathbf{\ ^{\shortparallel }}}}}%
g_{\alpha \beta }\rightarrow \mathbf{\mathbf{\mathbf{\mathbf{\
^{\shortparallel }}}}g}_{\alpha \beta }=[\mathbf{\mathbf{\mathbf{\mathbf{\
^{\shortparallel }}}}g}_{ij},\mathbf{\mathbf{\mathbf{\mathbf{\
^{\shortparallel }}}}g}^{ab}]$ etc. Such transforms can be performed for a a
nonholonomic frame structure (\ref{nhframtr}) which allows us to N-elongate
partial derivative operators and differentials and N-adapt formulas.
Applying such rules, we can nonholonomically deform the formula (5.80) in 
\cite{aschieri17} and compute the $\hbar \kappa =\ell _{s}^{3}/6$
contribution to the nonassociative canonical Ricci d-tensor (\ref%
{driccicanonstar}), 
\begin{eqnarray}
\mathbf{\mathbf{\mathbf{\mathbf{\ \ }}}}_{[11]}^{\shortparallel }\widehat{%
\mathbf{\mathbf{\mathbf{\mathbf{R}}}}}ic_{\beta \gamma }^{\star } &=&\mathbf{%
\mathbf{\mathbf{\mathbf{\ ^{\shortparallel }}}}e}_{\alpha }\mathbf{\mathbf{%
\mathbf{\mathbf{\ }}}}_{[11]}^{\shortparallel }\widehat{\mathbf{\Gamma }}_{\
\beta \gamma }^{\alpha }-\mathbf{\mathbf{\mathbf{\mathbf{\ ^{\shortparallel }%
}}}e}_{\beta }\ _{[11]}^{\shortparallel }\widehat{\mathbf{\Gamma }}_{\
\gamma \alpha }^{\alpha }+\mathbf{\mathbf{\mathbf{\mathbf{\ }}}}%
_{[11]}^{\shortparallel }\widehat{\mathbf{\Gamma }}_{\ \beta \gamma }^{\mu }%
\mathbf{\mathbf{\mathbf{\mathbf{\ ^{\shortparallel }}}}}\widehat{\mathbf{%
\Gamma }}_{\ \mu \alpha }^{\alpha }+\mathbf{\mathbf{\mathbf{\mathbf{\ }}}}%
_{[10]}^{\shortparallel }\widehat{\mathbf{\Gamma }}_{\ \beta \gamma }^{\mu
}\ _{[01]}^{\shortparallel }\widehat{\mathbf{\Gamma }}_{\ \mu \alpha
}^{\alpha }+\mathbf{\mathbf{\mathbf{\mathbf{\ }}}}_{[01]}^{\shortparallel }%
\widehat{\mathbf{\Gamma }}_{\ \beta \gamma }^{\mu }\mathbf{\mathbf{\mathbf{%
\mathbf{\ }}}}_{[10]}^{\shortparallel }\widehat{\mathbf{\Gamma }}_{\ \mu
\alpha }^{\alpha }  \notag \\
&&+\mathbf{\mathbf{\mathbf{\mathbf{\ ^{\shortparallel }}}}}\widehat{\mathbf{%
\Gamma }}_{\ \beta \gamma }^{\mu }\mathbf{\mathbf{\mathbf{\mathbf{\ \ }}}}%
_{[11]}^{\shortparallel }\widehat{\mathbf{\Gamma }}_{\ \mu \alpha }^{\alpha
}-\mathbf{\mathbf{\mathbf{\mathbf{\ \ }}}}_{[11]}^{\shortparallel }\widehat{%
\mathbf{\Gamma }}_{\ \beta \alpha }^{\mu }\mathbf{\mathbf{\mathbf{\mathbf{\
^{\shortparallel }}}}}\widehat{\mathbf{\Gamma }}_{\ \mu \gamma }^{\alpha }-%
\mathbf{\mathbf{\mathbf{\mathbf{\ \ }}}}_{[10]}^{\shortparallel }\widehat{%
\mathbf{\Gamma }}_{\ \beta \alpha }^{\mu }\mathbf{\mathbf{\mathbf{\mathbf{\ }%
}}}_{[01]}^{\shortparallel }\widehat{\mathbf{\Gamma }}_{\ \mu \gamma
}^{\alpha }-\mathbf{\mathbf{\mathbf{\mathbf{\ \ }}}}_{[01]}^{\shortparallel }%
\widehat{\mathbf{\Gamma }}_{\ \beta \alpha }^{\mu }\mathbf{\mathbf{\mathbf{%
\mathbf{\ }}}}_{[10]}^{\shortparallel }\widehat{\mathbf{\Gamma }}_{\ \mu
\gamma }^{\alpha }  \notag \\
&&-\mathbf{\mathbf{\mathbf{\mathbf{\ ^{\shortparallel }}}}}\widehat{\mathbf{%
\Gamma }}_{\ \beta \alpha }^{\mu }\mathbf{\mathbf{\mathbf{\mathbf{\ }}}}%
_{[11]}^{\shortparallel }\widehat{\mathbf{\Gamma }}_{\ \mu \gamma }^{\alpha
}+\mathbf{\mathbf{\mathbf{\mathbf{\ ^{\shortparallel }}}}}w_{\ \gamma \alpha
}^{\mu }\mathbf{\mathbf{\mathbf{\mathbf{\ }}}}_{[11]}^{\shortparallel }%
\widehat{\mathbf{\Gamma }}_{\ \beta \mu }^{\alpha }  \label{ric53}
\end{eqnarray}%
\begin{eqnarray*}
&&+\frac{i\hbar }{2}[(\ \mathbf{^{\shortparallel }e}_{l}\
_{[10]}^{\shortparallel }\widehat{\mathbf{\Gamma }}_{\ \mu \alpha }^{\alpha
})(\ ^{\shortparallel }\partial ^{n+l}\mathbf{\mathbf{\mathbf{\mathbf{\ }}}}%
^{\shortparallel }\widehat{\mathbf{\Gamma }}_{\ \beta \gamma }^{\mu })+(\ 
\mathbf{^{\shortparallel }e}_{l}\ ^{\shortparallel }\widehat{\mathbf{\Gamma }%
}_{\ \mu \alpha }^{\alpha })(\ ^{\shortparallel }\partial ^{n+l}\mathbf{%
\mathbf{\mathbf{\mathbf{\ }}}}_{[10]}^{\shortparallel }\widehat{\mathbf{%
\Gamma }}_{\ \beta \gamma }^{\mu }) \\
&&-(\ ^{\shortparallel }\partial ^{n+l}\mathbf{\mathbf{\mathbf{\mathbf{\ }}}}%
\ _{[10]}^{\shortparallel }\widehat{\mathbf{\Gamma }}_{\ \mu \alpha
}^{\alpha })(\ \mathbf{^{\shortparallel }e}_{l}\mathbf{~}^{\shortparallel }%
\widehat{\mathbf{\Gamma }}_{\ \beta \gamma }^{\mu })-(\ ^{\shortparallel
}\partial ^{n+l}\ ^{\shortparallel }\widehat{\mathbf{\Gamma }}_{\ \mu \alpha
}^{\alpha })(\ \mathbf{^{\shortparallel }e}_{l}\mathbf{\mathbf{\mathbf{%
\mathbf{\ }}}}_{[10]}^{\shortparallel }\widehat{\mathbf{\Gamma }}_{\ \beta
\gamma }^{\mu })] \\
&&-\frac{i\hbar }{2}[(\ \mathbf{^{\shortparallel }e}_{l}\mathbf{\mathbf{%
\mathbf{\mathbf{\ }}}}_{[10]}^{\shortparallel }\widehat{\mathbf{\Gamma }}_{\
\mu \gamma }^{\alpha })(\ ^{\shortparallel }\partial ^{n+l}\mathbf{\mathbf{%
\mathbf{\mathbf{\ ^{\shortparallel }}}}}\widehat{\mathbf{\Gamma }}_{\ \beta
\alpha }^{\mu })+\ \mathbf{^{\shortparallel }e}_{l}\mathbf{\mathbf{\mathbf{%
\mathbf{\ }}}}^{\shortparallel }\widehat{\mathbf{\Gamma }}_{\ \mu \gamma
}^{\alpha })(\ ^{\shortparallel }\partial ^{n+l}\ _{[10]}^{\shortparallel }%
\widehat{\mathbf{\Gamma }}_{\ \beta \alpha }^{\mu }) \\
&&-(\ ^{\shortparallel }\partial ^{n+l}\mathbf{\mathbf{\mathbf{\mathbf{\ }}}}%
_{[10]}^{\shortparallel }\widehat{\mathbf{\Gamma }}_{\ \mu \gamma }^{\alpha
})(\ \mathbf{^{\shortparallel }e}_{l}\mathbf{\mathbf{\mathbf{\mathbf{\
^{\shortparallel }}}}}\widehat{\mathbf{\Gamma }}_{\ \beta \alpha }^{\mu
})-(\ ^{\shortparallel }\partial ^{n+l}\mathbf{\mathbf{\mathbf{\mathbf{\ }}}}%
^{\shortparallel }\widehat{\mathbf{\Gamma }}_{\ \mu \gamma }^{\alpha })(\ 
\mathbf{^{\shortparallel }e}_{l}\ _{[10]}^{\shortparallel }\widehat{\mathbf{%
\Gamma }}_{\ \beta \alpha }^{\mu })]
\end{eqnarray*}%
\begin{eqnarray*}
&&+i\kappa \overline{\mathcal{R}}_{\quad \quad \quad }^{n+i\ b\ n+k}\mathbf{%
\mathbf{\mathbf{\mathbf{\ }}}}\ ^{\shortparallel }p_{b}[(\ \mathbf{%
^{\shortparallel }e}_{k}\mathbf{\mathbf{\mathbf{\mathbf{\ }}}}\
_{[01]}^{\shortparallel }\widehat{\mathbf{\Gamma }}_{\ \mu \alpha }^{\alpha
})(\ \mathbf{^{\shortparallel }e}_{i}\mathbf{~}^{\shortparallel }\widehat{%
\mathbf{\Gamma }}_{\ \beta \gamma }^{\mu })+(\ \mathbf{^{\shortparallel }e}%
_{k}\mathbf{\mathbf{\mathbf{\mathbf{\ }}}}\ ^{\shortparallel }\widehat{%
\mathbf{\Gamma }}_{\ \mu \alpha }^{\alpha })(\ \mathbf{^{\shortparallel }e}%
_{i}\mathbf{~}_{[01]}^{\shortparallel }\widehat{\mathbf{\Gamma }}_{\ \beta
\gamma }^{\mu }) \\
&&-(\ \mathbf{^{\shortparallel }e}_{k}\mathbf{\mathbf{\mathbf{\mathbf{\ }}}}%
_{[01]}^{\shortparallel }\widehat{\mathbf{\Gamma }}_{\ \mu \gamma }^{\alpha
})(\ \ \mathbf{^{\shortparallel }e}_{i}\mathbf{\mathbf{\mathbf{\mathbf{\
^{\shortparallel }}}}}\widehat{\mathbf{\Gamma }}_{\ \beta \alpha }^{\mu
})-(\ \mathbf{^{\shortparallel }e}_{k}\mathbf{\mathbf{\mathbf{\mathbf{\ }}}}%
^{\shortparallel }\widehat{\mathbf{\Gamma }}_{\ \mu \gamma }^{\alpha })(\ \ 
\mathbf{^{\shortparallel }e}_{i}\mathbf{\mathbf{\mathbf{\mathbf{\ }}}}%
_{[01]}^{\shortparallel }\widehat{\mathbf{\Gamma }}_{\ \beta \alpha }^{\mu
})]
\end{eqnarray*}%
\begin{eqnarray*}
&&-\frac{\hbar \kappa }{2}\overline{\mathcal{R}}_{\quad \quad \quad }^{n+i\
b\ n+k}\mathbf{\mathbf{\mathbf{\mathbf{\ }}}}\ ^{\shortparallel }p_{b}[(\ 
\mathbf{^{\shortparallel }e}_{l}\mathbf{~}\ \mathbf{^{\shortparallel }e}_{k}%
\mathbf{\mathbf{\mathbf{\mathbf{\ }}}}\ ^{\shortparallel }\widehat{\mathbf{%
\Gamma }}_{\ \mu \alpha }^{\alpha })(\ ^{\shortparallel }\partial ^{n+l}\ 
\mathbf{^{\shortparallel }e}_{i}\mathbf{~}^{\shortparallel }\widehat{\mathbf{%
\Gamma }}_{\ \beta \gamma }^{\mu })-(\ ^{\shortparallel }\partial ^{n+l}\ 
\mathbf{^{\shortparallel }e}_{k}\mathbf{\mathbf{\mathbf{\mathbf{\ }}}}\
^{\shortparallel }\widehat{\mathbf{\Gamma }}_{\ \mu \alpha }^{\alpha })(\ 
\mathbf{^{\shortparallel }e}_{l}\mathbf{~}\ \mathbf{^{\shortparallel }e}_{i}%
\mathbf{~}^{\shortparallel }\widehat{\mathbf{\Gamma }}_{\ \beta \gamma
}^{\mu }) \\
&&-(\ \mathbf{^{\shortparallel }e}_{l}\mathbf{~}\ \mathbf{^{\shortparallel }e%
}_{k}\mathbf{\mathbf{\mathbf{\mathbf{\ }}}}^{\shortparallel }\widehat{%
\mathbf{\Gamma }}_{\ \mu \gamma }^{\alpha })(\ ^{\shortparallel }\partial
^{n+l}\ \ \mathbf{^{\shortparallel }e}_{i}\mathbf{\mathbf{\mathbf{\mathbf{\
^{\shortparallel }}}}}\widehat{\mathbf{\Gamma }}_{\ \beta \alpha }^{\mu
})+(\ ^{\shortparallel }\partial ^{n+l}\ \mathbf{^{\shortparallel }e}_{k}%
\mathbf{\mathbf{\mathbf{\mathbf{\ }}}}^{\shortparallel }\widehat{\mathbf{%
\Gamma }}_{\ \mu \gamma }^{\alpha })(\ \mathbf{^{\shortparallel }e}_{l}%
\mathbf{~}\ \ \mathbf{^{\shortparallel }e}_{i}\mathbf{\mathbf{\mathbf{%
\mathbf{\ ^{\shortparallel }}}}}\widehat{\mathbf{\Gamma }}_{\ \beta \alpha
}^{\mu })]
\end{eqnarray*}%
\begin{eqnarray*}
&&-i\kappa \overline{\mathcal{R}}_{\quad \quad \quad }^{n+i\ n+j\
n+k}[\delta _{\alpha k}\left( (\ \mathbf{^{\shortparallel }e}_{j}\mathbf{%
\mathbf{\mathbf{\mathbf{\ }}}}^{\shortparallel }\widehat{\mathbf{\Gamma }}%
_{\ \mu i}^{\alpha })\ _{[01]}^{\shortparallel }\widehat{\mathbf{\Gamma }}%
_{\ \beta \gamma }^{\mu }+(\ \mathbf{^{\shortparallel }e}_{j}\mathbf{\mathbf{%
\mathbf{\mathbf{\ }}}}_{[01]}^{\shortparallel }\widehat{\mathbf{\Gamma }}_{\
\mu i}^{\alpha })\mathbf{\mathbf{\mathbf{\mathbf{\ ^{\shortparallel }}}}}%
\widehat{\mathbf{\Gamma }}_{\ \beta \gamma }^{\mu }\right) \\
&&+\delta _{\gamma k}\left( \mathbf{\mathbf{\mathbf{\mathbf{\ }}}}%
^{\shortparallel }\widehat{\mathbf{\Gamma }}_{\ \mu i}^{\alpha }(\ \mathbf{%
^{\shortparallel }e}_{j}\ _{[01]}^{\shortparallel }\widehat{\mathbf{\Gamma }}%
_{\ \beta \alpha }^{\mu })+\mathbf{\mathbf{\mathbf{\mathbf{\ }}}}%
_{[01]}^{\shortparallel }\widehat{\mathbf{\Gamma }}_{\ \mu i}^{\alpha }(\ 
\mathbf{^{\shortparallel }e}_{j}\mathbf{\mathbf{\mathbf{\mathbf{\
^{\shortparallel }}}}}\widehat{\mathbf{\Gamma }}_{\ \beta \alpha }^{\mu
})\right) ]
\end{eqnarray*}%
\begin{eqnarray*}
&&+\frac{\hbar \kappa }{2}\overline{\mathcal{R}}_{\quad \quad \quad }^{n+i\
n+j\ n+k}[\delta _{\alpha k}\left( (\ \mathbf{^{\shortparallel }e}_{l}\ 
\mathbf{^{\shortparallel }e}_{j}\mathbf{\mathbf{\mathbf{\mathbf{\ }}}}%
^{\shortparallel }\widehat{\mathbf{\Gamma }}_{\ \mu i}^{\alpha })(\
^{\shortparallel }\partial ^{n+l}\ ^{\shortparallel }\widehat{\mathbf{\Gamma 
}}_{\ \beta \gamma }^{\mu })-(\ ^{\shortparallel }\partial ^{n+l}\ \mathbf{%
^{\shortparallel }e}_{j}\mathbf{\mathbf{\mathbf{\mathbf{\ }}}}%
^{\shortparallel }\widehat{\mathbf{\Gamma }}_{\ \mu i}^{\alpha })(\ \mathbf{%
^{\shortparallel }e}_{l}\mathbf{\mathbf{\mathbf{\mathbf{\ ^{\shortparallel }}%
}}}\widehat{\mathbf{\Gamma }}_{\ \beta \gamma }^{\mu })\right) \\
&&+\delta _{\gamma k}\left( (\ \mathbf{^{\shortparallel }e}_{l}\mathbf{~%
\mathbf{\mathbf{\mathbf{\ }}}}^{\shortparallel }\widehat{\mathbf{\Gamma }}%
_{\ \mu i}^{\alpha })(\ ^{\shortparallel }\partial ^{n+l}\ \mathbf{%
^{\shortparallel }e}_{j}\ ^{\shortparallel }\widehat{\mathbf{\Gamma }}_{\
\beta \alpha }^{\mu })-(\ ^{\shortparallel }\partial ^{n+l}\mathbf{\mathbf{%
\mathbf{\mathbf{\ }}}}^{\shortparallel }\widehat{\mathbf{\Gamma }}_{\ \mu
i}^{\alpha })(\ \mathbf{^{\shortparallel }e}_{l}\ \mathbf{^{\shortparallel }e%
}_{j}\mathbf{\mathbf{\mathbf{\mathbf{\ ^{\shortparallel }}}}}\widehat{%
\mathbf{\Gamma }}_{\ \beta \alpha }^{\mu })\right) ]
\end{eqnarray*}%
\begin{eqnarray*}
&&+i\kappa \overline{\mathcal{R}}_{\quad \quad \quad }^{n+i\ n+j\ n+k}\delta
_{\alpha k}[\left( (\ \mathbf{^{\shortparallel }e}_{j}\ \mathbf{%
^{\shortparallel }e}_{\gamma }\mathbf{\mathbf{\mathbf{\mathbf{\ }}}}%
_{[01]}^{\shortparallel }\widehat{\mathbf{\Gamma }}_{\ \mu i}^{\alpha })+(\ 
\mathbf{^{\shortparallel }e}_{j}\mathbf{\mathbf{\mathbf{\mathbf{\ }}}}%
^{\shortparallel }\widehat{\mathbf{\Gamma }}_{\ \mu \gamma }^{\alpha })\
_{[01]}^{\shortparallel }\widehat{\mathbf{\Gamma }}_{\ \beta i}^{\mu }+(\ 
\mathbf{^{\shortparallel }e}_{j}\mathbf{\mathbf{\mathbf{\mathbf{\ }}}}%
_{[01]}^{\shortparallel }\widehat{\mathbf{\Gamma }}_{\ \mu \gamma }^{\alpha
})\ ^{\shortparallel }\widehat{\mathbf{\Gamma }}_{\ \beta i}^{\mu }\right) \\
&&+\mathbf{\mathbf{\mathbf{\mathbf{\ }}}}^{\shortparallel }\widehat{\mathbf{%
\Gamma }}_{\ \mu \gamma }^{\alpha }(\ \mathbf{^{\shortparallel }e}_{j}\
_{[01]}^{\shortparallel }\widehat{\mathbf{\Gamma }}_{\ \beta i}^{\mu })+%
\mathbf{\mathbf{\mathbf{\mathbf{\ }}}}_{[01]}^{\shortparallel }\widehat{%
\mathbf{\Gamma }}_{\ \mu \gamma }^{\alpha }(\ \mathbf{^{\shortparallel }e}%
_{j}\ ^{\shortparallel }\widehat{\mathbf{\Gamma }}_{\ \beta i}^{\mu })]
\end{eqnarray*}%
\begin{eqnarray*}
&&-\frac{\hbar \kappa }{2}\overline{\mathcal{R}}_{\quad \quad \quad }^{n+i\
n+j\ n+k}\delta _{\alpha k}[(\ \mathbf{^{\shortparallel }e}_{l}\ \mathbf{%
^{\shortparallel }e}_{j}\mathbf{\mathbf{\mathbf{\mathbf{\ }}}}%
^{\shortparallel }\widehat{\mathbf{\Gamma }}_{\ \mu \gamma }^{\alpha })(\
^{\shortparallel }\partial ^{n+l}\ ^{\shortparallel }\widehat{\mathbf{\Gamma 
}}_{\ \beta i}^{\mu })-(\ ^{\shortparallel }\partial ^{n+l}\ \mathbf{%
^{\shortparallel }e}_{j}\mathbf{\mathbf{\mathbf{\mathbf{\ }}}}%
^{\shortparallel }\widehat{\mathbf{\Gamma }}_{\ \mu \gamma }^{\alpha })(\ 
\mathbf{^{\shortparallel }e}_{l}\mathbf{\mathbf{\mathbf{\mathbf{\
^{\shortparallel }}}}}\widehat{\mathbf{\Gamma }}_{\ \beta i}^{\mu }) \\
&&+(\ \mathbf{^{\shortparallel }e}_{l}\mathbf{~\mathbf{\mathbf{\mathbf{\ }}}}%
^{\shortparallel }\widehat{\mathbf{\Gamma }}_{\ \mu \gamma }^{\alpha })(\
^{\shortparallel }\partial ^{n+l}\ \mathbf{^{\shortparallel }e}_{j}\
^{\shortparallel }\widehat{\mathbf{\Gamma }}_{\ \beta i}^{\mu })-(\
^{\shortparallel }\partial ^{n+l}\mathbf{\mathbf{\mathbf{\mathbf{\ }}}}%
^{\shortparallel }\widehat{\mathbf{\Gamma }}_{\ \mu \gamma }^{\alpha })(\ 
\mathbf{^{\shortparallel }e}_{l}\ \mathbf{^{\shortparallel }e}_{j}\mathbf{%
\mathbf{\mathbf{\mathbf{\ ^{\shortparallel }}}}}\widehat{\mathbf{\Gamma }}%
_{\ \beta i}^{\mu })].
\end{eqnarray*}%
In an equivalent form, these formulas can be proven by performing a $\hbar
\kappa $ decomposition with respective contractions of the first and forth
indices in (\ref{driccicanonstar1}). To complete the computation of such
real nonassociative and noncommutative contributions we have to insert
formulas for coefficients of type $\ _{[01]}^{\shortparallel }\widehat{%
\Gamma }_{\star \alpha \beta }^{\nu },\ _{[10]}^{\shortparallel }\widehat{%
\Gamma }_{\star \alpha \beta }^{\nu }$ and $\ _{[11]}^{\shortparallel }%
\widehat{\Gamma }_{\star \alpha \beta }^{\nu },$ see (\ref{aux51}). We omit
such cumbersome formulas because they will be not used in this and partner
works. For additional nonholonomic constraints resulting in zero distortion
d-tensors and nonassociative LC-connections, when $\ ^{\shortparallel }%
\widehat{\mathbf{D}}_{\mid \ ^{\shortparallel }\widehat{\mathbf{T}}%
=0}^{\star }=\ ^{\shortparallel }\nabla ^{\star },$ see (\ref{lccondnonass}%
), in coordinate bases, we obtain the same (5.80) from \cite{aschieri17}. It
is important to have a prescription with rules 1)-4) in order to move
geometric d-objects from local coordinate to N-adapted bases (and
inversely), and to star deform them nonholonomicaly into configurations
determined by canonical data $(\ \mathbf{^{\shortparallel}e}_{\alpha},%
\mathbf{\mathbf{\ ^{\shortparallel }}g}_{\alpha \beta }, \mathbf{\mathbf{%
\mathbf{\mathbf{\ ^{\shortparallel }}}}}\widehat{\mathbf{\Gamma }}_{\ \beta
\gamma }^{\mu }).$ In such N-adapted variables, we can apply the AFCDM and
prove certain general decoupling and integrability properties of
nonassociative modified Einstein and Ricci flow equations. Such
constructions are not possible if we work only with (non) associative/
commutative LC-data.

For h- and c-splitting, we have%
\begin{equation}
\ \mathbf{\mathbf{\mathbf{\mathbf{^{\shortparallel }}}}}\widehat{\mathbf{%
\mathbf{\mathbf{\mathbf{R}}}}}ic_{\alpha \beta }^{\star }=\mathbf{\{}\ \
^{\shortparallel }\widehat{R}_{\ \star hj}=\ \ ^{\shortparallel }\widehat{%
\mathcal{\Re }}_{\ \star hji}^{i},\ \ ^{\shortparallel }\widehat{P}_{\star
j}^{\ a}=-\ ^{\shortparallel }\widehat{\mathcal{\Re }}_{\star \ ji}^{i\quad
a}\ ,\ ^{\shortparallel }\widehat{P}_{\star \ k}^{b\quad }=\
^{\shortparallel }\widehat{\mathcal{\Re }}_{\star c\ k}^{\ b\ c},\
^{\shortparallel }\widehat{S}_{\star \ }^{bc\quad }=\ ^{\shortparallel }%
\widehat{\mathcal{\Re }}_{\star a\ }^{\ bca}\ \},  \label{hcnonassocrcan}
\end{equation}%
where each component splits also in respective $[00],[01],[10]$ and $[11]$
terms, for instance,%
\begin{eqnarray}
\ \ _{[00]}^{\shortparallel }\widehat{\mathbf{\mathbf{\mathbf{\mathbf{R}}}}}%
ic_{\alpha \beta }^{\star } &=&\mathbf{\{}\ \ \ _{[00]}^{\shortparallel }%
\widehat{R}_{\ \star hj}=\ \ \ _{[00]}^{\shortparallel }\widehat{\mathcal{%
\Re }}_{\ \star hji}^{i},\ _{[00]}^{\shortparallel }\widehat{P}_{\star j}^{\
a}=-\ \ _{[00]}^{\shortparallel }\widehat{\mathcal{\Re }}_{\star \
ji}^{i\quad a}\ ,\ \ _{[00]}^{\shortparallel }\widehat{P}_{\star \
k}^{b\quad }=\ \ _{[00]}^{\shortparallel }\widehat{\mathcal{\Re }}_{\star c\
k}^{\ b\ c},  \notag \\
&&\ _{[00]}^{\shortparallel }\widehat{S}_{\star \ }^{bc\quad }=\ \
_{[00]}^{\shortparallel }\widehat{\mathcal{\Re }}_{\star a\ }^{\ bca}\ \},
\label{hcnonassocrcana} \\
&&\ _{[01]}^{\shortparallel }\widehat{\mathbf{\mathbf{\mathbf{\mathbf{R}}}}}%
ic_{\alpha \beta }^{\star }=\mathbf{\{}\ \ \ _{[01]}^{\shortparallel }%
\widehat{R}_{\ \star hj}=\ _{[01]}^{\shortparallel }\widehat{\mathcal{\Re }}%
_{\ \star hji}^{i},\ _{[01]}^{\shortparallel }\widehat{P}_{\star j}^{\ a}=-\
\ _{[01]}^{\shortparallel }\widehat{\mathcal{\Re }}_{\star \ ji}^{i\quad a}\
,  \notag \\
&&\ _{[01]}^{\shortparallel }\widehat{P}_{\star \ k}^{b\quad }=\ \
_{[01]}^{\shortparallel }\widehat{\mathcal{\Re }}_{\star c\ k}^{\ b\ c},\
_{[01]}^{\shortparallel }\widehat{S}_{\star \ }^{bc\quad }=\ \
_{[01]}^{\shortparallel }\widehat{\mathcal{\Re }}_{\star a\ }^{\ bca}\ \}, 
\notag \\
&&\mbox{ and similarly for }\lbrack 1,0]\mbox{ and }\lbrack 1,1].  \notag
\end{eqnarray}%
We can arrange the nonholonomic distributions on phase space that $\
_{[00]}^{\shortparallel }\widehat{\mathbf{\mathbf{\mathbf{\mathbf{R}}}}}%
ic_{\alpha \beta }^{\star }=\ ^{\shortparallel }\widehat{\mathbf{R}}_{\alpha
\beta }$ with commutative h- and c- coefficients (\ref{candricci}) but other
components contains nonassociative and noncommutative contributions from
star product deformations.

Introducing star distortions of the canonical d-connection $,\
^{\shortparallel }\widehat{\mathbf{D}}^{\star }=\ ^{\shortparallel }\nabla
^{\star }+\ _{\star }^{\shortparallel }\widehat{\mathbf{Z}}$ (\ref%
{candistrnas}) into formulas for $\ ^{\shortparallel }\widehat{\mathcal{\Re }%
}ic^{\star }$ (\ref{driccicanonstar}), we can compute the distortions of the
nonassocitative canonical Ricci scalar and Ricci d-tensors from
corresponding values determined by the LC-connection, 
\begin{equation}
\ ^{\shortparallel }\widehat{\mathcal{R}}sc^{\star }=\ \ _{\nabla
}^{\shortparallel }\mathcal{R}sc^{\star }+\ \ \ _{\nabla }^{\shortparallel }%
\widehat{\mathcal{Z}}sc^{\star }\mbox{
and }\ \mathbf{\mathbf{\mathbf{\mathbf{\ ^{\shortparallel }}}}}\widehat{%
\mathcal{\Re }}ic^{\star }=\ \ _{\nabla }^{\shortparallel }\mathcal{\Re }%
ic^{\star }+\ \ _{\nabla }^{\shortparallel }\widehat{\mathcal{Z}}ic^{\star },
\label{candriccidistna}
\end{equation}%
with corresponding distortion tensors $\ _{\nabla }^{\shortparallel}\widehat{%
\mathcal{Z}}$ and $\ _{\nabla }^{\shortparallel }\widehat{\mathcal{Z}}ic.$
Similar values for commutative configuration are determined by formulas (\ref%
{candriccidist}). Here, we use the nonsymmetric inverse d-metric $\
_{\star}^{\shortparallel}\mathfrak{g}^{\alpha \beta }$ (\ref{aux37}) and the
nonsymmetric canonical Ricci d-tensor $\ ^{\shortparallel}\widehat{\mathbf{R}%
}ic_{\alpha \beta }^{\star }$ (\ref{driccicanonstar}) for definition of the
nonassociative canonical Ricci d-scalar, $\ ^{\shortparallel }\widehat{%
\mathcal{R}}sc^{\star }:=\ _{\star }^{\shortparallel }\mathfrak{g}^{\alpha
\beta }\star _{N}\mathbf{\mathbf{\mathbf{\mathbf{\ ^{\shortparallel }}}}} 
\widehat{\mathbf{\mathbf{\mathbf{\mathbf{R}}}}}ic_{\alpha \beta }^{\star }.$
In section 5.6 of \cite{aschieri17}, there are provided the coordinate
formulas for the nonsymmetric Ricci tensor of a star deformed LC-connection,
i.e. for $\ _{\nabla}^{\shortparallel }\mathcal{\Re }ic^{\star }$ and
respective decompositions on parameters $\hbar $ and $\kappa .$ In
principle, those formulas can be used for formulating nonassociative
generalizations of vacuum Einstein equations, for instance, in the form $\
_{\nabla }^{\shortparallel }\mathcal{\Re }ic^{\star }=0,$ see also formulas
(5.66), (5.75) and (5.90) in that work. Unfortunately, it is not possible to
prove a general decoupling property of such systems of nonlinear PDEs
involving the (non) associative LC-connection ( $\ ^{\shortparallel }\nabla
^{\star }) \ ^{\shortparallel }\nabla $ because the zero-torsion conditions
contract the indices in certain forms which do not allow decoupling. In
another turn, the nonassociative gravitational field equations with $\
^{\shortparallel }\widehat{\mathcal{\Re }}ic^{\star }$ can be decoupled and
solved similarly to the commutative versions, in both case using the AFCDM.
Such geometric constructions are with N-connection splitting and requests
certain additional diadic decompositions both in h- and c-subspaces and can
be elaborated both in nonassociative / noncommutative / commutative
generalizations of the Einstein gravity.

Let us consider some important properties of on $\hbar $ and $\kappa $
decompositions of nonassociative star d-metrics and canonical d-connections
presented above:

\begin{enumerate}
\item We can consider nonholonomic distributions on a phase space $\mathcal{M%
}=\mathbf{T}_{\shortparallel }^{\ast }\mathbf{V}$ with N-adapted quasi-Hopf
structure that both the nonassociative LC-connection and respective
distortion to the canonical d-connection are characterized by \textbf{real}
terms with labels $[00],$ with components as in the commutative cotangent
Lorentz bundle; with\textbf{\ imaginary} (complex) terms determined with
labels $[01]$, which are proportional to $\hbar ,$ and with labels $[10]$
being proportional to $\kappa ;$ and with \textbf{real} terms $[11]$
proportional to $\hbar \kappa =\ell _{s}^{3}/6$ representing nontrivial
contributions from nonassocitative geometry.

\item Here we note that nontrivial nonassociative contributions (both in
real and imaginary forms) can be also encoded in the nontrivial N-connection
coefficients $\ ^{\shortparallel }N_{je}(x^{i},\ ^{\shortparallel }p_{a})$
in (\ref{nadapdc}) and related off-diagonal components of the metric (\ref%
{offd}) being considered star deformations and quasi-Hopf algebra valued
coefficients. A number of imaginary or real terms in d-metric and linear
connection structures vanish if we prescribe some subclasses of (non)
symmetric metrics with coefficients defined in such frames of reference/
coordinates which do not depend, or instance, on momentum like variables,
and/or are described by diagonal matrices etc. Nevertheless, even in such
cases there are certain terms proportional to  $\ell _{s}^{3}\mathcal{R}%
_{\quad \tau }^{\mu \nu }$ and linear on momenta $p_{a}$ reflecting a
nontrivial R-flux dependence due to nonassociativity (in the simplest case,
due to the associator acting on some coefficients of metric and a linear
connection). But a general covariant and self-consistent geometric treatment
of geometric and physical theories $\mathcal{M}=\mathbf{T}_{\shortparallel
}^{\ast }\mathbf{V}$ request nonassociative formalism with d-objects and
respective coefficients depending on all phase space coordinates $(x^{i},\
^{\shortparallel }p_{a})$ even for certain models we consider projections,
lifts, and pull-backs on a base spacetime $\mathbf{V}$ with local
coordinates $(x^{i}).$ All additional algebraic commutative / noncommutative
/ nonassociative and nonholonomic structures with star deformation
contribute on $\mathbf{V}$ even the dependence on $\ ^{\shortparallel }p_{a}$
is not considered.

\item Keeping arbitrary the dependence on momentum like variables allows us
not only to formulate a rigorous geometrization of nonassociative physical
models with nontrivial contributions to the Ricci d-tensor $\mathbf{\mathbf{%
\mathbf{\mathbf{\ ^{\shortparallel }}}}}\mathcal{\Re}ic^{\star }$ (\ref%
{driccina}) and related modified Einstein equations (to be derived in next
section using $\mathbf{\mathbf{\mathbf{\mathbf{\ ^{\shortparallel }}}}}%
\widehat{\mathcal{\Re }}ic^{\star }$ constructed for $^{\shortparallel }%
\widehat{\mathbf{D}}^{\star }$ (\ref{twoconnonas})). This give us the
possibility to apply also the AFCDM for constructing exact and parametric
solutions with general off-diagonal dependence on all phase space
coordinates $(x^{i},\ ^{\shortparallel }p_{a}).$ Further nonholonomic
constraints, or certain smooth limits, to nonassociative / noncommutative /
commutative configurations depending only on $(x^{i}),$ in diagonal or not
diagonal forms of (non) symmetric metrics, can be considered. This way we
can study, for instance, nonassociative off-diagonal star deformations of
the black hole metrics and possible implications in modern cosmology. A
general decoupling and integration property of nonassociative versions of
gravitational and matter field equations is possible, for instance, for
so-called diadic decompositions for the canonical d-connection  $\
^{\shortparallel }\widehat{\mathbf{D}}^{\star }$. Imposing additional
constraints (\ref{lccondnonass}) for $\ ^{\shortparallel }\widehat{\mathbf{D}%
}_{\mid \ ^{\shortparallel }\widehat{\mathbf{T}}=0}^{\star }=\
^{\shortparallel }\nabla ^{\star },$ we can extract zero-torsion
LC-configurations with contributions from nonassociative geometry.
\end{enumerate}

\section{Nonassociative nonholonomic vacuum Einstein equations}

\label{sec4} In this section, we address the point how to formulate star
nonholonomic deformations of vacuum Einstein equations with nontrivial
cosmological constant and effective sources determined by real R-flux
contributions on phase space and spacetime projections. For nonholonomic
configurations, there are involved both types of symmetric and nonsymmetric
components of a d-metric structure which can be connected respectively with
the symmetric and nonsymmetric components of corresponding Ricci d-tensor.
This allows us to perform such geometric constructions on total
nonassociative phase spaces and do consider decompositions into h- and
c(v)-components. We can project all geometric and physical objects, and
respective fundamental equations, on commutative nonholonomic spacetimes
which may contain nontrivial nonassociative term contributions. In result,
we obtain a geometric proof of vacuum nonholonomic Einstein equations
generalizing to nonassociative and nonholonomic configurations the ideas
from \cite{misner,vacaru18,bubuianu18a}:\ Physically important systems of
nonlinear PDEs (for dynamical interactions and flow evolution gravitational
and matter fields in various types models and theories) can be defined
following "pure" geometric principles considering metrics, covariant
derivative operators, and respective Ricci tensors and curvature scalars.
More advanced topics with generalized nonassociative variational calculus
star nonholonomic deformations of Einstein equations (in string gravity with
R-fluxes and/or for other type modifications) with nontrivial effective and
matter field sources are considered in our partner works.\footnote{%
We shall study also models with nonassociative (effective) energy-momentum
sources and/or in the context of nonholonomic variants of
Einstein-Eisenhart-Moffat \cite%
{einstein25,einstein45,eisenhart51,eisenhart52,
moffat79,moffat95,moffat95a,moffat96,clayton96,clayton96a,moffat00}, see
nonholonomic generalizations in \cite{vacaru08aa,vacaru08bb,vacaru08cc}, and
related Finsler-Lagrange-Hamilton theories \cite%
{vacaru05a,vacaru18,bubuianu18a}, which are elaborated for star deformations
and nontrivial R-flux contributions in our partner works.}

\subsection{Nonassociative vacuum Einstein equations with nonsymmetric
metrics}

To derive geometrically such equations we consider star nonholonomic
deformations of geometric data,%
\begin{eqnarray}
&&(\mathbf{\mathbf{\mathbf{\mathbf{\ ^{\shortparallel }}}}}\partial _{\alpha
},\mathbf{\mathbf{\mathbf{\mathbf{\ ^{\shortparallel }}}}}g_{\alpha \beta },%
\mathbf{\mathbf{\mathbf{\mathbf{\ ^{\shortparallel }}}}}\nabla _{\beta },%
\mathbf{\mathbf{\mathbf{\mathbf{\ _{\nabla }^{\shortparallel }}}}}R_{\mu \nu
})%
\begin{array}{c}
\mbox{ commutative phase }\Longrightarrow \\ 
\mbox{ canonical nonholonomic deforms}%
\end{array}%
(\mathbf{\mathbf{\mathbf{\mathbf{\ ^{\shortparallel }}}}e}_{\alpha },\mathbf{%
\mathbf{\mathbf{\mathbf{\ ^{\shortparallel }}}}g}_{\alpha \beta },\mathbf{%
\mathbf{\mathbf{\mathbf{\ ^{\shortparallel }}}}}\widehat{\mathbf{D}}_{\beta
},\mathbf{\mathbf{\mathbf{\mathbf{\ ^{\shortparallel }}}}}\widehat{\mathbf{R}%
}_{\mu \nu }) \Longrightarrow  \notag \\
&& \Longrightarrow 
\begin{array}{c}
\mbox{nonassociative }\Longrightarrow \\ 
\mbox{N-adapted star deforms }%
\end{array}%
(\mathbf{\mathbf{\mathbf{\mathbf{\ ^{\shortparallel }}}}e}_{\alpha },\
_{\star }^{\shortparallel }\mathfrak{g}_{\alpha \beta },\ ^{\shortparallel }%
\widehat{\mathbf{D}}_{\beta }^{\star }=\ ^{\shortparallel }\nabla _{\beta
}^{\star }+\ _{\star }^{\shortparallel }\widehat{\mathbf{Z}}_{\beta },%
\mathbf{\mathbf{\mathbf{\mathbf{\ ^{\shortparallel }}}}}\widehat{\mathbf{%
\mathbf{\mathbf{\mathbf{R}}}}}ic_{\mu \nu }^{\star }).  \label{nonholstardef}
\end{eqnarray}%
Any nonsymmetric d-metric $\ _{\star }^{\shortparallel }\mathfrak{g}_{\alpha
\beta }=\ _{\star }^{\shortparallel }\mathfrak{g}_{\alpha \beta }^{[0]}+\
_{\star }^{\shortparallel }\mathfrak{g}_{\alpha \beta }^{[1]}(\kappa )$ (\ref%
{aux37}) can be decomposed into the symmetric part, 
\begin{eqnarray}
\ _{\star }^{\shortparallel }\mathfrak{\check{g}}_{\alpha \beta } &=&\frac{1%
}{2}(\ _{\star }^{\shortparallel }\mathfrak{g}_{\alpha \beta }+\ _{\star
}^{\shortparallel }\mathfrak{g}_{\beta \alpha })=\ _{\star }^{\shortparallel
}\mathbf{g}_{\alpha \beta }-\frac{i\kappa }{2}\left( \overline{\mathcal{R}}%
_{\quad \beta }^{\tau \xi }\ \mathbf{^{\shortparallel }e}_{\xi }\ _{\star
}^{\shortparallel }\mathbf{g}_{\tau \alpha }+\overline{\mathcal{R}}_{\quad
\alpha }^{\tau \xi }\ \mathbf{^{\shortparallel }e}_{\xi }\ _{\star
}^{\shortparallel }\mathbf{g}_{\beta \tau }\right) =\ _{\star
}^{\shortparallel }\mathfrak{\check{g}}_{\alpha \beta }^{[0]}+\ _{\star
}^{\shortparallel }\mathfrak{\check{g}}_{\alpha \beta }^{[1]}(\kappa ),
\label{aux40b} \\
&&\mbox{ for }\ _{\star }^{\shortparallel }\mathfrak{\check{g}}_{\alpha
\beta }^{[0]}=\ _{\star }^{\shortparallel }\mathbf{g}_{\alpha \beta }%
\mbox{ and
}\ _{\star }^{\shortparallel }\mathfrak{\check{g}}_{\alpha \beta
}^{[1]}(\kappa )=-\frac{i\kappa }{2}\left( \overline{\mathcal{R}}_{\quad
\beta }^{\tau \xi }\ \mathbf{^{\shortparallel }e}_{\xi }\ _{\star
}^{\shortparallel }\mathbf{g}_{\tau \alpha }+\overline{\mathcal{R}}_{\quad
\alpha }^{\tau \xi }\ \mathbf{^{\shortparallel }e}_{\xi }\ _{\star
}^{\shortparallel }\mathbf{g}_{\beta \tau }\right) ,  \notag
\end{eqnarray}%
and the anti-symmetric part, 
\begin{eqnarray}
\ _{\star }^{\shortparallel }\mathfrak{a}_{\alpha \beta }&:= &\frac{1}{2}(\
_{\star }^{\shortparallel }\mathfrak{g}_{\alpha \beta }-\ _{\star
}^{\shortparallel }\mathfrak{g}_{\beta \alpha })=\frac{i\kappa }{2}\left( 
\overline{\mathcal{R}}_{\quad \beta }^{\tau \xi }\ \mathbf{^{\shortparallel
}e}_{\xi }\ _{\star }^{\shortparallel }\mathbf{g}_{\tau \alpha }-\overline{%
\mathcal{R}}_{\quad \alpha }^{\tau \xi }\ \mathbf{^{\shortparallel }e}_{\xi
}\ _{\star }^{\shortparallel }\mathbf{g}_{\beta \tau }\right)  \notag \\
&=&\ _{\star }^{\shortparallel }\mathfrak{a}_{\alpha \beta }^{[1]}(\kappa )=%
\frac{1}{2}(\ _{\star }^{\shortparallel }\mathfrak{g}_{\alpha \beta
}^{[1]}(\kappa )-\ _{\star }^{\shortparallel }\mathfrak{g}_{\beta \alpha
}^{[1]}(\kappa )),  \label{aux40a}
\end{eqnarray}%
where $\ _{\star }^{\shortparallel }\mathfrak{a}_{\alpha \beta }^{[0]}=0$
for nonassociative star deformations of commutative theories with symmetric
metrics,%
\begin{equation}
\ _{\star }^{\shortparallel }\mathfrak{g}_{\alpha \beta }=\ _{\star
}^{\shortparallel }\mathfrak{\check{g}}_{\alpha \beta }+\ _{\star
}^{\shortparallel }\mathfrak{a}_{\alpha \beta }.  \label{splitdmetr}
\end{equation}%
In a similar form, we can decompose into symmetric and nonsymmetric parts
the inverse d-metric (\ref{aux38a}),%
\begin{equation*}
\ _{\star }^{\shortparallel }\mathfrak{g}^{\alpha \beta }=\ _{\star
}^{\shortparallel }\mathfrak{\check{g}}^{\alpha \beta }+\ _{\star
}^{\shortparallel }\mathfrak{a}^{\alpha \beta },
\end{equation*}%
where 
\begin{eqnarray}
\ _{\star }^{\shortparallel }\mathfrak{\check{g}}^{\alpha \beta } &=&\frac{1%
}{2}(\ _{\star }^{\shortparallel }\mathfrak{\check{g}}^{\alpha \beta }+\
_{\star }^{\shortparallel }\mathfrak{\check{g}}^{\beta \alpha })  \notag \\
&=&\frac{1}{2}(\ _{\star }^{\shortparallel }\mathbf{g}^{\alpha \beta }+\
_{\star }^{\shortparallel }\mathbf{g}^{\beta \alpha })-\frac{i\kappa }{2}(\
_{\star }^{\shortparallel }\mathbf{g}^{\alpha \tau }\overline{\mathcal{R}}%
_{\quad \tau }^{\mu \nu }(\ \mathbf{^{\shortparallel }e}_{\mu }\ \ _{\star
}^{\shortparallel }\mathbf{g}_{\nu \varepsilon })\ \ _{\star
}^{\shortparallel }\mathbf{g}^{\varepsilon \beta }+\ _{\star
}^{\shortparallel }\mathbf{g}^{\beta \tau }\overline{\mathcal{R}}_{\quad
\tau }^{\mu \nu }(\ \mathbf{^{\shortparallel }e}_{\mu }\ \ _{\star
}^{\shortparallel }\mathbf{g}_{\nu \varepsilon })\ \ _{\star
}^{\shortparallel }\mathbf{g}^{\varepsilon \alpha })+O(\kappa ^{2})\   \notag
\\
&=&\frac{1}{2}(\ _{\ast }^{\shortparallel }\mathfrak{g}_{[0]}^{\alpha \beta
}+\ _{\star }^{\shortparallel }\mathfrak{g}_{[1]}^{\alpha \beta }(\kappa )+\
_{\ast }^{\shortparallel }\mathfrak{g}_{[0]}^{\beta \alpha }+\ _{\star
}^{\shortparallel }\mathfrak{g}_{[1]}^{\beta \alpha }(\kappa ))+O(\kappa
^{2})  \label{aux40} \\
&=&\ _{\ast }^{\shortparallel }\mathfrak{g}_{[0]}^{\alpha \beta }+\frac{1}{2}%
(\ _{\star }^{\shortparallel }\mathfrak{g}_{[1]}^{\alpha \beta }(\kappa )+\
_{\star }^{\shortparallel }\mathfrak{g}_{[1]}^{\beta \alpha }(\kappa
))+O(\kappa ^{2}),  \notag \\
&=&\ _{\ast }^{\shortparallel }\mathfrak{g}_{[0]}^{\alpha \beta }+\ _{\ast
}^{\shortparallel }\mathfrak{\check{g}}_{[1]}^{\alpha \beta }(\kappa
)+O(\kappa ^{2})=\ _{\ast }^{\shortparallel }\mathfrak{g}_{[00]}^{\alpha
\beta }+\ _{\ast }^{\shortparallel }\mathfrak{g}_{[01]}^{\alpha \beta
}(\hbar )+\ _{\ast }^{\shortparallel }\mathfrak{\check{g}}_{[10]}^{\alpha
\beta }(\kappa )+\ _{\ast }^{\shortparallel }\mathfrak{\check{g}}%
_{[11]}^{\alpha \beta }(\hbar \kappa )+O(\hbar ^{2},\kappa ^{2}),  \notag \\
&&\mbox{ for }\ _{\ast }^{\shortparallel }\mathfrak{\check{g}}_{[0]}^{\alpha
\beta }=\ _{\ast }^{\shortparallel }\mathfrak{g}_{[0]}^{\alpha \beta },\
_{\ast }^{\shortparallel }\mathfrak{\check{g}}_{[1]}^{\alpha \beta }(\kappa
)=\ _{\ast }^{\shortparallel }\mathfrak{g}_{[1]}^{\alpha \beta }(\kappa ),\
_{\ast }^{\shortparallel }\mathfrak{g}_{[0]}^{\alpha \beta }=\ _{\ast
}^{\shortparallel }\mathfrak{g}_{[0]}^{\beta \alpha },%
\mbox{ see
(\ref{aux38a})},  \notag
\end{eqnarray}%
and 
\begin{eqnarray*}
\ _{\star }^{\shortparallel }\mathfrak{a}^{\alpha \beta } &=&\frac{1}{2}(\
_{\star }^{\shortparallel }\mathfrak{\check{g}}^{\alpha \beta }-\ _{\star
}^{\shortparallel }\mathfrak{\check{g}}^{\beta \alpha }) \\
&=&\frac{1}{2}(\ _{\star }^{\shortparallel }\mathbf{g}^{\alpha \beta }-\
_{\star }^{\shortparallel }\mathbf{g}^{\beta \alpha })-\frac{i\kappa }{2}(\
_{\star }^{\shortparallel }\mathbf{g}^{\alpha \tau }\overline{\mathcal{R}}%
_{\quad \tau }^{\mu \nu }(\ \mathbf{^{\shortparallel }e}_{\mu }\ \ _{\star
}^{\shortparallel }\mathbf{g}_{\nu \varepsilon })\ \ _{\star
}^{\shortparallel }\mathbf{g}^{\varepsilon \beta }-\ _{\star
}^{\shortparallel }\mathbf{g}^{\beta \tau }\overline{\mathcal{R}}_{\quad
\tau }^{\mu \nu }(\ \mathbf{^{\shortparallel }e}_{\mu }\ \ _{\star
}^{\shortparallel }\mathbf{g}_{\nu \varepsilon })\ \ _{\star
}^{\shortparallel }\mathbf{g}^{\varepsilon \alpha })+O(\kappa ^{2})\  \\
&=&\frac{1}{2}(\ _{\ast }^{\shortparallel }\mathfrak{g}_{[0]}^{\alpha \beta
}+\ _{\star }^{\shortparallel }\mathfrak{g}_{[1]}^{\alpha \beta }(\kappa )-\
_{\ast }^{\shortparallel }\mathfrak{g}_{[0]}^{\beta \alpha }-\ _{\star
}^{\shortparallel }\mathfrak{g}_{[1]}^{\beta \alpha }(\kappa ))+O(\kappa
^{2}) \\
&=&\frac{1}{2}(\ _{\star }^{\shortparallel }\mathfrak{g}_{[1]}^{\alpha \beta
}(\kappa )-\ _{\star }^{\shortparallel }\mathfrak{g}_{[1]}^{\beta \alpha
}(\kappa ))+O(\kappa ^{2})=\ _{\star }^{\shortparallel }\mathfrak{a}%
_{[1]}^{\alpha \beta }(\kappa )+O(\kappa ^{2})=\ _{\ast }^{\shortparallel }%
\mathfrak{a}_{[10]}^{\alpha \beta }(\kappa )+\ _{\ast }^{\shortparallel }%
\mathfrak{a}_{[11]}^{\alpha \beta }(\hbar \kappa )+O(\hbar ^{2},\kappa ^{2}),
\\
&&\mbox{ for }\ _{\ast }^{\shortparallel }\mathfrak{g}_{[0]}^{\alpha \beta
}=\ _{\ast }^{\shortparallel }\mathfrak{g}_{[0]}^{\beta \alpha },\ _{\star
}^{\shortparallel }\mathfrak{a}_{[1]}^{\alpha \beta }(\kappa )=\frac{1}{2}(\
_{\star }^{\shortparallel }\mathfrak{g}_{[1]}^{\alpha \beta }(\kappa )-\
_{\star }^{\shortparallel }\mathfrak{g}_{[1]}^{\beta \alpha }(\kappa )),\
_{\ast }^{\shortparallel }\mathfrak{a}_{[00]}^{\alpha \beta }=0,\quad _{\ast
}^{\shortparallel }\mathfrak{a}_{[01]}^{\alpha \beta }(\hbar )=0,%
\mbox{ see
(\ref{aux38a})},
\end{eqnarray*}%
where the commutative and noncommutative d-metric and their inverses are
considered to be symmetric but with nontrivial antisymmetric contributions
from R-fluxes. The decompositions of the inverse nonsymmetric d-metric
allows us to define and compute the nonassociative nonholonomic canonical
Ricci scalar curvature%
\begin{eqnarray}
\ ^{\shortparallel }\widehat{\mathbf{R}}sc^{\star }&:= &\ _{\star
}^{\shortparallel }\mathfrak{g}^{\mu \nu }\mathbf{\mathbf{\mathbf{\mathbf{\
^{\shortparallel }}}}}\widehat{\mathbf{\mathbf{\mathbf{\mathbf{R}}}}}ic_{\mu
\nu }^{\star }=\left( \ _{\star }^{\shortparallel }\mathfrak{\check{g}}^{\mu
\nu }+\ _{\star }^{\shortparallel }\mathfrak{a}^{\mu \nu }\right) \left( 
\mathbf{\mathbf{\mathbf{\mathbf{\ ^{\shortparallel }}}}}\widehat{\mathbf{%
\mathbf{\mathbf{\mathbf{R}}}}}ic_{(\mu \nu )}^{\star }+\mathbf{\mathbf{%
\mathbf{\mathbf{\ ^{\shortparallel }}}}}\widehat{\mathbf{\mathbf{\mathbf{%
\mathbf{R}}}}}ic_{[\mu \nu ]}^{\star }\right) =\mathbf{\mathbf{\mathbf{%
\mathbf{\ ^{\shortparallel }}}}}\widehat{\mathbf{\mathbf{\mathbf{\mathbf{R}}}%
}}ss^{\star }+\mathbf{\mathbf{\mathbf{\mathbf{\ ^{\shortparallel }}}}}%
\widehat{\mathbf{\mathbf{\mathbf{\mathbf{R}}}}}sa^{\star },  \notag \\
&&\mbox{ where }\mathbf{\mathbf{\mathbf{\mathbf{\ ^{\shortparallel }}}}}%
\widehat{\mathbf{\mathbf{\mathbf{\mathbf{R}}}}}ss^{\star }=:\ _{\star
}^{\shortparallel }\mathfrak{\check{g}}^{\mu \nu }\mathbf{\mathbf{\mathbf{%
\mathbf{\ ^{\shortparallel }}}}}\widehat{\mathbf{\mathbf{\mathbf{\mathbf{R}}}%
}}ic_{(\mu \nu )}^{\star }\mbox{ and }\mathbf{\mathbf{\mathbf{\mathbf{\
^{\shortparallel }}}}}\widehat{\mathbf{\mathbf{\mathbf{\mathbf{R}}}}}%
sa^{\star }:=\ _{\star }^{\shortparallel }\mathfrak{a}^{\mu \nu }\mathbf{%
\mathbf{\mathbf{\mathbf{\ ^{\shortparallel }}}}}\widehat{\mathbf{\mathbf{%
\mathbf{\mathbf{R}}}}}ic_{[\mu \nu ]}^{\star }.  \label{ricciscsymnonsym}
\end{eqnarray}%
In these formulas, it is considered a respective symmetric $\left(
...\right) $ and anti-symmetric $\left[ ...\right] $ decomposition of the
canonical Ricci d-tensor, 
\begin{equation*}
\mathbf{\mathbf{\mathbf{\mathbf{\ ^{\shortparallel }}}}}\widehat{\mathbf{%
\mathbf{\mathbf{\mathbf{R}}}}}ic_{\mu \nu }^{\star }=\mathbf{\mathbf{\mathbf{%
\mathbf{\ ^{\shortparallel }}}}}\widehat{\mathbf{\mathbf{\mathbf{\mathbf{R}}}%
}}ic_{(\mu \nu )}^{\star }+\mathbf{\mathbf{\mathbf{\mathbf{\
^{\shortparallel }}}}}\widehat{\mathbf{\mathbf{\mathbf{\mathbf{R}}}}}%
ic_{[\mu \nu ]}^{\star },
\end{equation*}%
where symmetrization and anti-symmetrization are performed using the
multiple $1/2.$

The commutative vacuum Einstein equations on phase space with cosmological
constant $^{\shortparallel }\lambda$ can be postulated in such equivalent
8-d forms,%
\begin{eqnarray}
\ _{\nabla }^{\shortparallel }E_{\mu \nu } &:=&\ _{\nabla }^{\shortparallel
}Ric_{\mu \nu }-\frac{1}{2}\mathbf{\mathbf{\mathbf{\mathbf{\
^{\shortparallel }}}}}g_{\mu \nu }\ \ _{\nabla }^{\shortparallel }Rsc_{\mu
\nu }=\mathbf{\mathbf{\mathbf{\mathbf{\ ^{\shortparallel }}}}}\lambda 
\mathbf{\mathbf{\mathbf{\mathbf{\ ^{\shortparallel }}}}}g_{\mu \nu },%
\mbox{
or }  \notag \\
\ ^{\shortparallel }\widehat{\mathbf{E}}_{\mu \nu } &:=& \mathbf{\mathbf{%
\mathbf{\mathbf{\ ^{\shortparallel }}}}}\widehat{\mathbf{\mathbf{\mathbf{%
\mathbf{R}}}}}ic_{\mu \nu }-\frac{1}{2}\mathbf{\mathbf{\mathbf{\mathbf{\
^{\shortparallel }}}}g}_{\mu \nu }\mathbf{\mathbf{\mathbf{\mathbf{\
^{\shortparallel }}}}}\widehat{\mathbf{\mathbf{\mathbf{\mathbf{R}}}}}sc=%
\mathbf{\mathbf{\mathbf{\mathbf{\ ^{\shortparallel }}}}}\lambda \mathbf{%
\mathbf{\mathbf{\mathbf{\ ^{\shortparallel }}}}g}_{\mu \nu },
\label{commutvacdeinst}
\end{eqnarray}%
see formulas (\ref{candricci}) and (\ref{candriccidist}), where $\ _{\nabla
}^{\shortparallel }E_{\mu \nu }$ and $\ ^{\shortparallel }\widehat{\mathbf{E}%
}_{\mu \nu }$ are the respective Einstein tensor and d-tensor. We note that
using the Bianchi identities for the LC-connection we obtain%
\begin{equation*}
\mathbf{\mathbf{\mathbf{\mathbf{\ ^{\shortparallel }}}}}\nabla ^{\mu }\ \
_{\nabla }^{\shortparallel }E_{\mu \nu }=0,\mbox{ but }\mathbf{\mathbf{%
\mathbf{\mathbf{\ ^{\shortparallel }}}}}\widehat{\mathbf{D}}^{\mu }(\mathbf{%
\mathbf{\mathbf{\mathbf{\ ^{\shortparallel }}}}}\widehat{\mathbf{E}}_{\mu
\nu })=\mathbf{\mathbf{\mathbf{\mathbf{\ ^{\shortparallel }}}}}\widehat{%
\mathbf{S}}_{\nu }[\ ^{\shortparallel }\widehat{\mathbf{Z}}_{\beta }]\neq 0,
\end{equation*}%
where $\ ^{\shortparallel }\widehat{\mathbf{S}}_{\nu }$ is computed using
the canonical distortion relations (\ref{candistr}). For nonholonomic
structures in commutative classical geometrical mechanics and modified
gravity theories, in general, $\ ^{\shortparallel }\widehat{\mathbf{S}}%
_{\nu}\neq 0$ which is a consequence of non-integrability conditions.

Applying star nonholonomic deformations (\ref{nonholstardef}) to (\ref%
{commutvacdeinst}), we can define and compute N-adapted components of
nonassociative vacuum Einstein equations, 
\begin{equation}
\ ^{\shortparallel }\widehat{\mathbf{\mathbf{\mathbf{\mathbf{R}}}}}%
ic_{\alpha \beta }^{\star }-\frac{1}{2}\ _{\star }^{\shortparallel }%
\mathfrak{g}_{\alpha \beta }\mathbf{\mathbf{\mathbf{\mathbf{\
^{\shortparallel }}}}}\widehat{\mathbf{\mathbf{\mathbf{\mathbf{R}}}}}%
sc^{\star }=\mathbf{\mathbf{\mathbf{\mathbf{\ ^{\shortparallel }}}}}\lambda 
\mathbf{\mathbf{\mathbf{\mathbf{\ }}}}\ _{\star }^{\shortparallel }\mathfrak{%
g}_{\alpha \beta }.  \label{nonassocdeinst1}
\end{equation}%
We can write these equations emphasizing explicitly the symmetric and
nonsymmetric components of d-metric (\ref{splitdmetr}),%
\begin{eqnarray}
\ _{\star }^{\shortparallel }\mathfrak{\check{g}}_{\mu \nu }(\mathbf{\mathbf{%
\mathbf{\mathbf{\ ^{\shortparallel }}}}}\lambda +\frac{1}{2}\mathbf{\mathbf{%
\mathbf{\mathbf{\ ^{\shortparallel }}}}}\widehat{\mathbf{\mathbf{\mathbf{%
\mathbf{R}}}}}sc^{\star }) &=&\mathbf{\mathbf{\mathbf{\mathbf{\
^{\shortparallel }}}}}\widehat{\mathbf{\mathbf{\mathbf{\mathbf{R}}}}}%
ic_{(\mu \nu )}^{\star }\mbox{ and }  \label{nonassocdeinst2a} \\
\ _{\star }^{\shortparallel }\mathfrak{a}_{\mu \nu }(\mathbf{\mathbf{\mathbf{%
\mathbf{\ ^{\shortparallel }}}}}\lambda +\frac{1}{2}\mathbf{\mathbf{\mathbf{%
\mathbf{\ ^{\shortparallel }}}}}\widehat{\mathbf{\mathbf{\mathbf{\mathbf{R}}}%
}}sc^{\star }) &=&\mathbf{\mathbf{\mathbf{\mathbf{\ ^{\shortparallel }}}}}%
\widehat{\mathbf{\mathbf{\mathbf{\mathbf{R}}}}}ic_{[\mu \nu ]}^{\star }.
\label{nonassocdeinst2b}
\end{eqnarray}%
From solutions of such systems of nonlinear PDEs, we can extract
LC-configurations imposing at the end the conditions (\ref{lccondnonass})
for $\ ^{\shortparallel }\widehat{\mathbf{D}}_{\mid \ ^{\shortparallel }%
\widehat{\mathbf{T}}=0}^{\star }=\ ^{\shortparallel }\nabla ^{\star }.$ The
equations (\ref{nonassocdeinst2a}) and/or (\ref{nonassocdeinst2b}) can be
re-written equivalently for $\ ^{\shortparallel }\nabla ^{\star }$ (when the
Ricci d-tensor is symmetric) using canonical star distortions (\ref%
{candistrnas}) which transforms such equations in a more coupled
nonholonomic and nonlinear system.

The systems of nonlinear PDEs (\ref{nonassocdeinst1}) and, respectively, (%
\ref{nonassocdeinst2a}) and (\ref{nonassocdeinst2b}) are self-consistent
because we can always start with a solution of commutative vacuum Einstein
equations (in general, such a solution is generic off-diagonal) and compute
any term for nonassociative models of gravity as N-adapted star deformations
for a symmetric $\ ^{\shortparallel }\mathbf{g}_{\alpha \beta }$ and
respective N-adapted configurations. In a more general approach, we can
prescribe an ansatz for $\ _{\star}^{\shortparallel }\mathfrak{\check{g}}%
_{\alpha \beta }$ and search for a nonsymmetric star deformation $\ _{\star
}^{\shortparallel }\mathfrak{a}_{\mu \nu }$ for an N-adapted $\
^{\shortparallel }\mathbf{e}_{\mu }$ which is subjected to the condition to
determine a solution of vacuum nonassociative gravitational field equations
with geometric objects and nontrivial sources induced by a R-flux.

If $\ ^{\shortparallel }\lambda =0,$ we get systems of nonlinear PDEs of
type 
\begin{equation}
\mathbf{\mathbf{\mathbf{\mathbf{\ ^{\shortparallel }}}}}\widehat{\mathbf{%
\mathbf{\mathbf{\mathbf{R}}}}}ic_{\mu \nu }^{\star }=\mathbf{\mathbf{\mathbf{%
\mathbf{\ ^{\shortparallel }}}}}\widehat{\mathbf{\mathbf{\mathbf{\mathbf{R}}}%
}}ic_{(\mu \nu )}^{\star }+\mathbf{\mathbf{\mathbf{\mathbf{\
^{\shortparallel }}}}}\widehat{\mathbf{\mathbf{\mathbf{\mathbf{R}}}}}%
ic_{[\mu \nu ]}^{\star }=0  \label{nonassocdeinst3}
\end{equation}%
which for LC-configurations and ansatz of metrics depending only on
spacetime coordinates transform into nonassociative vacuum Einstein
equations considered in \cite{aschieri17}. In our partner works (see, for
instance, \cite{partner02}), we prove that the equations (\ref%
{nonassocdeinst3}) (and above nonassociative gravitational equations with
cosmological constant) can be decoupled and integrated in very general forms
with generic off-diagonal symmetric and nonsymmetric metrics and generalized
connections, in particular, for LC-configurations.

In a similar form, we can provide geometric proofs for canonical
d-connections $\ _{\flat }^{\shortparallel }\widehat{\mathbf{D}}=\
^{\shortparallel }\nabla +\mathbf{\mathbf{\mathbf{\ _{\flat
}^{\shortparallel }}}}\widehat{\mathbf{Z}}$ and corresponding distortions of
type (\ref{distdtens}), for nonassociative models not considering quasi-Hopf
structures, which for LC-configurations and respective ansatz of phase space
metrics result in nonassociative vacuum Einstein equations considered in 
\cite{blumenhagen16}. We omit such constructions in this work.

\subsection{Nonassociative vacuum gravitational equations in hc-variables}

The N-connection splitting defines a h- and c-decomposition of (\ref%
{nonassocdeinst1}); (\ref{nonassocdeinst2a}) and (\ref{nonassocdeinst2b});
and (\ref{nonassocdeinst3}) using formulas for the nonassociative canonical
Ricci d-tensor (\ref{hcnonassocrcan}). We have 
\begin{eqnarray*}
\ ^{\shortparallel }\widehat{\mathbf{\mathbf{\mathbf{\mathbf{R}}}}}ic_{(\mu
\nu )}^{\star } &=&\{\ \frac{1}{2}(\ ^{\shortparallel }\widehat{R}_{\ \star
hj}+\ ^{\shortparallel }\widehat{R}_{\ \star jh}),\frac{1}{2}(\
^{\shortparallel }\widehat{P}_{\star j}^{\quad a}+\ ^{\shortparallel }%
\widehat{P}_{\star \ j}^{a\quad }),\frac{1}{2}(\ ^{\shortparallel }\widehat{P%
}_{\star k}^{\quad b}+\ ^{\shortparallel }\widehat{P}_{\star \ k}^{b\quad }),%
\frac{1}{2}(\ ^{\shortparallel }\widehat{S}_{\star }^{bc\quad }+\
^{\shortparallel }\widehat{S}_{\star}^{cb})\} \mbox{ and } \\
\mathbf{\mathbf{\mathbf{\mathbf{\ ^{\shortparallel }}}}}\widehat{\mathbf{%
\mathbf{\mathbf{\mathbf{R}}}}}ic_{[\mu \nu ]}^{\star } &=&\{\frac{1}{2}(\
^{\shortparallel }\widehat{R}_{\ \star hj}-\ ^{\shortparallel }\widehat{R}%
_{\ \star jh}),\frac{1}{2}(\ ^{\shortparallel }\widehat{P}_{\star j}^{\quad
a}-\ ^{\shortparallel }\widehat{P}_{\star \ j}^{a\quad }),\frac{1}{2}(\
^{\shortparallel }\widehat{P}_{\star k}^{\quad b}-\ ^{\shortparallel }%
\widehat{P}_{\star \ k}^{b\quad }),\frac{1}{2}(\ ^{\shortparallel} \widehat{S%
}_{\star}^{bc}-\ ^{\shortparallel }\widehat{S}_{\star}^{cb})\},
\end{eqnarray*}%
where explicit computation of N-adapted coefficients involved contacting of
indices in (\ref{nadriemhopfcan}). All such values are computed for the
nonassociative canonical d-connection with $\ ^{\shortparallel }\widehat{%
\Gamma }_{\star \alpha \gamma }^{\mu }$ determined by the star d-metric $\
_{\star }^{\shortparallel }\mathbf{g}_{\alpha \beta } =(\
_{\star}^{\shortparallel }g_{jr}, \ _{\star }^{\shortparallel }g^{ab})$ (\ref%
{dmss}) via formulas (\ref{0canconnonas}), (\ref{aux311}) and (\ref{aux311a}%
).

Following formulas (\ref{aux40b}), with $\ _{\star }^{\shortparallel }%
\mathfrak{\check{g}}_{\alpha \beta }=\ _{\star }^{\shortparallel }\mathfrak{%
\check{g}}_{\alpha \beta }^{[0]}+\ _{\star }^{\shortparallel }\mathfrak{%
\check{g}}_{\alpha \beta }^{[1]}(\kappa ),$ and (\ref{aux40a}), we find 
\begin{equation}
\ _{\star }^{\shortparallel }\mathfrak{a}_{\alpha \beta }:=\left( 0,0,0,\
_{\star }^{\shortparallel }\mathfrak{a}^{ab}=\frac{i\kappa }{2}(\overline{%
\mathcal{R}}_{c\quad }^{\ n+ka}\ \mathbf{^{\shortparallel }e}_{k}\ _{\star
}^{\shortparallel }\mathbf{g}^{cb}-\overline{\mathcal{R}}_{c\quad }^{\
n+kb}\ \mathbf{^{\shortparallel }e}_{k}\ _{\star }^{\shortparallel }\mathbf{g%
}^{ca})=\ _{\star }^{\shortparallel }\mathfrak{a}_{[1]}^{ab}(\kappa )=\frac{1%
}{2}(\ _{\star }^{\shortparallel }\mathfrak{g}_{[1]}^{ab}-\ _{\star
}^{\shortparallel }\mathfrak{g}_{[1]}^{ba})(\kappa )\right) .  \notag
\end{equation}%
The canonical Ricci scalar is computed using block $(4\times 4)+(4\times 4)$
parametrization of respective matrices, $\ _{\star }^{\shortparallel }%
\mathfrak{\check{g}}^{\mu \nu }=(\ _{\star }^{\shortparallel }\mathfrak{%
\check{g}}^{ij},0,0,\ _{\star }^{\shortparallel }\mathfrak{\check{g}}_{ab})$
and $\ _{\star }^{\shortparallel }\mathfrak{a}^{\mu \nu }=(0,0,0,\ _{\star
}^{\shortparallel }\mathfrak{a}_{cb}),$%
\begin{eqnarray*}
\ \mathbf{\mathbf{\mathbf{\mathbf{^{\shortparallel }}}}}\widehat{\mathbf{%
\mathbf{\mathbf{\mathbf{R}}}}}ss^{\star } & =: &\ _{\star }^{\shortparallel }%
\mathfrak{\check{g}}^{\mu \nu }\mathbf{\mathbf{\mathbf{\mathbf{\
^{\shortparallel }}}}}\widehat{\mathbf{\mathbf{\mathbf{\mathbf{R}}}}}%
ic_{(\mu \nu )}^{\star }=\mathbf{\mathbf{\mathbf{\mathbf{^{\shortparallel }}}%
}}\widehat{\mathbf{\mathbf{\mathbf{\mathbf{R}}}}}sh^{\star }+\ \mathbf{%
\mathbf{\mathbf{\mathbf{^{\shortparallel }}}}}\widehat{\mathbf{\mathbf{%
\mathbf{\mathbf{R}}}}}sc^{\star }\mbox{ and }\mathbf{\mathbf{\mathbf{\mathbf{%
\ ^{\shortparallel }}}}}\widehat{\mathbf{\mathbf{\mathbf{\mathbf{R}}}}}%
sa^{\star }:=\ _{\star }^{\shortparallel }\mathfrak{a}_{cb}\mathbf{\mathbf{%
\mathbf{\mathbf{\ ^{\shortparallel }}}}}\widehat{\mathbf{\mathbf{\mathbf{%
\mathbf{R}}}}}ic^{\star \lbrack cb]},\mbox{ for } \\
&& \ ^{\shortparallel }\widehat{\mathbf{\mathbf{\mathbf{\mathbf{R}}}}}%
sh^{\star } = \ _{\star }^{\shortparallel }\mathfrak{\check{g}}^{ij}\mathbf{%
\mathbf{\mathbf{\mathbf{\ ^{\shortparallel }}}}}\widehat{\mathbf{\mathbf{%
\mathbf{\mathbf{R}}}}}ic_{ij}^{\star },\ \mathbf{\mathbf{\mathbf{\mathbf{%
^{\shortparallel }}}}}\widehat{\mathbf{\mathbf{\mathbf{\mathbf{R}}}}}%
sc^{\star }=\ _{\star }^{\shortparallel }\mathfrak{\check{g}}_{ab}\mathbf{%
\mathbf{\mathbf{\mathbf{\ ^{\shortparallel }}}}}\widehat{\mathbf{\mathbf{%
\mathbf{\mathbf{R}}}}}ic^{\star ab}.
\end{eqnarray*}%
For such N-adapted nonassociative vacuum configurations, the system of
nonlinear PDEs (\ref{nonassocdeinst2a}) and (\ref{nonassocdeinst2b}) splits
into such h- and c-components 
\begin{eqnarray*}
\ _{\star }^{\shortparallel }\mathfrak{\check{g}}_{ij}(\mathbf{\mathbf{%
\mathbf{\mathbf{\ }}}}2\mathbf{\mathbf{\mathbf{\mathbf{^{\shortparallel }}}}}%
\lambda +\mathbf{\mathbf{\mathbf{\mathbf{^{\shortparallel }}}}}\widehat{%
\mathbf{\mathbf{\mathbf{\mathbf{R}}}}}sh^{\star }+\ \mathbf{\mathbf{\mathbf{%
\mathbf{^{\shortparallel }}}}}\widehat{\mathbf{\mathbf{\mathbf{\mathbf{R}}}}}%
sc^{\star }+\mathbf{\mathbf{\mathbf{\mathbf{\ ^{\shortparallel }}}}}\widehat{%
\mathbf{\mathbf{\mathbf{\mathbf{R}}}}}sa^{\star }) &=&2\ ^{\shortparallel }%
\widehat{R}_{\ \star ij}=2\ ^{\shortparallel }\widehat{R}_{\ \star ji}, \\
\ _{\star }^{\shortparallel }\mathfrak{\check{g}}^{bc}(\mathbf{\mathbf{%
\mathbf{\mathbf{\ }}}}2\mathbf{\mathbf{\mathbf{\mathbf{^{\shortparallel }}}}}%
\lambda +\mathbf{\mathbf{\mathbf{\mathbf{^{\shortparallel }}}}}\widehat{%
\mathbf{\mathbf{\mathbf{\mathbf{R}}}}}sh^{\star }+\ \mathbf{\mathbf{\mathbf{%
\mathbf{^{\shortparallel }}}}}\widehat{\mathbf{\mathbf{\mathbf{\mathbf{R}}}}}%
sc^{\star }+\mathbf{\mathbf{\mathbf{\mathbf{\ ^{\shortparallel }}}}}\widehat{%
\mathbf{\mathbf{\mathbf{\mathbf{R}}}}}sa^{\star }) &=&\ ^{\shortparallel }%
\widehat{S}_{\star \ }^{bc\quad }+\ ^{\shortparallel }\widehat{S}_{\star \
}^{cb\quad },\ ^{\shortparallel }\widehat{P}_{\star j}^{\quad a}=\
^{\shortparallel }\widehat{P}_{\star \ j}^{a\quad }=0, \\
\ _{\star }^{\shortparallel }\mathfrak{a}^{bc}(\mathbf{\mathbf{\mathbf{%
\mathbf{\ }}}}2\mathbf{\mathbf{\mathbf{\mathbf{^{\shortparallel }}}}}\lambda
+\mathbf{\mathbf{\mathbf{\mathbf{^{\shortparallel }}}}}\widehat{\mathbf{%
\mathbf{\mathbf{\mathbf{R}}}}}sh^{\star }+\ \mathbf{\mathbf{\mathbf{\mathbf{%
^{\shortparallel }}}}}\widehat{\mathbf{\mathbf{\mathbf{\mathbf{R}}}}}%
sc^{\star }+\mathbf{\mathbf{\mathbf{\mathbf{\ ^{\shortparallel }}}}}\widehat{%
\mathbf{\mathbf{\mathbf{\mathbf{R}}}}}sa^{\star }) &=&\ ^{\shortparallel }%
\widehat{S}_{\star \ }^{bc\quad }-\ ^{\shortparallel }\widehat{S}_{\star \
}^{cb\quad }.
\end{eqnarray*}%
Such a decoupling property is very important for extending the AFCDM and
constructing exact and parametric solutions in nonassociative and (non)
commutative gravity.

\subsection{Parametric decomposition of nonassociative vacuum gravitational
equations}

Using decompositions (\ref{hcnonassocrcana}), we can decompose above
formulas on small parameters $\hbar $ and $\kappa ,$%
\begin{eqnarray*}
\mathbf{\mathbf{\mathbf{\mathbf{\ ^{\shortparallel }}}}}\widehat{\mathbf{%
\mathbf{\mathbf{\mathbf{R}}}}}ic_{(\mu \nu )}^{\star } &=&\mathbf{\mathbf{%
\mathbf{\mathbf{\ \mathbf{\mathbf{\mathbf{\mathbf{\ }}}}}}}}\
_{[00]}^{\shortparallel }\widehat{\mathbf{\mathbf{\mathbf{\mathbf{R}}}}}%
ic_{(\mu \nu )}^{\star }+\mathbf{\mathbf{\mathbf{\mathbf{\ \ }}}}%
_{[01]}^{\shortparallel }\widehat{\mathbf{\mathbf{\mathbf{\mathbf{R}}}}}%
ic_{(\mu \nu )}^{\star }(\hbar )+\mathbf{\mathbf{\mathbf{\mathbf{\ }}}}%
_{[10]}^{\shortparallel }\widehat{\mathbf{\mathbf{\mathbf{\mathbf{R}}}}}%
ic_{(\mu \nu )}^{\star }(\kappa )+\mathbf{\mathbf{\mathbf{\mathbf{\ }}}}%
_{[11]}^{\shortparallel }\widehat{\mathbf{\mathbf{\mathbf{\mathbf{R}}}}}%
ic_{(\mu \nu )}^{\star }(\hbar \kappa )+O(\hbar ^{2},\kappa ^{2},...);%
\mbox{
and } \\
\mathbf{\mathbf{\mathbf{\mathbf{\ ^{\shortparallel }}}}}\widehat{\mathbf{%
\mathbf{\mathbf{\mathbf{R}}}}}ic_{[\mu \nu ]}^{\star } &=&\mathbf{\mathbf{%
\mathbf{\mathbf{\ \mathbf{\mathbf{\mathbf{\mathbf{\ }}}}}}}}\
_{[00]}^{\shortparallel }\widehat{\mathbf{\mathbf{\mathbf{\mathbf{R}}}}}%
ic_{[\mu \nu ]}^{\star }+\mathbf{\mathbf{\mathbf{\mathbf{\ \ }}}}%
_{[01]}^{\shortparallel }\widehat{\mathbf{\mathbf{\mathbf{\mathbf{R}}}}}%
ic_{[\mu \nu ]}^{\star }(\hbar )+\mathbf{\mathbf{\mathbf{\mathbf{\ }}}}%
_{[10]}^{\shortparallel }\widehat{\mathbf{\mathbf{\mathbf{\mathbf{R}}}}}%
ic_{[\mu \nu ]}^{\star }(\kappa )+\mathbf{\mathbf{\mathbf{\mathbf{\ }}}}%
_{[11]}^{\shortparallel }\widehat{\mathbf{\mathbf{\mathbf{\mathbf{R}}}}}%
ic_{[\mu \nu ]}^{\star }(\hbar \kappa )+O(\hbar ^{2},\kappa ^{2},...),
\end{eqnarray*}%
where, for $\ \ _{[00]}^{\shortparallel }\widehat{R}_{\ \star hj}=\ \
_{[00]}^{\shortparallel }\widehat{R}_{\ \star jh}$ and $\
_{[00]}^{\shortparallel }\widehat{S}_{\star \ }^{bc\quad }=\
_{[00]}^{\shortparallel }\widehat{S}_{\star \ }^{cb\quad },$ 
\begin{eqnarray*}
\ _{[00]}^{\shortparallel }\widehat{\mathbf{\mathbf{\mathbf{\mathbf{R}}}}}%
ic_{(\mu \nu )}^{\star } &=&\{\ _{[00]}^{\shortparallel }\widehat{R}_{\
\star hj},\frac{1}{2}(\ _{[00]}^{\shortparallel }\widehat{P}_{\star
j}^{\quad a}+\ _{[00]}^{\shortparallel }\widehat{P}_{\star \ j}^{a\quad }),%
\frac{1}{2}(\ _{[00]}^{\shortparallel }\widehat{P}_{\star k}^{\quad b}+\
_{[00]}^{\shortparallel }\widehat{P}_{\star \ k}^{b\quad }),\
_{[00]}^{\shortparallel }\widehat{S}_{\star \ }^{bc\quad }\}, \\
\ _{[01]}^{\shortparallel }\widehat{\mathbf{\mathbf{\mathbf{\mathbf{R}}}}}%
ic_{(\mu \nu )}^{\star } &=&\mathbf{\{}\frac{1}{2}(\ \
_{[01]}^{\shortparallel }\widehat{R}_{\ \star hj}+\ \
_{[01]}^{\shortparallel }\widehat{R}_{\ \star jh}),\frac{1}{2}(\
_{[01]}^{\shortparallel }\widehat{P}_{\star j}^{\quad a}+\
_{[01]}^{\shortparallel }\widehat{P}_{\star \ j}^{a\quad }), \\
&&\frac{1}{2}(\ _{[01]}^{\shortparallel }\widehat{P}_{\star k}^{\quad b}+\
_{[01]}^{\shortparallel }\widehat{P}_{\star \ k}^{b\quad }),\frac{1}{2}(\
_{[01]}^{\shortparallel }\widehat{S}_{\star \ }^{bc\quad }+\
_{[01]}^{\shortparallel }\widehat{S}_{\star \ }^{cb\quad })\}, \\
&&\mbox{ and similarly for }\lbrack 1,0]\mbox{ and }\lbrack 1,1];
\end{eqnarray*}%
and%
\begin{eqnarray*}
\ _{[00]}^{\shortparallel }\widehat{\mathbf{\mathbf{\mathbf{\mathbf{R}}}}}%
ic_{[\mu \nu ]}^{\star } &=&\{\ \ 0,\frac{1}{2}(\ _{[00]}^{\shortparallel }%
\widehat{P}_{\star j}^{\quad a}-\ _{[00]}^{\shortparallel }\widehat{P}%
_{\star \ j}^{a\quad }),-\frac{1}{2}(\ _{[00]}^{\shortparallel }\widehat{P}%
_{\star j}^{\quad a}-\ _{[00]}^{\shortparallel }\widehat{P}_{\star \
j}^{a\quad }),\ 0\}, \\
\ _{[01]}^{\shortparallel }\widehat{\mathbf{\mathbf{\mathbf{\mathbf{R}}}}}%
ic_{[\mu \nu ]}^{\star } &=&\{\ \frac{1}{2}(\ \ _{[01]}^{\shortparallel }%
\widehat{R}_{\ \star hj}-\ \ _{[01]}^{\shortparallel }\widehat{R}_{\ \star
jh}),\frac{1}{2}(\ _{[01]}^{\shortparallel }\widehat{P}_{\star j}^{\quad
a}-\ _{[01]}^{\shortparallel }\widehat{P}_{\star \ j}^{a\quad }), \\
&&-\frac{1}{2}(\ _{[01]}^{\shortparallel }\widehat{P}_{\star j}^{\quad a}-\
_{[01]}^{\shortparallel }\widehat{P}_{\star \ j}^{a\quad }),\frac{1}{2}(\
_{[01]}^{\shortparallel }\widehat{S}_{\star \ }^{bc\quad }-\
_{[01]}^{\shortparallel }\widehat{S}_{\star \ }^{cb\quad })\}, \\
&&\mbox{ and similarly for }\lbrack 1,0]\mbox{ and }\lbrack 1,1].
\end{eqnarray*}%
Further computations are possible for parametric decomposition of the
nonassociative canonical Ricci d-tensor following formulas (\ref{ric50})-(%
\ref{ric53}).

Considering such formulas and (\ref{aux40}), we compute the parametric
decomposition on $\hbar $ and $\kappa $ of the nonassociative Ricci scalar (%
\ref{ricciscsymnonsym}),%
\begin{equation*}
\ ^{\shortparallel }\widehat{\mathbf{\mathbf{\mathbf{\mathbf{R}}}}}sc^{\star
}:=\ _{[00]}^{\shortparallel }\widehat{\mathbf{\mathbf{\mathbf{\mathbf{R}}}}}%
sc^{\star }+\ _{[01]}^{\shortparallel }\widehat{\mathbf{\mathbf{\mathbf{%
\mathbf{R}}}}}sc^{\star }(\hbar )+\mathbf{\mathbf{\mathbf{\mathbf{\ }}}}\
_{[10]}^{\shortparallel }\widehat{\mathbf{\mathbf{\mathbf{\mathbf{R}}}}}%
sc^{\star }(\kappa )+\mathbf{\mathbf{\mathbf{\mathbf{\ }}}}\
_{[11]}^{\shortparallel }\widehat{\mathbf{\mathbf{\mathbf{\mathbf{R}}}}}%
sc^{\star }(\hbar \kappa )+O(\hbar ^{2},\kappa ^{2},...),
\end{equation*}%
where%
\begin{eqnarray*}
\ _{[00]}^{\shortparallel }\widehat{\mathbf{\mathbf{\mathbf{\mathbf{R}}}}}%
sc^{\star } &=&\mathbf{\mathbf{\mathbf{\mathbf{\mathbf{\mathbf{\mathbf{%
\mathbf{\ }}}}}}}}\ _{\ast }^{\shortparallel }\mathfrak{g}_{[00]}^{\mu \nu
}\ _{[00]}^{\shortparallel }\widehat{\mathbf{\mathbf{\mathbf{\mathbf{R}}}}}%
ic_{(\mu \nu )}^{\star }, \\
\mathbf{\mathbf{\mathbf{\mathbf{\ }}}}\ _{[01]}^{\shortparallel }\widehat{%
\mathbf{\mathbf{\mathbf{\mathbf{R}}}}}sc^{\star } &=&\ _{\ast
}^{\shortparallel }\mathfrak{g}_{[00]}^{\mu \nu }\mathbf{\mathbf{\mathbf{%
\mathbf{\ \ }}}}_{[01]}^{\shortparallel }\widehat{\mathbf{\mathbf{\mathbf{%
\mathbf{R}}}}}ic_{(\mu \nu )}^{\star }+\ _{\ast }^{\shortparallel }\mathfrak{%
g}_{[01]}^{\mu \nu }\ _{[00]}^{\shortparallel }\widehat{\mathbf{\mathbf{%
\mathbf{\mathbf{R}}}}}ic_{(\mu \nu )}^{\star }, \\
\mathbf{\mathbf{\mathbf{\mathbf{\ }}}}\ _{[10]}^{\shortparallel }\widehat{%
\mathbf{\mathbf{\mathbf{\mathbf{R}}}}}sc^{\star } &=&\ _{\ast
}^{\shortparallel }\mathfrak{g}_{[00]}^{\mu \nu }\mathbf{\mathbf{\mathbf{%
\mathbf{\ }}}}_{[10]}^{\shortparallel }\widehat{\mathbf{\mathbf{\mathbf{%
\mathbf{R}}}}}ic_{(\mu \nu )}^{\star }+\ _{\ast }^{\shortparallel }\mathfrak{%
\check{g}}_{[10]}^{\mu \nu }\ _{[00]}^{\shortparallel }\widehat{\mathbf{%
\mathbf{\mathbf{\mathbf{R}}}}}ic_{(\mu \nu )}^{\star }+\ _{\ast
}^{\shortparallel }\mathfrak{a}_{[10]}^{\alpha \beta }\ \mathbf{\mathbf{%
\mathbf{\mathbf{\mathbf{\mathbf{\mathbf{\mathbf{\ }}}}}}}}\
_{[00]}^{\shortparallel }\widehat{\mathbf{\mathbf{\mathbf{\mathbf{R}}}}}%
ic_{[\mu \nu ]}^{\star }, \\
\mathbf{\mathbf{\mathbf{\mathbf{\ }}}}\ _{[11]}^{\shortparallel }\widehat{%
\mathbf{\mathbf{\mathbf{\mathbf{R}}}}}sc^{\star } &=&\ _{\ast
}^{\shortparallel }\mathfrak{g}_{[00]}^{\mu \nu }\mathbf{\mathbf{\mathbf{%
\mathbf{\ }}}}_{[11]}^{\shortparallel }\widehat{\mathbf{\mathbf{\mathbf{%
\mathbf{R}}}}}ic_{(\mu \nu )}^{\star }+\ _{\ast }^{\shortparallel }\mathfrak{%
g}_{[01]}^{\mu \nu }\mathbf{\mathbf{\mathbf{\mathbf{\ }}}}%
_{[10]}^{\shortparallel }\widehat{\mathbf{\mathbf{\mathbf{\mathbf{R}}}}}%
ic_{(\mu \nu )}^{\star }+\ _{\ast }^{\shortparallel }\mathfrak{\check{g}}%
_{[10]}^{\mu \nu }\mathbf{\mathbf{\mathbf{\mathbf{\ \ }}}}%
_{[01]}^{\shortparallel }\widehat{\mathbf{\mathbf{\mathbf{\mathbf{R}}}}}%
ic_{(\mu \nu )}^{\star }+\ _{\ast }^{\shortparallel }\mathfrak{\check{g}}%
_{[11]}^{\mu \nu }\mathbf{\mathbf{\mathbf{\mathbf{\mathbf{\mathbf{\mathbf{%
\mathbf{\ }}}}}}}}\ _{[00]}^{\shortparallel }\widehat{\mathbf{\mathbf{%
\mathbf{\mathbf{R}}}}}ic_{(\mu \nu )}^{\star } \\
&&+\ _{\ast }^{\shortparallel }\mathfrak{a}_{[10]}^{\alpha \beta }\mathbf{%
\mathbf{\mathbf{\mathbf{\ \ }}}}_{[01]}^{\shortparallel }\widehat{\mathbf{%
\mathbf{\mathbf{\mathbf{R}}}}}ic_{[\mu \nu ]}^{\star }+\ _{\ast
}^{\shortparallel }\mathfrak{a}_{[11]}^{\alpha \beta }\mathbf{\mathbf{%
\mathbf{\mathbf{\mathbf{\mathbf{\mathbf{\mathbf{\ }}}}}}}}\
_{[00]}^{\shortparallel }\widehat{\mathbf{\mathbf{\mathbf{\mathbf{R}}}}}%
ic_{[\mu \nu ]}^{\star },
\end{eqnarray*}%
where a complete parametric decomposition is possible if we apply (\ref%
{ric50})-(\ref{ric53}).

The parametric decomposition of the nonassociative system (\ref%
{nonassocdeinst2a}) and (\ref{nonassocdeinst2b}) for $\ _{\star
}^{\shortparallel }\mathfrak{g}_{\alpha \beta }=\ _{\star }^{\shortparallel }%
\mathfrak{g}_{\alpha \beta }^{[0]}+\ _{\star }^{\shortparallel }\mathfrak{g}%
_{\alpha \beta }^{[1]}(\kappa )$ (\ref{aux37}) and $\ _{\star
}^{\shortparallel }\mathfrak{a}_{\alpha \beta }=\ _{\star }^{\shortparallel }%
\mathfrak{a}_{\alpha \beta }^{[1]}(\kappa )$ is defined by such recurrent
formulas 
\begin{eqnarray}
\ \ _{\star }^{\shortparallel }\mathfrak{g}_{\mu \nu }^{[0]}(\ 2\mathbf{%
\mathbf{\mathbf{\mathbf{\ ^{\shortparallel }}}}}\lambda +\mathbf{\mathbf{%
\mathbf{\mathbf{\ }}}}\ _{[00]}^{\shortparallel }\widehat{\mathbf{\mathbf{%
\mathbf{\mathbf{R}}}}}sc^{\star }) &=&2\mathbf{\mathbf{\mathbf{\mathbf{\ \ 
\mathbf{\mathbf{\mathbf{\mathbf{\ }}}}}}}}\ _{[00]}^{\shortparallel }%
\widehat{\mathbf{\mathbf{\mathbf{\mathbf{R}}}}}ic_{(\mu \nu )}^{\star },\ 
\label{nonassocdeinst4ab} \\
\ _{\star }^{\shortparallel }\mathfrak{g}_{\mu \nu }^{[0]}\mathbf{\mathbf{%
\mathbf{\mathbf{\ }}}}\ _{[01]}^{\shortparallel }\widehat{\mathbf{\mathbf{%
\mathbf{\mathbf{R}}}}}sc^{\star } &=&2\mathbf{\mathbf{\mathbf{\mathbf{\ \ }}}%
}_{[01]}^{\shortparallel }\widehat{\mathbf{\mathbf{\mathbf{\mathbf{R}}}}}%
ic_{(\mu \nu )}^{\star },  \notag \\
\ _{\star }^{\shortparallel }\mathfrak{g}_{\mu \nu }^{[0]}\mathbf{\mathbf{%
\mathbf{\mathbf{\ }}}}\ _{[10]}^{\shortparallel }\widehat{\mathbf{\mathbf{%
\mathbf{\mathbf{R}}}}}sc^{\star }+\ _{\star }^{\shortparallel }\mathfrak{g}%
_{\mu \nu }^{[1]}(\ 2\mathbf{\mathbf{\mathbf{\mathbf{\ ^{\shortparallel }}}}}%
\lambda +\mathbf{\mathbf{\mathbf{\mathbf{\ }}}}\ _{[00]}^{\shortparallel }%
\widehat{\mathbf{\mathbf{\mathbf{\mathbf{R}}}}}sc^{\star }) &=&2\mathbf{%
\mathbf{\mathbf{\mathbf{\ \ }}}}_{[10]}^{\shortparallel }\widehat{\mathbf{%
\mathbf{\mathbf{\mathbf{R}}}}}ic_{(\mu \nu )}^{\star },  \notag \\
\ _{\star }^{\shortparallel }\mathfrak{g}_{\mu \nu }^{[0]}\mathbf{\mathbf{%
\mathbf{\mathbf{\ }}}}\ _{[11]}^{\shortparallel }\widehat{\mathbf{\mathbf{%
\mathbf{\mathbf{R}}}}}sc^{\star }+\ _{\star }^{\shortparallel }\mathfrak{g}%
_{\mu \nu }^{[1]}\mathbf{\mathbf{\mathbf{\mathbf{\ }}}}\
_{[01]}^{\shortparallel }\widehat{\mathbf{\mathbf{\mathbf{\mathbf{R}}}}}%
sc^{\star } &=&2\mathbf{\mathbf{\mathbf{\mathbf{\ \ }}}}_{[11]}^{%
\shortparallel }\widehat{\mathbf{\mathbf{\mathbf{\mathbf{R}}}}}ic_{(\mu \nu
)}^{\star }\mbox{ and }  \notag \\
\mathbf{\mathbf{\mathbf{\mathbf{\ \mathbf{\mathbf{\mathbf{\mathbf{\ }}}}}}}}%
\ _{[00]}^{\shortparallel }\widehat{\mathbf{\mathbf{\mathbf{\mathbf{R}}}}}%
ic_{[\mu \nu ]}^{\star } &=&0,\mathbf{\mathbf{\mathbf{\mathbf{\ \ }}}}%
_{[01]}^{\shortparallel }\widehat{\mathbf{\mathbf{\mathbf{\mathbf{R}}}}}%
ic_{[\mu \nu ]}^{\star }=0,  \notag \\
\ _{\star }^{\shortparallel }\mathfrak{a}_{\mu \nu }^{[1]}(\ 2\mathbf{%
\mathbf{\mathbf{\mathbf{\ ^{\shortparallel }}}}}\lambda +\mathbf{\mathbf{%
\mathbf{\mathbf{\ }}}}\ _{[00]}^{\shortparallel }\widehat{\mathbf{\mathbf{%
\mathbf{\mathbf{R}}}}}sc^{\star }) &=&2\mathbf{\mathbf{\mathbf{\mathbf{\ }}}}%
_{[10]}^{\shortparallel }\widehat{\mathbf{\mathbf{\mathbf{\mathbf{R}}}}}%
ic_{[\mu \nu ]}^{\star },  \notag \\
\ _{\star }^{\shortparallel }\mathfrak{a}_{\mu \nu }^{[1]}\mathbf{\mathbf{%
\mathbf{\mathbf{\ }}}}\ _{[01]}^{\shortparallel }\widehat{\mathbf{\mathbf{%
\mathbf{\mathbf{R}}}}}sc^{\star } &=&2\mathbf{\mathbf{\mathbf{\mathbf{\ }}}}%
_{[11]}^{\shortparallel }\widehat{\mathbf{\mathbf{\mathbf{\mathbf{R}}}}}%
ic_{[\mu \nu ]}^{\star }.  \notag
\end{eqnarray}

These formulas involve h- and c-decompositions for block parameterizations
which can be written in explicit form using additionally formulas from
previous subsection. In such a case, it is convenient to use block
parameterizations of the symmetric and anti-symmetric d-metric components, 
\begin{equation*}
\ _{\star }^{\shortparallel }\mathfrak{\check{g}}_{\mu \nu }=(\ _{\star
}^{\shortparallel }\mathfrak{\check{g}}_{ij},0,0,\ _{\star }^{\shortparallel
}\mathfrak{\check{g}}^{ab}),\ _{\star }^{\shortparallel }\mathfrak{a}_{\mu
\nu }=(0,0,0,\ _{\star }^{\shortparallel }\mathfrak{a}^{cb})\mbox{ and }\
_{\star }^{\shortparallel }\mathfrak{\check{g}}^{\mu \nu }=(\ _{\star
}^{\shortparallel }\mathfrak{\check{g}}^{ij},0,0,\ _{\star }^{\shortparallel
}\mathfrak{\check{g}}_{ab}),\ _{\star }^{\shortparallel }\mathfrak{a}^{\mu
\nu }=(0,0,0,\ _{\star }^{\shortparallel }\mathfrak{a}_{cb}),
\end{equation*}%
when the inverse matrices are computed not directly but using in N-adapted
form respective formulas of type (\ref{aux34}). We omit such formulas in
this work but we prove a general decoupling property and provide examples of
parametric solutions in our partner works.

The values $\ _{\star }^{\shortparallel }\mathfrak{g}_{\mu \nu }^{[0]}$ in
the first equation in (\ref{nonassocdeinst4ab}) can be taken to determine a
solution of vacuum Einstein equations (in general, with nontrivial
cosmological constant) in a commutative gravity theory. Other terms are
computed recurrently for respective parametric star deformations. The
N-adapted structure should be chosen in such forms as the next step
equations in that system of PDEs are self-consistent and define parametric
solutions in respective noncommutative and nonassociative extensions. This
should be checked always for explicit classes of solutions and their
symmetries. We can not elaborate such geometric constructions of solutions
if we work only with local coordinate frames and diagonal metric structures.
Even we star with a commutative diagonal metric, further steps will result
in certain generic off-diagonal symmetric and nonsymmetric contributions
which involve also nonsymmetric Ricci d-tensor star deformations. In our
partner work we show that such a nonsymmetric extension of the AFCDM can be
applied in certain self-consistent forms resulting in explicit classes of
exact/ parametric solutions.

\subsection{Horizontal N-adapted projections and lifts of nonassociative
Ricci d-tensors}

In this work, our main goals is to prove that nonassociative gravity models
determined by "non-geometric" R-fluxes in string theory \cite%
{mylonas13,blumenhagen16,aschieri17} can be completely geometrized on
nonholonomic phase spaces modeled as cotangent Lorentz bundles enabled with
N-connection structure. In such an approach star nonholonomic deformations
of commutative spacetime geometries result both in symmetric and
nonsymmetric metrics which in the vacuum case can be related to respective
symmetric and nonsymmetric components of some Ricci d-tensors. The
constructions can be performed for certain canonical d-connections which
allow a general decoupling and integration (for certain well-defined
off-diagonal metric and distorted linear connection structures) of
nonassociative nonholonomic Einstein equations and then re-definitions for
LC-configurations.

For holonomic configurations, we start from tensors on a Lorentz manifold $%
V, $ lift them on $\mathcal{M}=T^{\ast }V$ and/or $\ ^{\shortparallel }%
\mathcal{M}$ with complex momenta coordinates, and then construct new
composite tensor using nonassociative star deformations of the geometry of $%
\mathcal{M};$ the results can be reordered using the associator and
projected back to $V.$ Functions and forms can be lifted from $V$ to $%
\mathcal{M}$ as pullbacks of forms using the canonical projection $\pi :\
T^{\ast }V\rightarrow V.$ In the opposite direction, we can work with an
embedding $\sigma :\ V\rightarrow \mathcal{M}$, which is given by zero
section $x\rightarrow \sigma (x)=(x,0),$ we pull back forms on $\mathcal{M}$
to forms on $V.$ Lifts of vectors are obtained by considering a foliation of 
$\mathcal{M}$ with constant momentum leaves (each such leaf is chosen to be
diffeomorphic to $V$). Using coordinated bases, such lifts of $\partial _{i}$
on $V$ a transformed into vector on $\mathcal{M}$, when $b^{i}(x)\partial
_{i}\rightarrow \pi ^{\ast }(b^{i})(x,p)\partial _{i},$ where $\pi ^{\ast
}(b^{i})(x,p)\partial _{i}=(b^{i})(\pi (x,p))=b^{i}(x).$ We can act also in
opposite direction when vector fields on $\mathcal{M}$ are projected to
vector fields on $V$ via zero section of $\sigma .$ In coordinate form, this
is written in the form $b^{i}(x,p)\partial _{i}+b_{a}(x,p)\partial
^{a}\rightarrow b^{i}(x,0)\partial _{i}.$

In our approach, we work on nonholonomic manifolds and (co) tangent Lorentz
bundles and perform the constructions in N-adapted form, when, for instance, 
$\ ^{\shortparallel }b^{i}(x,\ ^{\shortparallel }p)\ ^{\shortparallel }%
\mathbf{e}_{i}+\ ^{\shortparallel }b_{a}(x,\ ^{\shortparallel }p)\
^{\shortparallel }\partial ^{a}\rightarrow \ ^{\shortparallel }b^{i}(x,0)\
^{\shortparallel }\mathbf{e}_{i},$ see formulas (\ref{nadapdc}). Any metric
tensor $\widehat{g}_{jk}(x)dx^{j}\otimes dx^{k}$ on $V$ can be lifted into
an off-diagonal metric and/or equivalent d--metric $\ ^{\shortparallel }%
\mathbf{g}$ (\ref{commetr}) on $\mathbf{T}_{\shortparallel }^{\ast }\mathbf{V%
}$. With respect to tensor product of N-adapted cobases $\ ^{\shortparallel }%
\mathbf{e}^{\beta }=(dx^{i},\ ^{\shortparallel }\mathbf{e}^{a}),$ we can
consider a matrix is parametrization for symmetric d-tensors,%
\begin{eqnarray*}
\ \widehat{g}_{jk}(x) &\rightarrow &\ \ ^{\shortparallel }\widehat{g}_{%
\underline{\alpha }\underline{\beta }}(x,\ ^{\shortparallel }p)=\left[ 
\begin{array}{cc}
\ \ ^{\shortparallel }\widehat{g}_{ij}(x)+\ ^{\shortparallel }\widehat{g}%
^{ab}(x,\ ^{\shortparallel }p)\ ^{\shortparallel }N_{ia}(x,^{\shortparallel
}p)\ \ ^{\shortparallel }N_{jb}(x,\ ^{\shortparallel }p) & \ \
^{\shortparallel }\widehat{g}^{ae}(x)\ ^{\shortparallel }N_{je}(x,\
^{\shortparallel }p) \\ 
\ \ ^{\shortparallel }\widehat{g}^{be}(x)\ ^{\shortparallel }N_{ie}(x,\
^{\shortparallel }p) & \ \ ^{\shortparallel }\widehat{g}^{ab}(x))\ 
\end{array}%
\right] \\
&\simeq &\left[ \ \widehat{\mathbf{g}}_{jk}(x,\ ^{\shortparallel }p),\
^{\shortparallel }\widehat{\mathbf{g}}^{ab}(x,\ ^{\shortparallel }p))\right]
=\ ^{\shortparallel }\widehat{\mathbf{g}}_{\mu \nu }(x,\ ^{\shortparallel
}p)=\left[ 
\begin{array}{cc}
\ \widehat{\mathbf{g}}_{jk}(x,\ ^{\shortparallel }p) & 0 \\ 
0 & \ ^{\shortparallel }\widehat{\mathbf{g}}^{ab}(x,\ ^{\shortparallel }p)%
\end{array}%
\right] \\
&\longrightarrow &\ ^{\shortparallel }\mathbf{g}_{\mu \nu }=\left[ 
\begin{array}{cc}
\mathbf{g}_{jk}(x,\ ^{\shortparallel }p) & 0 \\ 
0 & \ ^{\shortparallel }\mathbf{g}^{ab}(x,\ ^{\shortparallel }p)%
\end{array}%
\right] ,
\end{eqnarray*}%
where extensions to dependencies on all phase space coordinates $(x^{i},\
^{\shortparallel }p_{a})$ are defined by general frame and coordinate
transforms on $\mathcal{M}.$ N-adapted star deformations $\ ^{\shortparallel
}\mathbf{g}_{\alpha \beta }\rightarrow \ _{\star }^{\shortparallel }%
\mathfrak{g}_{\alpha \beta }$ (\ref{dmss1}) are introduced as 
\begin{equation*}
\ ^{\shortparallel }\mathbf{g=}\ ^{\shortparallel }\mathbf{g}_{\alpha \beta
}\ ^{\shortparallel }\mathbf{e}^{\alpha }\otimes _{\star N}\
^{\shortparallel }\mathbf{e}^{\beta }\rightarrow \ _{\star }^{\shortparallel
}\mathfrak{g}=\ _{\star }^{\shortparallel }\mathfrak{g}_{\alpha \beta }\star
_{N}(\ ^{\shortparallel }\mathbf{e}^{\alpha }\otimes _{\star N}\
^{\shortparallel }\mathbf{e}^{\beta }),
\end{equation*}%
when the linear R-flux corrections can be considered as off-diagonal/ block
terms, 
\begin{eqnarray}
\left[ 
\begin{array}{cc}
\mathbf{g}_{jk}(x,\ ^{\shortparallel }p) & 0 \\ 
0 & \ ^{\shortparallel }\mathbf{g}^{ab}(x,\ ^{\shortparallel }p)%
\end{array}%
\right] &\rightarrow &\left[ 
\begin{array}{cc}
\mathbf{g}_{jk}(x,\ ^{\shortparallel }p) & -\frac{i\kappa }{2}\overline{%
\mathcal{R}}_{\quad \quad \quad n+k}^{n+i\ n+l}\ ^{\shortparallel }\mathbf{e}%
_{i}\mathbf{g}_{jl} \\ 
-\frac{i\kappa }{2}\overline{\mathcal{R}}_{\quad \quad \quad n+j}^{n+i\
n+l}\ ^{\shortparallel }\mathbf{e}_{i}\mathbf{g}_{kl} & \ ^{\shortparallel }%
\mathbf{g}^{ab}(x,\ ^{\shortparallel }p)%
\end{array}%
\right]  \label{auxm61} \\
&\rightarrow &\left[ 
\begin{array}{cc}
\mathbf{g}_{jk}(x) & -\frac{i\kappa }{2}\overline{\mathcal{R}}_{\quad \quad
\quad n+k}^{n+i\ n+l}\mathbf{\partial }_{i}\mathbf{g}_{jl} \\ 
-\frac{i\kappa }{2}\overline{\mathcal{R}}_{\quad \quad \quad n+j}^{n+i\ n+l}%
\mathbf{\partial }_{i}\mathbf{g}_{kl} & \ ^{\shortparallel }\mathbf{g}%
^{ab}(x,\ ^{\shortparallel }p)%
\end{array}%
\right] ,\mbox{ if }\mathbf{g}_{jk}=\mathbf{g}_{jk}(x).  \notag
\end{eqnarray}

The h-projection of the Ricci d-tensor $h\ \mathbf{\mathbf{\mathbf{\mathbf{%
^{\shortparallel }}}}}\widehat{\mathbf{\mathbf{\mathbf{\mathbf{R}}}}}%
ic_{\alpha \beta }^{\star }=\mathbf{\{}\ \ ^{\shortparallel }\widehat{R}_{\
\star hj}\},$ see formulas (\ref{hcnonassocrcan}) and (\ref{hcnonassocrcana}%
), is a canonical d-connection analog of the nonassociative Ricci tensor for
LC-connfiguration and projections on $V.$ For zero distortions, $\ _{\star
}^{\shortparallel }\widehat{\mathbf{Z}}=0,$ i. e.$\ ^{\shortparallel }%
\widehat{\mathbf{D}}_{\mid \ ^{\shortparallel }\widehat{\mathbf{T}}%
=0}^{\star }=\ ^{\shortparallel }\nabla ^{\star },$ see formulas (\ref%
{lccondnonass}) and (\ref{candistrnas}), in a correspondingly defined local
basis, restricting the components of the Ricci tensors and d-tensors, $%
Ric_{\alpha \beta },$ to spacetime directions and setting the momentum
dependence to zero, 
\begin{equation*}
\ \ ^{\shortparallel }\widehat{R}_{\ \star hj}(x,\ ^{\shortparallel
}p)\rightarrow \ \ _{\nabla }^{\shortparallel }\mathcal{\Re }ic_{hj}^{\star
}(x)=Ric_{hj}^{\circ }(x)=\sigma ^{\ast }(\ \ _{\nabla }^{\shortparallel }%
\mathcal{\Re }ic_{hj}^{\star }(x,p))=R_{hj}(x,0).
\end{equation*}%
In \cite{aschieri17}, the nonassociative vacuum solutions on phase space are
postulated in the form%
\begin{equation}
\ \ _{\nabla }^{\shortparallel }\mathcal{\Re }ic_{\alpha \beta }^{\star
}(x,p)=0  \label{veinstlc}
\end{equation}%
with an assumption that every solution of such a system of nonlinear PDEs
result in a solution of 
\begin{equation}
Ric_{hj}^{\circ }(x)=0.  \label{veinsthlc}
\end{equation}%
It was emphasized in that paper that not all solutions of (\ref{veinsthlc})
can be lifted to solutions of (\ref{veinstlc}) on the phase space. Such a
property was a consequence of the non-geometric R-flux contributions in
string theory when it is not clear why the dynamics should be completely
determined on phase spaces. In our approach, R-flux contributions can be
geometrized in the framework of a nonassociative nonholonomic models with
N-connections formulated on (co) tangent Lorentz bundles. We can
nonholonomically deform any solution of (\ref{veinsthlc}) to solutions of (%
\ref{nonassocdeinst3}) and inversely, we can consider respective
nonholonomic constraints of type (\ref{lccondnonass}) and generate exact
solutions for LC-configurations both in phase spaces and Lorentz manifold
spacetimes. Such constructions involve, in general, generic off-diagonal
components of symmetric and nonsymmetric metrics and even at the end, in
some adapted frames/coordinates, the dependence on momentum like variables
is not considered, the solution contain contributions from the nonholonomic
dynamics on $\mathcal{M}=T^{\ast }V$ and/or $\ ^{\shortparallel }\mathcal{M}$
with respective $\hbar $ and $\kappa $ terms. This conclusion is supported
by various classes of exact and parametric solutions in nonassociative/
noncommutative / commutative gravity and geometric flow theory constructed
by applying the AFCDM.

\subsection{First order nonassociative nonholonomic corrections}

Let us analyse some important physical properties of h-components of
nonassociative and nonholonomic vacuum Einstein equations (\ref%
{nonassocdeinst3}). Such formulas projected on a base spacetime $V$ can be
used for computing possible R-flux contributions. We consider a nontrivial
dependence on momentum coordinates for the h-component of d-metric $\mathbf{g%
}_{jk}(x,\ ^{\shortparallel }p)$ for an arbitrary value of a c-component $\
^{\shortparallel }\mathbf{g}^{ab}(x,\ ^{\shortparallel }p)$ in (\ref{auxm61}%
). A nonholonomic frame structure (\ref{nhframtr}) can be prescribed with
"small" canonical distortions (\ref{candistrnas}) when 
\begin{eqnarray*}
\ _{[00]}^{\shortparallel }\widehat{\mathbf{\Gamma }}_{\ jk}^{i} &=&\
^{\shortparallel }\widehat{\mathbf{\Gamma }}_{\ jk}^{i}=\ ^{\shortparallel }%
\widehat{L}_{\ jk}^{i},\ _{[01]}^{\shortparallel }\widehat{\mathbf{\Gamma }}%
_{\ jk}^{i}=0, \\
\ _{[10]}^{\shortparallel }\widehat{\mathbf{\Gamma }}_{\ jk}^{i} &=&-i\kappa 
\overline{\mathcal{R}}_{\quad \quad \quad }^{n+o\ a~n+l}\ p_{a}\
^{\shortparallel }\mathbf{g}^{iq}(\ ^{\shortparallel }\mathbf{e}_{l}\
^{\shortparallel }\mathbf{g}_{qr})(\ ^{\shortparallel }\mathbf{e}_{o}\
^{\shortparallel }\widehat{L}_{\ jk}^{r}),\ _{[10]}^{\shortparallel }%
\widehat{\mathbf{\Gamma }}_{\ jk}^{a}=-\frac{i\kappa }{2}\overline{\mathcal{R%
}}_{\quad \quad \quad }^{n+i\ a~n+q}\ ^{\shortparallel }\mathbf{g}_{lq}\
^{\shortparallel }\mathbf{e}_{i}\ ^{\shortparallel }\widehat{L}_{\ jk}^{q};
\\
\ _{[10]}^{\shortparallel }\widehat{\mathbf{\Gamma }}_{\quad k}^{ia} &=&%
\frac{i\kappa }{2}\overline{\mathcal{R}}_{\quad \quad \quad }^{n+o\ a~n+l}\
\ ^{\shortparallel }\mathbf{g}^{qi}\ ^{\shortparallel }\mathbf{e}_{o}(\
^{\shortparallel }\mathbf{g}_{qr}\ ^{\shortparallel }\widehat{L}_{\
lk}^{r}),\ _{[10]}^{\shortparallel }\widehat{\mathbf{\Gamma }}_{~k}^{i~a}=%
\frac{i\kappa }{2}\overline{\mathcal{R}}_{\quad \quad \quad }^{n+o\ a~n+l}\
\ ^{\shortparallel }\mathbf{g}^{iq}\ ^{\shortparallel }\mathbf{e}_{o}(\
^{\shortparallel }\mathbf{g}_{qr}\ ^{\shortparallel }\widehat{L}_{\ lk}^{r}),
\\
\ _{[11]}^{\shortparallel }\widehat{\mathbf{\Gamma }}_{\ jk}^{i} &=&\frac{%
\hbar \kappa }{2}\overline{\mathcal{R}}_{\quad \quad \quad }^{n+o\ n+q\
~n+l}\ (\ ^{\shortparallel }\mathbf{e}_{o}\ ^{\shortparallel }\mathbf{g}%
^{is})(\ ^{\shortparallel }\mathbf{e}_{q}\ ^{\shortparallel }\mathbf{g}%
_{sr})(\ ^{\shortparallel }\mathbf{e}_{l}\ ^{\shortparallel }\widehat{L}_{\
jk}^{r}),
\end{eqnarray*}%
see respective formulas (\ref{canhc}), (\ref{0canconnonas}) and (\ref%
{aux311a}) for the canonical d-connections, their star deformations and
parametric decompositions. We note that for general nonholonomic
distributions $\ _{[01]}^{\shortparallel }\widehat{\mathbf{\Gamma }}_{\
jk}^{i}$ (\ref{aux51}) and $\ _{[01]}^{\shortparallel }\widehat{\mathbf{R}}%
ic_{\beta \gamma }^{\star }$ (\ref{ric51}) being proportional to $\hbar $
may be not zero even respective values for the LC-connection are zero in
certain coordinate frames, see formulas (5.82)-(5.85) in \cite{aschieri17}.

R-fluxes define a nontrivial imaginary $[10]$ h-term of the canonical Ricci
d-tensor (\ref{ric52}) which for a corresponding N-adapting with "small"
canonical distortions,%
\begin{eqnarray*}
\mathbf{\mathbf{\mathbf{\mathbf{\ \ }}}}_{[10]}^{\shortparallel }\widehat{%
\mathbf{\mathbf{\mathbf{\mathbf{R}}}}}ic_{jk}^{\star } &=&i\kappa \overline{%
\mathcal{R}}_{\quad \quad \quad }^{n+o\ a~n+l}\ p_{a}\{-\ ^{\shortparallel }%
\mathbf{e}_{i}[~^{\shortparallel }\mathbf{g}^{iq}(\ ^{\shortparallel }%
\mathbf{e}_{l}\ ^{\shortparallel }\mathbf{g}_{qr})(\ ^{\shortparallel }%
\mathbf{e}_{o}\ ^{\shortparallel }\widehat{L}_{\ jk}^{r})]+\
^{\shortparallel }\mathbf{e}_{k}[~^{\shortparallel }\mathbf{g}^{iq}(\
^{\shortparallel }\mathbf{e}_{l}\ ^{\shortparallel }\mathbf{g}_{qr})(\
^{\shortparallel }\mathbf{e}_{o}\ ^{\shortparallel }\widehat{L}_{\ ji}^{r})]
\\
&&-\ ^{\shortparallel }\widehat{L}_{\ jk}^{m}~^{\shortparallel }\mathbf{g}%
^{iq}(\ ^{\shortparallel }\mathbf{e}_{l}\ ^{\shortparallel }\mathbf{g}%
_{qr})(\ ^{\shortparallel }\mathbf{e}_{o}\ ^{\shortparallel }\widehat{L}%
_{mi}^{r})-\ ^{\shortparallel }\widehat{L}_{\ im}^{i}~^{\shortparallel }%
\mathbf{g}^{mq}(\ ^{\shortparallel }\mathbf{e}_{l}\ ^{\shortparallel }%
\mathbf{g}_{qr})(\ ^{\shortparallel }\mathbf{e}_{o}\ ^{\shortparallel }%
\widehat{L}_{jk}^{r}) \\
&&+\ ^{\shortparallel }\widehat{L}_{\ ji}^{m}~^{\shortparallel }\mathbf{g}%
^{iq}(\ ^{\shortparallel }\mathbf{e}_{l}\ ^{\shortparallel }\mathbf{g}%
_{qr})(\ ^{\shortparallel }\mathbf{e}_{o}\ ^{\shortparallel }\widehat{L}%
_{mk}^{r})+\ ^{\shortparallel }\widehat{L}_{\ mk}^{i}~^{\shortparallel }%
\mathbf{g}^{mq}(\ ^{\shortparallel }\mathbf{e}_{l}\ ^{\shortparallel }%
\mathbf{g}_{qr})(\ ^{\shortparallel }\mathbf{e}_{o}\ ^{\shortparallel }%
\widehat{L}_{ji}^{r}) \\
&&+(~^{\shortparallel }\mathbf{e}_{l}\ ^{\shortparallel }\widehat{L}_{\
qi}^{i})(\ ^{\shortparallel }\mathbf{g}_{qr}\ ^{\shortparallel }\widehat{L}%
_{\ lk}^{r})-(~^{\shortparallel }\mathbf{e}_{l}\ ^{\shortparallel }\widehat{L%
}_{\ qk}^{i})(\ ^{\shortparallel }\mathbf{g}_{qr}\ ^{\shortparallel }%
\widehat{L}_{\ li}^{r})\}.
\end{eqnarray*}%
This term is imaginary but there is a real nontrivial contribution via%
\begin{eqnarray*}
\mathbf{\mathbf{\mathbf{\mathbf{\ \ }}}}_{[11]}^{\shortparallel }\widehat{%
\mathbf{\mathbf{\mathbf{\mathbf{R}}}}}ic_{jk}^{\star } &=&\frac{\hbar \kappa 
}{2}\overline{\mathcal{R}}_{\quad \quad \quad }^{n+o\ n+q\ ~n+l}\{\
^{\shortparallel }\mathbf{e}_{i}[(\ ^{\shortparallel }\mathbf{e}%
_{o}~^{\shortparallel }\mathbf{g}^{im})(\ ^{\shortparallel }\mathbf{e}_{q}\
^{\shortparallel }\mathbf{g}_{mr})(\ ^{\shortparallel }\mathbf{e}_{l}\
^{\shortparallel }\widehat{L}_{\ jk}^{r})] \\
&&-\ ^{\shortparallel }\mathbf{e}_{k}[(\ ^{\shortparallel }\mathbf{e}%
_{o}~^{\shortparallel }\mathbf{g}^{im})(\ ^{\shortparallel }\mathbf{e}_{q}\
^{\shortparallel }\mathbf{g}_{mr})(\ ^{\shortparallel }\mathbf{e}_{l}\
^{\shortparallel }\widehat{L}_{\ ji}^{r})] \\
&&+(\ ^{\shortparallel }\mathbf{e}_{l}~^{\shortparallel }\mathbf{g}_{qr})[\
^{\shortparallel }\mathbf{e}_{o}(\ ~^{\shortparallel }\mathbf{g}^{iq}\
^{\shortparallel }\widehat{L}_{\ ik}^{m})(\ ^{\shortparallel }\mathbf{e}%
_{q}\ ^{\shortparallel }\widehat{L}_{jm}^{r})-\ ^{\shortparallel }\mathbf{e}%
_{o}(\ ~^{\shortparallel }\mathbf{g}^{iq}\ ^{\shortparallel }\widehat{L}_{\
im}^{m})(\ ^{\shortparallel }\mathbf{e}_{q}\ ^{\shortparallel }\widehat{L}%
_{jk}^{r}) \\
&&+\left( \ ^{\shortparallel }\widehat{L}_{\ ji}^{m}\ ^{\shortparallel }%
\mathbf{e}_{o}(\ ~^{\shortparallel }\mathbf{g}^{ir})-\ ^{\shortparallel }%
\mathbf{e}_{o}(\ ^{\shortparallel }\widehat{L}_{\ ji}^{m})\
~^{\shortparallel }\mathbf{g}^{ir}\right) (\ ^{\shortparallel }\mathbf{e}%
_{q}\ ^{\shortparallel }\widehat{L}_{\ mk}^{r}) \\
&&-\left( \ ^{\shortparallel }\widehat{L}_{\ jk}^{m}\ ^{\shortparallel }%
\mathbf{e}_{o}(\ ~^{\shortparallel }\mathbf{g}^{ir})-\ ^{\shortparallel }%
\mathbf{e}_{o}(\ ^{\shortparallel }\widehat{L}_{\ jk}^{m})\
~^{\shortparallel }\mathbf{g}^{ir}\right) (\ ^{\shortparallel }\mathbf{e}%
_{q}\ ^{\shortparallel }\widehat{L}_{\ mi}^{r})]\}.
\end{eqnarray*}%
In result, the horizontal N--adapted components of nonassociative vacuum
Einstein equations (\ref{nonassocdeinst3}) can be written in a form
containing real contributions from string gravity with R-fluxes, 
\begin{equation}
\ \ ^{\shortparallel }\widehat{R}_{\ \star hj}(x,\ ^{\shortparallel }p)=\ \
^{\shortparallel }\widehat{R}_{\ hj}(x,\ ^{\shortparallel }p)+\mathbf{%
\mathbf{\mathbf{\mathbf{\ \ }}}}_{[11]}^{\shortparallel }\widehat{\mathbf{%
\mathbf{\mathbf{\mathbf{R}}}}}ic_{jk}^{\star }(x,\ ^{\shortparallel }p)=0.
\label{aux61a}
\end{equation}%
Applying the AFCDM, we can decouple and integrate such equations in
arbitrary form for various classes of h-metric components $\mathbf{g}%
_{jk}(x,\ ^{\shortparallel }p)$ of a d-metric (\ref{auxm61}).

For additional nonholonomic constraints $\ ^{\shortparallel }\widehat{%
\mathbf{D}}_{\mid \ ^{\shortparallel }\widehat{\mathbf{T}}=0}^{\star }=\
^{\shortparallel }\nabla ^{\star }$ (\ref{lccondnonass}), we extract from (%
\ref{aux61a}) horizontal LC-configurations of type (\ref{veinstlc}),%
\begin{equation*}
\ \ _{\nabla }^{\shortparallel }\mathcal{\Re }ic_{hj}^{\star }(x,p)=\ \
_{\nabla }^{\shortparallel }R_{\ hj}(x,\ ^{\shortparallel }p)+\mathbf{%
\mathbf{\mathbf{\mathbf{\ \ }}}}_{[11]}^{\shortparallel \nabla
}R_{jk}^{\star }(x,\ ^{\shortparallel }p)=0.
\end{equation*}%
Solutions of such systems can be extracted for zero canonical distortions.
For holonomic structures and $\mathbf{g}_{jk}(x,\ ^{\shortparallel }p)=%
\mathbf{g}_{jk}(x),$ we generate configurations of type (\ref{veinsthlc}), 
\begin{equation*}
Ric_{hj}^{\circ }(x)=\ \ _{\nabla }Ric_{hj}(x)+\mathbf{\mathbf{\mathbf{%
\mathbf{\ \ }}}}_{[11]}^{\shortparallel \nabla }R_{jk}^{\star }(x)=0.
\end{equation*}%
In coordinate frames, these equations are equivalent to (5.90) from \cite%
{aschieri17}, when R-flux corrections to Ricci tensors determined on
spacetime by $\ ^{\shortparallel }\nabla $ are independent on $\hbar $. This
property holds true for some special classes of metrics $\mathbf{g}_{jk}(x).$
In general, nonassociative gravity theories formulated on phase spaces
involve contributions from the total space dynamics and geometric evolution
processes. Such systems are generic nonlinear and off-diagonal when
imaginary parts determine also real terms for nontrivial gravitational and
matter field interactions on modified Einstein spaces. We construct such
exact and parametric solutions and study possible physical implications in
our partner works.

\section{Summary of results and conclusions}

\label{sec5} In this work we have elaborated a formalism which allows us to
geometrize on star deformed nonholonomic cotangent Lorentz bundles
(considered as phase spaces with possible complexified momentum like
coordinates) the nonassociative gravity models determined by non-geometric
R-flux backgrounds of string theory. There are three important motivations
to work on such curved phase spaces both in general nonholonomic forms and
with associated nonlinear connection, N-connection structure:

\begin{enumerate}
\item Nonassociative R-flux contributions and respective star deformations
involve momentum like variables $p_a$ which in relativistic commutative
limits are related to geometric objects on a dual 4-d Lorentz manifold (with
cotangent Minkowski space), for instance, of signature (+++-). We work in
explicit form with 4-d and (4+4)-d configurations following the assumption
that in certain commutative limits our nonassociative gravity and geometric
flow models transforms into respective Einstein gravity (or possible $f(R)$,
metric-affine and similar modifications). For more general constructions in
string / M-theory and supergravity, we can elaborate on 5-d and 6-d (super)
manifolds and respective (5+5)-d and (6+6)-d (co) tangent (super) bundles.
We omit such considerations in our series of works but emphasize that a
rigorous geometric (frame and coordinate free) formalism with nonassociative
and noncommutative star products can be developed only using certain classes
of geometric objects on (co) tangent [and (co) vector, with possible, for
instance, Lie algebroid (dual) structures, or other type fibered structures,
in order to work with different spacetime and (co) fiber dimensions]. In
self-consistent geometric form, such constructions are adapted to certain
classes of nonholonomic distributions which for (co) tangent bundles encode
respective horizontal and (co) vertical splitting determined by
N-connections.

\item Former works on nonassociative geometry and gravity are based on
coordinate base star deforming formalism adapted to certain commutative/
noncommutative / nonassociative structures like in \cite%
{blumenhagen16,aschieri17}, see also references therein. This follows a
principle of deforming certain commutative algebras into nonassociative/
noncommutative ones and then to elaborate on plethora of nonassocitative
geometric models when the explicit constructions on the type of adapting to
respective nonholonomic structure and conventions on multiple local and
covariant derivatives, symmetric and nonsymmetric metrics etc. In our
approach, the main assumption is to begin with commutative configurations
(for well-defined geometric and physical models) with general frame, metric
and (non) linear connection structures and then to elaborate on
nonassociative and noncommutative star nonholonomic star deformations in
certain N-adapted and prescribed algebraic structure forms.

\item Building over this, we can move to constructions for star nonholonomic
deforming of commutative models of differential geometry and gravity (and
other types of geometrized matter field interactions). Such star
deformations are determined by R-flux contributions and may encode
commutative effects of some (commutative / modified) Einstein like equations
if the definition of star products involves not just partial derivatives but
certain general nonholonomic and respectively N-elongated partial
derivative. In nonholonomic form, we can consider 2 (or 3)+2+2+... diadic
decompositions which allow to generate exact and parametric solutions of
nonassociative geometric and physically important systems nonlinear partial
derivative equations, PDEs, using respective commutative analogs and their
solutions. Such a procedure can be performed self-consistently at least for
nonassocitaive/ noncommutative corrections with terms proportional to
parameters $\hbar ,\kappa $ and $\hbar \kappa $ terms. Decoupling and
integration properties of such systems can be proven for certain canonical
nonholonomic configurations \cite{partner02} but not in the case of
holonomic systems with partial derivatives and their duals.
\end{enumerate}

Above stated geometric constructions have been performed on full phase
spaces endowed with canonical nonholonomic (nonintegrable, equivalently,
anholonomic) structures when physically important systems of nonlinear PDEs
in nonassociative gravity and geometric flow theories can be decoupled and
integrated in very general forms. Such models involve star deformations of
generalized metrics and nonlinear and linear connections like in
(complexified and/or almost complex) relativistic Finsler-Lagrange-Hamilton
geometry \cite{vacaru96a,vacaru96b,vacaru07b,vacaru18,bubuianu18a} and
extensions to Einstein-Eisenhart-Moffat theories \cite%
{einstein25,einstein45,eisenhart51,
eisenhart52,moffat79,moffat95,moffat95a,moffat00,vacaru08aa,vacaru08bb,vacaru08cc}
with nonsymmetric metrics. For additional nonholonomic constraints and small
parametric limits, we can extract Levi-Civita configurations and reproduce
in coordinate bases the nonassociative geometric objects and vacuum
gravitational equations from \cite{blumenhagen16,aschieri17}. It should be
noted here that there are alternative models of nonassociative geometry, for
instance, with octonionic structure, see \cite%
{kurdgelaidze,okubo,castro1,castro2,castro3,gunaydin} and references
therein. It would be interesting to see how such theories can be formulated
in general nonholonomic forms and to see if the anholonomic frame
deformation method, AFCDM, works for such configurations.

Our two main principles which we follow in this and partner papers are that

\begin{itemize}
\item 1] a self-consistent geometric approach to nonassociative physical
models requests general frame formulations for certain commutative
relativistic well-defined theories which for nontrivial R-flux and other
type string/ quantum are completely star deformed following Convention 2 and
formulas (\ref{conv2}); and

\item 2] the nonassociative generalized gravitational and matter fields
dynamics, and various models of geometric evolution with quantum and
classical information encoding must be completely determined on star
nonholonomic deformed phase space.
\end{itemize}

Such a program can be performed only for geometric constructions adapted in
canonical forms to respective N-connection structures. The nonassociative
nonholonomic vacuum gravitational equations can formulated on the total
phase space, projected on horizontal and cofiber components, and computed
analytically on all order corrections on R-flux and noncommutative
parameters. Following additional geometric and/or nonassociative generalized
variational principles such nonassociative modified Einstein equations can
be generalized for nontrivial (effective) matter field sources when
decoupling properties can be preserved (proofs are similar to those in \cite%
{vacaru18,bubuianu18a,bubuianu17,bubuianu19}).

We can pull back the vacuum and nonvacuum nonassociative nonholonomic
Einstein equations from phase space to a commutative spacetime with R-flux
real corrections, for instance, via the zero momentum section $\sigma :\
V\rightarrow \mathcal{M}$ as it is considered at the end of previous section
and in \ \cite{aschieri17}. The nonassociativity survives on the action of
diffeormorphisms on commutative base spacetime but following our
nonholonomic geometric approach we can compute additional generic
off-diagonal and N-connections effects which is not possible if we work only
in local coordinate (co) frames. Such geometric and analytic constructions
and applications in modern astrophysics and cosmology were considered in
noncommutative and commutative modified gravity and string theories \cite%
{vacaru01,vacaru03,vacaru09a,vacaru16,bubuianu17b,bubuianu17,
bubuianu19,vacaru19,vacaru20}.

Finally,we emphasize that the main goal of this work was to provide a
nonholonomic geometric background which will allow to develop our AFCDM and
quantum deformation methods for nonassociative classical and quantum
geometric gravitational and information theories.

\vskip6pt

\textbf{Acknowledgments:} This work develops for nonassociative geometry and
gravity some research programs on geometry and physics, during 2006-2015,
supported by a project IDEI--Romania and senior fellowships at the Perimeter
Institute and Fields Institute (Ontario, Canada), CERN (Geneva, Switzerland)
and Max Planck Institut f\"{u}r Physik / Werner Heisenberg Institut, M\"{u}%
nchen (Germany). SV is grateful to professors V. G. Kupriyanov, D. L\"{u}st,
N. Mavromatos, J. Moffat, D. Singleton, P. Stavrinos, and other host
professors, for respective hosting of short/ long terms visits, seminars,
and/or discussing important ideas and preliminary results.

\end{document}